\documentclass[fleqn,usenatbib]{mnras}

\usepackage{newtxtext,newtxmath}

\usepackage[T1]{fontenc}
\DeclareRobustCommand{\VAN}[3]{#2}
\let\VANthebibliography\thebibliography
\def\thebibliography{\DeclareRobustCommand{\VAN}[3]{##3}\VANthebibliography}

\usepackage{graphicx}	
\usepackage{amsmath}	
\usepackage{amssymb}	
\usepackage{pdflscape}

\graphicspath{{./}{Figures/}}

\newcommand{\um}{\ensuremath{\mu{\textrm{m}}}}
\newcommand{\sfr}{M$_\odot\,{\textrm{yr}}^{-1}$}

\title[A 1.1 mm LMT/AzTEC Survey of Red-\textit{Herschel} galaxies.]{Early Science with the Large Millimeter Telescope: a 1.1 mm AzTEC Survey of Red-\textit{Herschel} dusty star-forming galaxies}

\author[A. Monta\~na et al.]{A. Monta\~na,$^{1,2}$\thanks{E-mail: amontana@inaoep.mx}
J. A. Zavala,$^{3}$
I. Aretxaga,$^{2}$
D. H. Hughes,$^{2}$
R. J. Ivison,$^{4}$
A. Pope,$^{5}$
D. S\'anchez-Arg\"uelles,$^{1,2}$
\newauthor
G. W. Wilson,$^{5}$
M. Yun,$^{5}$
O. A. Cantua,$^{3}$
M. McCrackan,$^{5}$
M. J. Micha\l owski,$^{6}$
E. Valiante,$^{7}$
V. Arumugam,$^{8}$
\newauthor
C. M. Casey,$^{3}$
R. Ch\'avez,$^{1,9}$
E. Col\'in-Beltr\'an,$^{1,2}$
H. Dannerbauer,$^{10,11}$
J. S. Dunlop,$^{12}$
L. Dunne,$^{13}$
S. Eales,$^{13}$
\newauthor
D. Ferrusca,$^{2}$
V. G\'omez-Rivera,$^{2}$
A. I. G\'omez-Ruiz,$^{1,2}$
V. H. de la Luz,$^{14}$
S. J.  Maddox,$^{13,12}$
G. Narayanan,$^{5}$
\newauthor
A. Omont,$^{15}$
I. Rodr\'iguez-Montoya,$^{1,2}$
S. Serjeant,$^{16}$
F. P. Schloerb,$^{5}$
M. Vel\'azquez,$^{2}$
S. Ventura-Gonz\'alez,$^{17}$
\newauthor
P. van der Werf,$^{18}$
M. Zeballos$^{19,2}$
\\
$^{1}$Consejo Nacional de Ciencia y Tecnolog\'ia, Av. Insurgentes Sur 1582, Col. Cr\'edito Constructor, Alcald\'ia Benito Ju\'arez, C.P. 03940, Ciudad de M\'exico, M\'exico\\
$^{2}$Instituto Nacional de Astrof\'isica \'Optica y Electr\'onica, Luis Enrique Erro 1, CP 72840, Tonantzintla, Puebla, M\'exico\\
$^{3}$Department of Astronomy, The University of Texas at Austin, 2515 Speedway Blvd Stop C1400, Austin, TX 78712, USA\\
$^{4}$European Southern Observatory, Karl-Schwarzschild-Stra\ss e 2, D-85748 Garching, Germany \\
$^{5}$Department of Astronomy, University of Massachusetts, Amherst, MA 01003, USA \\
$^{6}$Astronomical Observatory Institute, Faculty of Physics, Adam
Mickiewicz University, ul.~S{\l}oneczna 36, 60-286 Pozna{\'n}, Poland \\
$^{7}$1QB Information Technologies (1QBit), 1285 W Pender St Unit 200, Vancouver, BC V6E 4B1, Canada \\
$^{8}$ Institut de Radioastronomie Millim\'etrique, 300 rue de la Piscine, Domaine Universitaire, 38406 Saint-Martin-d'Hères, France \\
$^{9}$Instituto de Radioastronom\'ia y Astrof\'isica, UNAM, Campus Morelia, Morelia C.P. 58089, M\'exico \\
$^{10}$Instituto de Astrof\'isica de Canarias (IAC), E-38205 La Laguna, Tenerife, Spain. \\
$^{11}$Universidad de La Laguna, Dpto. Astrof\'isica, E-38206 La Laguna, Tenerife, Spain \\
$^{12}$Institute for Astronomy, University of Edinburgh, Royal Observatory, Edinburgh, EH9 3HJ \\
$^{13}$School of Physics and Astronomy, Cardiff University, Queens Buildings, The Parade, Cardiff, CF24 3AA, UK \\
$^{14}$Escuela Nacional de Estudios Superiores Unidad Morelia, Universidad
Nacional Aut\'onoma de M\'exico, Morelia, 58190, M\'exico \\
$^{15}$Institut d'Astrophysique de Paris, 98bis Bd Arago,75014, Paris, France \\
$^{16}$School of Physical Sciences, The Open University, Milton Keynes, MK7 6AA, UK \\
$^{17}$University of Hawai‘i at Manoa, Hawai'i 96822, USA \\
$^{18}$Leiden Observatory, Leiden University, P.O. Box 9513, NL-2300 RA Leiden, The Netherlands \\
$^{19}$Universidad de las Am\'ericas Puebla. Ex Hacienda Sta. Catarina M\'artir S/N. San Andr\'es Cholula, Puebla 72810, M\'exico
}

\date{Accepted 2021 June 2. Received 2021 June 2; in original form 2020 November 2.}

\pubyear{2021}

\begin{document}
\label{firstpage}
\pagerange{\pageref{firstpage}--\pageref{lastpage}}
\maketitle

\begin{abstract}
We present LMT/AzTEC\,1.1\,mm observations of $\sim100$ luminous high-redshift dusty star-forming galaxy candidates from the $\sim600\,$sq.deg {\it Herschel}-ATLAS survey, selected on the basis of their SPIRE red far-infrared colours and with $S_{500\mu\rm m}=35-80$ mJy. With an effective $\theta_{\rm FWHM}\approx9.5\,$\,arcsec angular resolution, our observations reveal that at least 9 per cent of the targets break into multiple systems with SNR $\geq 4$ members (i.e. without considering close mergers). The fraction of multiple systems increases to $\sim23$\, per cent (or more) if some non-detected targets are multiples, as suggested by the data. Combining the new AzTEC and deblended {\it Herschel} photometry we derive photometric redshifts, IR luminosities, and star formation rates. While the median redshifts of the multiple and single systems are similar ($z_{\rm med}\approx3.6$), the latter are skewed towards higher redshifts. Of the AzTEC sources $\sim85$\,per cent lie at $z_{\rm phot}>3$ while $\sim33$\, per cent are at $z_{\rm phot}>4$. This corresponds to a lower limit on the space density of ultra-red sources at $4<z<6$ of $\sim3\times10^{-7}\, \textrm{Mpc}^{-3}$ and a contribution to the obscured star-formation $\gtrsim 8\times10^{-4}\, \textrm{M}_\odot \textrm{yr}^{-1} \textrm{Mpc}^{-3}$. Some of the multiple systems have members with photometric redshifts consistent among them suggesting possible physical associations. Given their angular separations, these systems are most likely galaxy over-densities and/or early-stage pre-coalescence mergers. Finally, we present 3mm LMT/RSR spectroscopic redshifts of six red-{\it Herschel} galaxies at $z_{\rm spec}=3.85-6.03$, two of them (at $z \sim 4.7$) representing new redshift confirmations. Here we release the AzTEC and deblended {\it Herschel} photometry as well as catalogues of the most promising interacting systems and $z>4$ galaxies.\\
\end{abstract}

\begin{keywords}
submillimetre: galaxies -- galaxies: high-redshift -- galaxies: starburst -- galaxies: interactions
\end{keywords}



\section{Introduction}

Taking advantage of a narrow atmospheric window at $\lambda\approx850\,\mu\rm m$, around two decades ago, the first surveys taken at submillimeter wavelengths with the SCUBA camera ---which now sits in the National Museum of Scotland in Edinburgh--- confirmed the existence of a  population of high-redshift dust-enshrouded star-forming galaxies (e.g. \citealt{Smail1997a,Barger1998a,Hughes1998a}).

Thanks to the significant two-decade effort poured into determining the physical properties of these galaxies, it is now known that they show the most extreme star formation rates (SFR from 100's to 1000's \sfr) in the Universe (modulo the possibility of a very top heavy stellar initial
mass function, e.g.  \citealt{Zhang2018}), have large stellar and dust masses ($\sim10^{10-11}$ and $\sim10^{8-9}$ M$_\odot$, respectively) with large gas mass reservoirs ($\sim10^{10-11}$ M$_\odot$), and contribute significantly to the cosmic star formation rate density (see reviews by \citealt*{Casey2014a} and \citealt*{Hodge2020}). These sources are also considered to be the progenitors of massive, quiescent galaxies observed at $z\approx2-3$, which ultimately lead to the assembly of the massive elliptical galaxies observed in the local Universe (\citealt{Toft2014a, Simpson2014}).

Nevertheless, despite their recognized importance in our understanding of galaxy formation and evolution, fundamental questions remain unanswered. For example, although the bulk of the population is known to lie at $z\approx2-4$ (e.g. \citealt{Aretxaga2003,Chapman2005,Aretxaga2007,Yun2012,Michal2012,Simpson2014,Koprowski2016,Brisbin2017,Zavala2018b,Simpson2020,Dudzeviciute2020}), their distribution at high-redshifts ($z>4$) and its dependence with flux density remains unclear. Constraining the prevalence of these galaxies is crucial, for instance, to derive a complete census of the cosmic star formation rate density and to test our current models of cosmic structure formation, since these galaxies are expected to trace the assembly of the first massive dark matter halos in the Universe (\citealt{Marrone2017a}). 

An important step towards understanding the formation processes that built up these extreme galaxies relies on determining their triggering mechanisms and their star formation modes. Pioneering observational and theoretical studies concluded that the formation scenario of Sub-Millimeter Galaxies (SMGs) involves major and minor gas-rich mergers (e.g., \citealt{Tacconi2006a,Ivison2007,Bothwell2010a,Engel2010a,Narayanan2010a}). Nevertheless, subsequent theoretical work showed that early-stage mergers (pre-coalescence galaxy pairs), isolated star-forming disks, and even line-of-sight projections or gravitational lensing can also lead to these bright submm fluxes (e.g., \citealt{Dave2010a,Hayward2011a,Narayanan2015a}). Although it is now clear that the population of dusty star-forming galaxies (DSFGs) may be rather heterogeneous (e.g. \citealt{Hayward2018,Jimenez-Andrade2020}), the relative importance of each component remains uncertain.

Characterizing these specifics requires sensitive and wide enough surveys to capture the rarest systems, which allow us to test the predictions from galaxy-formation models (e.g. \citealt{Hayward2013,Gruppioni2015,Lacey2016,Lagos2019a,McAlpine2019}). 

The large-area surveys conducted with the {\it Herschel} Space Observatory (as well as the South Pole Telescope -- SPT; \citealt{Vieira2010a}) have already identified remarkable examples of such systems, including some of the most distant dusty galaxies currently known at $z\approx6-7$ (\citealt{Riechers2013a,Fudamoto2017,Zavala2018a}) and extreme galaxy over-densities (i.e. proto-cluster structures) in the early Universe (e.g. \citealt{Ivison2013,Oteo2018a}; see \citealt{Strandet2017a} and \citealt{Miller2018a} for similar systems selected by the SPT).
Part of this success relies on the availability of simultaneous observations at 250, 350, and $500\,\mu\rm m$ with SPIRE, which enable a straightforward selection criteria ($S_{250\mu\rm m}<S_{350\mu\rm m}<S_{500\mu\rm m}$) of high-redshift candidates known as `$500\mu\rm m$ risers' or `red-{\it Herschel} galaxies' (e.g. \citealt{Pope2010,Cox2011,Dowell2014a,Asboth2016,Ivison2016,Donevski2018,Duivenvoorden2018, Ma2019,Bakx2020b}).

Follow-up observations with higher angular resolutions at (ideally) longer wavelengths than those used to select these galaxies are, however, necessary as an intermediate step to identify and filter out possible contaminants, whilst providing more accurate positions for spectroscopic surveys. 
Previous works, as those discussed in more detail below, have focused on samples of red-{\it Herschel} sources followed-up with the SCUBA-2 camera at $850\,\micron$ ($\theta_{\rm FWHM}\approx15\,$\,arcsec), LABOCA at $870\,\micron$ ($\theta_{\rm FWHM}\approx19\,$\,arcsec; \citealt{Dowell2014a,Ivison2016,Asboth2016,Duivenvoorden2018,Donevski2018}), or with higher angular resolution interferometric observations with ALMA, NOEMA and the SMA \citep[e.g.][]{Ma2019,Greenslade2020}.

Here, we present $1.1\rm\,mm$ imaging, using the AzTEC camera \citep{Wilson2008a} on the Large Millimeter Telescope (LMT\footnote{\url{www.lmtgtm.org}}, \citealt{Hughes2010a}) of a relatively large sample of 100 {\it Herschel}-selected galaxies. Additionally, we present 3mm spectra of 6 red-{\it Herschel} sources using the Redshift Search Receiver (RSR). We provide new redshifts for 2 of these sources (with $z\gtrsim4.7$) and confirm the redshift of the other 4, which were already known \citep{Fudamoto2017,Zavala2018a}.

The 32-m illuminated surface of the telescope, at the time of the observations, provides an effective angular resolution of $\theta_{\rm FWHM}\approx9.5\,$\,arcsec, a factor of 4 better than {\it Herschel} at $500\,\mu\rm m$ ($\theta_{\rm FWHM}\approx36\,$\,arcsec). 
This  angular resolution enables  the identification of not only the most promising high-redshift candidates but also of physically interacting galaxies blended within the {\it Herschel} beam, as discussed below.

This paper is structured as follows: \S\ref{secc:observations} describes the sample selection and AzTEC/RSR observations. The analysis of these images and the bulk of the results are presented in \S\ref{secc:results}. This includes constraints on the multiplicity, as well as photometric redshift, luminosity, and SFR estimations. In \S\ref{secc:subsamples} we identify and present sub-samples of the most promising high redshift candidates and physically interacting galaxies. Finally,  the implications of these results in our general understanding of the properties of this population of galaxies are discussed in \S\ref{secc:discussion}, where our conclusions are also summarized. 

Throughout this paper, we assume a flat $\Lambda$CDM cosmology with $H_0 = 68\rm\,km\,s^{-1}\,Mpc^{-1}$ and $\Omega_\Lambda= 0.69$ 
(\citealt{Planck-Collaboration2016a}), and the \citet{Chabrier2003a} initial mass function (IMF) for SFR estimations.

\section{Sample and observations} \label{secc:observations}

\subsection{Sample Selection}
\label{secc:sample_selection}

The {\it Herschel} Astrophysical Terahertz Large Area Survey \citep[\textit{H}-ATLAS,][]{Eales2010,Valiante2016} is one of the largest surveys ($\sim600\rm\,deg^2$) carried out with the {\it Herschel} Space Observatory. The thousands of sources detected at 250, 350 and $500\,\um$ in the South Galactic Pole (SGP), North Galactic Pole (NGP) and the Galaxy and Mass Assembly 9hr (G09), 12hr (G12) and 15hr (G15) fields, make \textit{H}-ATLAS an ideal survey to search for rare high-redshift ($z\gtrsim4$) dusty star-forming galaxies.

Our sample was taken from a parent sample of ultrared DSFGs obtained by \citet{Ivison2016}, where a detailed description of the selection process is presented and summarized here. First, candidates at $>2.2\sigma$  were identified in the 250$\mu$m maps using the Multi-band Algorithm for source eXtraction \citep[MADX;][]{MADX2020}. Then,  PSFs with scaled flux densities at each band  were subtracted from the SPIRE maps. Subsequently, a second and third set of candidates were generated by repeating the process and searching for  $>2.4\sigma$ and $>3.5\sigma$ peaks in the 350$\mu$m and 500$\mu$m residual maps respectively. The final catalog of 7961 ultra-red sources includes only those detected at $\geq3.5\sigma$ at 500$\mu$m and with $S_{500\mu\textrm{m}}/S_{250\mu\textrm{m}} \geq 1.5$ and $S_{500\mu\textrm{m}}/S_{350\mu\textrm{m}} \geq 0.85$. A sub-sample of 2725 ultrared candidates was eyeballed to find a reliable sample for ground-based observations. It is important to note that the eyeballing process rejected 22\,per cent of the candidates which were heavily confused and whose flux densities (and therefore colours) were unreliable (see \citealt{Ivison2016} for details).

This colour selected sources with spectral energy distributions (SEDs) that rise from 250 to $350\,\um$ and continue rising onwards to $500\um$ ($S_{250\rm\um} < S_{350\rm\um} < S_{500\um} $) are called  `$500\um$ risers' or red-{\it Herschel} sources. This technique enabled the identification of DSFGs up to $z\sim6$ (e.g. \citealt{Riechers2013a,Asboth2016,Zavala2018a}).

\begin{figure}\hspace{-0.75cm}\vspace{-0.5cm}
 \includegraphics[width=0.53\textwidth]{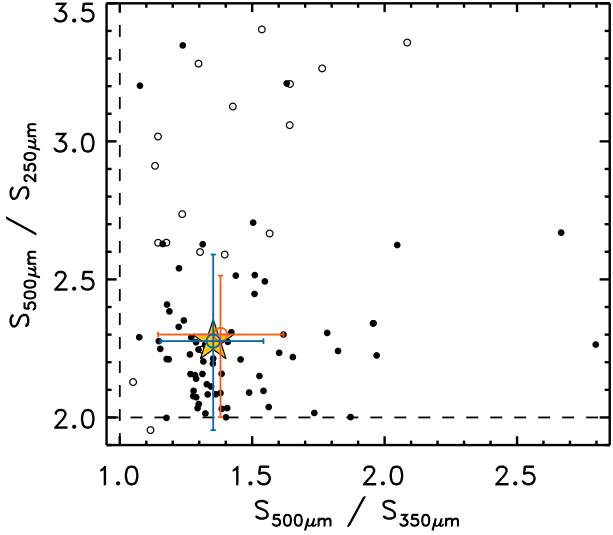}
 \caption{Colour selection ($S_{500\um}/S_{250\um}>2$ and $S_{500\um}/S_{350\um}>1$) of the \textit{H}-ATLAS targets observed with AzTEC: filled black circles identified in the $250\mu\textrm{m}$ map and empty circles in the $350\mu\textrm{m}$ (i.e. BANDFLAG 1 and 2 in \citet{Ivison2016}, respectively). All targets were detected above $5\sigma$ at $500\mu\textrm{m}$. The yellow star indicates the average colour of the sample, while the blue and orange circles indicate, respectively, the median colour of the targets with and without AzTEC detections above a 4$\sigma$ level. The similar colours shown by the two sub-samples suggest that differences in SEDs cannot explain the bulk of AzTEC non-detections discussed in \S \ref{secc:non_detections}.}
 \label{fig:sample_colours}
\end{figure}

The final sample of 108 ultra-red {\it Herschel} sources selected for LMT follow-up observations has signal-to-noise ratios at 500$\mu$m $\gtrsim 5$ and colour cuts $S_{500\um}/S_{250\um}>2$ and $S_{500\um}/S_{350\um}>1$ (see Figure \ref{fig:sample_colours}). Most of the sample ($\sim 80$ per cent) was identified in the $250\mu$m map, with the remaining ones identified in the $350\mu$m residual maps (Figure \ref{fig:sample_colours}). An additional selection criteria of $ 35\rm\,mJy < S_{500\um} < 80\rm\,mJy$ was imposed to minimize the contamination by false detections and by gravitationally lensed sources, since the fraction of lensed galaxies falls with decreasing $500\um$ flux density (with $\sim100$\, per cent of lensed galaxies expected at $S_{500\um}>100\,$mJy, and $\sim 50$\, per cent at $S_{500\um}>55\,$mJy; \citealt{Negrello2010}; see also \citealt{Wardlow2013,Negrello2017, Bakx2020a}). The sample was also checked not to have contamination by radio-loud AGN and it was correlated with optical/near-IR imaging \citep{Bourne2016} to reject nearby galaxies and any obvious lenses that could have entered the sample. Note that, while the contamination by radio-loud AGN is expected to be negligible, that from gravitationally lensed systems is not (e.g. \citealt{Donevski2018}). This is because the selection of high redshift candidates increases the probability of line of sight alignment with lower redshift massive structures. Indeed, such lensed systems have been confirmed in the sample (e.g. \citealt{Zavala2018a}).

\subsection{AzTEC Observations}
\label{secc:AzTEC_observations}

Of the 108 targets proposed for the LMT 2014-ES3 campaign (Project 2014AHUGD011, PI: D.H. Hughes), we obtained AzTEC data for 100 sources. AzTEC observations, using the photometry map mode which covers a $\sim 3.5$ arcmin diameter area, were conducted between November 2014 and June 2015 under optimal atmospheric conditions with $\langle{\tau}_{225\rm GHz}\rangle \approx 0.06 \pm 0.02$ ($0.03 \leq \tau_{225\rm GHz} \leq 0.11$). The integration times devoted to each target were in the range of 3 -- 50 min (21.3 hr in total), with a median of 11 min (Table \ref{table:HAFields}). Pointing measurements on known quasars close to the targets were made before and after science observations, and were used in the data reduction process to compensate for any pointing drifts, resulting in a r.m.s pointing accuracy $\lesssim1$\,arcsec.
 
The data was reduced using MACANA, the C++ version of the standard AzTEC data reduction pipeline \citep[e.g.][]{Scott2008}, with a Wiener-filter applied to improve the detection of point-like sources, at the expense of increasing the nominal FWHM by $\sim 11$\, per cent (from $\approx 8.6$\,arcsec to $\approx9.5$\,arcsec). The AzTEC pipeline produces four main outputs: signal and signal-to-noise maps, a weight map representative of the noise in each pixel of the map, and a 2D transfer function which tracks the effects of the reduction process on the shape of a synthetic 1 Jy point source (i.e. the Point Spread Function, PSF). Additionally, the pipeline can generate a set of simulated `jackknifed' noise maps by randomly multiplying the clean time-stream data by $\pm1$. In \S\ref{secc:source_detection} we take advantage of these simulations to measure false-detection-rate (FDR) probabilities in our AzTEC maps.

\begin{table*}
\centering
\caption{Sample of the 93 \textit{H}-ATLAS fields targeted at 1.1\,mm with AzTEC (Figure \ref{fig:pstamps}). Right Ascensions and Declinations (J2000) correspond to the \textit{H}-ATLAS source centroids where the AzTEC maps were centered. The quoted  $\sigma_\textrm{r.m.s}$ values correspond to the 85\,per cent coverage area of the AzTEC maps (i.e. the $\sim 36.6$\,arcsec diameter central area used for our multiplicity analysis). \label{table:HAFields}}
\begin{tabular}{lrrrcc|lrrrc}
\hline
ID & R.A. & Dec. & $t_\textrm{int}$ & $\sigma_\textrm{r.m.s}$ & \hspace{2cm} & ID & R.A. & Dec. & $t_\textrm{int}$ & $\sigma_\textrm{r.m.s}$ \\
    &
[deg.] &
[deg.] &
[min] &
[mJy] &
    &
    &
[deg.] &
[deg.] &
[min] &
[mJy] \\
\hline

 G09-12469 & $140.422082$ & $0.895139$  & 30 &  $1.15$      &   & NGP-203484$^a$ & $204.917078$ & $31.369083$  & 22 &  $1.49$ \\
 G09-44907 & $129.345002$ & $1.429361$  & 20 &  $0.98$      &   & NGP-211862 & $193.669996$ & $26.823668$  & 11 &  $1.62$     \\
 G09-58643 & $127.822924$ & $1.282944$  & 15 &  $1.20$      &   & NGP-222757 & $199.172916$ & $23.675194$  & 11 &  $1.35$     \\
 G09-62610 & $137.354579$ & $1.928361$  & 30 &  $0.72$      &   & NGP-235542 & $195.640411$ & $26.673582$  & 11 &  $1.78$ \\
 G09-64894 & $138.189998$ & $1.195028$  & 30 &  $1.27$      &   & NGP-240219 & $192.798743$ & $30.998196$  &  3 &  $2.67$ \\
 G09-71054 & $137.194161$ & $1.891806$  & 11 &  $0.89$      &   & NGP-244082 & $199.487500$ & $34.490723$  & 11 &  $1.73$ \\
 G09-75817 & $130.181251$ & $1.948556$  & 11 &  $1.38$      &   & NGP-246114$^a$ & $205.309166$ & $33.992964$  & 10 &  $1.64$ \\
 G09-80523 & $132.810001$ & $1.034833$  & 15 &  $1.16$      &   & NGP-248192 & $193.215837$ & $34.549946$  & 11 &  $1.55$ \\
 G09-81106$^a$ & $132.404541$ & $0.248816$  & 45 &  $0.82$  &   & NGP-248712 & $197.039995$ & $22.836945$  & 11 &  $1.37$ \\
 G09-83808$^a$ & $135.189171$ & $0.690306$  & 11 &  $0.89$  &   & NGP-248948 & $192.160006$ & $29.628666$  &  3 &  $2.78$ \\
 G12-23831 & $179.638753$ & $-1.823000$  & 15 &  $1.40$     &   & NGP-249138 & $193.611245$ & $24.625444$  &  3 &  $2.06$ \\
 G12-26926$^a$ & $183.489161$ & $-1.372833$  & 11 &  $1.68$ &   & NGP-249475 & $206.234994$ & $31.441082$  & 22 &  $1.55$ \\
 G12-31529 & $181.649580$ & $1.549667$  & 60 &  $1.00$      &   & NGP-284357$^a$ & $203.214926$ & $33.394179$  & 10 &  $1.54$ \\
 G12-42911 & $175.811248$ & $0.479556$  & 10 &  $1.60$      &   &  NGP-49609 & $203.213339$ & $32.936916$  &  3 &  $2.48$ \\
 G12-47416 & $182.763333$ & $0.215972$  & 22 &  $1.12$      &   &  NGP-55628 & $206.604581$ & $34.270832$  & 22 &  $1.54$ \\
 G12-49632 & $179.715414$ & $-0.164528$  & 22 &  $1.17$     &   &  NGP-78659 & $207.216253$ & $26.899221$  &  6 &  $2.26$ \\
 G12-53832 & $184.296255$ & $-0.128139$  &  3 &  $2.79$     &   &  NGP-87226 & $205.289998$ & $32.933891$  &  6 &  $2.59$ \\
 G12-58719 & $185.308757$ & $0.930917$  & 11 &  $1.41$      &   &  NGP-94843 & $204.695005$ & $25.669167$  & 22 &  $1.44$ \\
 G12-73303 & $183.509173$ & $-1.932833$  & 11 &  $1.26$     &   & SGP-101187 & $16.970000$ & $-30.297194$ & 11 &  $1.13$ \\
 G12-77419 & $182.271252$ & $-1.089417$  &  6 &  $1.67$     &   & SGP-106123 & $13.851250$ & $-28.010111$ & 26 &  $0.99$ \\
 G12-78868 & $179.059167$ & $1.651611$  &  3 &  $2.61$      &   & SGP-211713 & $22.584168$ & $-31.662361$ & 15 &  $1.98 $ \\
 G15-23358 & $214.860005$ & $0.193861$  & 11 &  $1.11$      &   & SGP-215925 & $351.983328$ & $-32.768387$ & 11 &  $2.38 $ \\
 G15-26675 & $221.138749$ & $0.277694$  &  6 &  $1.86$      &   & SGP-238944 & $0.077917$ & $-33.636837$ & 11 &  $2.86 $ \\
 G15-29728 & $211.117501$ & $1.582861$  & 11 &  $1.10$      &   & SGP-267200 & $350.531673$ & $-34.577248$ & 11 &  $1.73 $ \\
 G15-48916 & $214.615831$ & $0.789056$  & 11 &  $1.12$      &   & SGP-272197 & $1.531667$ & $-32.444195$ & 15 &  $1.62 $ \\
 G15-57401 & $214.029579$ & $1.159361$  & 11 &  $1.18$      &   & SGP-280787 & $350.393744$ & $-33.081249$ & 11 &  $1.99 $ \\
 G15-63483 & $221.295834$ & $0.017833$  & 11 &  $1.43$      &   & SGP-284969 & $15.858334$ & $-30.059082$ & 26 &  $0.91 $ \\
 G15-68998 & $222.726660$ & $-0.595778$  & 11 &  $1.27$     &   & SGP-289463 & $25.536667$ & $-32.574276$ & 15 &  $2.07 $ \\
 G15-72333 & $217.463751$ & $1.003722$  & 11 &  $1.07$      &   & SGP-293180 & $18.160833$ & $-30.783361$ & 11 &  $1.22 $ \\
 G15-78944 & $221.771665$ & $1.024111$  & 11 &  $1.28$      &   & SGP-316248 & $354.393339$ & $-34.839191$ & 11 &  $1.43 $ \\
 G15-82597 & $220.157919$ & $0.801556$  & 11 &  $1.31$      &   & SGP-322449 & $344.887505$ & $-34.446335$ & 11 &  $1.34 $ \\
 G15-82610 & $220.729165$ & $1.162305$  & 11 &  $1.45$      &    & SGP-323041 & $18.501250$ & $-32.517056$ &  5 &  $2.16 $ \\
 G15-82660 & $215.634999$ & $0.505694$  & 11 &  $1.08$      &   & SGP-340137 & $344.271240$ & $-33.839111$ & 11 &  $1.24 $ \\
 G15-83272 & $213.802500$ & $-0.276806$  &  3 &  $1.70$     &   & SGP-348040 & $350.404987$ & $-35.024223$ & 11 &  $1.39 $ \\
NGP-112775 & $204.170423$ & $26.266306$  & 22 &  $1.38$     &   & SGP-352624 & $19.664167$ & $-27.638889$ & 15 &  $1.47 $ \\
NGP-113203 & $196.662912$ & $28.464140$  & 22 &  $1.43$      &  & SGP-359921 & $16.919584$ & $-28.448500$ & 11 &  $1.21 $ \\
NGP-115876 & $204.650416$ & $27.546473$  & 22 &  $1.50$     &   & SGP-379994 & $11.356249$ & $-32.554695$ & 15 &  $1.17 $ \\
NGP-124539 & $195.212088$ & $32.776001$  & 11 &  $1.67$     &   & SGP-384367 & $21.078333$ & $-32.980335$ & 15 &  $1.13 $ \\
NGP-131281 & $193.201246$ & $34.404278$  & 18 &  $1.36$     &    & SGP-396540 & $10.270833$ & $-28.222473$ & 11 &  $1.37 $ \\
NGP-139851 & $196.441255$ & $25.498222$  &  3 &  $2.60$     &    & SGP-396663 & $12.548333$ & $-32.482918$ & 11 &  $1.52 $ \\
NGP-145039 & $194.372921$ & $29.280277$  & 22 &  $1.32$     &   & SGP-396921 & $16.555417$ & $-28.231222$ & 26 &  $0.95 $ \\
NGP-149267 & $203.001666$ & $26.422222$  & 22 &  $1.36$     &   & SGP-396966 & $8.878333$ & $-31.504917$ & 15 &  $1.27 $ \\
NGP-157992 & $193.517504$ & $27.177750$  &  3 &  $2.60$      &  & SGP-399383 & $20.202917$ & $-30.976528$ & 15 &  $1.32 $ \\
NGP-168019 & $205.405412$ & $32.476780$  & 22 &  $1.59$      &  & SGP-400082 & $8.386249$ & $-30.080584$ &  6 &  $1.39 $ \\
NGP-172727 & $196.315827$ & $25.516083$  &  3 &  $2.50$     &   & SGP-403579 & $355.064993$ & $-30.445110$ &  6 &  $1.28 $ \\
NGP-176261 & $199.242082$ & $33.915947$  &  9 &  $1.70$     &   &  SGP-68123 & $341.992092$ & $-29.945444$ & 11 &  $1.24 $ \\
NGP-194548 & $203.407087$ & $24.261639$  & 11 &  $1.32$     &   &            &              &              &    &          \\

\hline
\multicolumn{9}{l}{\footnotesize$^a$Selected for spectroscopic follow-up with the 3mm RSR (see \autoref{table:RSR}).} \\
\end{tabular}
\end{table*}

\begin{figure}\hspace{-1cm}
\includegraphics[width=0.52\textwidth]{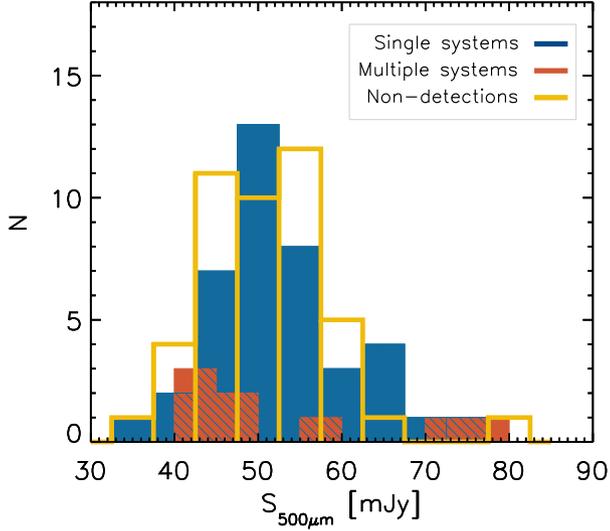}
\caption{Number of sources as a function of original (before deblending) {\it Herschel} $500\rm\,\um$ flux density. AzTEC sources classified as single and multiple systems (see \S\ref{secc:results}) are illustrated by the blue and orange histograms, while AzTEC non-detections (SNR $< 4$) are represented by the yellow histogram. As shown, the three different sets include sources with similar flux densities, discarding possible selection biases in the classification. The bins are slightly shifted for better visualization.
\label{fig:sample}}
\end{figure}

\begin{figure}\hspace{-1cm}
\includegraphics[width=0.52\textwidth]{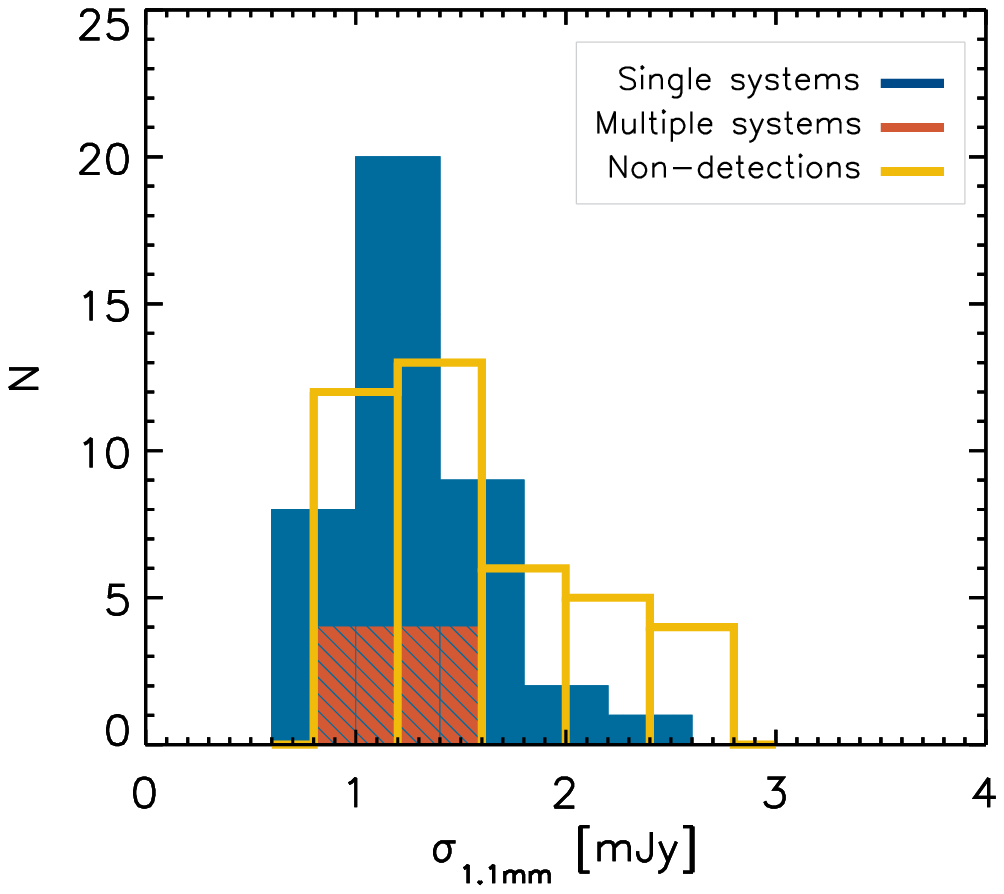}
\caption{Histogram of achieved sensitivity in the AzTEC observations for the three different classifications: Single systems (in blue), multiple systems (orange), and non-detections (yellow, SNR $< 4$). Our survey shows an homogeneous sensitivity, with all of the maps reaching a depth within $0.7<\sigma_{1.1\rm mm}<2.8\,\rm mJy$, therefore ruling out any observational bias in our classification.}
\label{fig:sample2}
\end{figure}

Seven of the targets in the SGP field, observed in the poorest weather conditions ($\tau_{225\rm GHz} \approx 0.11$), did not reach the target sensitivity (with $\sigma_{1.1\textrm{mm}} > 3.0$ mJy r.m.s.) and, therefore, were removed from the analysis. Thus, we focus on the remaining 93 sources observed with AzTEC (Table \ref{table:HAFields}). 

The  maps of the final sample have an average depth of $\langle{\sigma}_{1.1\textrm{mm}}\rangle = 1.5\pm0.5\,\textrm{mJy}$ in the central region used for the counterpart analysis (i.e. within the $\sim 85$\, per cent coverage area) and $\langle{\sigma}_{1.1\textrm{mm}}\rangle =2.1\pm0.7\,\textrm{mJy}$ over the 50\,per cent coverage area used to detect sources (see \S\ref{secc:source_detection}). The filtered maps have PSFs with $\langle{\theta}_\textrm{FWHM}\rangle = 9.6 \pm 0.5\,$\,arcsec. Figures \ref{fig:sample} and \ref{fig:sample2} show the $500\,\um$-flux density distribution of our sample and the attained $\sigma_{1.1\rm mm}\approx0.7-2.8\,\rm mJy$ r.m.s. distribution of the AzTEC observations respectively.

\subsection{RSR observations}
\label{secc:RSR}
Six red-\textit{Herschel} sources, confirmed with AzTEC to be single systems at high-redshift ($z_\textrm{phot} \gtrsim 4$), were selected for spectroscopic follow-up observations with the Redshift Search Receiver on the LMT (Table \ref{table:RSR}). The RSR is a broadband spectrometer covering the 3\,mm window (73-111 GHz) with four detectors in a dual–beam dual–polarization configuration \citep[RSR, ][]{Erikson2007}. The observations were done using both the 32m and 50m (since 2018) configurations of the LMT.

The RSR data was reduced using the \textsc{Dreampy} package (Data REduction and Analysis Methods in PYthon, written by G. Narayanan) and following the standard procedure \citep[e.g. ][]{Yun2015, Cybulski2016,Wong2017}. A careful visual inspection of individual scans was performed to identify and remove those with the noisiest spectral features. 

\begin{figure*}
\includegraphics[width=0.92\textwidth]{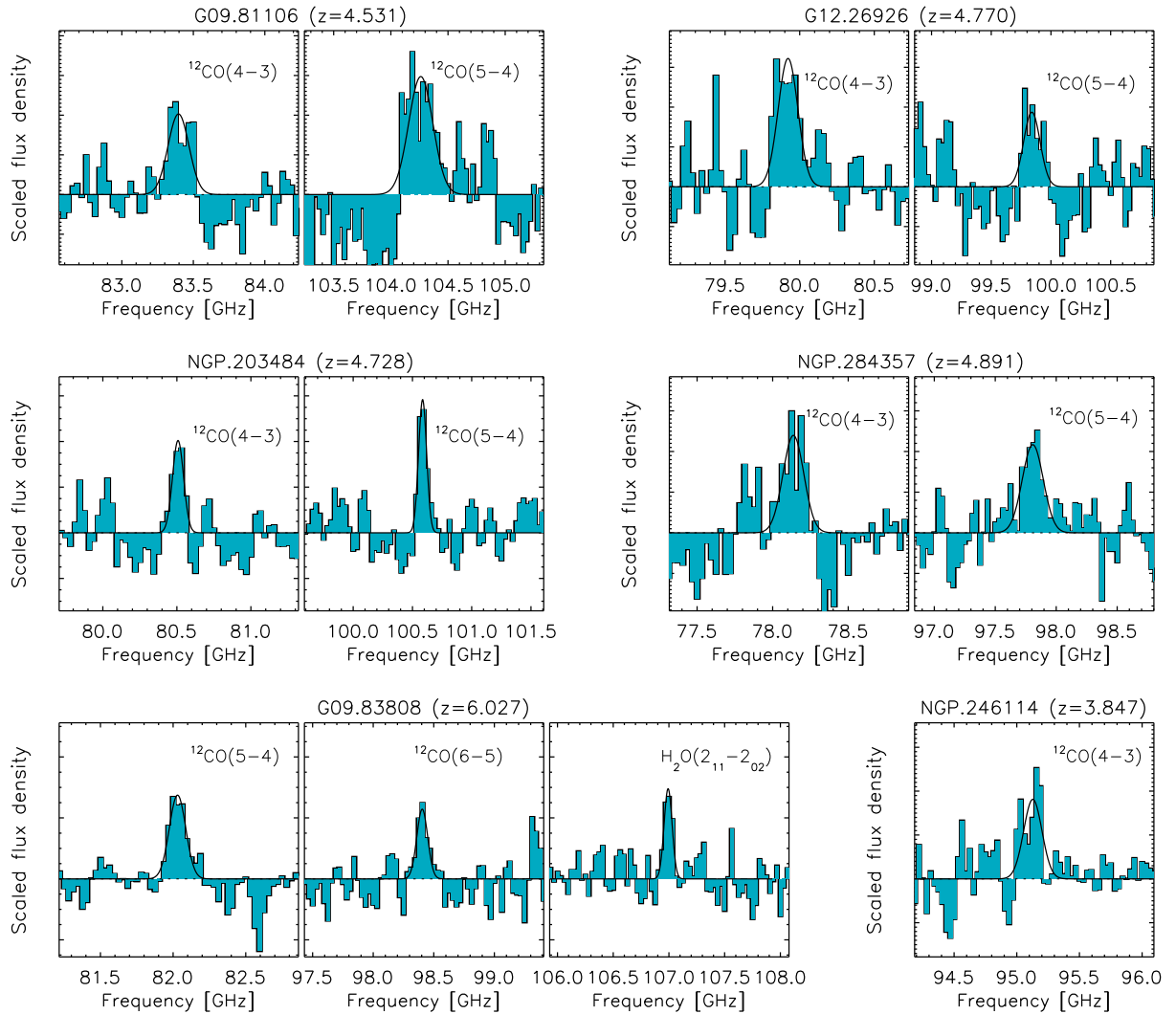}
\caption{Identified lines in the RSR spectra along with the best-fit Gaussian functions. The redshift of each source is identified on the top and the individual transitions are labeled close to the lines. Each panel has a total width of 6,000\,km/s and is centered on the central frequency of the respective line. The identified lines in G09.83808 are the same as in \citet{Zavala2018a}. The fitted parameters to each line are listed in Table \ref{table:RSR}}.
\label{fig:lines}
\end{figure*}

We confirm the redshifts presented in  \citet{Fudamoto2017} and \citet{Zavala2018a} for four of these sources, and provide two new determinations at $z=4.768$ and 4.728 (see Table \ref{table:RSR}). The latter two correspond to G12-26926 and NGP-203484 respectively, whose redshifts are unambiguously identified from at least two emission lines detected with $\textrm{SNR}\ge5$ in each RSR spectrum. Their redshifts are independently confirmed using the template cross-correlation analysis described by \citet{Yun2015}. The same template cross-correlation method yields the unique redshifts of three other objects with two or more emission lines in the RSR spectrum (G09-81106, G09-83808, NGP-284347). Only one emission line is detected in the RSR spectrum of NGP-246114, but it is the same CO(4-3) line at $z=3.847$ previously reported by \citet{Fudamoto2017} who also detected a CO(6-5) transition. Figure \ref{fig:lines} shows the identified lines in the RSR spectra. Their fitted parameters are summarized in Table \ref{table:RSR}. We use these spectroscopic redshifts in Figure \ref{fig:redshift_comparison} to characterize the accuracy of our photometric-redshift determinations.

\begin{table*}
\centering
\caption{Measured properties of the detected CO transitions in the 3mm RSR spectra. The spectroscopic redshifts estimated in this work from the LMT/RSR spectra are given in column 7 and, for comparison, those previously published are listed in column 8.}
\label{table:RSR}
\begin{tabular}{lcrccccc}
\hline
ID &
Transition &
$\nu_\textrm{line}$ & Peak flux & Integrated flux &
FWHM &
\multicolumn{2}{c}{$z_\textrm{spec}$}  \\
  & 
  &
[GHz] &
[mJy] & 
[Jy km s$^{-1}$] & 
[km s$^{-1}$] & 
LMT/RSR & 
Literature  \\
\hline

G09-81106  &  CO$(4-3)$  & $83.40\pm0.01$   & $2.50\pm0.41$  & $1.36\pm0.17$  & $727\pm128$  & $4.531\pm0.006$ & $4.531\pm 0.001^a$  \\
           &  CO$(5-4)$  & $104.28\pm0.02$  & $4.06\pm0.33$  & $2.21\pm0.32$  & $899\pm70$   & &   \\
G09-83808  &  CO$(5-4)$  & $82.03\pm0.01$   & $2.75\pm0.29$  & $1.27\pm0.17$  & $584\pm120$  & $6.026\pm0.005$ & $6.0269\pm 0.0006^b$  \\
           &  CO$(6-5)$  & $98.41\pm 0.01$  & $2.29\pm0.41$  & $0.90\pm0.17$  & $418\pm115$  & &   \\
           &H$_2$O($2_{11}-2_{02}$) &  $106.99\pm0.01$  & $2.96\pm0.48$  & $0.79\pm0.11$ & $227\pm49$ & &   \\
G12-26926  & CO$(4-3)^d$ & $79.93\pm0.01$   & $8.53\pm0.45$ & $1.98\pm0.18$  & $641\pm38$   & $4.768\pm0.002^c$ &   \\
           & CO$(5-4)^d$ & $99.86\pm0.01$   & $4.58\pm0.40$ & $0.93\pm0.22$  & $520\pm41$   & &   \\
NGP-203484 & CO$(4-3)$   & $80.51\pm0.01$   & $5.39\pm0.54$ & $1.53\pm0.25$  & $376\pm 45$  & $4.728\pm0.002^c$ &  \\
           & CO$(5-4)$   & $100.59\pm0.01$  & $7.03\pm0.68$ & $1.59\pm0.17$  & $243\pm 31$  & &   \\
NGP-246114 & CO$(4-3)$   & $95.13\pm0.03$   & $2.05\pm0.52$ & $1.04\pm0.13$  & $476\pm168$  & $3.847\pm 0.002^e$ & $3.847\pm 0.002^a$ \\
NGP-284357 & CO$(4-3)^d$ & $78.16\pm0.01$   & $8.83\pm0.41$ & $1.60\pm0.21$  & $626\pm36$  & $4.891\pm0.006$ & $4.894\pm 0.003^a$ \\
           & CO$(5-4)$   & $97.81\pm0.01$   & $3.31\pm0.48$ & $1.69\pm0.22$  & $535\pm100$  & &   \\
\hline
\multicolumn{8}{l}{$^a$ \citet{Fudamoto2017}. $^b$ \citet{Zavala2018a} including an additional [CII] with SMA. $^c$ New LMT/RSR determinations derived in this work.} \\
\multicolumn{8}{l}{$^d$ First published CO transitions using the 50m-LMT. $^e$ Estimated including the CO(6-5) transition of \citet{Fudamoto2017}.}\\
\end{tabular}
\end{table*}

An alternative reduction was produced using a new \textsc{Python} wrapper script developed by D. O. S\'anchez-Arg\"uelles for the \textsc{Dreampy} package, known as \textsc{rsr\_driver}\footnote{The \textsc{rsr\_driver} and its documentation is publicly available at the {\scshape{LMT devs}} github repository \url{https://github.com/LMTdevs/RSR_driver}.}. This script aims to provide the LMT user community with a front-end interface to generate RSR scientific-quality data from raw observations. The \textsc{rsr\_driver} reduction procedure is very similar to the standard process, and below a brief description is presented. 

\begin{figure}\hspace{-0.35cm}
 \includegraphics[width=0.5\textwidth]{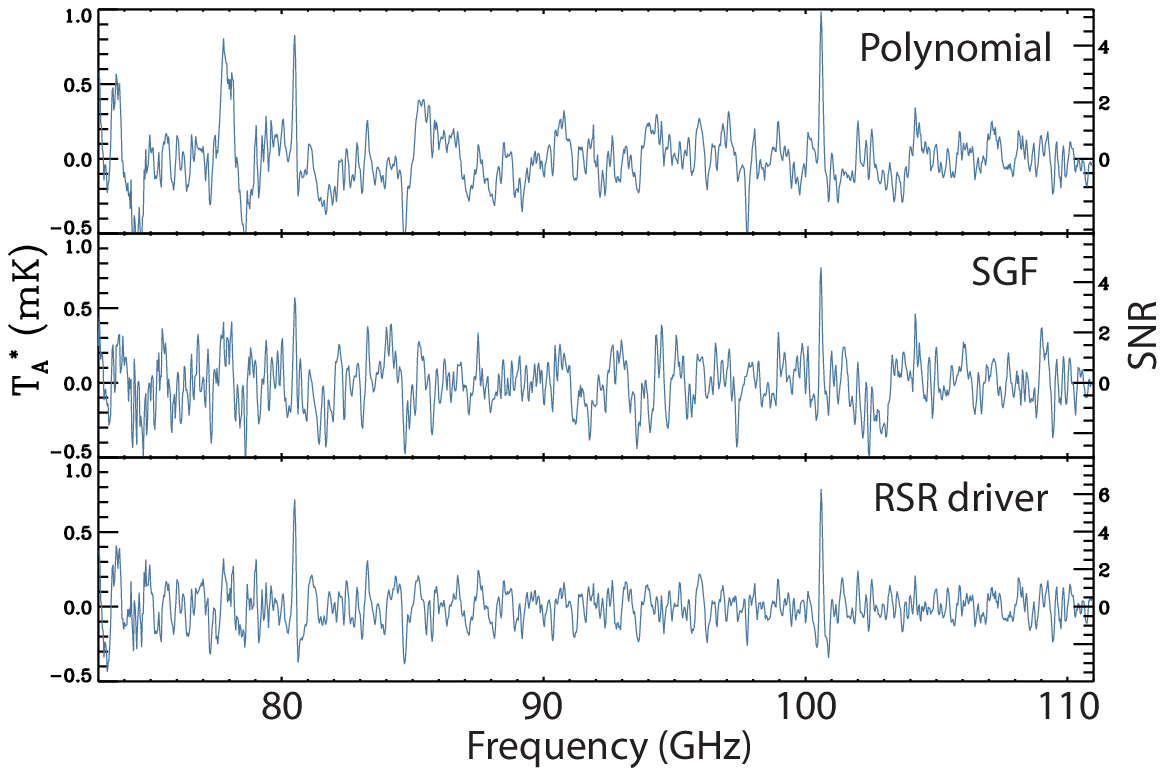}
 \caption{RSR spectra of source NGP-203484 ($z=4.728$) obtained with the different baseline-subtraction techniques available within the pipeline. Top: the standard output of the \textsc{Dreampy} package using a low order polynomial baseline subtraction. Middle: the resulting spectrum by visually inspecting and removing the noisy data, and subtracting a Savistzky-Golay filter (SGF). Bottom: the output produced by the recently developed \textsc{rsr\_driver}, including both, a low order polynomial baseline subtraction and a SGF. The output of the \textsc{rsr\_driver} wrapper produces a $\sim\times1.5$ improvement on the line-peak SNR of the observed CO transitions.}
 \label{fig:rsr_pipeline_demo}
\end{figure}

For each detector the RSR backend records the autocorrelation function (ACF) of the observed sky brightness. It is important to notice that the broad 73-111 GHz bandwidth of the RSR is achieved by dividing it into six bands. The raw data is therefore comprised of six ACFs. The pipeline starts by processing the ACFs using a Fourier transform matrix to reconstruct the spectrum of the astronomical source. At this stage, a user defined low order ($\leq 3$) polynomial baseline is computed across each band and subtracted from the spectrum. The relatively fast internal switching ($\simeq 1$ kHz) between the RSR beams allows the minimization of the contribution from atmospheric noise into the ACF; nevertheless, small differences in the switching duty-cycle can introduce large baseline artifacts. To increase the detectability of spectral lines, the \textsc{rsr\_driver} can calculate and subtract a Savitzky-Golay filter (SGF) for each band, analogous to performing a high-pass filter on the observed data\footnote{If the SGF is applied, the polynomial baseline subtraction from the previous step is not performed.}. The number of points used to simultaneously fit the filter determines its cut-off frequency. In this work we used a length of 55 frequency channels to cut out all the features broader than $\simeq 1.71$ GHz ($\Delta V \simeq 5500 $ km/s), which is much larger than the expected CO line widths from high-redshift SMGs. The \textsc{rsr\_driver} allows us to automatically remove noisy data from the spectrum. A typical five minutes integration yields a $\sigma_{T_{A^\ast}} \simeq 2$ -- $4$ mK. All bands with $\sigma_{T_{A^\ast}} > 5$ mK are therefore ignored by the pipeline. The construction of the final spectrum is achieved by a weighted average of all the observations available for an astronomical source. Figure \ref{fig:rsr_pipeline_demo} shows one example of the resulting RSR spectrum and a comparison between the results obtained with the different baseline improvement techniques. It is important to notice that the output of the \textsc{rsr\_driver} wrapper produces a significant ($\times1.5$) improvement on the line-peak SNR of the observed CO transitions, which would be important for the identification of fainter transitions with lower SNRs. This semi-automatic procedure provides an user-friendly tool to reduce RSR data in an homogeneous and efficient way.

\section{Analysis and Results} \label{secc:results}
\subsection{Source detection and flux measurement}\label{secc:source_detection}
\subsubsection{Detection algorithm}
Sources were identified using the AzTEC signal-to-noise ratio (SNR) map of each observation and adopting a SNR threshold. If a source is detected in a map, we measure the 1.1\,mm flux density and noise at the position of the pixel with the maximum SNR value as well as its celestial coordinates. A mask of 1.5 times the size of the AzTEC beam ($\theta_{1.1\rm mm}\approx9.5$\,arcsec) is applied to the source before repeating the process again until no more sources above our adopted threshold are detected in the map. This process is conducted within the 50 per cent coverage of the maximum depth, which corresponds to typical areas of $6.1 - 10.8\,\rm arcmin^2$ per map.

A $4\sigma$ detection threshold was adopted to minimize the contamination from false detections due to the noise in the maps. False detection probabilities are estimated in three different ways. In method a) the noise simulations generated by the AzTEC pipeline (jackknife maps) are used to identify positive noise peaks (i.e. the false detections), which are then divided by the number of detections in the real maps. In method b), the number of false sources estimated above are divided by the expected number of sources in the map, which we calculate by adding the false sources to the number counts from blank fields \citep{Scott12} plus 1 (to compensate for the fact that we are targeting biased fields where we expect to find at least one source). Finally, in method c) the number of negative peaks in the SNR map (representative of the noise in the map) are divided by the number of positive detections.

\begin{figure}\hspace{-1cm}
\includegraphics[width=0.53\textwidth]{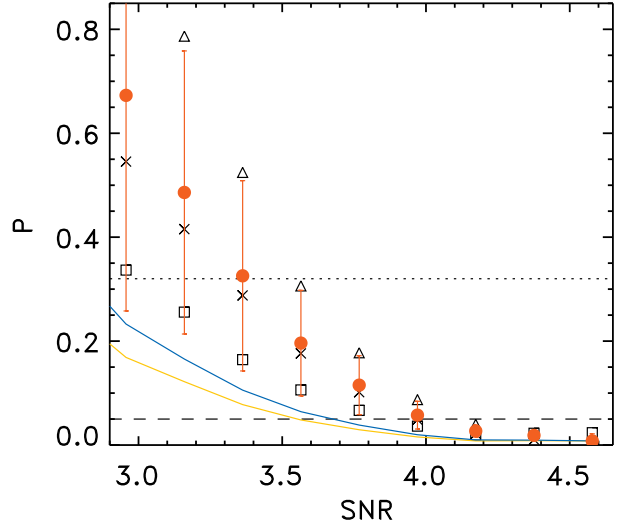}
\caption{False detection rate (P) estimated within the adopted search area ($r=36.6\,$ arcsec) as a function of signal to noise ratio. Triangles, crosses, and squares correspond to methods a), b), and c), respectively, to estimate the false detection rates (see text). Filled red circles indicate the mean of the three methods. Dashed and dotted lines indicate 5 and 32\,per cent contamination, respectively. The blue and yellow curves correspond to the expected false detection rate for sources detected within the deeper central $r=15\,$ and $r=10\,$ arcsec region of the maps.} 
\label{fig:FDRprobaility}
\end{figure}

Figure \ref{fig:FDRprobaility} shows the results of our false detection rate analysis and how, for our adopted search radius of $r=36.6$\,arcsec, the three methods converge at $\rm{SNR} \gtrsim 4$, where the contamination due to false detections is $\lesssim 5$\, per cent. This contamination drops to $\sim2$ per cent for $4\sigma$ sources detected within the central and deeper $r = 15$ arcsec area of the maps, where the reliability of $3.5\sigma$ and $3\sigma$ detections increases to $\gtrsim$ 90 and 75 per cent respectively.

\begin{figure}\hspace{-1cm}
\includegraphics[width=0.53\textwidth]{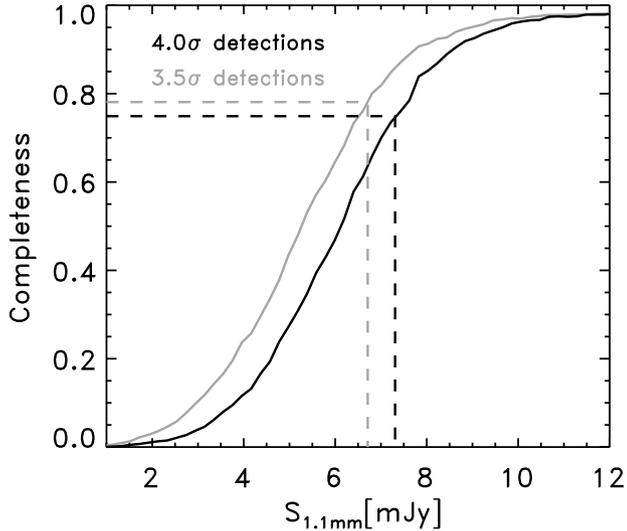}
\caption{Completeness fraction as a function of 1.1 mm flux density for our AzTEC survey of red-\textit{Herschel} targets, estimated by inserting synthetic sources in different flux density bins and attempting to recover them above a 4.0$\sigma$ (black) and 3.5$\sigma$ (grey) detection threshold. If we extrapolate the $500 \mu\textrm{m}$ flux density using typical SMG SED templates (e.g. \citealt{Michalowski2010a,daCunha2015}; or modified black bodies with $T_\textrm{dust} = 40 - 50\,$K and $\beta = 1 - 2$ ) at $z = 3$, the expected median 1.1 mm flux densities are $\gtrsim 9 \,$mJy, corresponding to a completeness $\gtrsim 90$ per cent. The median flux density of the AzTEC detections, however, suggests a more conservative $\sim 75$ per cent completeness (dashed lines).
}
\label{fig:completeness}
\end{figure}

We have also estimated the completeness of our survey as a function of flux density by inserting synthetic sources (1000 sources per flux density bin) in the AzTEC maps and quantifying the recovery efficiency. The sources are inserted within the 85 per cent coverage region considered in our analysis. Nevertheless, if a source is detected within 5 arcsec from a real detection, it is excluded from the completeness calculation. Figure \ref{fig:completeness} summarizes the completeness of our survey. 
Assuming typical SMG SED templates (e.g. \citealt{Michalowski2010a,daCunha2015}; or modified black-bodies with $T_\textrm{dust}= 40 - 50\,$K and $\beta = 1 - 2$) at $z \geq 3$, scaled to the average $500 \mu\textrm{m}$ flux density of our red-\textit{Herschel} targets, we infer a completeness $\gtrsim 90$ per cent given the expected 1.1\,mm flux densities $\gtrsim 9 \,$mJy. Nevertheless, the median flux density of our AzTEC 4$\sigma$ detections of $S_{1.1\textrm{mm}} = 7\, \textrm{mJy}$ (see below) suggests a lower completeness of around 75 per cent, which should be considered more reliable.

The same set of simulations are then used to explore the impact of flux boosting, meaning sources’ flux densities systematically biased upwards by noise and the presence of unresolved astronomical sources below the detection threshold. Given the relatively low number of sources at the depth of our observations (around 0.006 sources per AzTEC beam at our typical $1\sigma$ RMS depth), we infer an average flux boosting factor of $\sim1.15$ for those sources detected at our detection threshold of SNR=4. The average flux boosting decreases with flux density (or similarly with SNR) and it is almost negligible at SNR$\gtrsim5.5$\footnote{Note that our observations are far from being confusion noise limited. Assuming the most recent $1.1\,$mm number counts from \citet{Zavala2021a} and defining confusion noise at the level of 1/30 source per beam, we estimate the confusion noise to be around $0.35\,$mJy for the 32-m LMT, which is a factor of $\approx4-6\times$ deeper than the typical noise in our observations.}. This value is not taken into account given the relatively larger uncertainties of the sources' flux densities (25 per cent for a source detected at SNR=4).

\begin{figure}\hspace{-1cm}
\includegraphics[width=0.53\textwidth]{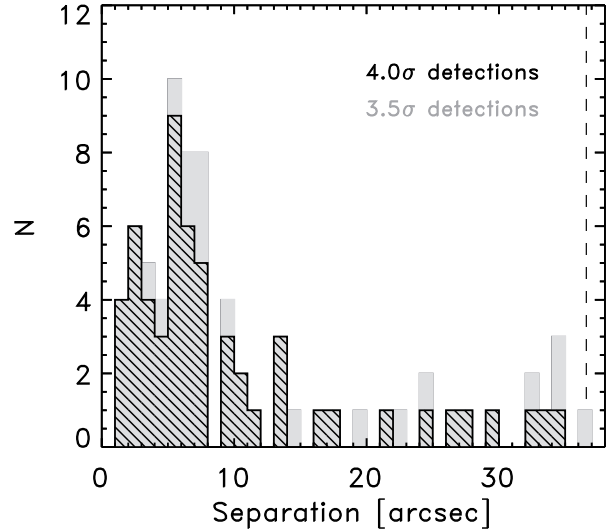}
\caption{Histogram of the separations between the AzTEC detection and \textit{H}-ATLAS targeted positions. Although a radius $\theta_{500\um}=36.6\,$arcsec (vertical dashed line) is used for our counterpart analysis, $\sim 82$\, per cent ($75$\, per cent) of the $4\sigma$ ($3.5\sigma$) AzTEC sources are separated by $\lesssim 15\,$arcsec off the \textit{H}-ATLAS target, were the reliability of the detections is $\gtrsim 98\,$ per cent (90 per cent; see Fig. \ref{fig:FDRprobaility}).
}
\label{fig:separations}
\end{figure}

In the 93 analyzed maps, we find a  total of 79 AzTEC detections above our adopted threshold (SNR>4) within the 50\,per cent coverage area. The counterpart matching between the {\it Herschel} and these AzTEC sources was then performed using the $500\,\um$ {\it Herschel} beamsize as a reference. An AzTEC source is associated with a {\it Herschel} source if its AzTEC position lies within $\theta_{500\um}=36.6\,$arcsec of the {\it Herschel} position (although most of them lie within $\sim15\,$arcsec; see Figure \ref{fig:separations}). Out of the 93 red-{\it Herschel} targets, 40 are associated with individual AzTEC sources, while eight break into multiple AzTEC components (comprising a total of 16 AzTEC detections) and therefore are classified as multiple systems (see \S\ref{secc:multiplicity}). 
This leaves 45 {\it Herschel} targets with no AzTEC detections at the $>4\sigma$ level, which are discussed in \S\ref{secc:non_detections}.

\subsubsection{Serendipitous sources}

Additionally, we find 23 AzTEC ``serendipitous'' detections that lie outside the adopted search radius, i.e. they are not directly associated with any red-{\it Herschel} source in the sample and do not contribute to the {\it Herschel}-500$\mu$m flux density. Given the total mapped area ($\sim720$ square arcmin without considering the central regions of the maps) and the flux detection threshold assumed in our analysis ($S_{\rm{1.1mm}} \gtrsim 8$ mJy in the outer region of the maps), the number of serendipitous detections (23) is much larger than the number of sources ($\sim4$) predicted from AzTEC blank-field number counts \citep{Scott12}. The probability of finding this number of serendipitous sources within the mapped area by chance is $\sim 8\times10^{-12}$. Furthermore, moderate resolution single-dish number counts as those reported in \citet{Scott12} are known to over-estimate the number of bright sources due to source blending \citep[e.g.][]{Lindner2011,Karim2013,Bethermin2017a,Stach2018}. Taking this into account would further increase the discrepancy we report. 

The estimated overdensity parameter of 4.75 ($\delta(>S) \equiv N(>S)/N(>S)_\textrm{blankfield} - 1$ = 4.75) is conservative since we are excluding all sources within the multiplicity search radius (i.e. $r = 36.6$ arcsec). It is also consistent with the results of \citet{Lewis2018} from LABOCA 870$\mu$m follow-up observations of 22 ultra-red \textit{Herschel} sources, who found $\delta(>S_{870\mu\textrm{m}}) \sim 4 - 30$ for $S_{870\mu\textrm{m}} \sim 13 - 16$ mJy (i.e. equivalent to our $4\sigma$ detection threshold of $S_{\rm{1.1mm}} \gtrsim 8$ mJy). This excess suggests that some of these red-{\it Herschel} sources are associated with galaxy overdensities. This deserves further analysis which is beyond the scope of this work; therefore, these serendipitous sources are not discussed in the rest of the paper since they are not directly associated with the originally targeted red-{\it Herschel} sources.

\subsubsection{Deblending Herschel observations}

We estimate deblended flux densities in the {\it Herschel} bands for all the AzTEC detections in a similar way to that presented in \citet[][see details of the method therein]{michalowski17}. Briefly, we extract a square 120\,arcsec wide around the position of a given AzTEC source and simultaneously fit 2-dimensional Gaussian functions at the positions of all AzTEC sources within this square patch \citep[the fitting is performed using the IDL {\sc mpfit} package,][]{mpfit}. The normalization of each Gaussian function is kept as a free parameter, whereas its FWHM is fixed at the size of the respective {\it Herschel} beam. 
The errors on the deblended flux densities are calculated from the covariance matrix in order to take into account the possible degeneracies in the fitting. This is especially important for close sources that lie within the beam at a given band, whose fluxes are highly degenerate. 
The confusion limit of the SPIRE data, reported as 5.8, 6.3, and 6.8\,mJy\,beam$^{-1}$ at 250, 350 and $500\,\mu\rm m$ \citep{nguyen10}, are also added in quadrature. The AzTEC $1.1\,$mm flux densities and the deblended {\it Herschel} flux densities derived in this work are reported in Table \ref{table:catalogue}.

\subsection{On the nature of non-detections}\label{secc:non_detections}

After performing the counterpart matching, 45 of the 93 analyzed red-{\it Herschel} targets do not have an associated AzTEC detection at $\geq 4\sigma$ significance within $36.6$\,arcsec. Since the incompleteness of our survey can not explain the bulk of these non-detections (see \S \ref{secc:source_detection}), here we explore four possible scenarios to explain them: (1) AzTEC observations do not reach the desired sensitivity; (2) these targets correspond to the faintest {\it Herschel} sources and thus deeper observations are needed;  (3) these sources are made up of multiple intrinsically fainter components blended within the {\it Herschel} beam; and (4) the sources show different SED properties.

As shown in Figure \ref{fig:sample2}, in general, the AzTEC observations on these non-detected targets have a similar r.m.s noise as those in which sources were detected (see yellow histogram in the figure). This confirms the homogeneity of our observations, ruling out the first scenario discussed above. Similarly, these AzTEC non-detections have similar {\it Herschel} flux densities to the detected galaxies (Figure \ref{fig:sample}), spanning a flux density range of  $S_{\rm 500\mu m}\approx40-80\,$mJy.
 Therefore, if those were single sources with SEDs similar to those of the detected galaxies, we would expect most of them to be detected above the adopted threshold, although a small fraction of them could be associated to the faintest sources. Actually, looking at the AzTEC maps individually, we find 8 (15) sources at SNR $\geq 3.5$ (3.0) close to the \textit{Herschel} position (at $r \lesssim 10$\,arcsec), which are consistent with being single systems but falling below our detection threshold (SNR$ = 4$).
 Note that the reliability of these $3.5\sigma$  and $3\sigma$ detections is $\gtrsim 93\,$ and $\gtrsim 80\,$ per cent, respectively (see FDR for the central $r=10\,$arcsec radius  region in Fig. \ref{fig:FDRprobaility}), which suggest that these single faint sources are real.
 
Considering that the percentage of spurious detections in the \textit{H}-ATLAS catalogues is reported to be $\simeq 0.2 $\, per cent\footnote{Given the complex selection process of our sample (see \S\ref{secc:sample_selection}) the false detection rate may be $> 0.2 $\, per cent. Nevertheless, we do not expect it to be large enough to justify all the AzTEC non detections, since all of our \textit{Herschel} sources were identified in the 250 and 350$\mu \textrm{m}$ maps, and detected above $5\sigma$ at $500\mu \textrm{m}$.} \citep{Valiante2016}, the remaining non-detections are therefore likely multiple systems with individual members' flux densities below our sensitivity limit or sources with different SED properties.  In fact, \citet{Valiante2016} explicitly suggest that ``a more important problem than spurious sources is likely to be  sources that are actually multiple sources''. After visual inspection, we  identify at least nine systems with multiple components at SNR$>3.5$, suggesting that multiplicity is indeed a main reason for the non-detection of these galaxies. 

However, although there are no significant differences between the {\it Herschel} colours of the detected and the non-detected systems (see Figure \ref{fig:sample_colours}), suggesting similar SED shapes, we cannot rule out the possibility that some of the non-detected sources have a higher dust emissivity spectral index, $\beta$. This would also decrease the expected flux density in the Rayleigh–Jeans regime probed by the AzTEC 1.1\,mm observations, potentially explaining the lack of detections in some of these targets. In fact the median 1.1\,mm flux density of the AzTEC detections seems a bit lower than what it is predicted by using typical SED templates (e.g. \citealt{Michalowski2010a,da-Cunha2015a}; see also \S \ref{secc:source_detection}). This might be in line with recent results reporting steeper $\beta$ values in $z>3$ galaxies (e.g. \citealt{Kato2018a,Jin2019}; Casey et al. in prep.), suggesting an evolution of the dust emissivity index with redshift and/or luminosity.

\subsection{Multiplicity fraction}\label{secc:multiplicity}

Our observations are sensitive to galaxies separated by $\Delta\theta>9.5\,$arcsec. Sources with such separation are hard to detect in the small field of view of interferometric observations as those achieved with ALMA\footnote{The ALMA primary beam at $850\,\mu\rm m$ has a half power beam width (HPBW) of $18$\,arcsec. Hence, at a radius larger than $\sim9$\,arcsec, the primary beam response drops below 0.5.} and NOEMA. Indeed, \citet{Ma2019} noted that, for some of their red-{\it Herschel} sources, the total flux densities measured by ALMA are systematically lower than those measured with single-dish telescopes, suggesting the presence of multiple components beyond their mapped area. Our observations thus probe multiplicity at a different scale from what has been studied so far with interferometers.\footnote{We note that the field-of-view of the SMA can probe angular scales < 27 arcsec (ignoring the drop in efficiency towards the edge of the beam). The relatively small samples observed so far, however, have limited the detection of systems separated by these larger angular scales.} 

\begin{figure}\hspace{-0.9cm}
\includegraphics[width=0.53\textwidth]{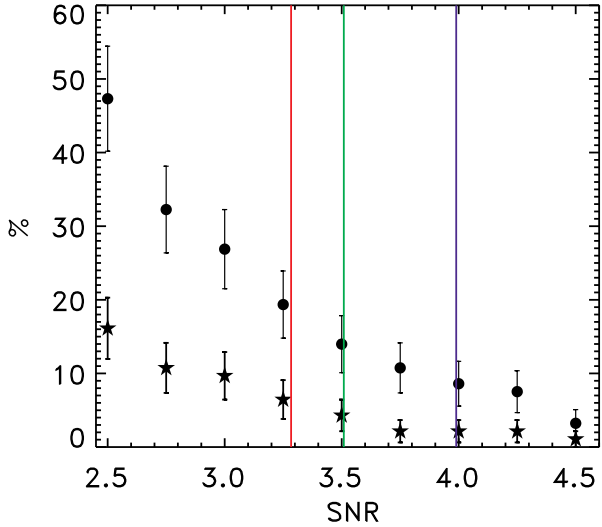}
\caption{Multiplicity fraction as a function of signal to noise detection threshold, and for 2 search radii: $36.6$\,arcsec (circles) and $18.3$\,arcsec (stars). Error bars correspond to $\sqrt{N}$. Vertical lines indicate false detection contamination of 5\,per cent (dark blue at SNR $\sim4$), 20\,per cent (green at SNR $\sim3.5$), and 32\,per cent (red at SNR $\sim3.3$), estimated within the central $r = 36.6$\,arcsec area of the AzTEC maps.}
\label{fig:multiplicytyfraction}
\end{figure}

Using our search radius of $36.6$\,arcsec (the size of the {\it Herschel} beam at $500\mu\textrm{m}$), we find that eight targets from the original sample show source multiplicity, comprising a total of 16 AzTEC detections. This implies that at least $\sim9$\, per cent (8/93) of the red-{\it Herschel} targets with AzTEC detections are composed of multiple systems. Additionally, nine of the non-detections are likely multiple systems made of intrinsically fainter galaxies with individual flux densities falling just below our detection threshold (see \S \ref{secc:non_detections}). Furthermore, four of the targets originally classified as single detections correspond to multiple systems if the detection threshold is reduced to include $\geq 3.5\sigma$ sources. These additional sources lie within the central ($r \lesssim 15$ arcsec) deeper region of the maps, where the reliability of $3.5\sigma$ detections is $> 90$ per cent. Including these, the multiplicity fraction is increased to $\sim23$\, per cent (21/93).
An extreme scenario would be if all of the AzTEC non-detections are also assumed to be multiple systems. In that case the  multiplicity fraction of the red {\it Herschel} sources would be as high as $50$\, per cent. Figure \ref{fig:multiplicytyfraction} shows, for two different search radii, how the multiplicity fraction increases as the detection threshold in our analysis is reduced.

The multiple fraction may be even larger if multiplicity at smaller scales than the AzTEC beam is also present within the sources classified here as single systems. 

To probe the multiplicity at smaller scales, we perform two different tests in which the measured PSF profile of single sources is comparable to that of an expected point source. Deviation on the width and shape of the PSF would be expected if two or more sources are blended within the AzTEC beam. 
First, we derive a radial profile for each detection by azimuthally averaging its flux density and compare its FWHM against that of the point-source PSF. Second, for each AzTEC source, we subtract the point-source PSF scaled to the corresponding measured flux value and quantify the residuals within a $1.5\times \textrm{FWHM}$ area. Then, those sources with broad  FWHMs ($\gtrsim 10$\, per cent than the ideal PSF -- i.e. $\sim 2\sigma_\textrm{PSF}$ the standard deviation of the PSFs in our sample) and/or  with residuals larger than the noise level are tagged as potential close multiples. Based on this analysis, we expect multiplicity in  $21-33$\, per cent of the AzTEC detections classified as singles. 

These estimations can be compared to the results from interferometric observations on samples of red-{\it Herschel} sources. For example, \citet{Ma2019} reported that $\sim27$\ of a compilation of 63 red-\textit{Herschel} galaxies observed with ALMA, NOEMA and the SMA 
are close multiple systems. \citet{Greenslade2020} have also recently reported $870\mu\textrm{m}$ and 1.1 mm SMA observation of 34 red $500\mu\rm{m}$-risers from the \textit{Herschel} Multi-tiered Extragalactic Survey \citep[HerMES,][]{HerMES}, and find a $\sim12$\, per cent multiplicity fraction. However, they argue that their 12 non-detections are most likely multiple systems with more than two members, in which case their multiplicity fraction increases to 47\,per cent. 

Our estimates of the fraction of close multiple systems are in broad agreement with the literature, implying that the multiplicity fraction of the whole sample might be larger than the values reported above when accounting for the multiplicity at smaller scales than the AzTEC beam. Nevertheless, combining these higher-resolution interferometic results with our multiplicity estimations is not straightforward. A careful visual inspection of the 63 red-\textit{Herschel} sources presented in \citet{Ma2019} indicates that 14 of them ($\sim$ 22 per cent) are multiples at scales below those probed by our AzTEC observations, and would therefore be classified as individual systems in our analysis. This implies that our multiplicity estimates should be increased by an additional $\sim$ 10 per cent due to multiple systems that are not resolved within the AzTEC beam. The resulting total multiplicity fraction of red-\textit{Herschel} sources would therefore be $\gtrsim 18$\, per cent in the conservative scenario and $\sim 60$\, per cent in the extreme one, in which most of the non-detections are also considered to be multiples.

Similar results are found if we instead adopt the results from our PSF modelling analysis, but the reader should keep in mind that different factors other than multiplicity (e.g. focus and astigmatism of the telescope, noise gradient in the maps, or even strong gravitational lensing effects) could distort and broaden the shape of the AzTEC beam. Therefore, we stress that follow-up higher angular resolution observations are necessary to derive a robust estimation of the total multiplicity fraction in our sample.

\subsection{Redshifts and luminosities}\label{secc:redshifts}

In order to estimate photometric redshifts, luminosities and SFRs, we follow the procedure described in \citet{Ivison2016}, in which a library of template SEDs is adopted in order to better characterize the diversity of the intrinsic SEDs and the uncertainties in the derived quantities. We use four SEDs which are representative of dusty star-forming galaxies, and particularly, of red-{\it Herschel} sources (see \citealt{Ivison2016}). This set includes Arp220 (\citealt{Silva1998a}), the Cosmic Eyelash (\citealt{Swinbank2010b,Ivison2010a}), and the two synthesized templates of \citet{Pope2008} and \citet{daCunha2015}. 

\begin{figure}\hspace{-1cm}
\includegraphics[width=0.53\textwidth]{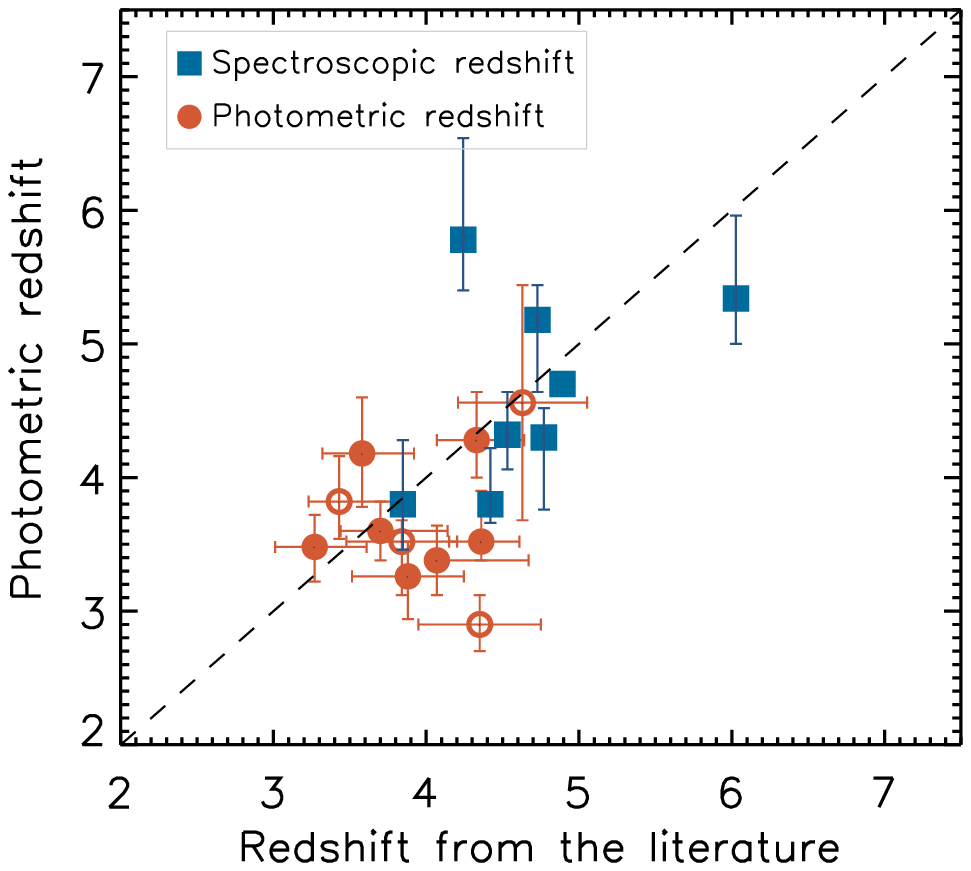}
\caption{Comparison between the photometric redshifts derived in this work and those reported in the literature (\citealt{Ivison2016,Fudamoto2017,Duivenvoorden2018,Zavala2018a,Ma2019}), including both photometric (orange circles) and spectroscopic redshifts (blue squares; see Table \ref{table:RSR} and \S\ref{secc:highz_candidates}). The open circles represent sources which are classified as multiple systems based on our AzTEC observations, while all the remaining sources correspond to single systems. In general, our estimated redshifts are in good agreement with those reported in the literature with a mean redshift deviation of $\Delta z/(1+z_{\rm ref})=0.09$ (considering only those sources with spectroscopic redshifts).
\label{fig:redshift_comparison}}
\end{figure}

Our SED fitting approach is based on a maximum-likelihood method which formally takes into account upper limits  in case of non-detections (e.g. \citealt{Aretxaga2007,Sawicki2012}). This is important since the {\it Herschel} flux densities of some of the sources lie below $2.5\sigma$ after using the AzTEC positions as priors to deblend the {\it Herschel} emission (see \S\ref{secc:source_detection}). We test our procedure combining the AzTEC photometry (including an additional 5 per cent calibration uncertainty) with all the {\it Herschel} data (PACS 100 and $160\,\um$ and, SPIRE 250, 350, and $500\,\um$), with only PACS $160\,\um$ plus all the SPIRE bands, and with only SPIRE photometry. Given the typical low SNR of PACS $100\,\um$ (plus the possible contribution from emission mechanisms not included in the adopted SED templates -- e.g. AGN, PAHs, etc.), the best fits are achieved when using only PACS $160\,\um$ in combination with the SPIRE and AzTEC photometry. We therefore discard the $100\,\um$ band during the SED fitting procedure. 

For each source in our catalog, a redshift probability distribution is calculated by combining the redshift distributions associated with the four different SED templates described above. Then, the best-fit photometric redshift is assumed to be the one with the maximum likelihood, and the 68 per cent confidence interval is estimated by integrating 
the combined redshift distribution.  
As shown in Figure \ref{fig:redshift_comparison}, the photometric redshifts derived by this method, which are reported in Table \ref{table:catalogue}, are in good agreement with those reported in the literature (\citealt{Ivison2016,Ma2019,Fudamoto2017,Zavala2018a,Duivenvoorden2018}); with eight being spectroscopic redshifts). The relative difference between our redshifts and those derived elsewhere is estimated to be $\Delta z/(1+z_{\rm ref}) = 0.10$, and 0.09 if only the spectroscopic redshifts are considered. These values are similar to the expected uncertainties for photometric redshifts \citep{Hughes2002}.  Additionally, we estimate the photometric redshifts with the MMPZ code (\citealt{Casey2020a}) and find consistent results (with a mean difference of $\Delta z/(1+z_{\rm ref}) = 0.004$ for the single sources), although with significantly  larger uncertainties.

\begin{figure}\hspace{-1cm}
\includegraphics[width=0.53\textwidth]{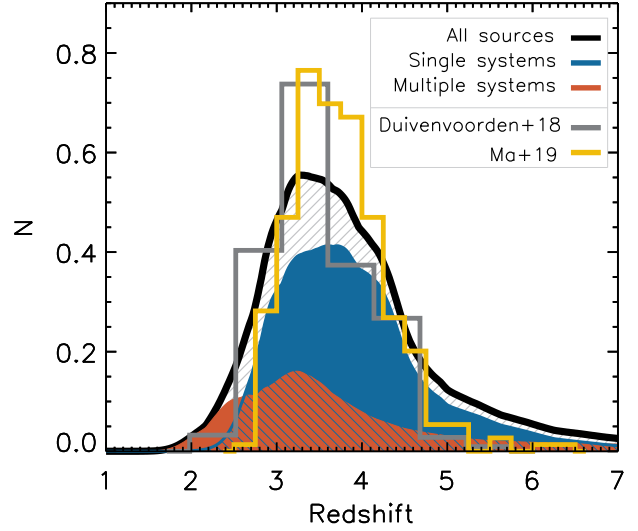}
\caption{Normalized redshift distribution of the AzTEC-detected red-{\it Herschel} sources derived from the stacking of the redshift probability distribution functions. The components of multiple systems are represented by the orange distribution while the single systems are illustrated by the blue distribution. Although both redshift distributions have a similar median value, that for multiple systems is more skewed towards lower redshifts. The redshift distribution of all the sources is plotted as the black solid curve. For comparison, previous estimations of similar red samples are also included (\citealt{Duivenvoorden2018,Ma2019}).
\label{fig:z_dist}}
\end{figure}

Figure \ref{fig:z_dist} shows the stacked probability redshift  distribution of all the AzTEC-detected red-{\it Herschel} sources, which has a median redshift of $z_{\rm med}\approx3.64$. We also plot the stacked redshift distribution of the single and multiple systems separately. Although both distributions have similar median redshifts (\textit{$z_{\rm med}\approx3.8$ vs $3.5$}), the multiple systems have a larger fraction of low-redshift sources (27\,per cent at $z_{\rm phot}<3$) compared to the single systems (10\,per cent). We highlight that, although only $\sim33$\, per cent of the total redshift distribution lie at $z_{\rm phot}>4$, the adopted colour selection criteria is efficient at selecting $z_{\rm phot}>3$ galaxies, where $\sim85$\, per cent of the sample lie. In Figure \ref{fig:z_dist} we also compare the redshift distribution of the AzTEC-{\it Herschel} sources with those from similar samples derived in \citet{Duivenvoorden2018} and \citet{Ma2019}, which have median photometric redshifts of 3.6 and 3.3, respectively. Similarly, \citet{Ivison2016} reported a median redshift of $3.66$, with $\sim32$\, per cent of the sources lying at $z>4$. Our results are therefore in general agreement with those previously reported.

The IR luminosity is then derived using the best-fit template and integrating from $8-1000\,\um$ (in the rest frame), from which the SFR is estimated assuming the \citet{Kennicutt2012} calibration for a \citet{Chabrier2003a} IMF,  SFR [\sfr] $=1.48\times10^{-10}\,L_{IR}\,[L_\odot]$. The uncertainties on the infrared luminosities (and hence SFRs) are propagated from the flux density and redshift errors using Monte Carlo simulations. 
The estimated IR luminosities and SFRs can be found  in Table \ref{table:catalogue}.

\begin{figure}\hspace{-1cm}
\includegraphics[width=0.53\textwidth]{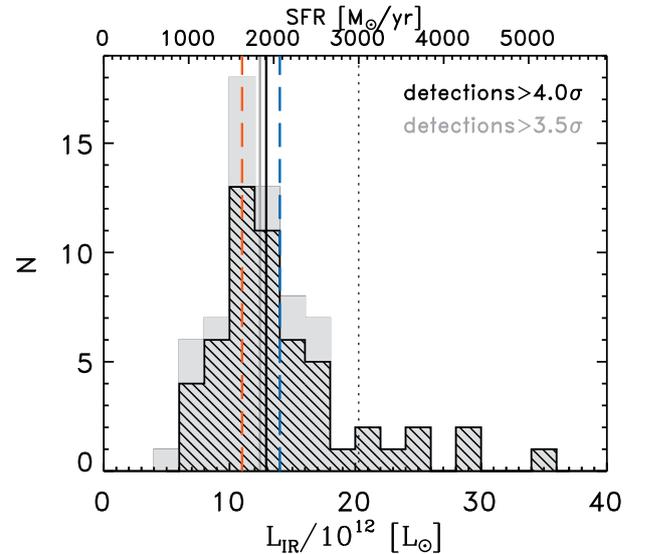}
\caption{Histogram of IR luminosities and SFRs estimated for the AzTEC detections above $4\sigma$ (black line filled) and $3.5\sigma$ (light grey). The median values of both distributions are indicated with vertical lines, including the median luminosity for the $4\sigma$ single systems ($1.4\times10^{13}\,L_\odot$; dashed blue) and the components of the multiple systems ($1.1\times10^{13}\,L_\odot$; dashed orange). Eight sources in our sample show $L_\textrm{IR} > 2\times10^{13}L_\odot$ and SFRs $> 3000$ \sfr\,(dotted vertical line), representing a population of extreme star-forming galaxies that cannot be explained within current physically plausible models, unless gravitational lensing effects are included.
\label{fig:LIR_dist}}
\end{figure}

Figure \ref{fig:LIR_dist} shows the IR luminosity and SFR histogram of our AzTEC sources, which have apparent IR luminosities in the range of $12.8 \leq \textrm{log}_{10}(L_\textrm{IR}/L_\odot) \leq 13.5$, with a median luminosity of $1.4\times10^{13}\,L_\odot$ for the single systems  and $1.1\times10^{13}\,L_\odot$ for the components of the multiple systems. 
Their SFRs span $\sim900$ \sfr to 5000 \sfr, representing some of the most extreme star-forming galaxies known (in the absence of gravitational lensing). These values are in good agreement with those reported in the literature for similar samples. For example, \citet{Ivison2016} reported apparent luminosities in the range of  $5\times10^{12}-6\times10^{13}\,L_\odot$ with a median of $1.3\times10^{13}\,L_\odot$, \citet{Ma2019} derived a  median luminosity of $9.0\times10^{12}\,L_\odot$, and \citet{Greenslade2020} report luminosities in the range of  $2\times10^{13}-6\times10^{13}\,L_\odot$. These values are also comparable to those estimated for the sample of DSFGs selected with the SPT, which have a median intrinsic luminosity of $L_{\rm IR}\approx1.5\times10^{13}\,L_\odot$ (\citealt{Reuter2020a}).

\section{Identification of interesting sub-samples}\label{secc:subsamples}
\subsection{High-redshift galaxy candidates}\label{secc:highz_candidates}

To isolate the most promising high-redshift galaxies, we select all those {\it Herschel}-AzTEC systems detected above the $4\sigma$ threshold and with $z_{\rm phot}>4$. The 18 sources which satisfy this criterion  are indentified in Table \ref{table:catalogue}, three of which are members of multiple systems. Their photometric redshifts span $z=4.0$ to $\sim5.8$ and have SFRs in the range of $\approx 1000-5000$ \sfr, representing some of the most luminous DSFGs known so far (in the absence of gravitational amplification).

Six of these candidates have already been spectroscopically confirmed at $z > 4$: SGP-272197 at $z=4.24$ (source SGP-261206 in \citealt{Fudamoto2017}) and five of the sources in Table \ref{table:RSR}. Interestingly, G09-838083 ($z = 6.03$) was found to be gravitationally lensed by a foreground elliptical galaxy, with a magnification factor of $\approx 9.3$ (\citealt{Zavala2018a}). This implies that, although the sources were selected to be preferentially non-lensed (see \S\ref{secc:observations}), there might be other amplified galaxies in the sample, and therefore, the SFRs quoted above would represent upper limits.

Regardless of their potential gravitational lensing amplification, these sources are ideal targets for future spectroscopic surveys aimed at identifying and characterizing dusty starburst galaxies in the early Universe. 

\citet{Ivison2016} developed robust simulations to estimate the different completeness factors affecting the selection of the ultrared \textit{Herschel} sample. Given the similar selection criteria between samples, we update their completeness estimates considering the 500$\mu$m flux density limit of our sample ($S_{500\mu\textrm{m}} \gtrsim 35$ mJy) and the number of sources in our analysis (93). Using this completeness correction, and considering the number of $z_{\rm phot}>4$ from our analysis, we estimate a lower limit for the space density of $4 < z < 6$ red DSFGs of $\approx 3 \times 10^{-7} \textrm{Mpc}^{-3}$. This is a factor of two lower than that found by \citet{Ivison2016}. However, \citeauthor{Ivison2016} assumed a SNR detection threshold $\geq 2.5$ for their SCUBA-2/LABOCA observations (FWHM $\sim 18.4$\,arcsec), and did not consider potential multiplicity effects. 

Combining our estimated space density and the median SFR of our $z_{\rm phot}>4$ sample ($\sim 2500\, \textrm{M}_\odot \textrm{yr}^{-1}$ not corrected for potential gravitational lensing effects), we conclude that luminous red \textit{Herschel} sources contribute $\gtrsim 8 \times 10^{-4}\, \textrm{M}_\odot \textrm{yr}^{-1} \textrm{Mpc}^{-3}$ to the obscured star formation at $4 < z < 6$.
This value is in very good agreement with the recent estimations of the dust-obscured star formation rate density presented by \citet{Zavala2021a} based on ALMA number counts at 1.2\,mm, 2\,mm, and 3\,mm. Their model predicts a dust-obscured star formation rate density of $\approx10^{+5}_{-6}\times10^{-4}\, \textrm{M}_\odot \textrm{yr}^{-1} \textrm{Mpc}^{-3}$ at $z=5$ from galaxies with IR luminosities in the range of our sources ($12.8 \leq \textrm{log}_{10}(L_\textrm{IR}/L_\odot) \leq 13.5$). 

\subsection{Physically Interacting galaxies}\label{secc:mergers}

With redshifts in hand, we can speculate the nature of the multiple systems in our sample. Are they chance projections at different redshifts (e.g. \citealt{Zavala2015a}) or physically interacting galaxies (e.g. \citealt{Oteo2016})? Examples of both systems have been reported in the literature and, indeed, it is likely that these multiple systems are composed by both physically associated galaxies and chance projections (e.g. \citealt{Wardlow2018,Hayward2018,Stach2018}). 

\begin{figure}\hspace{-0.6cm}
\begin{center} 
\includegraphics[width=0.45\textwidth]{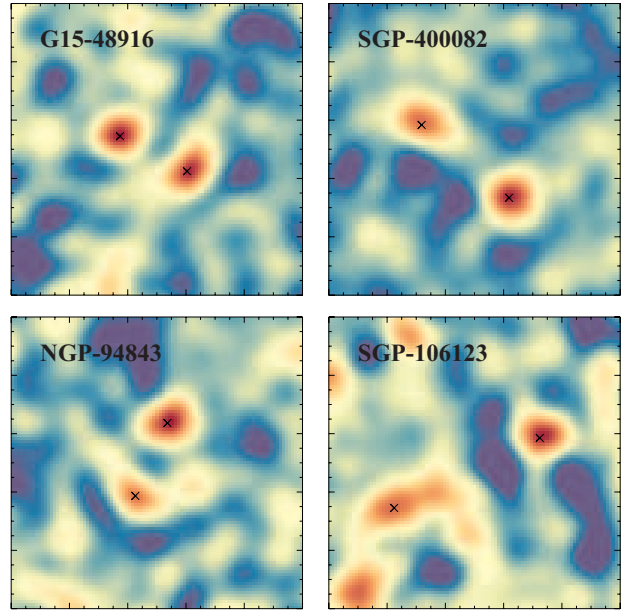}
\caption{ $\rm 76\,arcsec\times76\,arcsec$ AzTEC 1.1\,mm SNR maps of the four most likely physically interacting galaxy candidates in the sample. All the detected sources fall within  $36.6$\,arcsec of the original  {\it Herschel} position (the beamsize  at $500\,\um$) and have individual photometric redshifts in agreement with each other within the error bars. These sources, which are typically missed in intereferometric observations due to their small fields of view, are likely galaxy over-densities and/or pre-coalescence galaxy mergers.
\label{fig:multiple_example}}
\end{center}
\end{figure}

As discussed in \S\ref{secc:multiplicity}, we are only sensitive to sources separated by $\Delta\theta \gtrsim9.5$\,arcsec (which corresponds to $\Delta\gtrsim70\,$kpc at $z=4$). 
This prevents us from detecting late-stage mergers as those already identified by ALMA and the SMA (e.g. \citealt{Oteo2016}). Nevertheless, our observations enables the detection of pre-coalescence galaxy pairs and proto-cluster structures, whose identification by interferometers with small fields of view like ALMA or NOEMA is rather challenging.
Such complexes represent ideal laboratories to study the environmental effects on the star formation activity and
to understand the star formation process during the earliest stages of galaxy mergers. 

To identify the most promising physically interacting systems, we select those multiple galaxies with photometric redshift consistent with each other within $\Delta z<0.5$, since the typical photometric redshift uncertainty from SED-fitting methods is $\approx0.2-0.3$ (Table \ref{table:catalogue}; see also \citealt{Aretxaga2005}). Although this threshold might appear too relaxed, we highlight that the probability of finding a pair of two bright sources ($S_{\rm 1.1mm}\gtrsim5\,$mJy)  by chance line-of-sight alignment is very low since their surface density is estimated to be around $0.01\,\rm arcmin^{-2}$ \citep{Scott12}.

Out of the eight original targets that show multiplicity, only two fulfill this criterion. Two additional systems are identified if the detection threshold is reduced to $3.5\sigma$.
These sources are shown in Figure \ref{fig:multiple_example} and are also identified in Table \ref{table:catalogue}. As can be seen in the figure, some of these physically interacting candidates are well resolved into two separated sources. For three of these four systems, their redshifts agree within $\Delta z\lesssim0.06$. 

Although further observations are needed to  confirm their nature, they might represent observational evidence of the existence of early-stage (pre-coalescence) mergers  within the submillimeter galaxy population since their
angular separations ($\approx20-30$\,arcsec or $\approx150-200\,$kpc) and their flux ratios (1:2) are in very good agreement with the predictions from simulations (e.g. \citealt{Narayanan2010a,Hayward2012a}).

\section{Discussion and Conclusions}\label{secc:discussion}

As part of the Early Science Phase of the Large Millimeter Telescope, we obtained AzTEC 1.1 mm observations on a sample of 
 100 red-{\it Herschel} sources. Their red far-infrared colours ($S_{250\mu\rm m}<S_{350\mu\rm m}<S_{500\mu\rm m}$) and bright flux densities ($S_{500\mu\rm m}\approx35-80\,\rm mJy$) suggest that they are  high star formation rate galaxies ($\rm SFRs\gtrsim500\,$ \sfr) at high redshifts ($z\gtrsim3$). 
 
\begin{figure}\hspace{-0.5cm}
\includegraphics[width=0.5\textwidth]{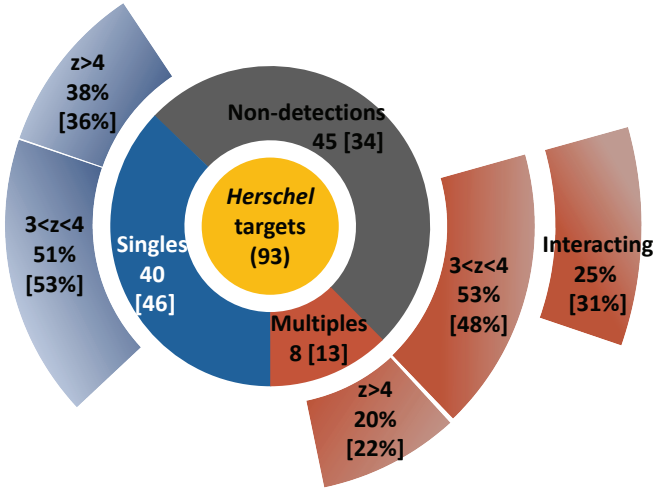}
\caption{Summary of the results derived from the LMT/AzTEC observations. From the 93 observed red-{\it Herschel} targets, and assuming a $4\sigma$ ($3.5\sigma$) AzTEC detection threshold, 40 (46) were associated with single detections, 8 (13) with multiple sources, and 45 (34) with non-detections, within a search radius of $36.6\,$arcsec. These non-detected galaxies are also likely multiple systems, although some of them might be explained by having steep Rayleigh-Jeans slopes (see \S\ref{secc:non_detections}). From the detected galaxies, around $85$\, per cent of all the AzTEC-{\it Herschel} sources lie at $z_{\rm phot}>3$, while only $33$\, per cent at $z_{\rm phot}>4$. From the 8 (13) multiple systems, $\sim25$\, per cent ($\sim31$\, per cent) are consistent with being physically associated galaxies ($\Delta z_{\rm phot}\lesssim0.5$), all of them lying at $z_{\rm phot}\gtrsim3$. The most promising $z>4$ single sources and the physically interacting galaxy candidates are identified in Table \ref{table:catalogue}. 
\label{fig:summary}}
\end{figure}

Combining the AzTEC data with our new deblended {\it Herschel} photometry, we constrained the multiplicity fraction in the sample and derived photometric redshifts, IR luminosities, and SFRs for all the sources in the catalog. Our main results are discussed below and are also summarized in Figure \ref{fig:summary}.  

Thanks to the $\theta_{\rm FWHM}\approx9.5$\,arcsec angular resolution provided by the 32\,m illuminated surface of the LMT (a factor of 4 better than {\it Herschel} at $500\,\mu\rm m$), we found that 8 of the red-\textit{Herschel} targets break into multiple components (with SNR $\geq 4$), which implies a multiplicity fraction of $\sim9$\, per cent. This value increases to $\sim23$\, per cent if we include those sources with  evidence  of  multiplicity  but slightly below our detection threshold (i.e. formally classified as non-detections). The multiplicity fraction can be even higher  (up to $\sim50$\, per cent) if some of the non-detected sources were also made of multiple systems\footnote{Note, however, that we cannot rule out the possibility that some of the non-detected sources might have a higher dust emissivity spectral index, $\beta$, which would decrease the expected flux density in the Rayleigh-Jeans regime probed by the AzTEC 1.1\,mm observations.} (see \S\ref{secc:multiplicity}). These multiple sources probe a different scale from what has been studied so far with smaller field of view interferometric observations. Hence, the multiplicity fraction quoted above might be larger if multiplicity at smaller scales \citep[e.g.][]{Ma2019,Greenslade2020} is also present within the sources classified here as single systems.

\begin{figure}\hspace{-1cm}
\includegraphics[width=0.53\textwidth]{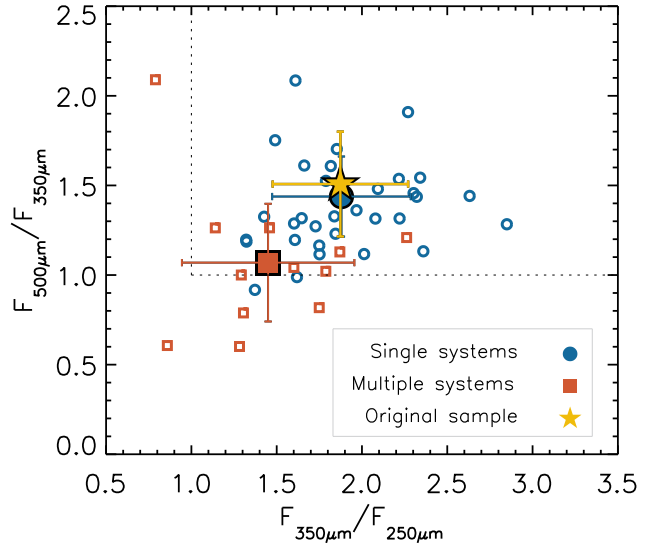}
\caption{FIR colour-colour plot showing the singles (blue empty circles) and the components of multiple systems (orange empty squares) after deblending the {\it Herschel} fluxes using the detected AzTEC positions as priors. The solid blue circle and orange square indicate the average value of the corresponding samples, which have slightly different colours, with the multiple systems showing the least ``red'' colours. Those sources with upper limits in any of the three bands are not plotted. For comparison, the average colour of the whole sample before deblending the fluxes is marked with a yellow star.
\label{fig:color_before_after}}
\end{figure}

Such high multiplicity should be taken into account when comparing the properties of these galaxies to results from theoretical models and simulations, particularly since source blending artificially increases the redness of the colour in the {\it Herschel} bands due to the larger beamsizes of the redder filters. This can be seen in Figure \ref{fig:color_before_after}, where the FIR colours of the single and multiple systems are plotted  after deblending the {\it Herschel} flux densities (blue circles and orange squares, respectively), along with the average colour of the whole sample before deblending the fluxes (yellow star).  In general, the multiple systems have individual colours which are less red than those of the single systems and the average original colour used for the selection of these galaxies. Indeed, as seen in Table \ref{table:catalogue}, 14 of the 56 AzTEC detections associated with {\it Herschel} sources ($\sim25$\, per cent) do not fulfill the condition $F_{250\rm\mu m}<F_{350\rm\mu m}<F_{500\rm\mu m}$ after the deblending of the {\it Herschel} fluxes\footnote{Note that in order to differentiate between the original and the deblended {\it Herschel} flux densities, we use the symbols $S_\lambda$ and $F_\lambda$, respectively.}, and 23 ($\sim41$\, per cent) would be excluded after our colour cut ($S_{500\rm\mu m}/S_{250\rm\mu m}>2$ and $S_{500\rm\mu m}/S_{350\rm\mu m}>1$; see \S\ref{secc:observations}). This is in line with \citet{Ma2019}, who suggest that $\sim20$\, per cent of their sources would not pass the selection criteria of $500\,\mu\rm m$-risers without blending, although lower than the $\sim60$\, per cent derived by \citet{Duivenvoorden2018} from mock observations using the \citet{Bethermin2017a} models (note that their sources are brighter with a flux density cut of $S_{500\mu\rm{m}}>63\,$mJy). 

\begin{figure}\hspace{-0.9cm}
\includegraphics[width=0.53\textwidth]{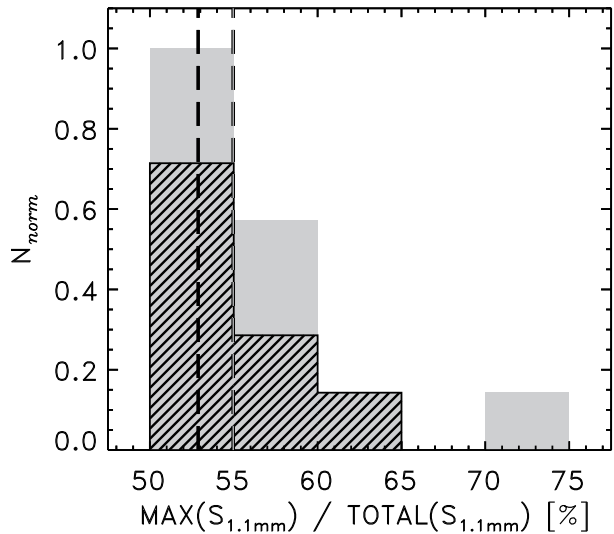}
\caption{Normalized distribution of the AzTEC 1.1mm flux density ratio between the brightest source of the multiple systems with respect to the total flux density of the system. Solid-gray and line-filled histograms correspond to a $3.5\sigma_{\textrm{1.1mm}}$ and $4.0\sigma_{\textrm{1.1mm}}$ detection thresholds, respectively. Vertical dashed lines indicate the median value for each distribution. Histograms have been normalized to the peak value of the $3.5\sigma_{\textrm{1.1mm}}$ distribution.
\label{fig:fluxratios}}
\end{figure}

Figure \ref{fig:fluxratios} shows the distribution of flux density ratios between the brightest component of the multiple system with respect to the total flux density of the system. Our analysis indicates that the brightest component contributes 50 - 75\, per cent (with a median $\approx 55$\, per cent) at 1.1mm. This is in agreement with results from previous works, using both interferometric and single-dish observations \citep[e.g.][]{Donevski2018,Ma2019,Greenslade2020}.

\begin{figure}\hspace{-1cm}
\includegraphics[width=0.53\textwidth]{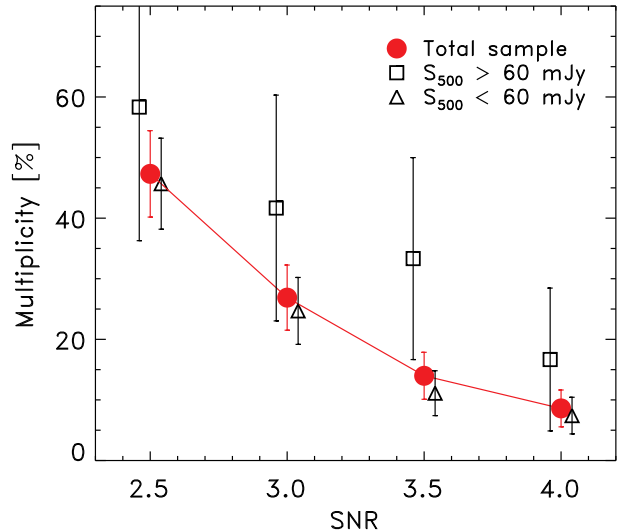}
\caption{Multiplicity fraction as a function of 1.1mm SNR detection threshold for: the whole sample (red circles), and sources fainter (triangles) and brighter (squares) than $S_{500 \mu\textrm{m}} = 60$ mJy (before deblending). The multiplicity fraction of the bright sample is $\sim 2$ times larger than that from the fainter sample, and should be considered by models and simulations that try to explain the bright-end of the 500$\mu$m \textit{Herschel} population of high-$z$ galaxies.
\label{fig:multi_fluxcut}}
\end{figure}

To shed light on the multiplicity as a function of the original $500 \mu\textrm{m}$ flux density (i.e. before deblending) and to compare to model predictions and other studies, we have divided our sample in two flux density bins: fainter and brighter than $S_{500 \mu\textrm{m}} = 60$ mJy. Figure \ref{fig:multi_fluxcut} shows the multiplicity fraction as a function of our AzTEC signal-to-noise detection threshold for the whole sample (as in Fig. \ref{fig:multiplicytyfraction}), compared to the faint and bright sub-samples. Of the twelve \textit{H}-ATLAS sources in our sample with $60 \leq S_{500 \mu\textrm{m}} \leq 80$ mJy, seven are identified as single systems, two break into multiple components, and three have no detections (assuming a SNR threshold $\geq 4$). This corresponds to a multiplicity fraction of $\sim 17$\, per cent , which is a factor of two larger than the multiplicity of the fainter sample. This is in agreement with the SMA results from \citet{Greenslade2020} who found, in a sample of 17 SPIRE $S_{500 \mu\textrm{m}} > 60$ mJy sources, twelve single systems, three multiples, and two non detections (i.e. a multiplicity fraction $\sim 18$\, per cent). We note that, although the sample from \citet{Greenslade2020} includes sources with a wider range of 500$\mu$m flux densities (up to $S_{500 \mu\textrm{m}} = 160$ mJy), they only find multiple systems or non-detections (that could potentially be associated with multiple systems) in sources with $S_{500 \mu\textrm{m}} \leq 82$ mJy. These results seem to disagree with models suggesting that \textit{Herschel} sources with $S_{500 \mu\textrm{m}} > 60$ mJy are most likely single galaxies, potentially magnified by gravitational lensing effects \citep[e.g.][]{Bethermin2017a}. 

The redshift distribution of all the {\it Herschel}-AzTEC sources (singles and multiples) shows that $\sim33$\, per cent of the objects lie at $z_{\rm phot}>4$ and $\sim85$\, per cent at $z_{\rm phot}>3$, with a median redshift of $z_{\rm med}\approx3.64$ (see Figure \ref{fig:z_dist}). 
These sources show high SFRs in the range of $\approx900-5000$ \sfr (in the absence of gravitational lensing). All of this confirms the high efficiency of the colour selection criterion to select luminous high-redshift ($z>3$) galaxies from the {\it Herschel} catalogs. 

In \S\ref{secc:subsamples}, we identified the most promising high redshift galaxies candidates, comprising 15 single sources and three members of multiple systems with $z_{\rm phot}>4$. Six of these sources have already been spectroscopically confirmed at $z=4.24-6.03$, including two new spectroscopic redshifts derived in this work using the complete 50\,m diameter aperture of the LMT (see \S \ref{secc:RSR}). The rest of the objects comprise ideal targets for future spectroscopic surveys aimed at identifying  the most distant dusty star-forming galaxies in the  Universe.

Given our $z_\textrm{phot} > 4$ sample, we estimate a lower limit for the space density of $4 < z < 6$ red DSFGs of $\approx 3 \times 10^{-7} \textrm{Mpc}^{-3}$ which, combined with their median SFR ($\approx 2500\, \textrm{M}_\odot \textrm{yr}^{-1}$ not corrected for potential gravitational lensing effects), results in a $\gtrsim 8 \times 10^{-4}\, \textrm{M}_\odot \textrm{yr}^{-1} \textrm{Mpc}^{-3}$ contribution to the obscured star formation of the Universe at these early epochs (1.5 - 0.9 Gyr after the Big Bang).

Similarly, we identified those multiple systems which could potentially be physically associated (rather than line-of-sight projections). The four candidates, whose members have consistent redshifts with each others within the error bars, are shown in Figure \ref{fig:multiple_example}. 
As discussed in \S\ref{secc:subsamples}, these sources might trace galaxy over-densities as those recently discovered within similar samples (e.g. \citealt{Oteo2018a}). Some of them are also in agreement with being galaxy pairs in an  early-stage (pre-coalescence) merger as those predicted by simulations (e.g. \citealt{Hayward2011a}). These systems, which given their component separations ($\Delta\gtrsim20\,$arcsec) are hard to identify with small fields-of-view interferometers, are hence ideal targets to study the environmental effects on the star formation activity  and  to  understand  the  star  formation  process during the earliest stages of galaxy mergers. 

The catalogue of AzTEC/{\it Herschel} sources is given in Table \ref{table:catalogue}, including their updated photometry, derived physical properties, and the best high-$z$ and physically interacting galaxy candidates.

Our results emphasize the importance of accounting for multiplicity in any conclusions derived from {\it Herschel}/SPIRE observations, particularly those that estimate number counts or the space density of DSFGs at high redshifts. 

The fast mapping speeds of a new generation of large format cameras for the 50m-LMT \citep{LMT2020}, e.g. MUSCAT \citep{MUSCAT2018} and TolTEC\footnote{http://toltec.astro.umass.edu/} \citep{TolTEC}, will result in thousands of DSFGs with better photometry and position accuracy for counterpart identification. 
The angular resolution provided by the 50m primary mirror of the LMT will allow the identification of multiple systems separated by, at least, angular scales $\gtrsim 5$\,arcsec (i.e. $\gtrsim 40$ kpc at $z = 3$), and reduce the confusion noise by an order of magnitude ($\sim 0.025$ mJy at 1.1 mm). This would be sufficient to resolve $~50$\, per cent of the multiple systems identified with interferometers \citep[e.g.][]{Ma2019} with enough sensitivity to explore the less extreme (and more abundant) population of Luminous Infrared Galaxies ($L_{\textrm{IR}} \gtrsim 10^{11} \textrm{L}_\odot$). All of these measurements combined will better constrain the space density of DSFGs and their contribution to the star formation history at the earliest stages of galaxy formation in the Universe.

\section*{Acknowledgements}

We would like to thank the support and assistance of all the LMT staff. We also thank the anonymous referee for a thorough reading of our paper and for the comments which helped to improve the clarity and robustness of our results. We thank I. R. Smail for insightful comments that improved the quality of the paper. A.M. thanks support from Consejo Nacional de Ciencia y Technolog\'ia (CONACYT) project A1-S-45680. This work was partially supported by CONACYT projects FDC-2016-1848 and CB-2016-281948. J.A.Z. and C.M.C. thank the University of Texas at Austin College of Natural Sciences for support. C.M.C. also thanks the National Science Foundation for support through grants AST-1814034 and AST-2009577, and the Research Corporation for Science Advancement from a 2019 Cottrell Scholar Award sponsored by IF/THEN, an initiative of Lyda Hill Philanthropies. R.J.I. is funded by the Deutsche Forschungsgemeinschaft (DFG, German Research Foundation) under Germany's Excellence Strategy --- EXC-2094 --- 390783311. M.J.M. acknowledges the support of the National Science Centre, Poland through the SONATA BIS grant 2018/30/E/ST9/00208.  H.D. acknowledges financial support from the Spanish Ministry of Science, Innovation and Universities (MICIU) under the 2014 Ram\'on y Cajal program RYC-2014-15686 and AYA2017-84061-P, the later one co-financed by FEDER (European Regional Development Funds). S.J.M. acknowledges support from the European Research Council (ERC) Consolidator Grant {\sc CosmicDust} (ERC-2014-CoG-647939, PI H\,L\,Gomez) from the ERC Advanced Investigator Program, COSMICISM (ERC-2012-ADG 20120216, PI R.J.Ivison). 
\section*{Data Availability}
 
The datasets generated and analysed during this study are available from the corresponding author on reasonable request.



\bibliographystyle{mnras}
\bibliography{biblio} 




\appendix

\section{Catalogue and postage stamps}
\label{appendix:catalog}

This Appendix presents $\rm 80\,arcsec \times 80\,arcsec$ postage stamps of the 93 \textit{H}-ATLAS targets included in our analysis (\autoref{fig:pstamps}), as well as the new photometry of the AzTEC detection (with SNR $\geq 3.5$) and their derived physical parameters \autoref{table:catalogue}.

\begin{figure*}\hspace{-0.6cm}
\begin{center}
\includegraphics[totalheight=4.25cm]{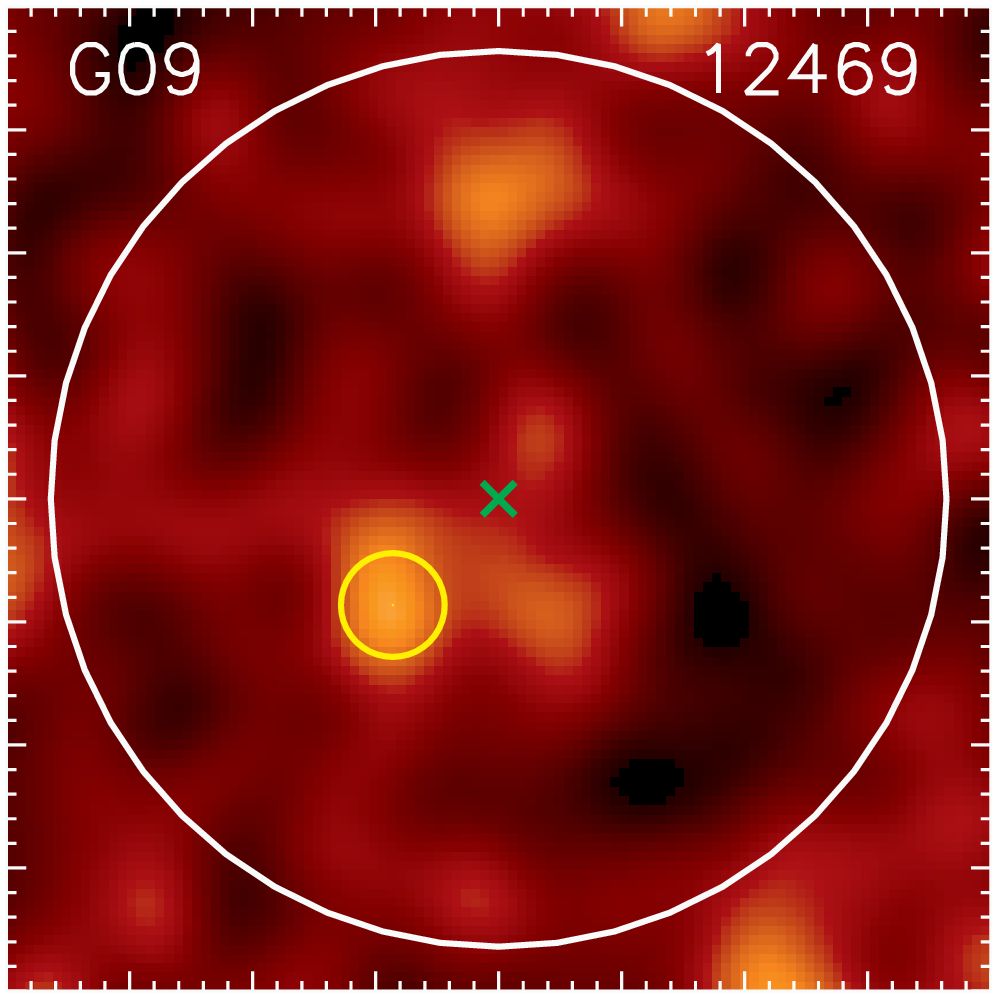}
\includegraphics[totalheight=4.25cm]{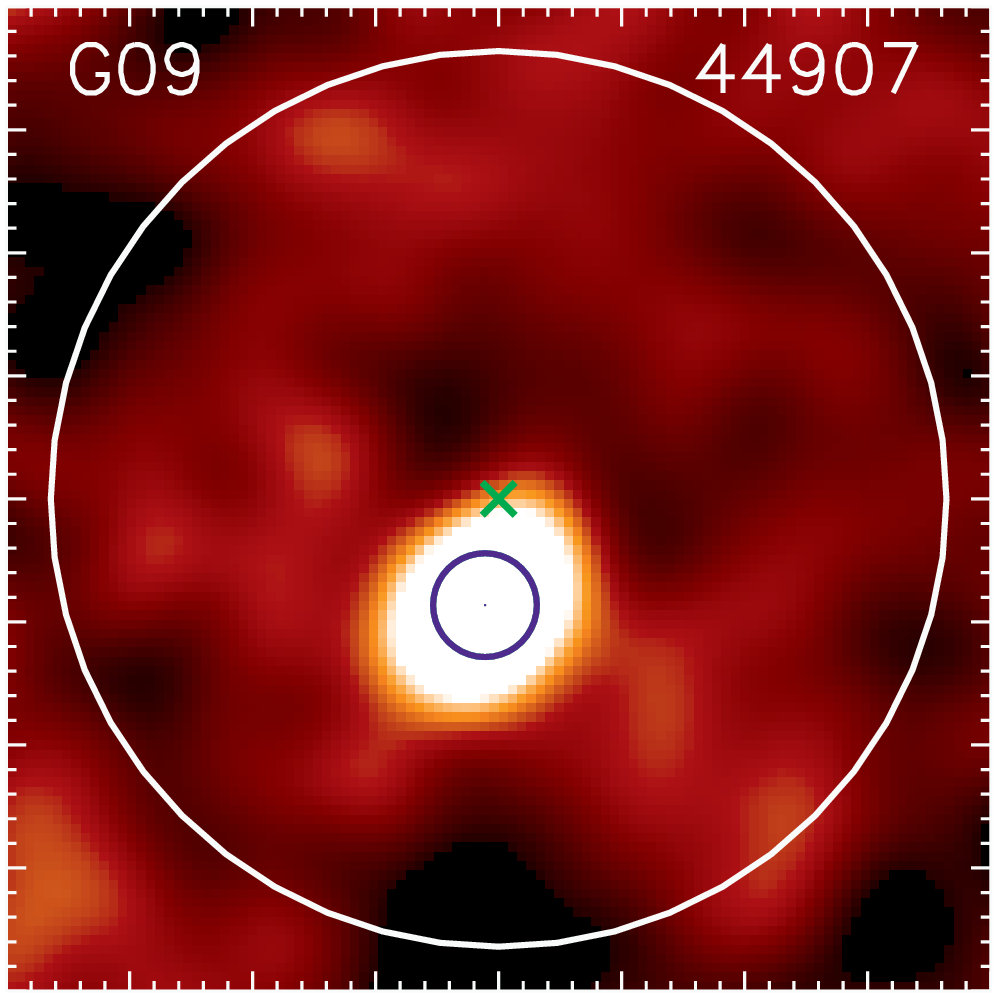}
\includegraphics[totalheight=4.25cm]{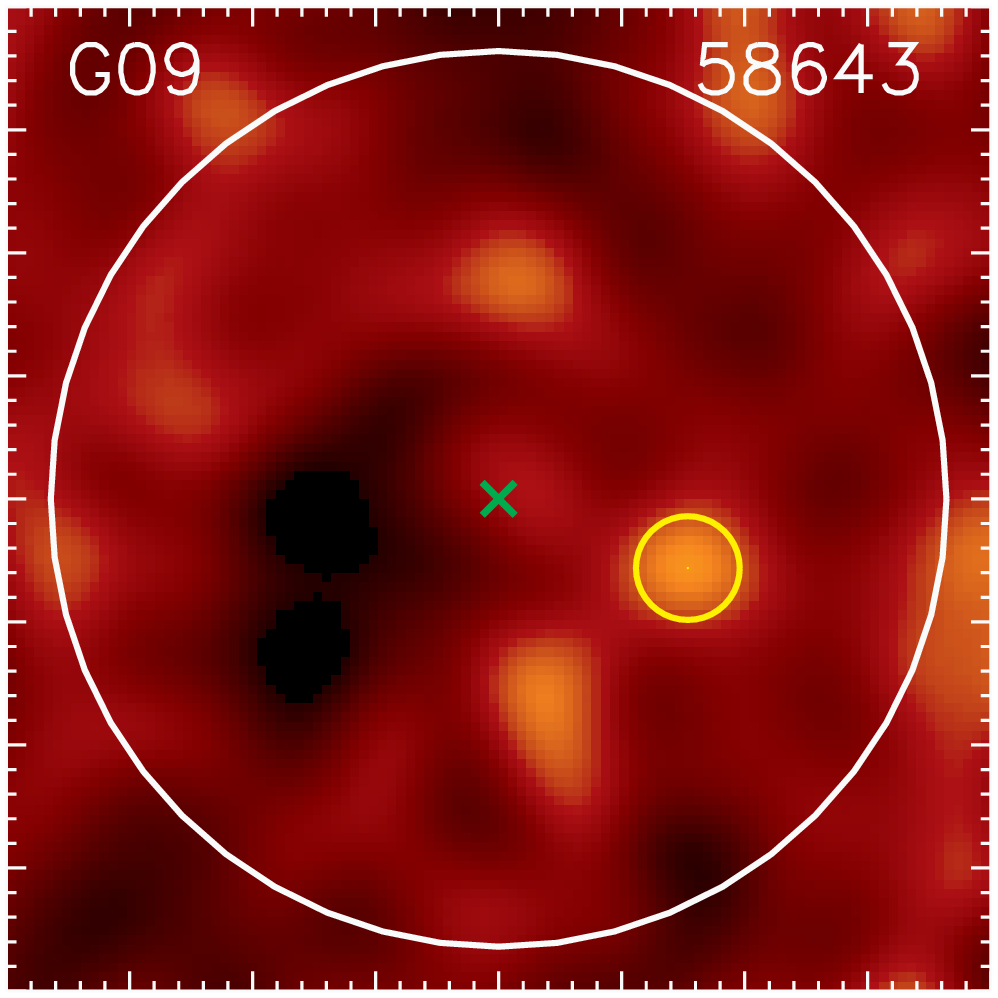}
\includegraphics[totalheight=4.25cm]{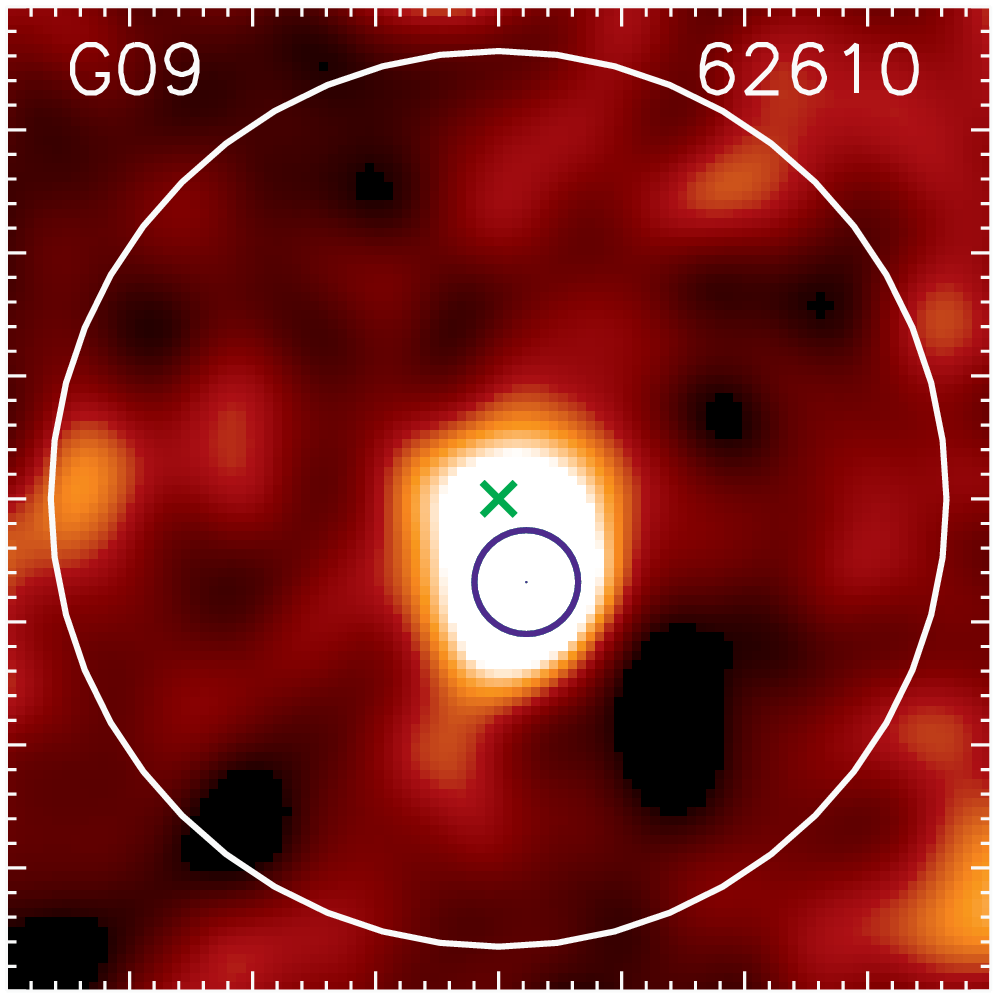} \\
\includegraphics[totalheight=4.25cm]{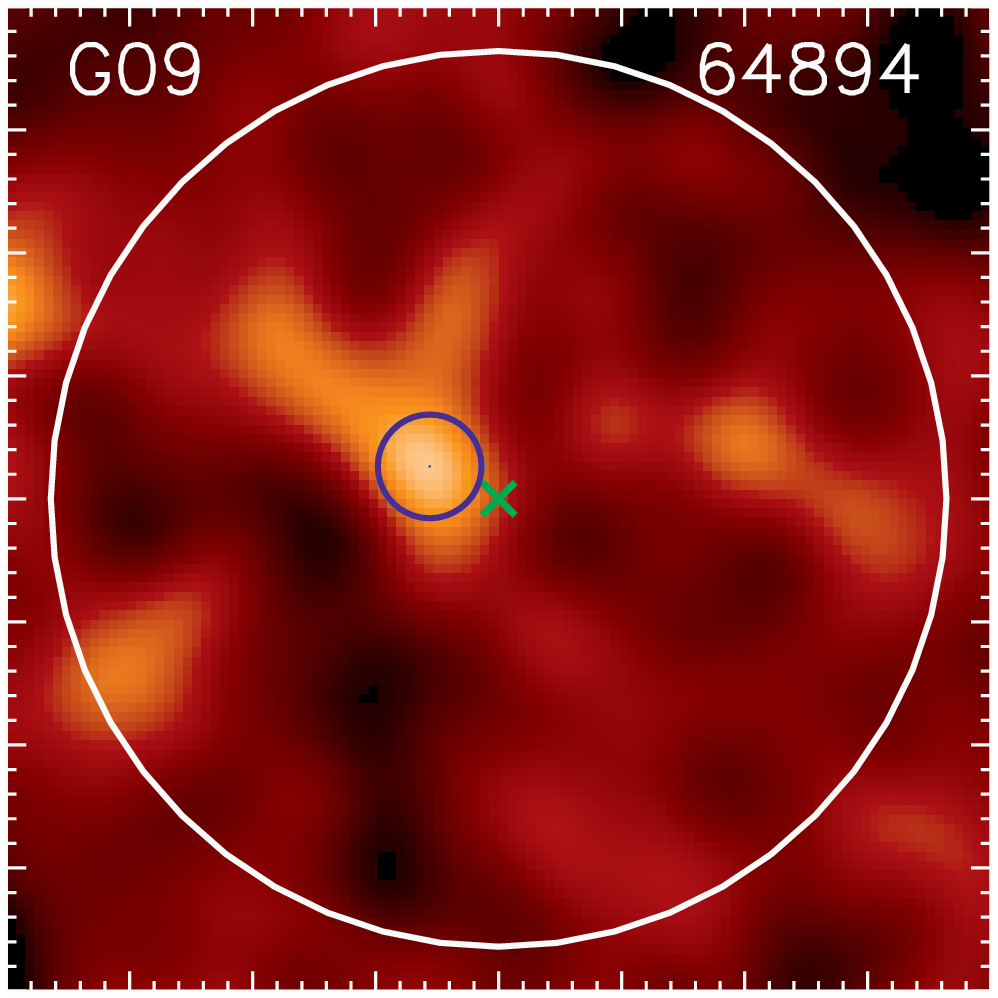}
\includegraphics[totalheight=4.25cm]{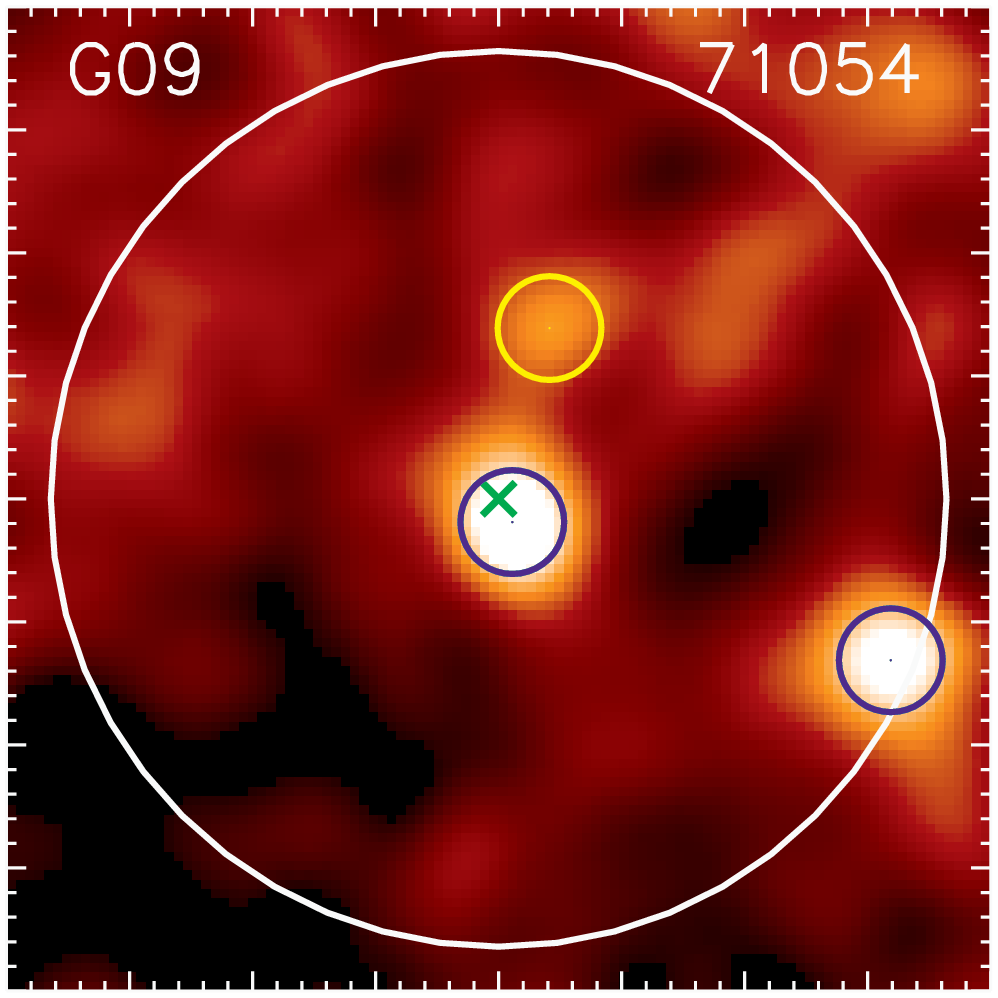}
\includegraphics[totalheight=4.25cm]{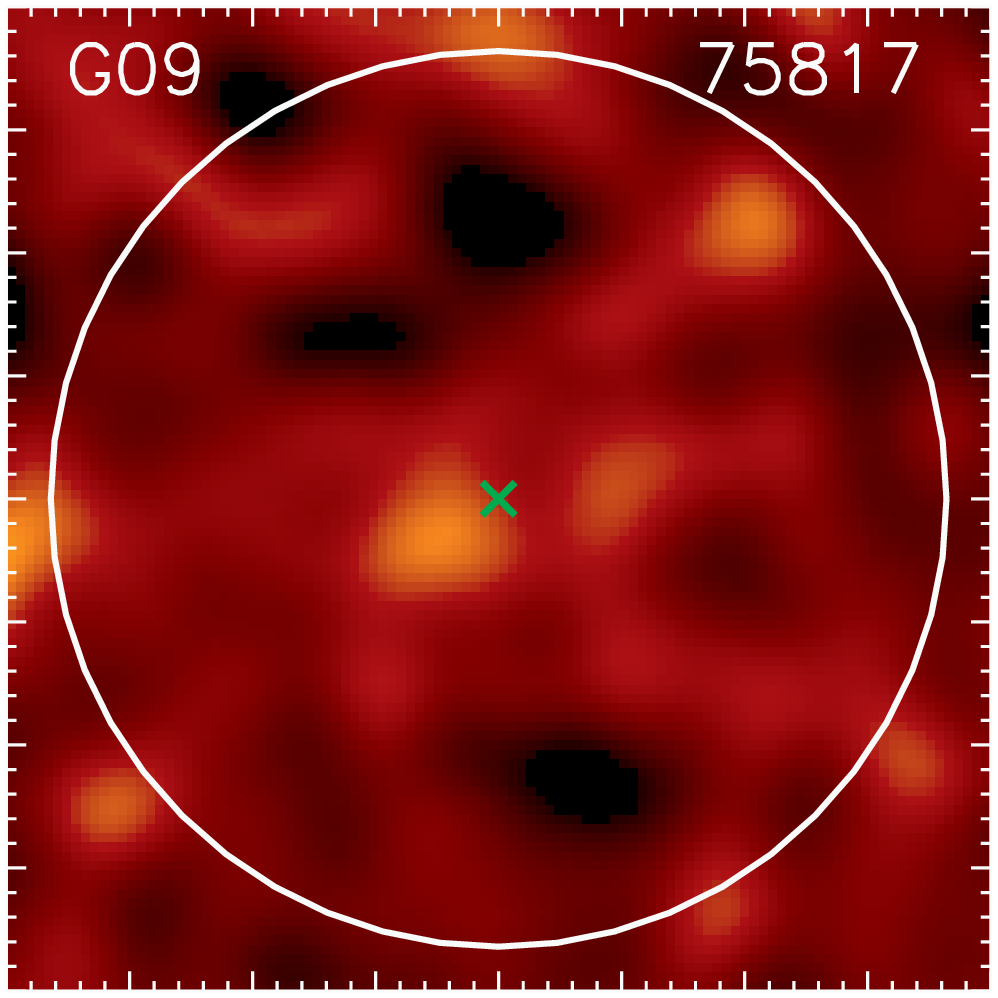}
\includegraphics[totalheight=4.25cm]{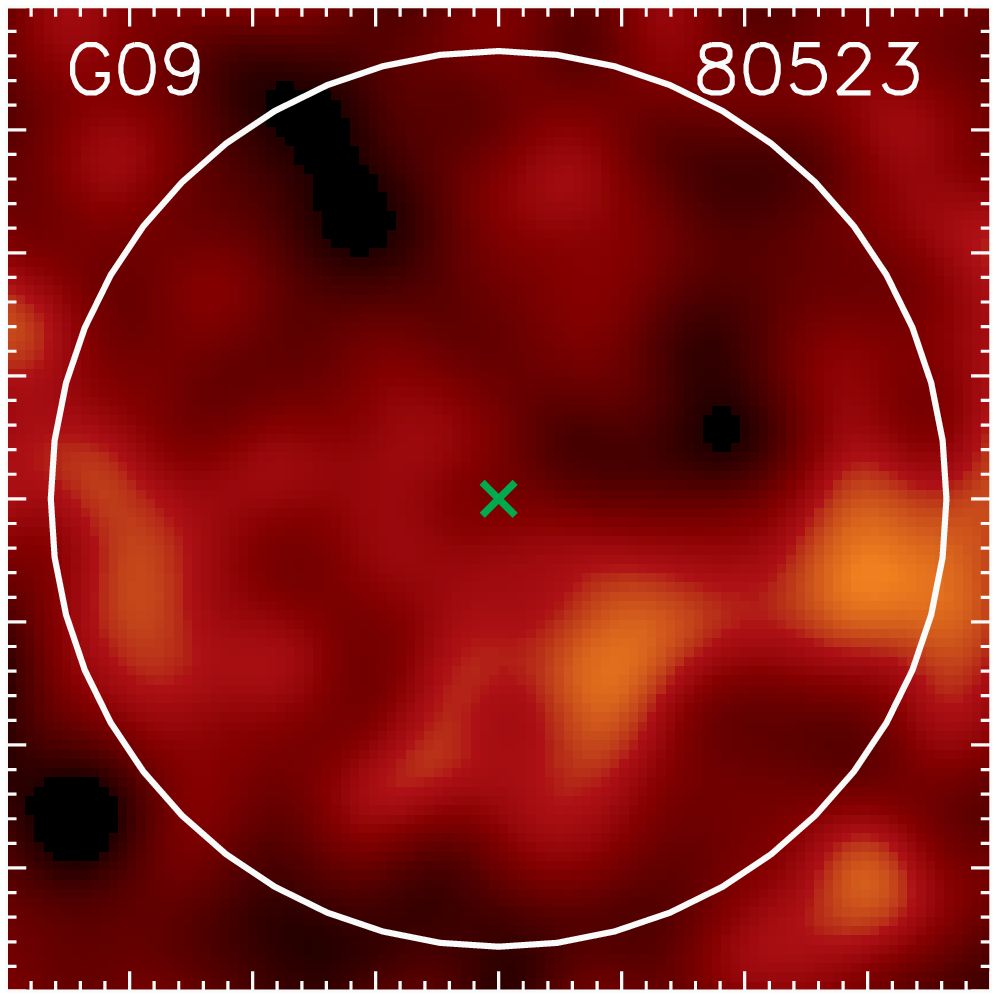} \\
\includegraphics[totalheight=4.25cm]{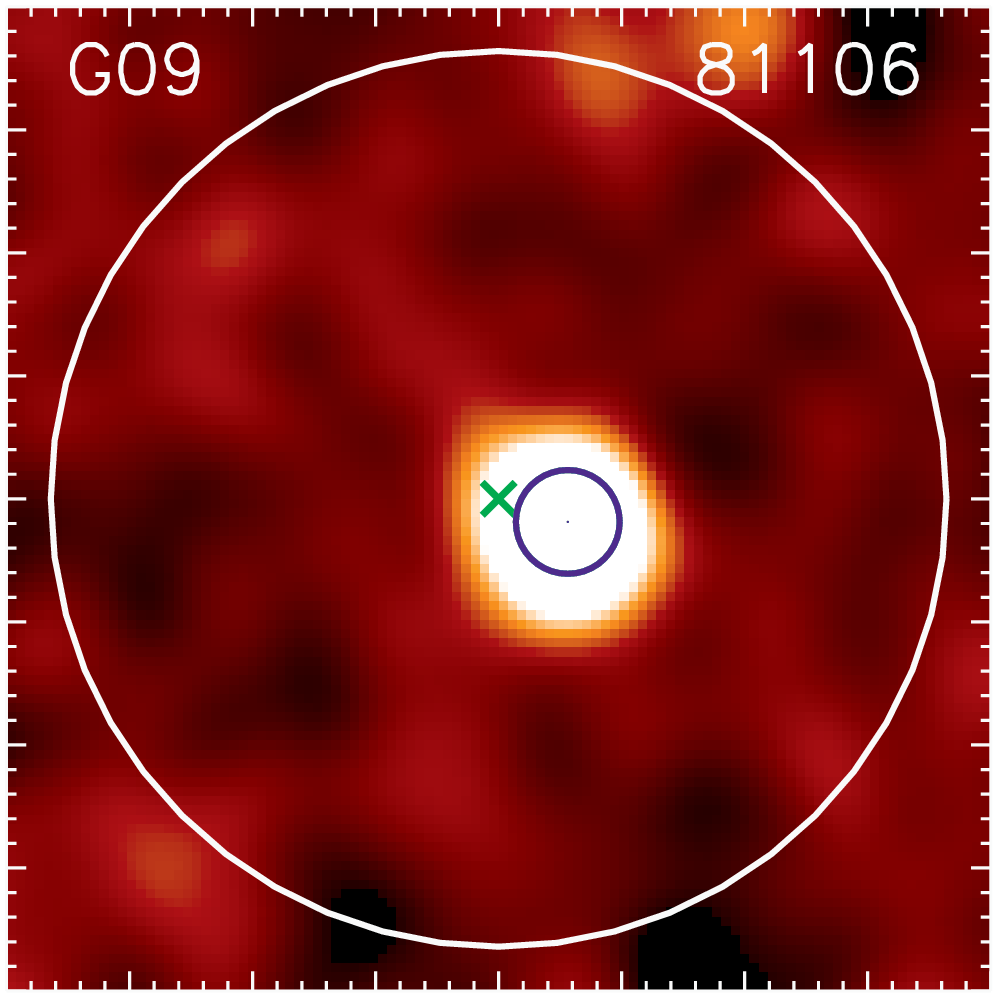}
\includegraphics[totalheight=4.25cm]{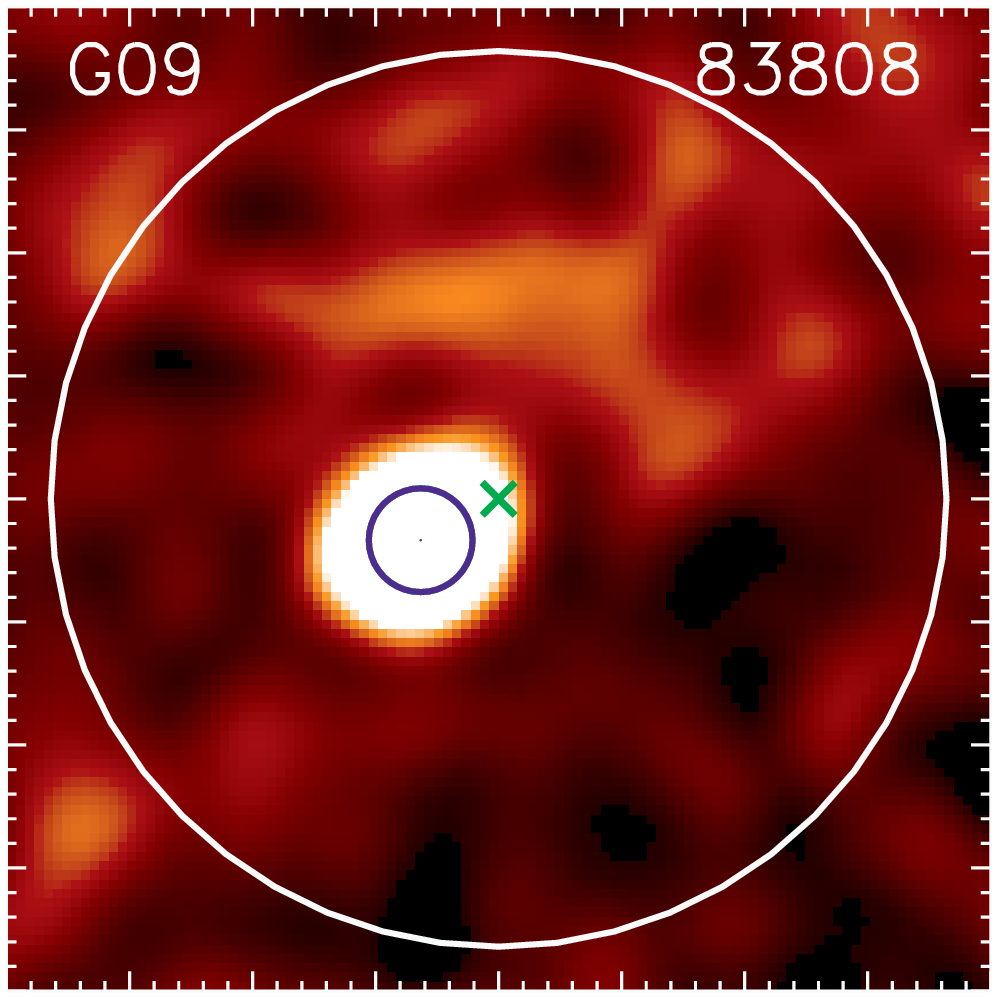}
\includegraphics[totalheight=4.25cm]{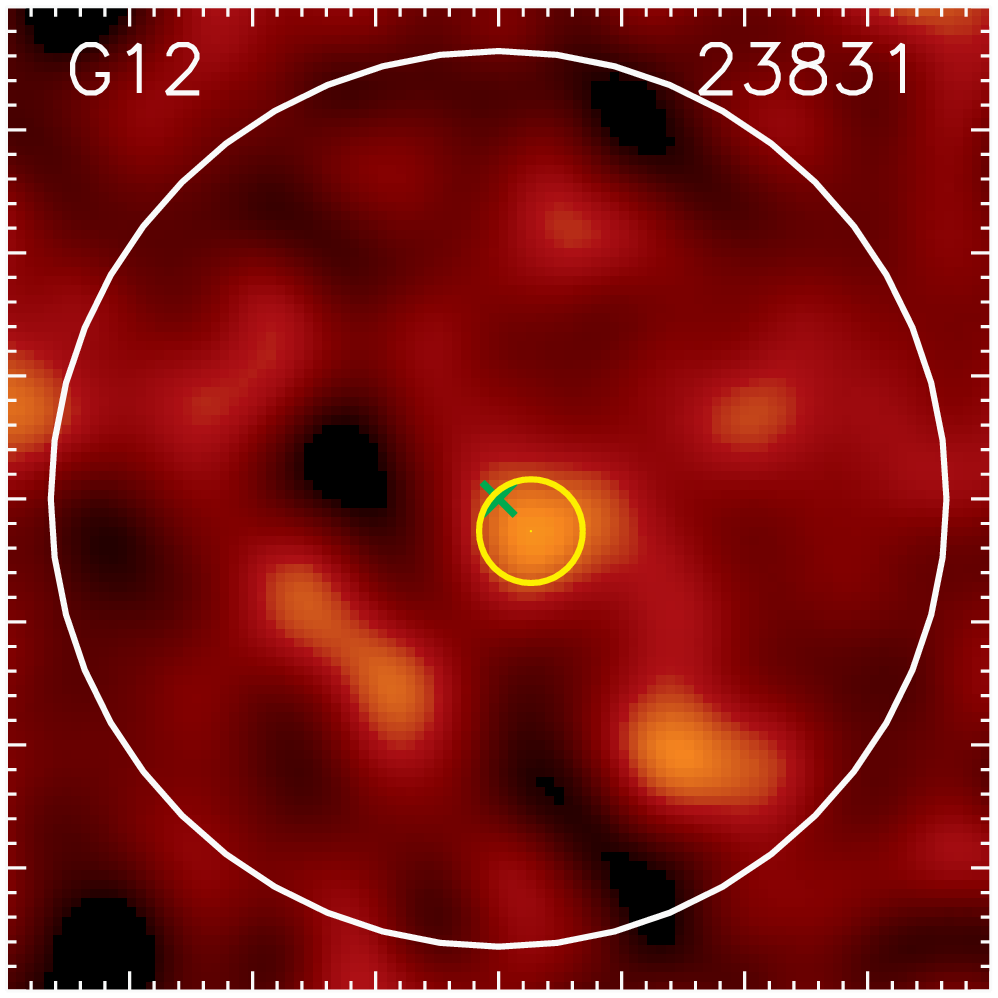}
\includegraphics[totalheight=4.25cm]{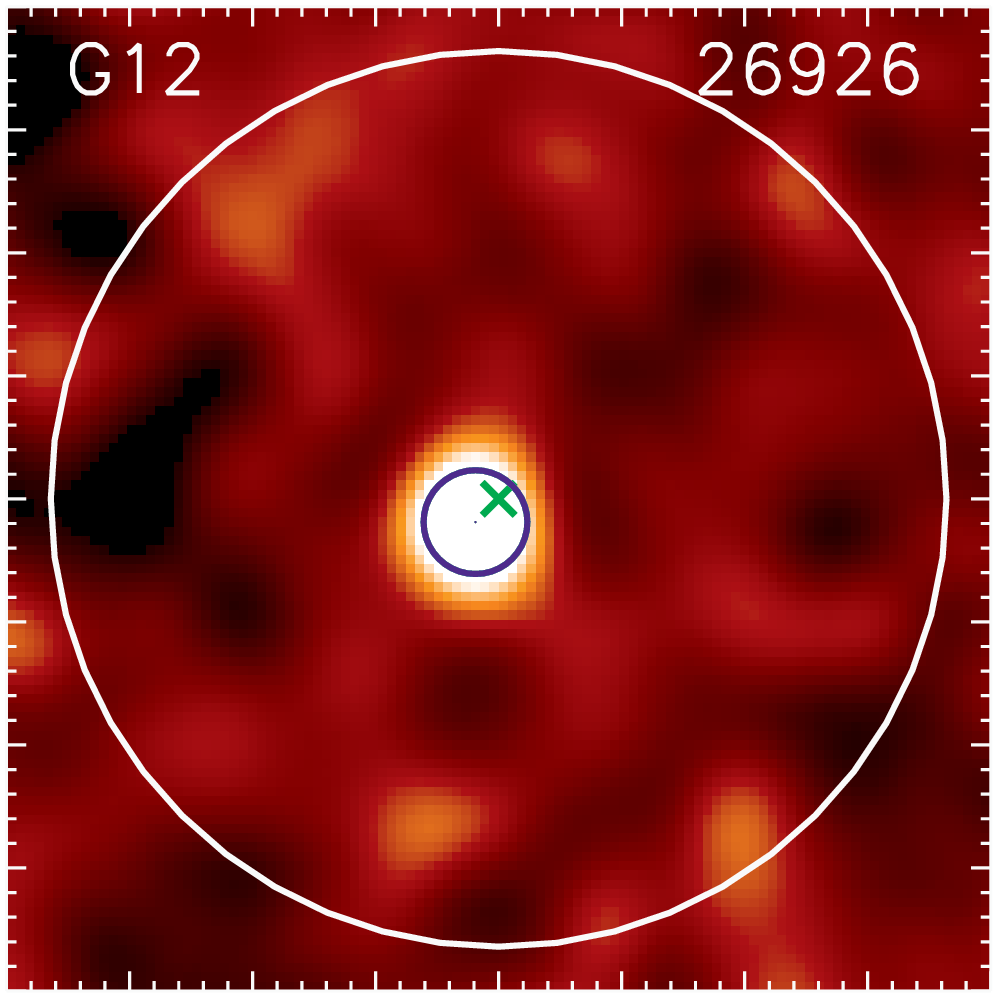} \\
\includegraphics[totalheight=4.25cm]{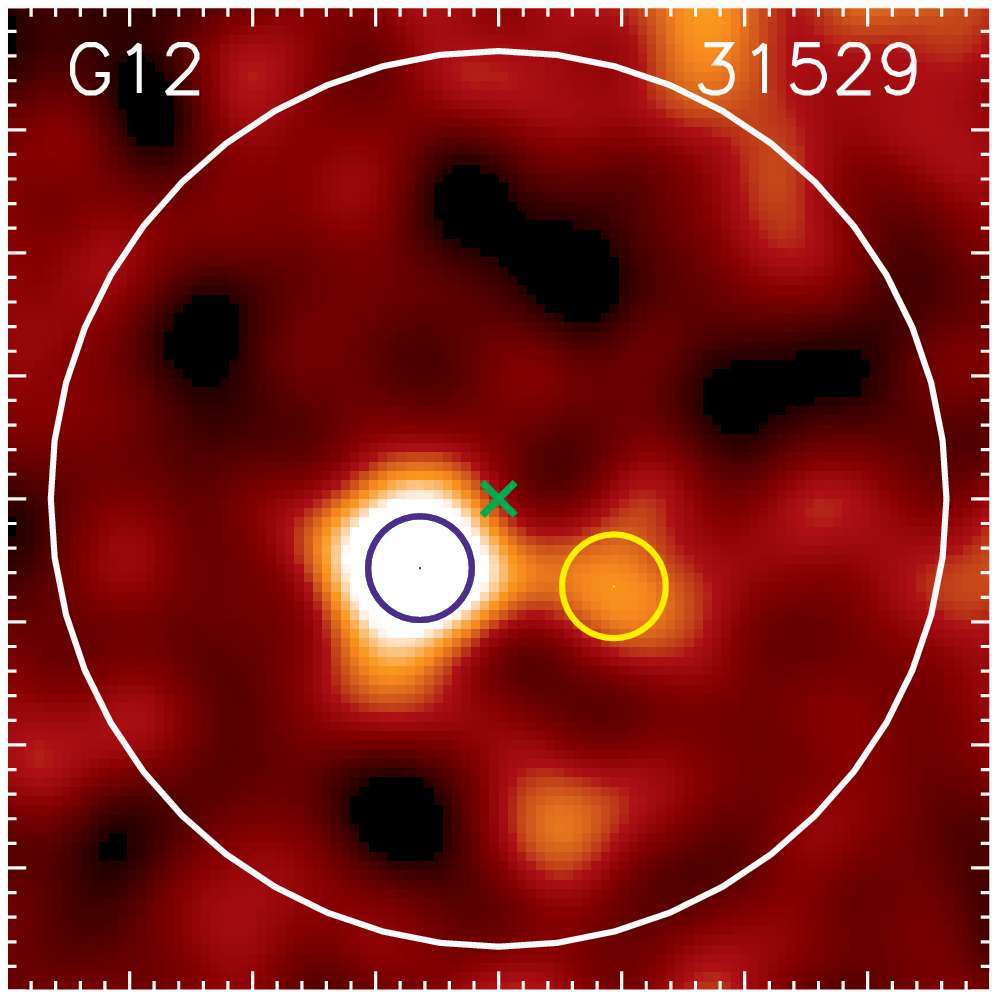}
\includegraphics[totalheight=4.25cm]{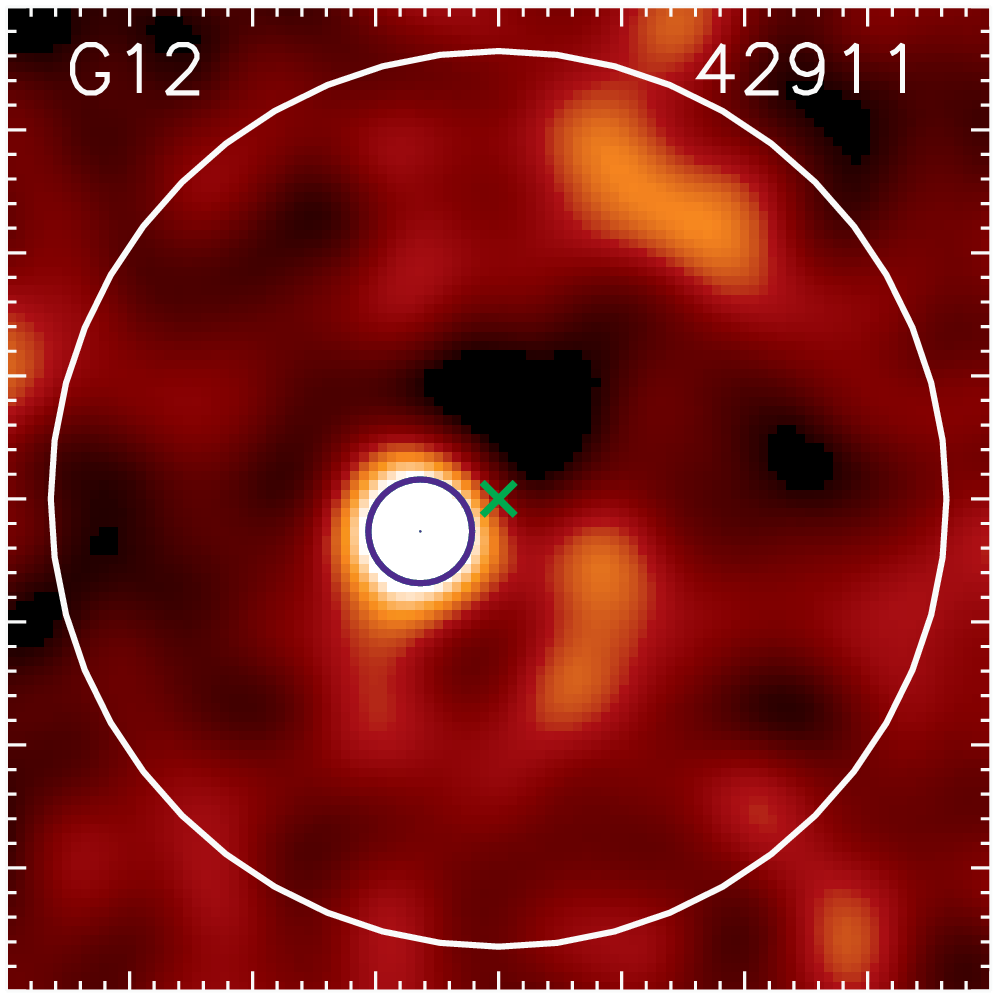}
\includegraphics[totalheight=4.25cm]{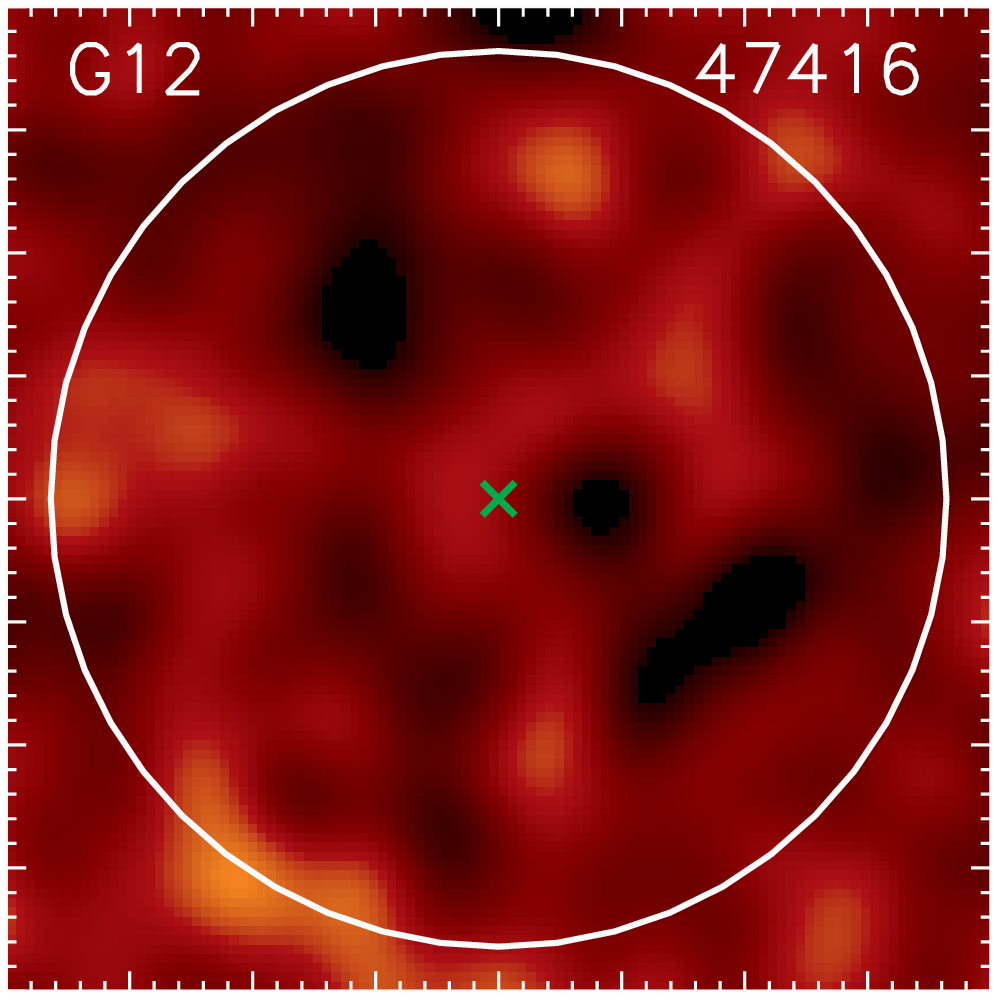}
\includegraphics[totalheight=4.25cm]{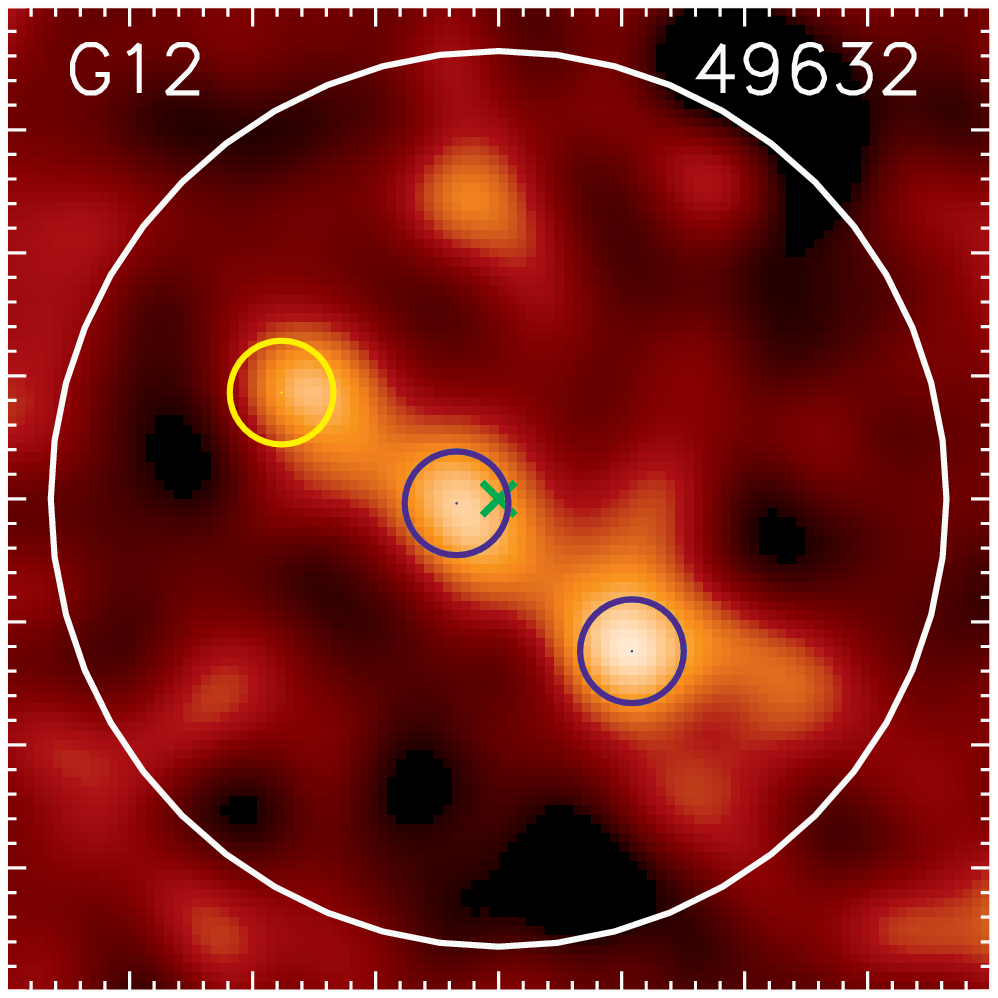} \\
\includegraphics[totalheight=4.25cm]{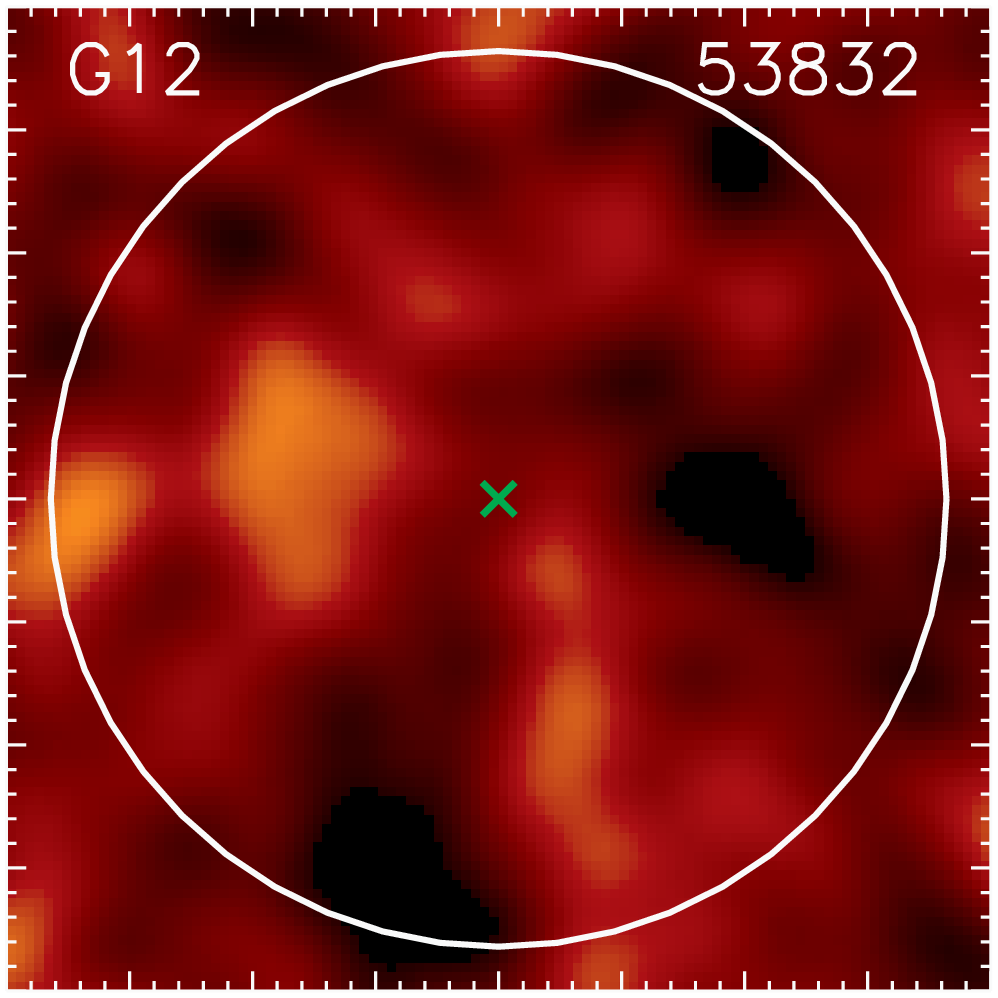}
\includegraphics[totalheight=4.25cm]{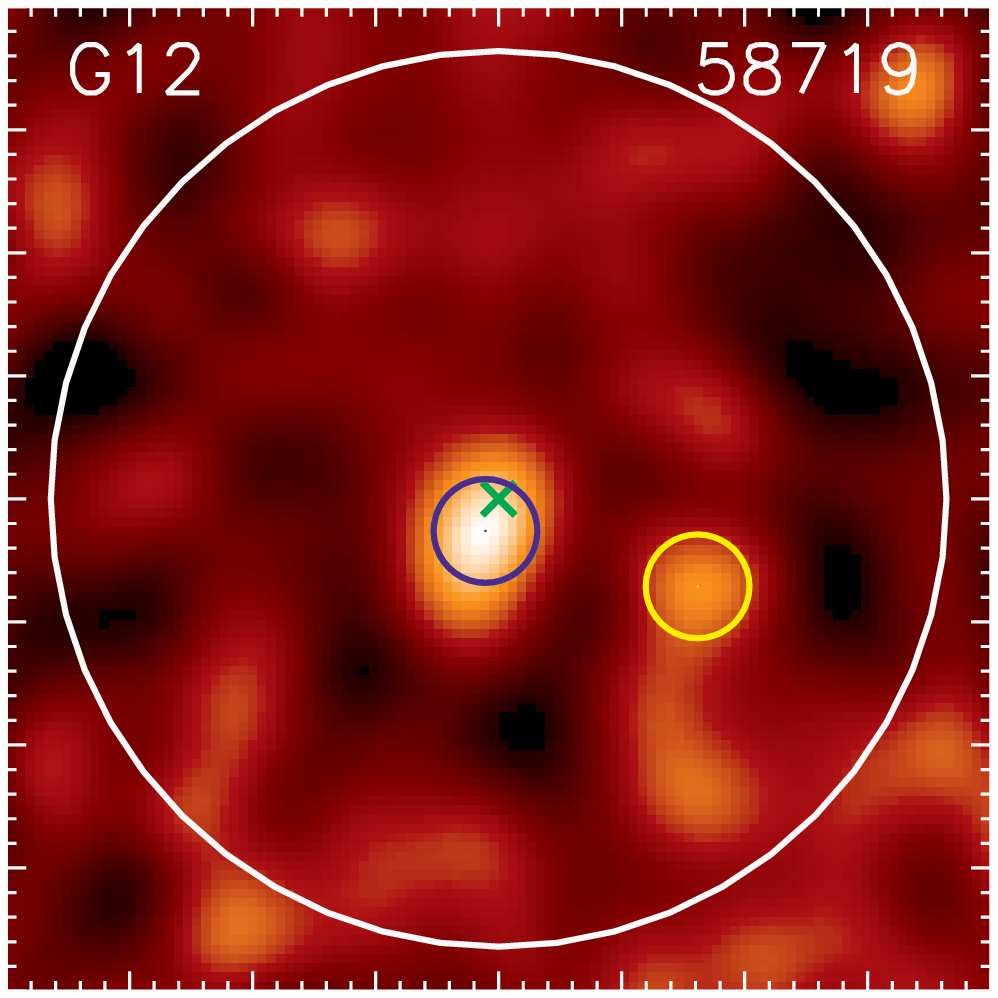}
\includegraphics[totalheight=4.25cm]{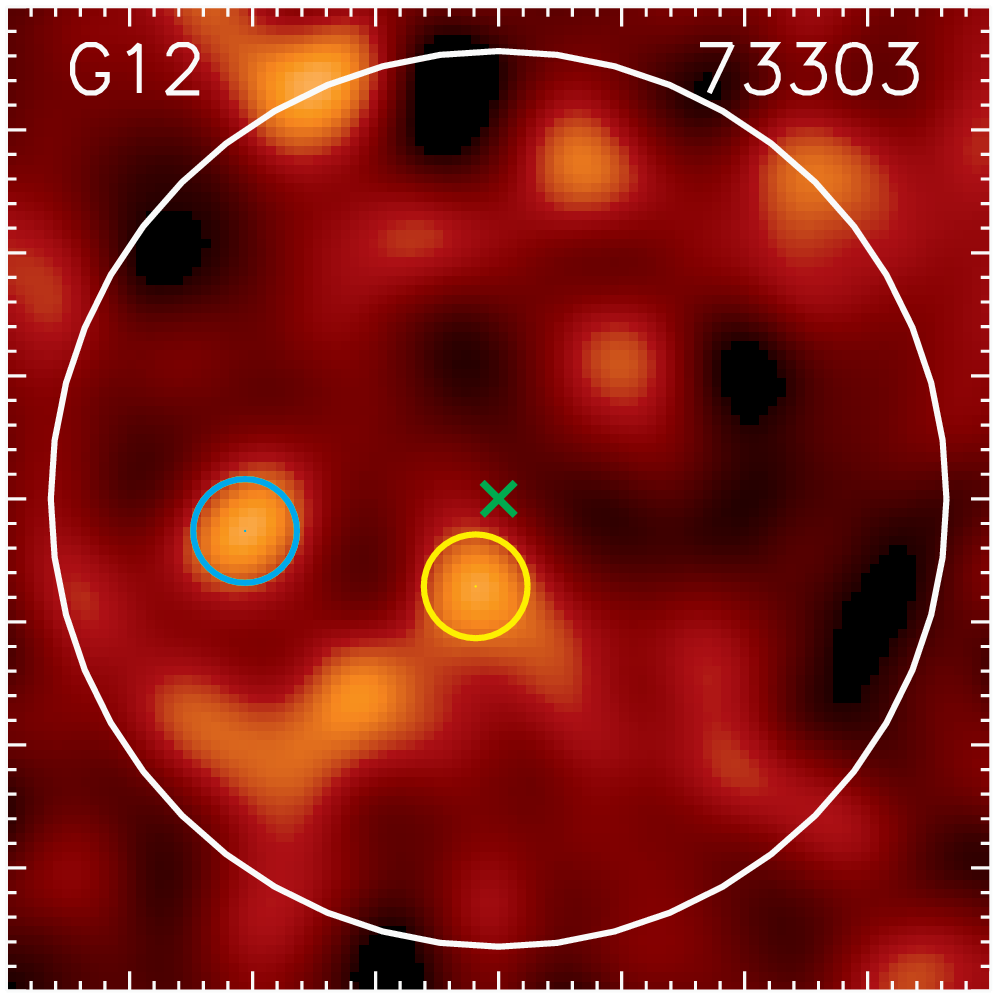}
\includegraphics[totalheight=4.25cm]{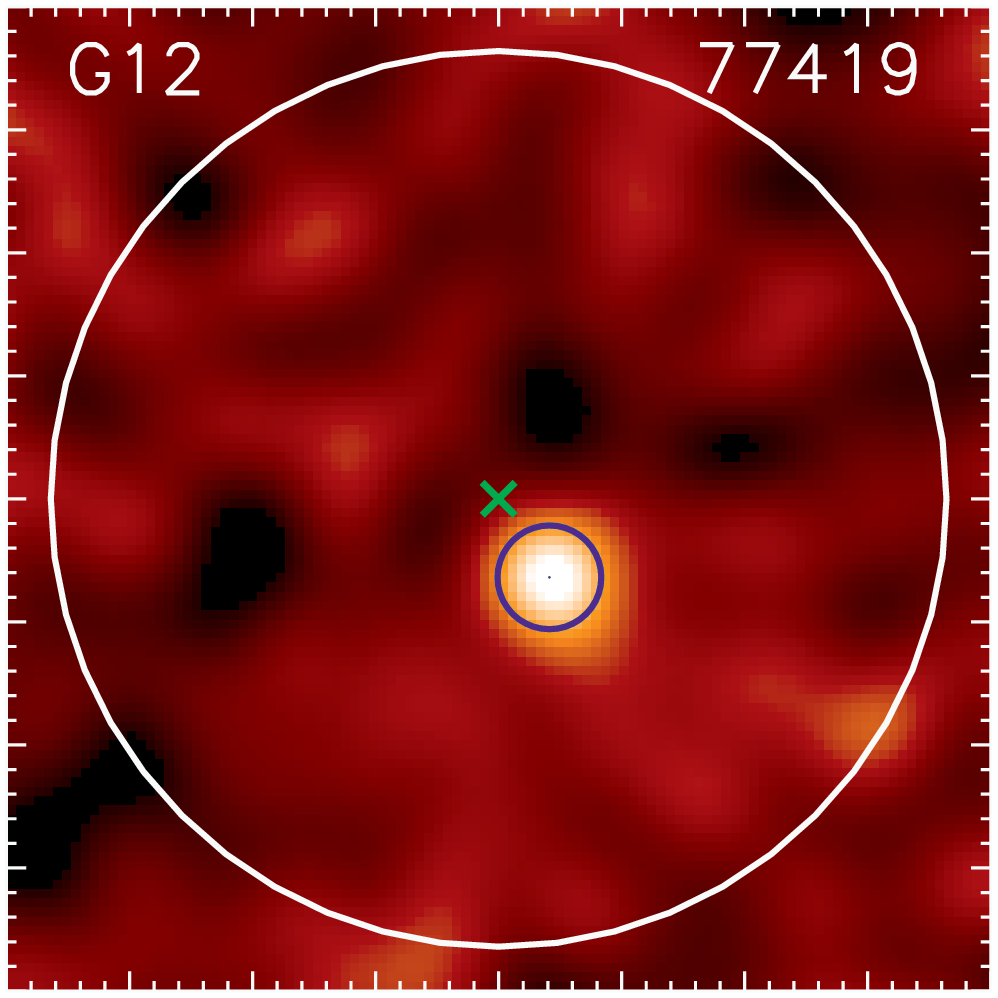} \\

\caption{$\rm 80\,arcsec \times 80\,arcsec$ postage stamps of the 93 \textit{H}-ATLAS targets included in our analysis. Images correspond to the 1.1 mm AzTEC SNR maps. 
Green crosses and white circles indicate the \textit{H}-ATLAS targeted position and the adopted search radius ($36.6$\,arcsec) respectively. Small circles ($9.5$\,arcsec in diameter) mark the position of AzTEC detections, with: yellow corresponding to $3.0 \leq \textrm{SNR} < 3.5$, light blue to $3.5 \leq \textrm{SNR} < 4.0$, and dark-blue to $\textrm{SNR} \geq$ 4.0.}
\label{fig:pstamps}
\end{center}
\end{figure*}
\begin{figure*}\hspace{-0.6cm}
\begin{center}
\includegraphics[totalheight=4.25cm]{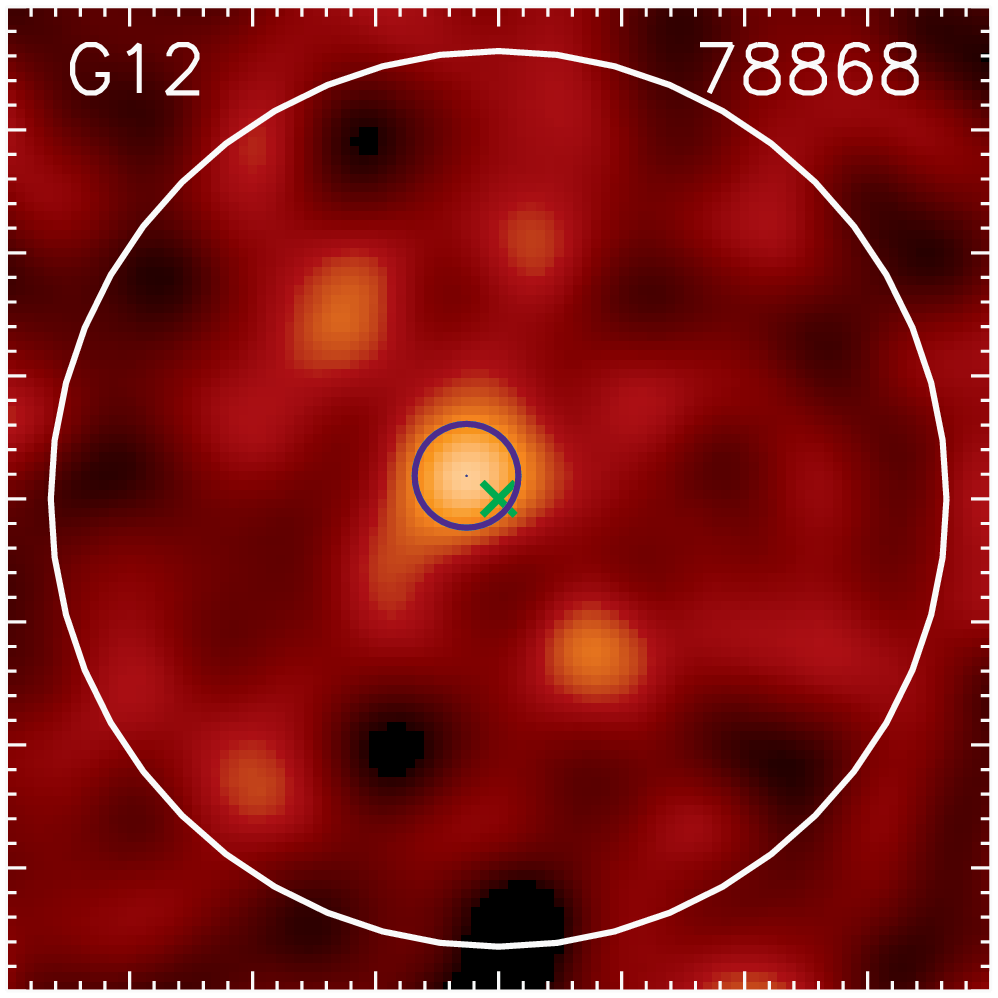}
\includegraphics[totalheight=4.25cm]{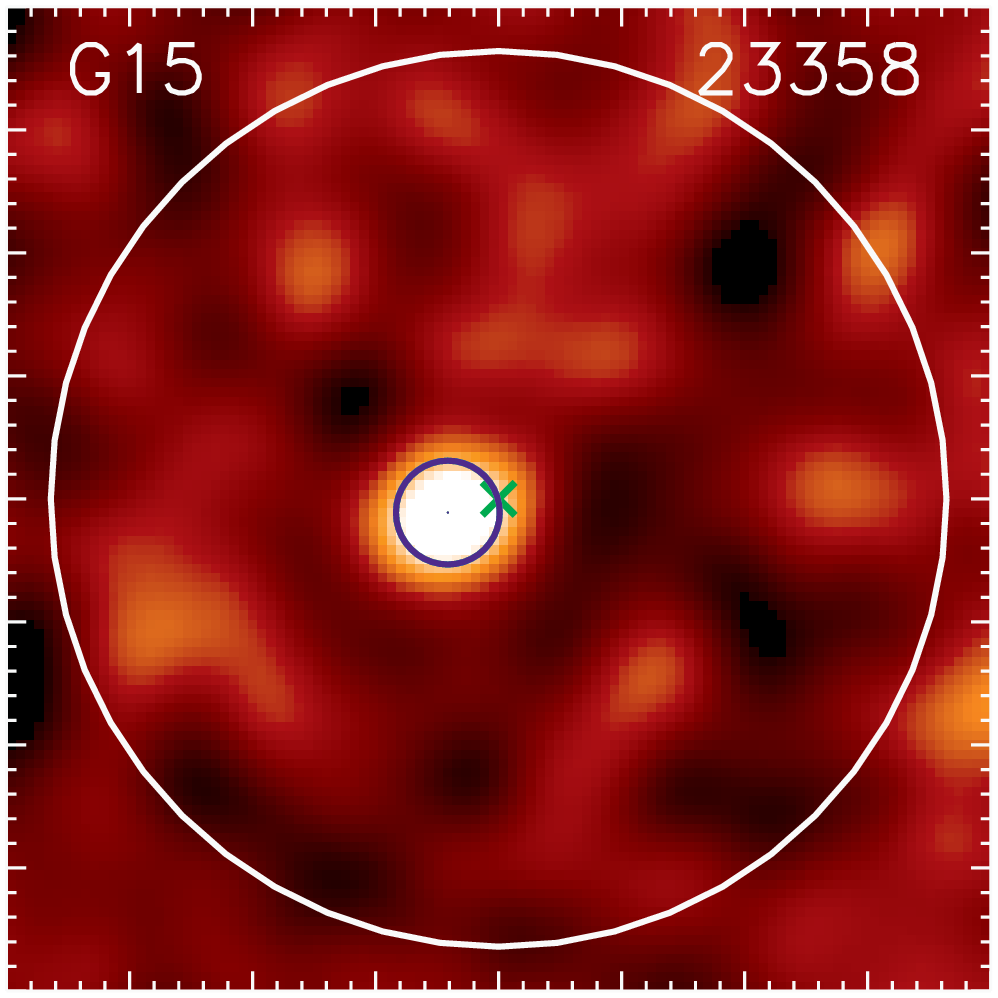}
\includegraphics[totalheight=4.25cm]{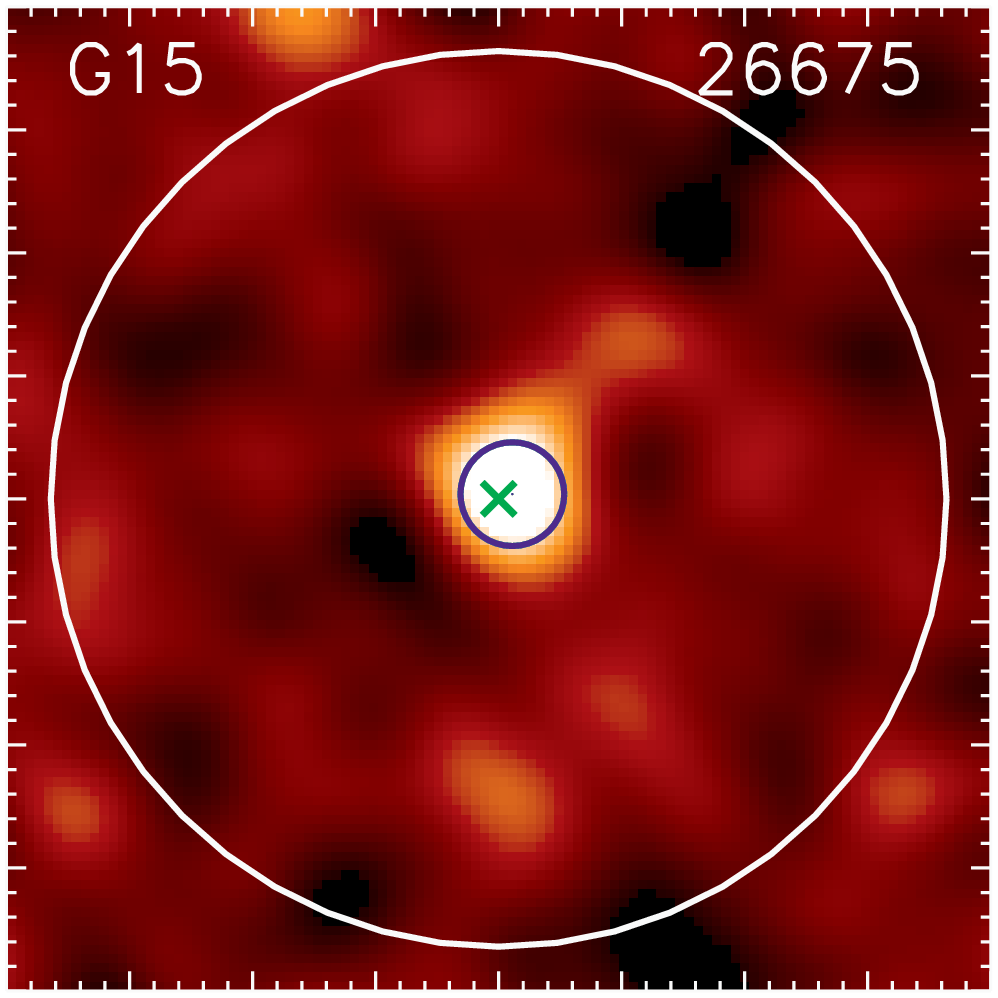}
\includegraphics[totalheight=4.25cm]{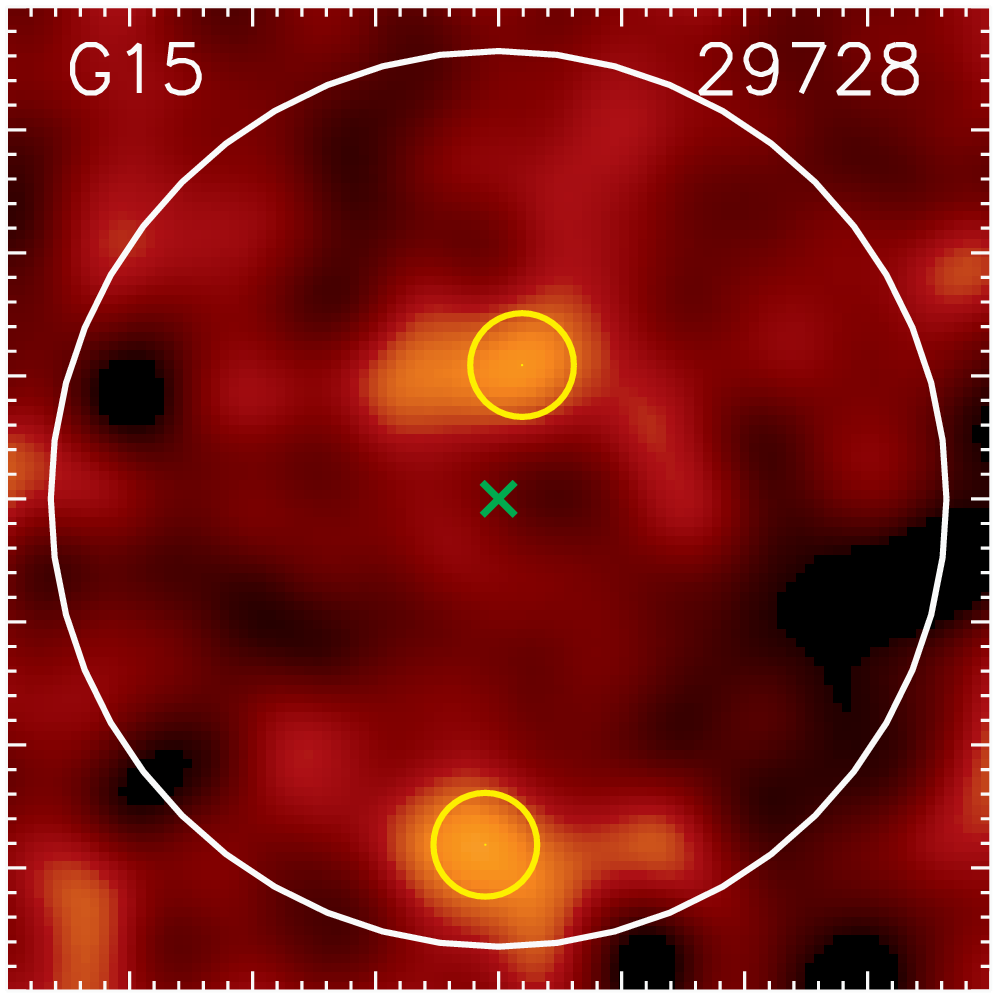} \\
\includegraphics[totalheight=4.25cm]{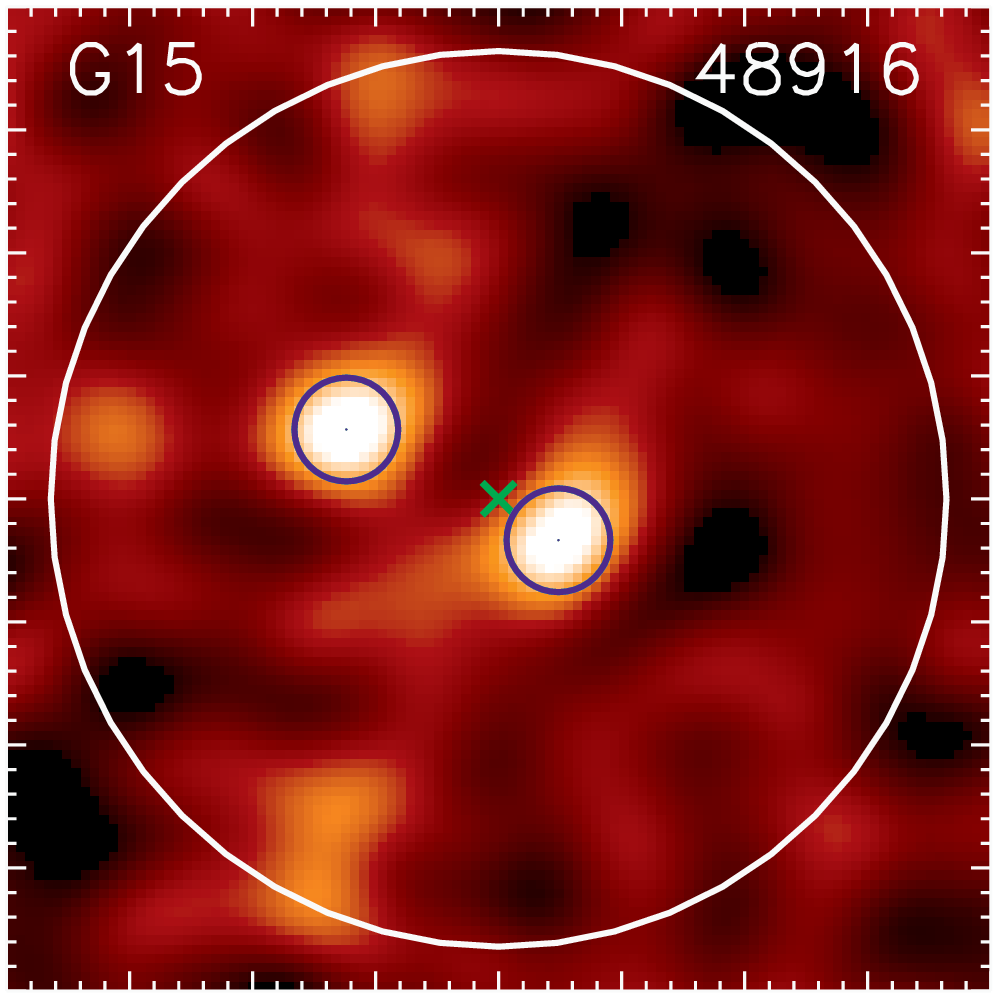}
\includegraphics[totalheight=4.25cm]{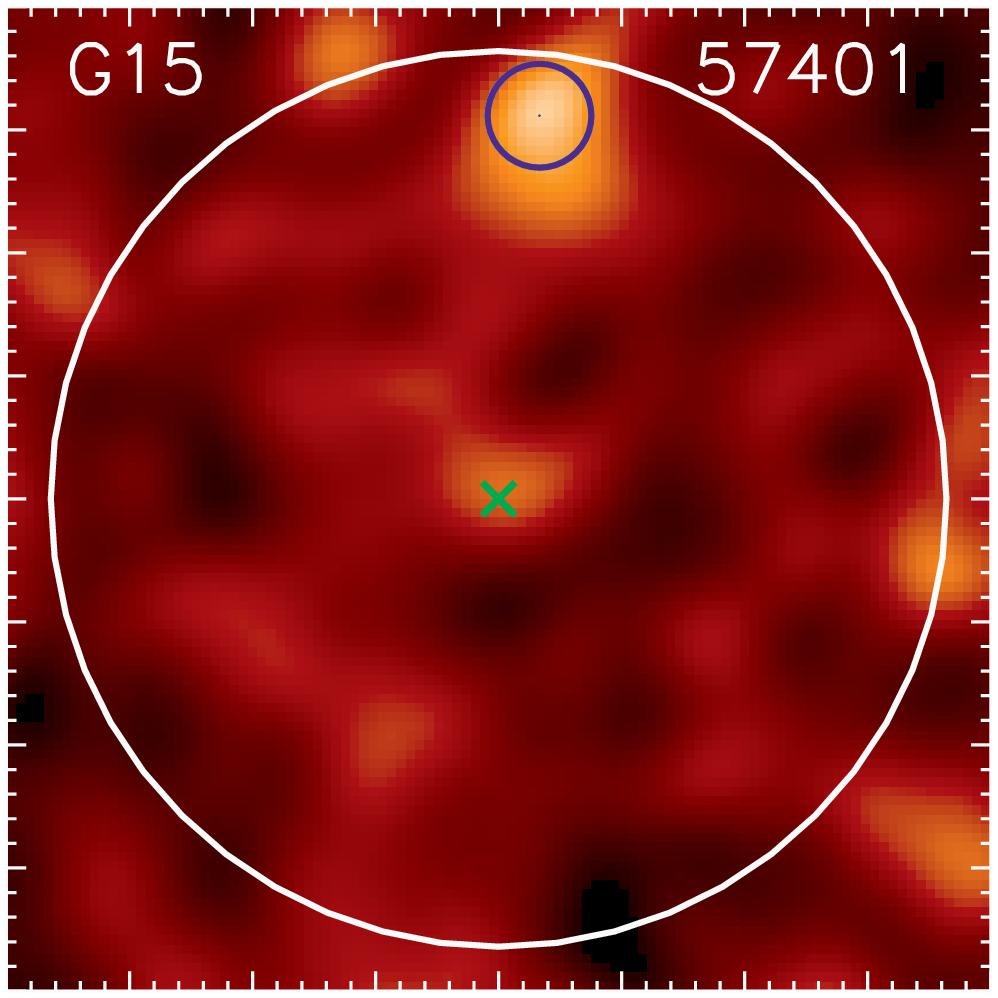}
\includegraphics[totalheight=4.25cm]{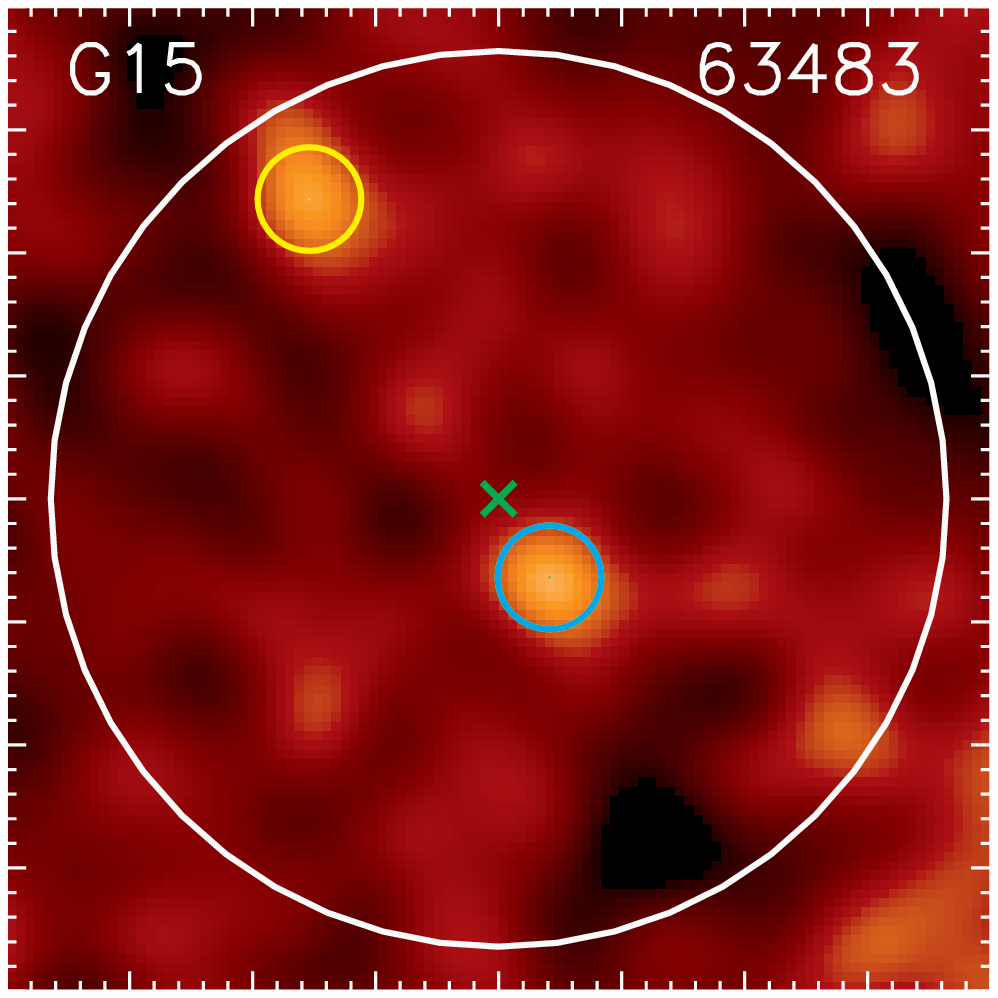}
\includegraphics[totalheight=4.25cm]{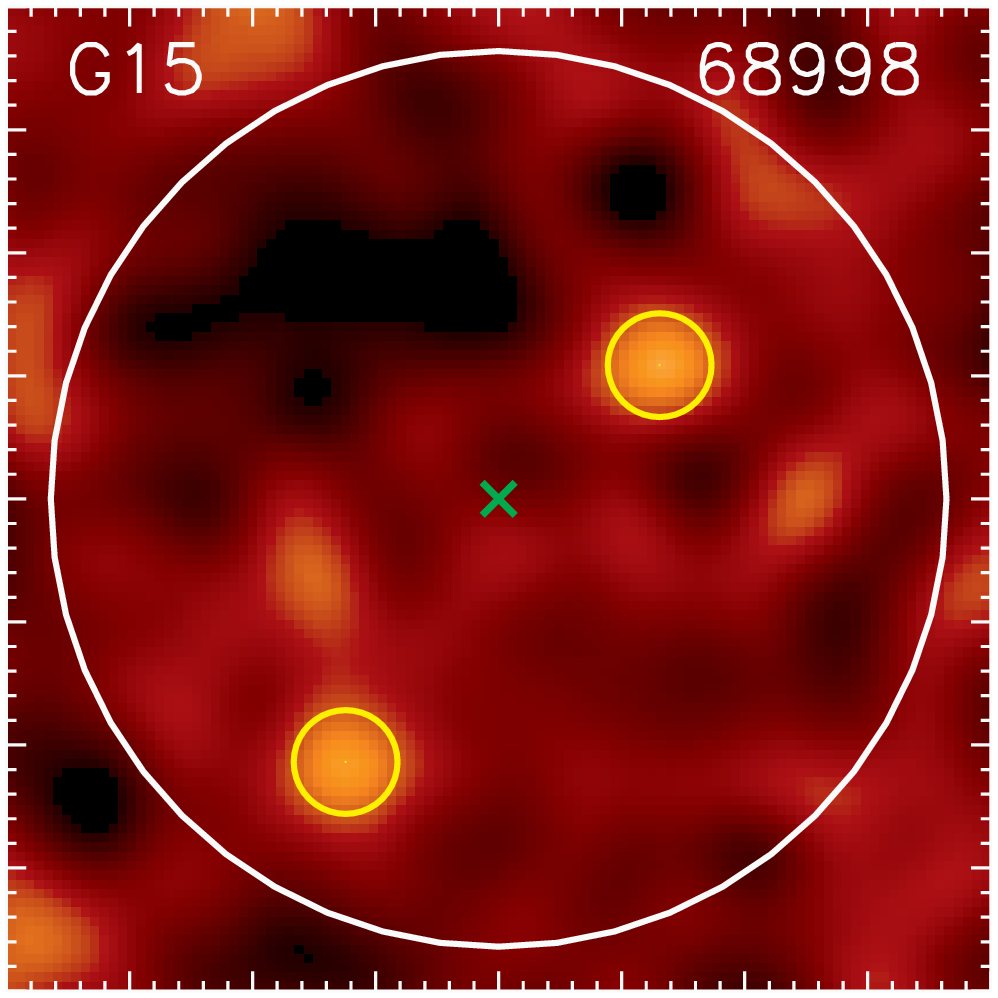} \\
\includegraphics[totalheight=4.25cm]{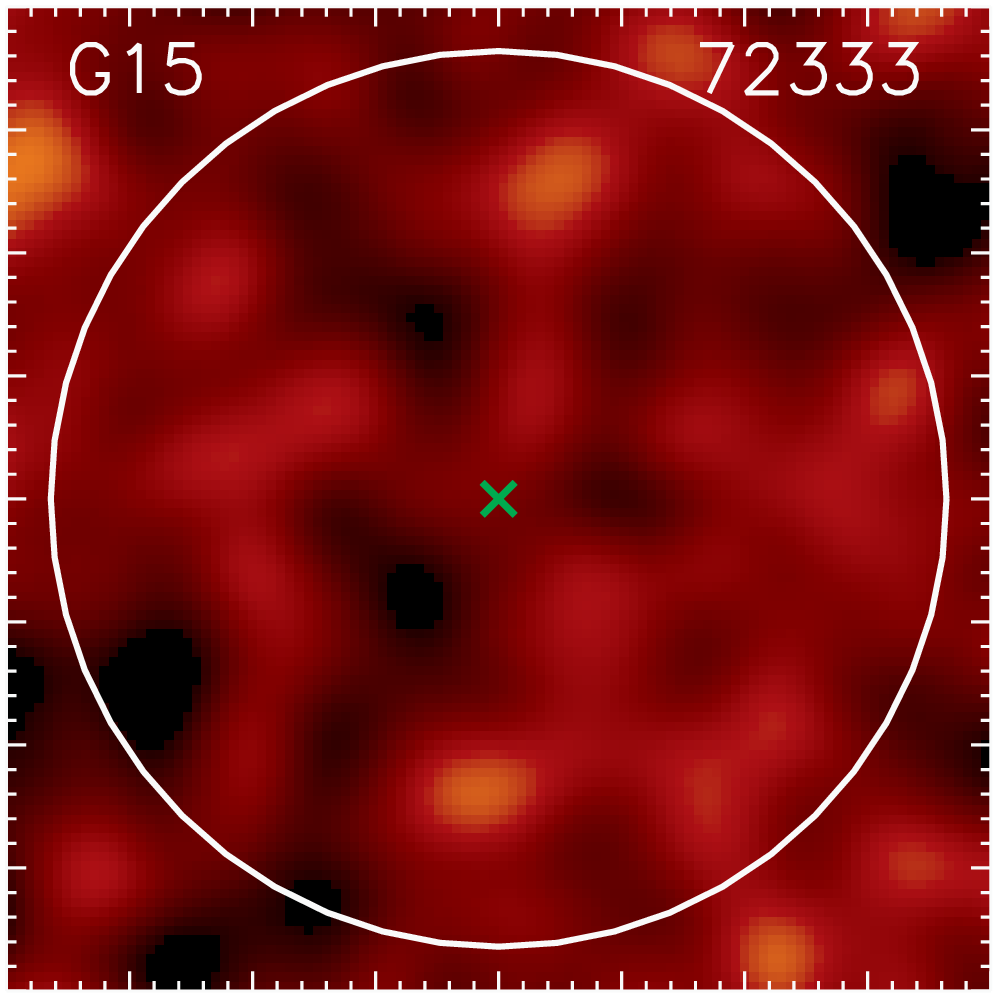}
\includegraphics[totalheight=4.25cm]{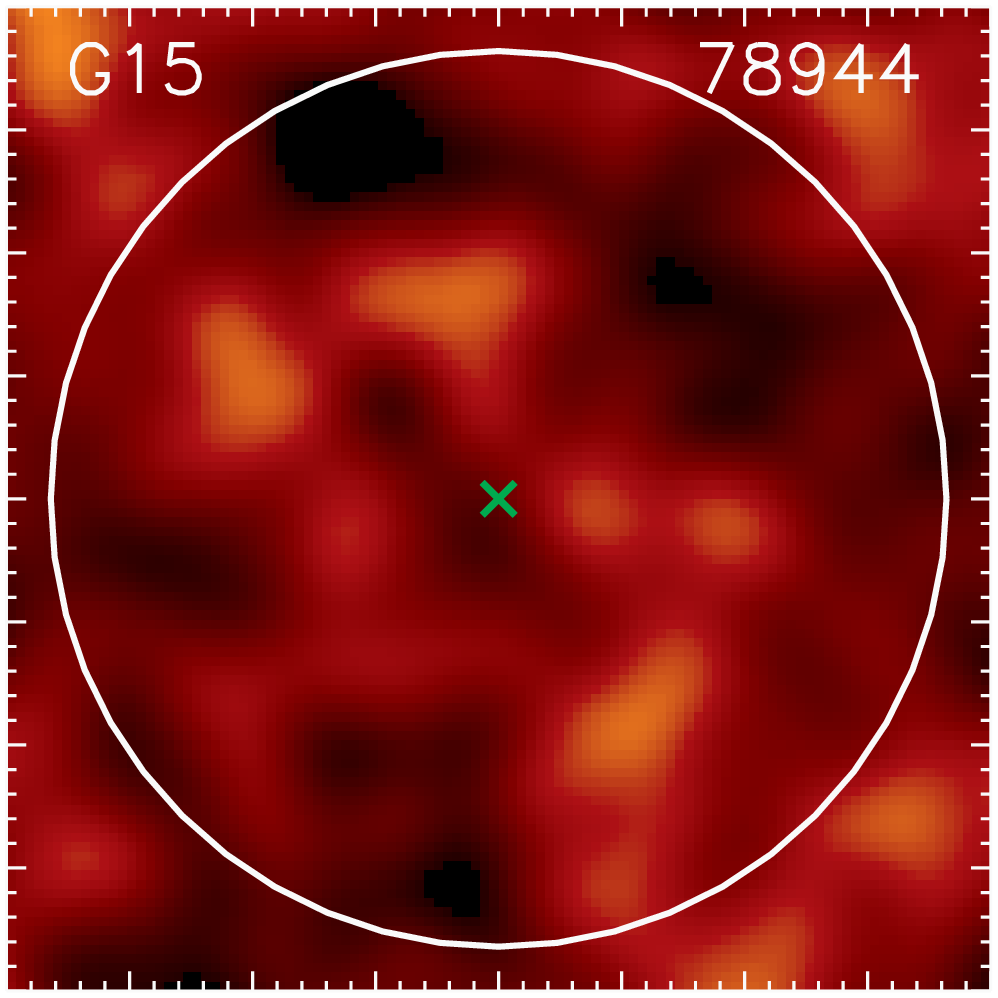}
\includegraphics[totalheight=4.25cm]{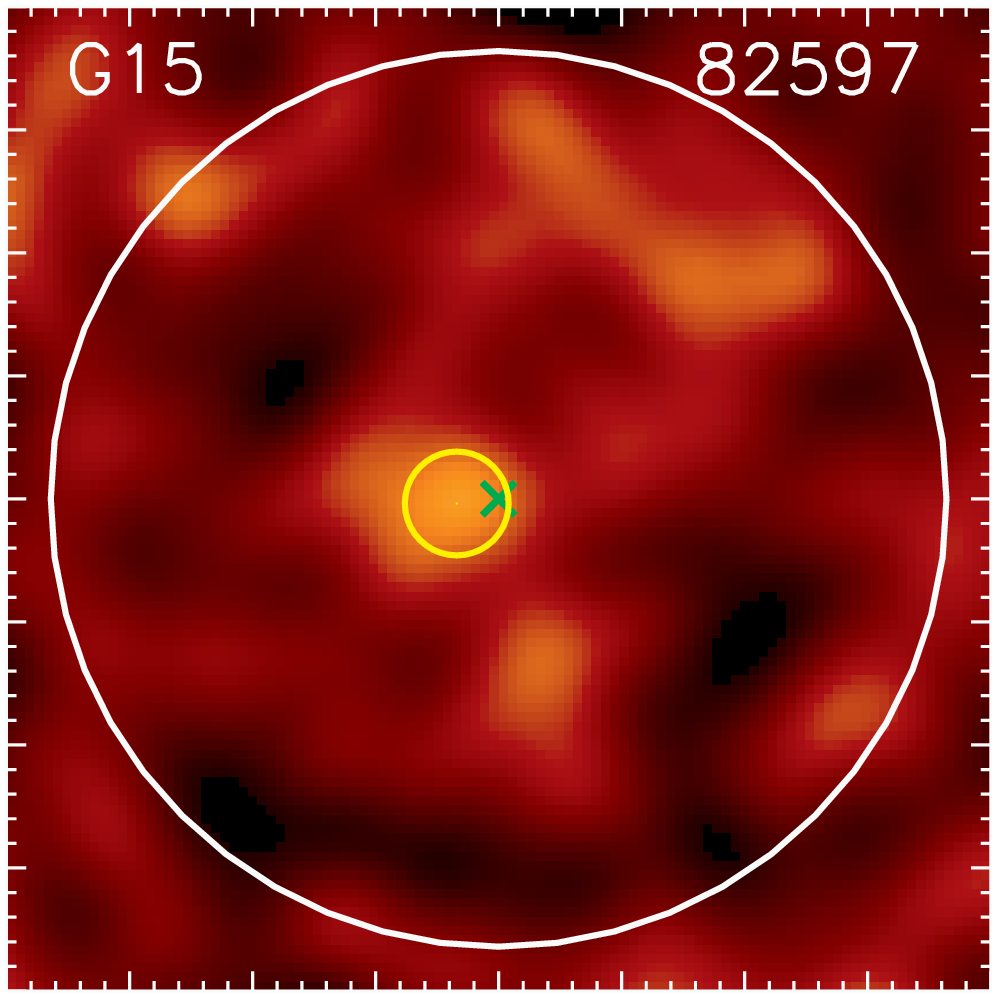}
\includegraphics[totalheight=4.25cm]{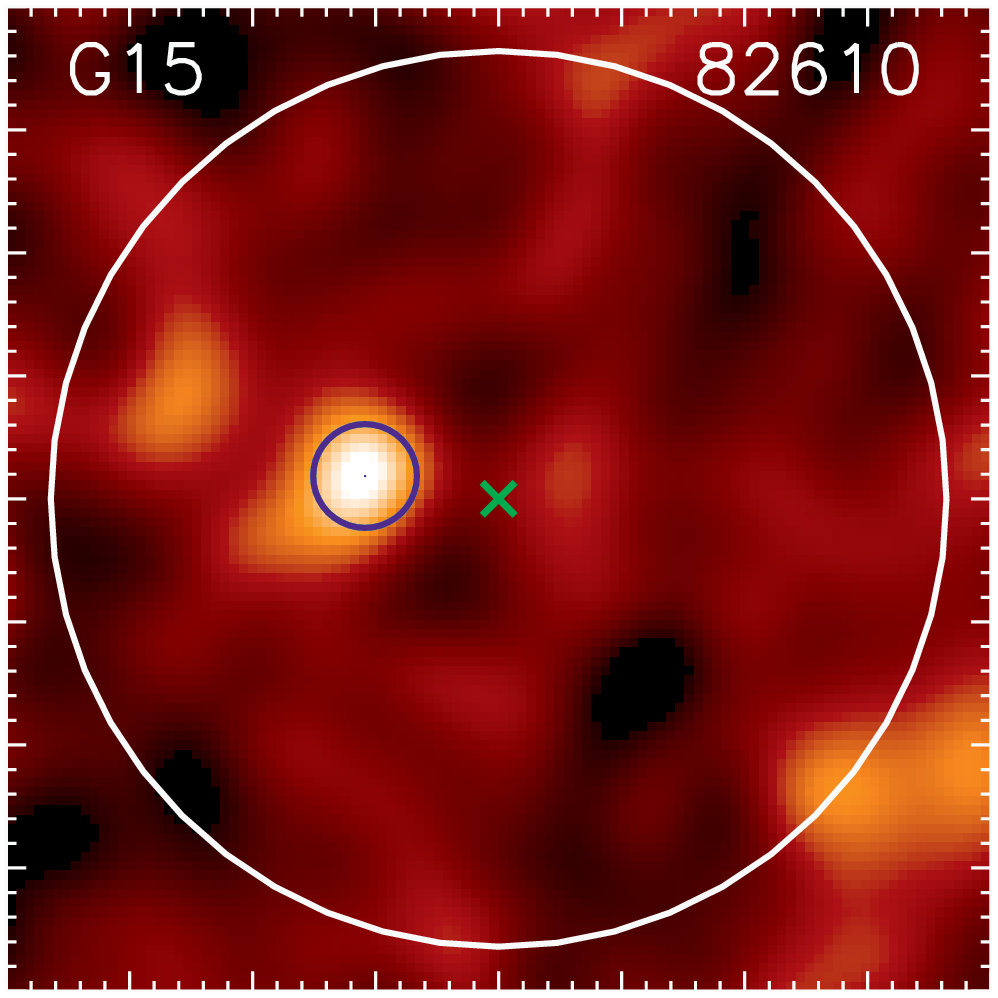} \\
\includegraphics[totalheight=4.25cm]{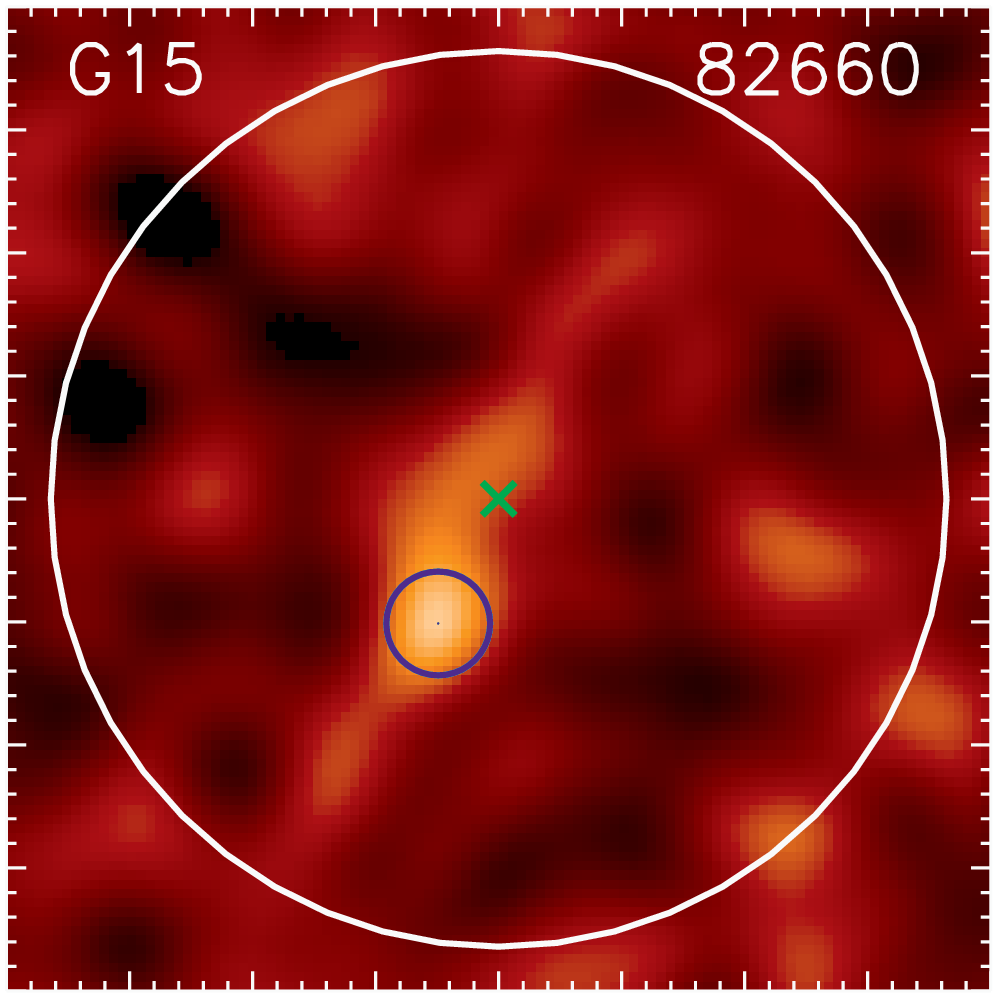}
\includegraphics[totalheight=4.25cm]{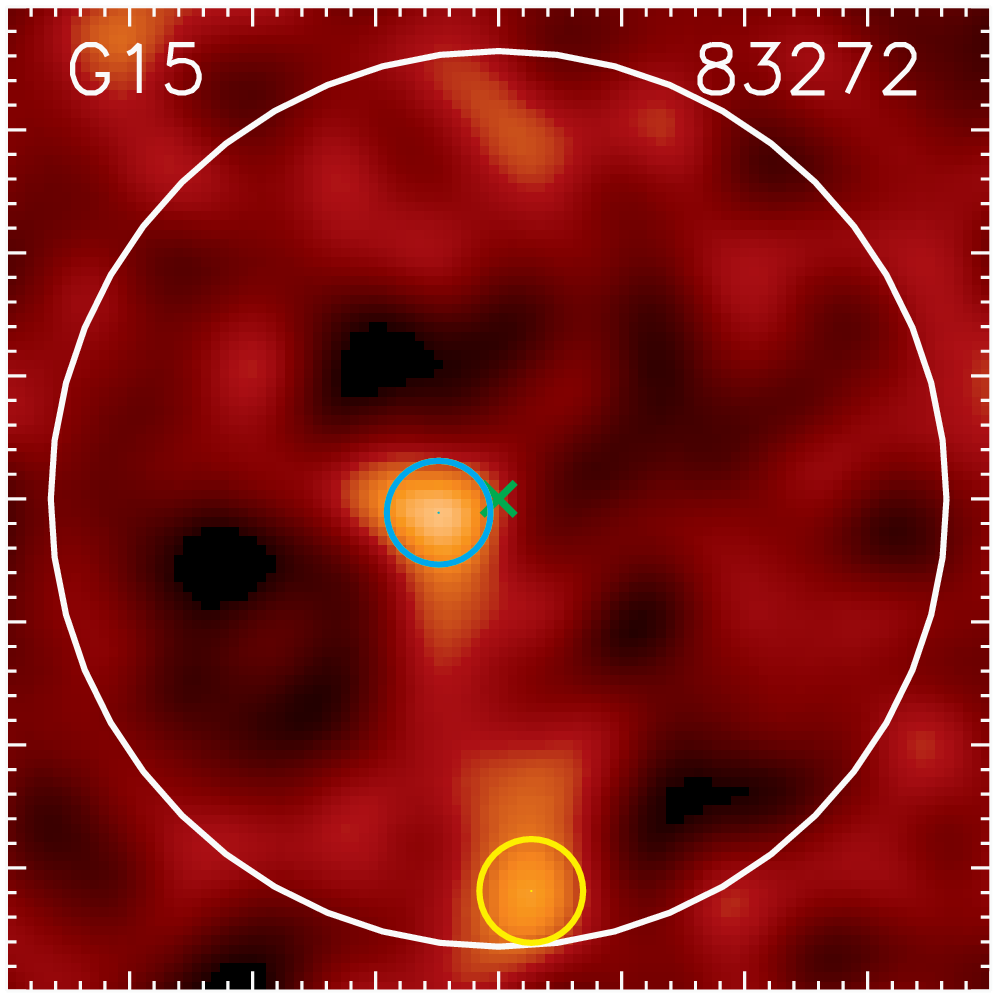}
\includegraphics[totalheight=4.25cm]{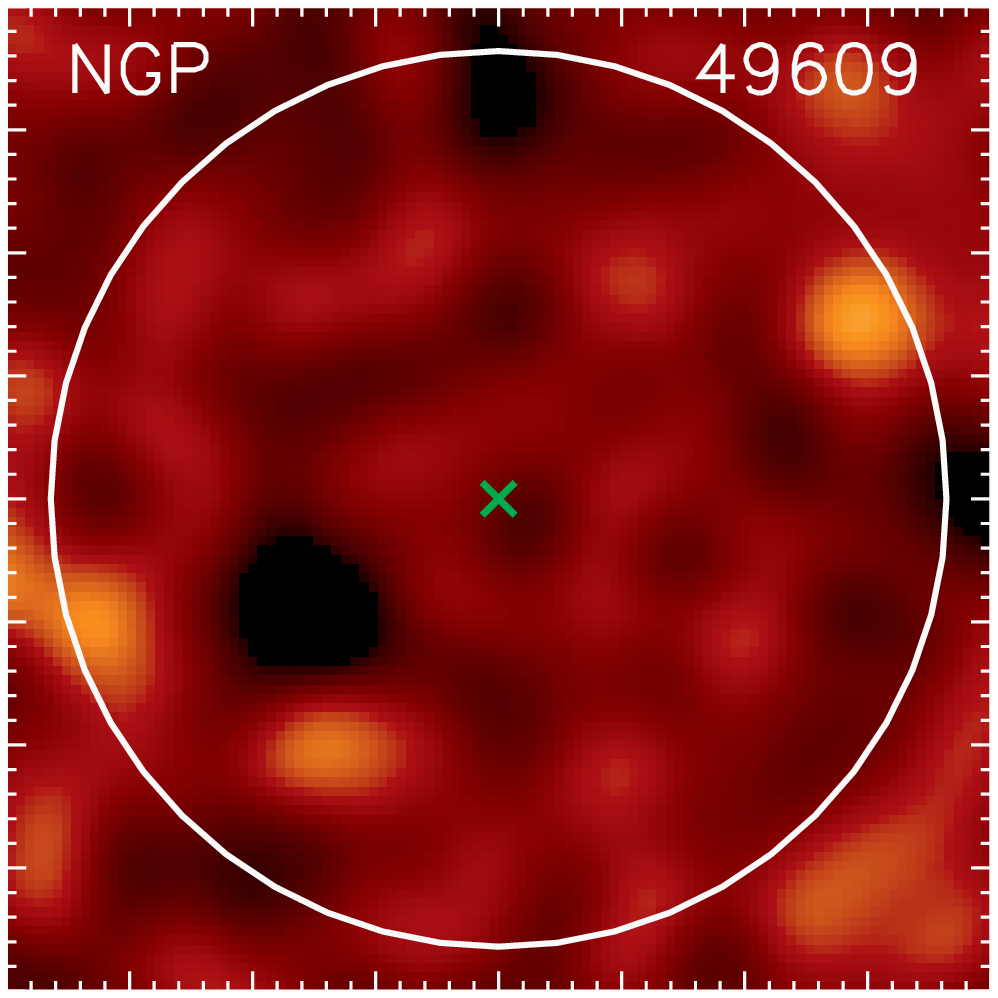}
\includegraphics[totalheight=4.25cm]{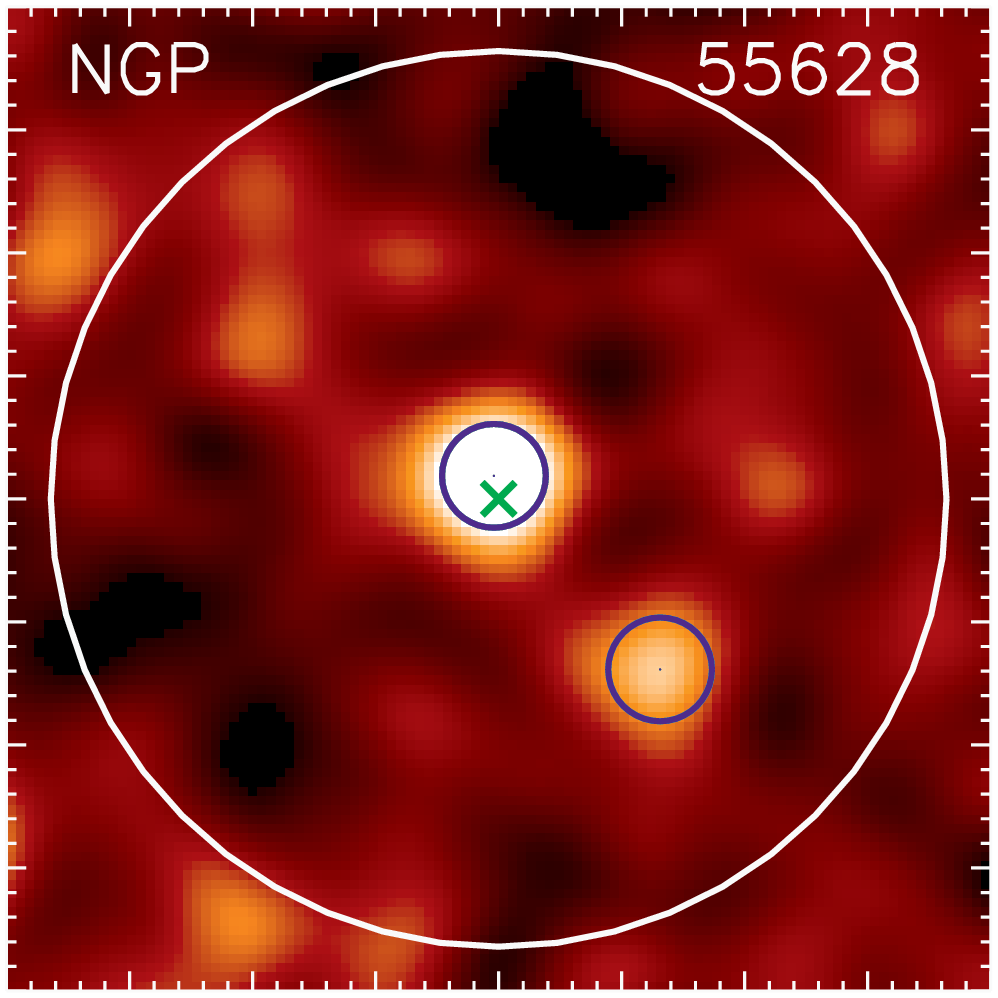} \\
\includegraphics[totalheight=4.25cm]{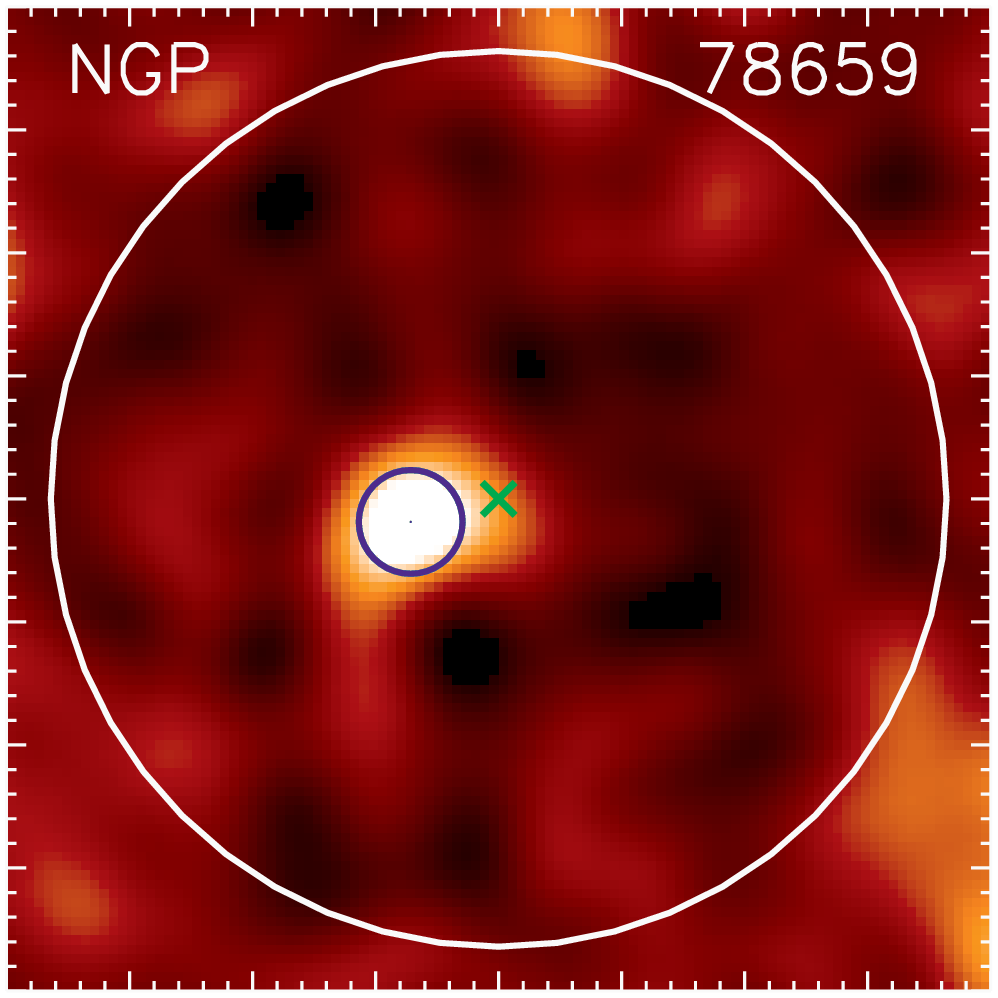}
\includegraphics[totalheight=4.25cm]{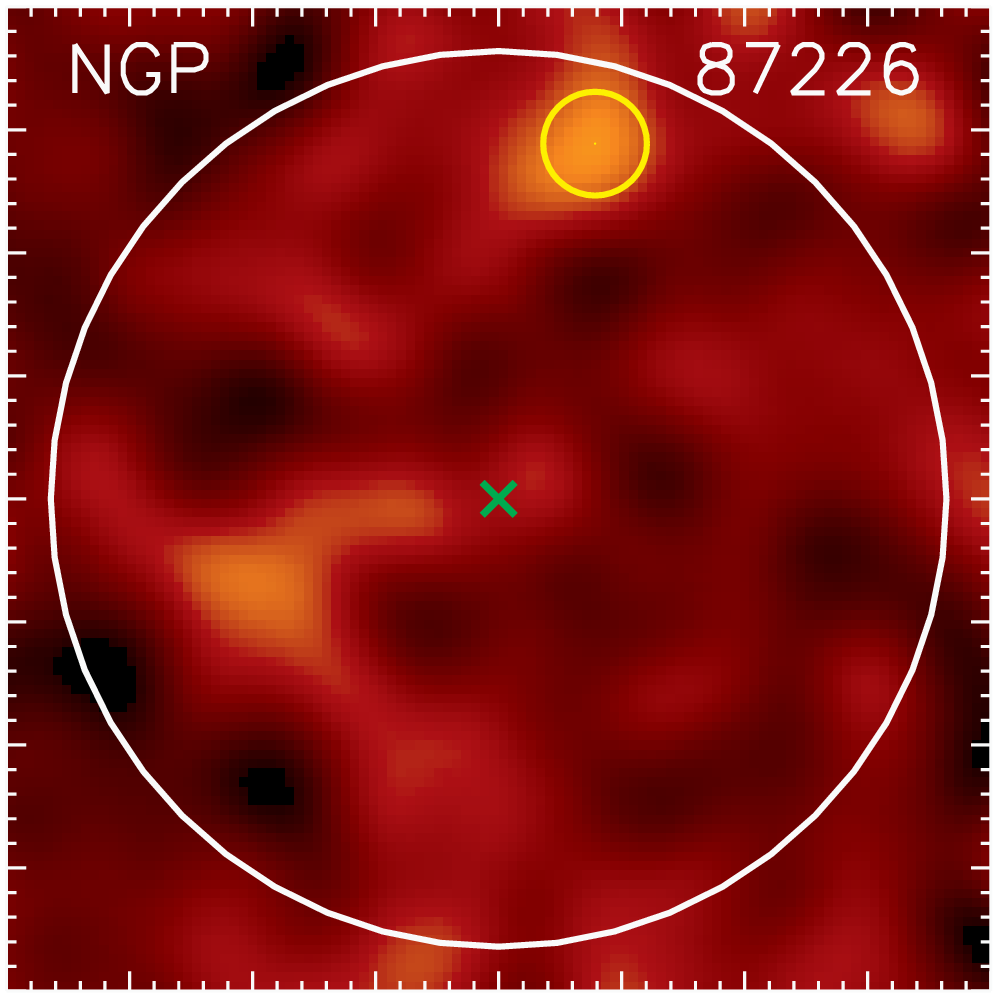}
\includegraphics[totalheight=4.25cm]{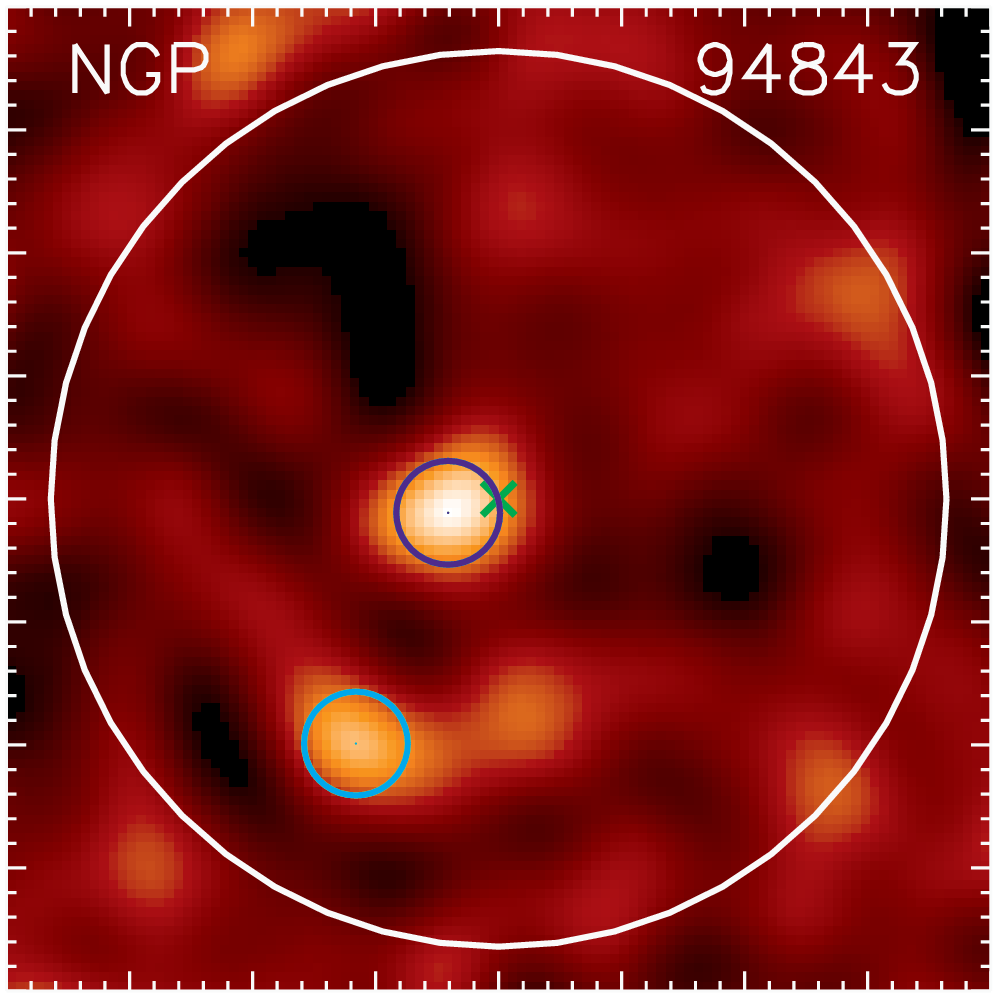}
\includegraphics[totalheight=4.25cm]{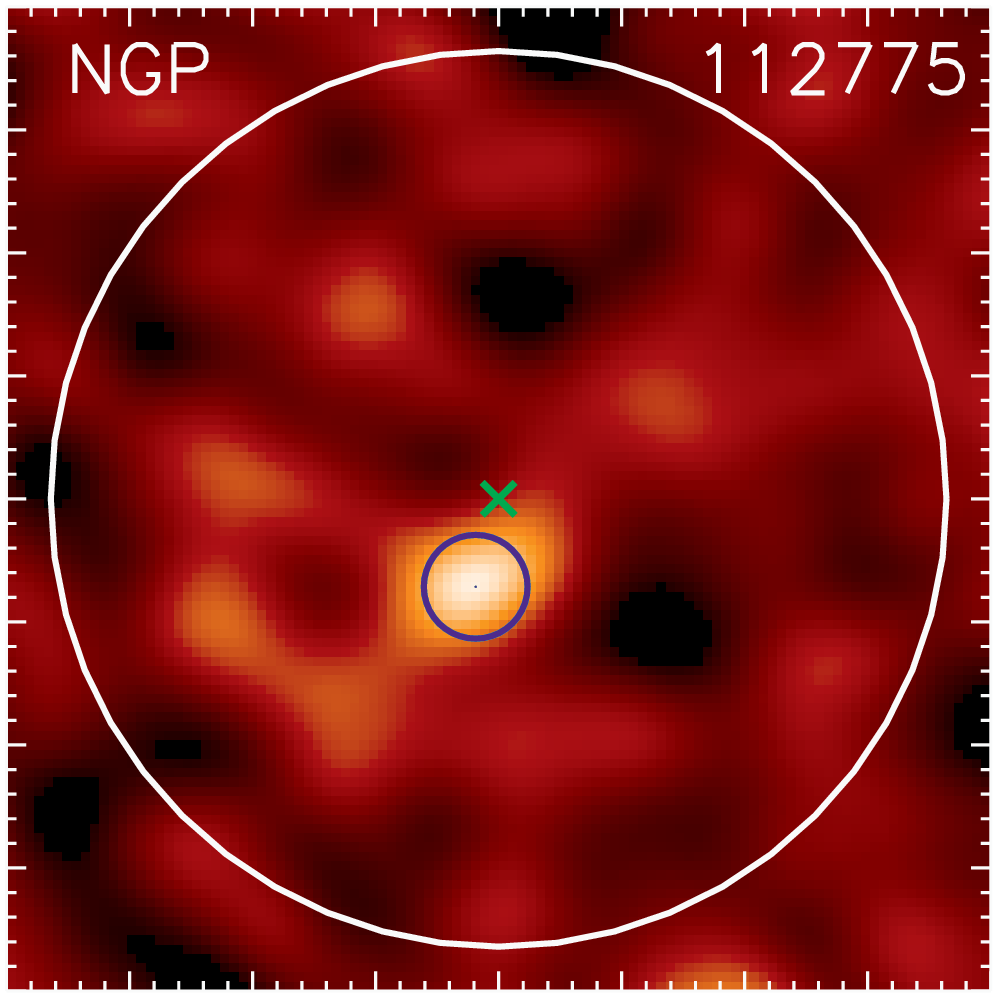} \\
\centering
\contcaption{ }
\end{center}
\end{figure*}
\begin{figure*}\hspace{-0.6cm}
\begin{center}
\includegraphics[totalheight=4.25cm]{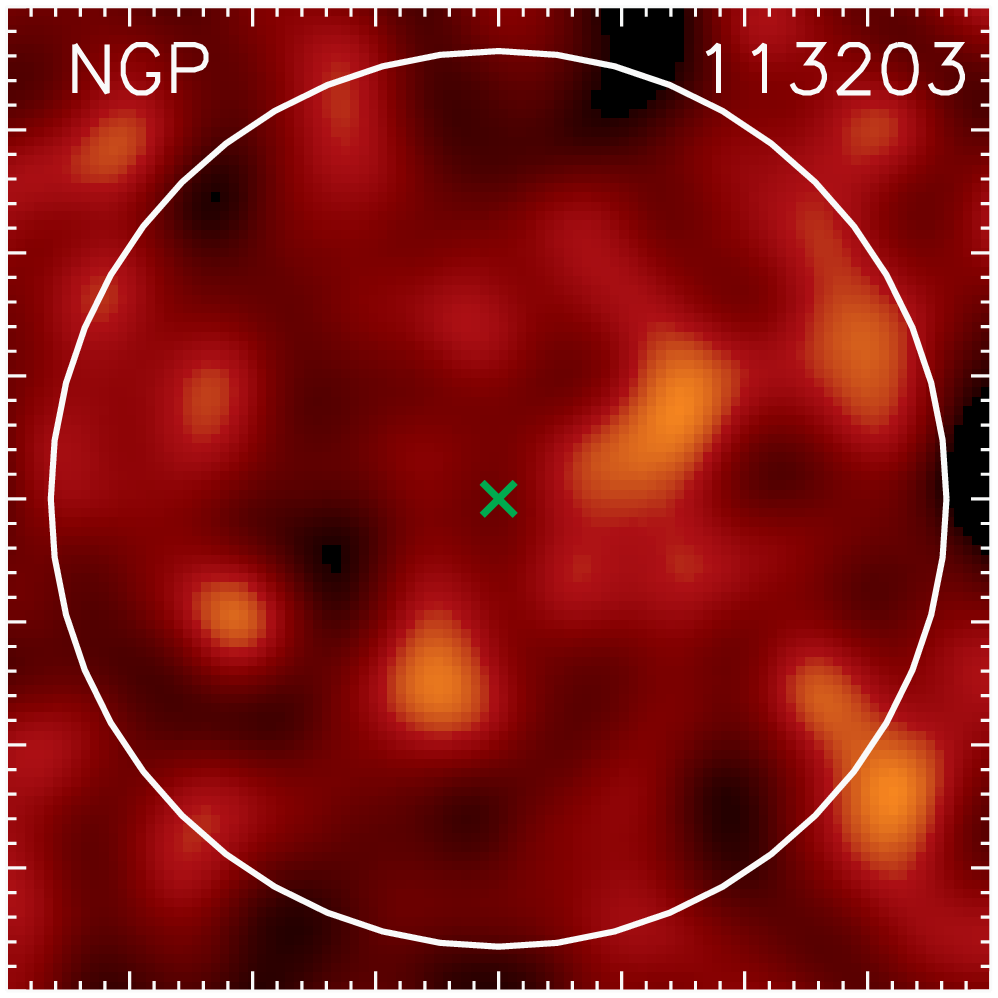}
\includegraphics[totalheight=4.25cm]{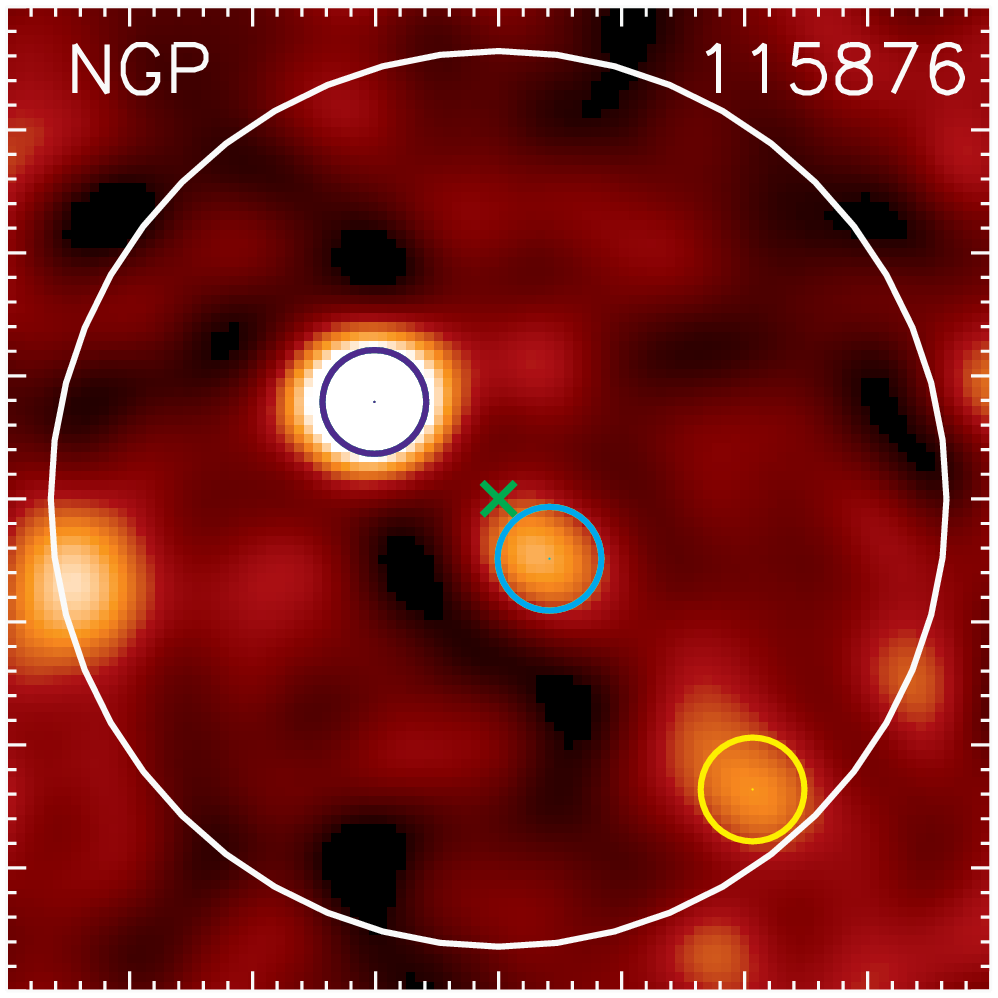}
\includegraphics[totalheight=4.25cm]{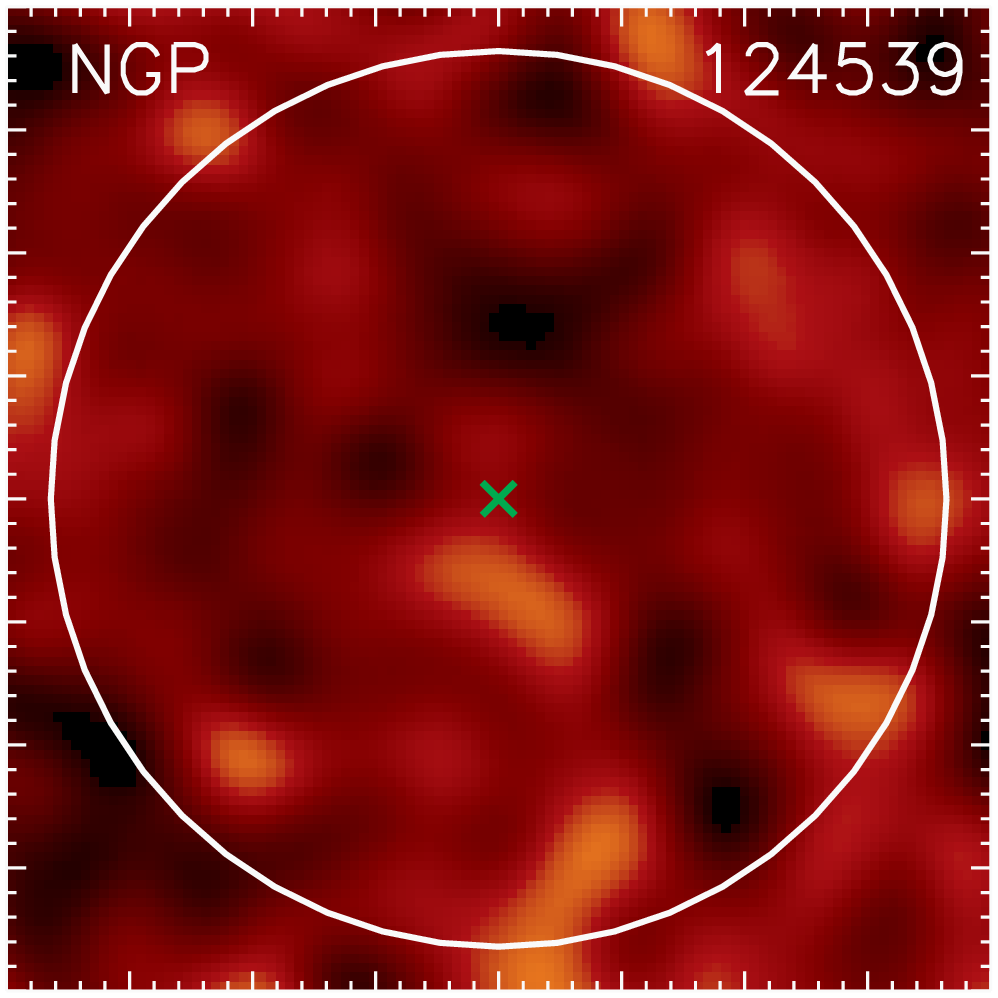}
\includegraphics[totalheight=4.25cm]{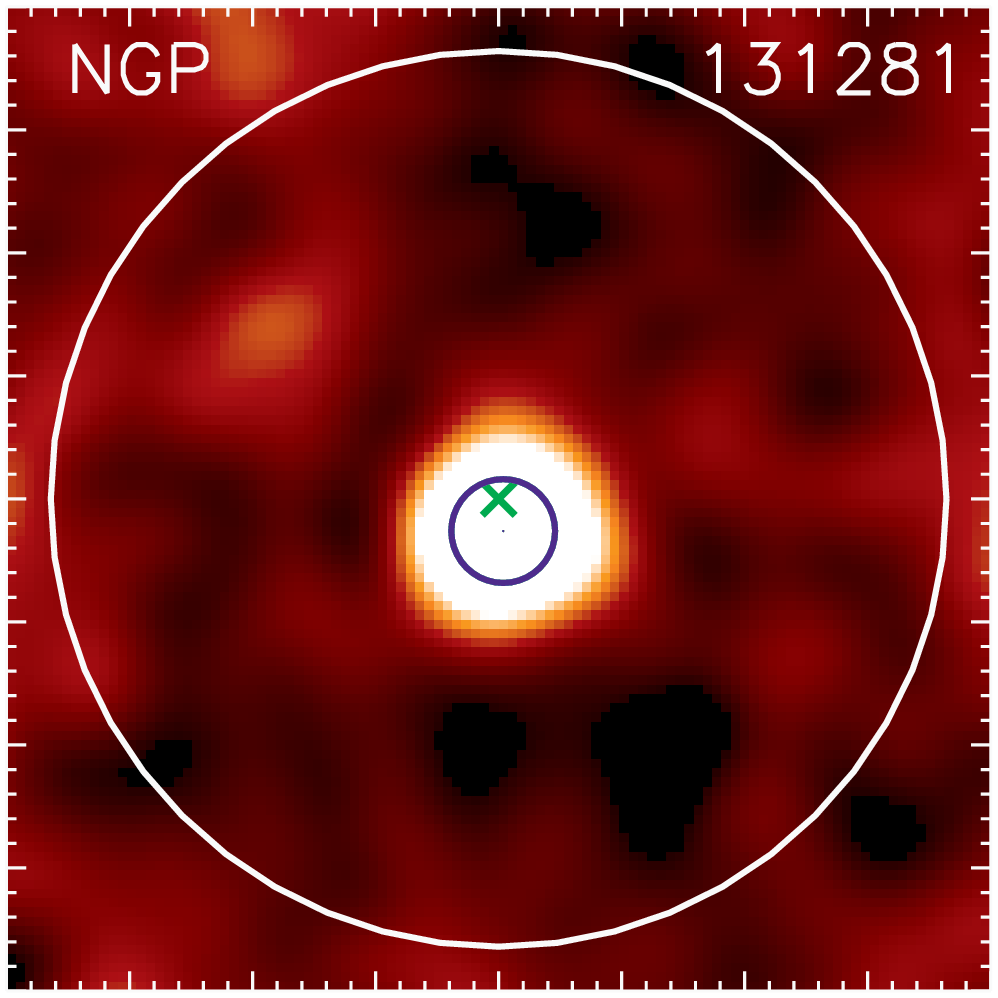} \\
\includegraphics[totalheight=4.25cm]{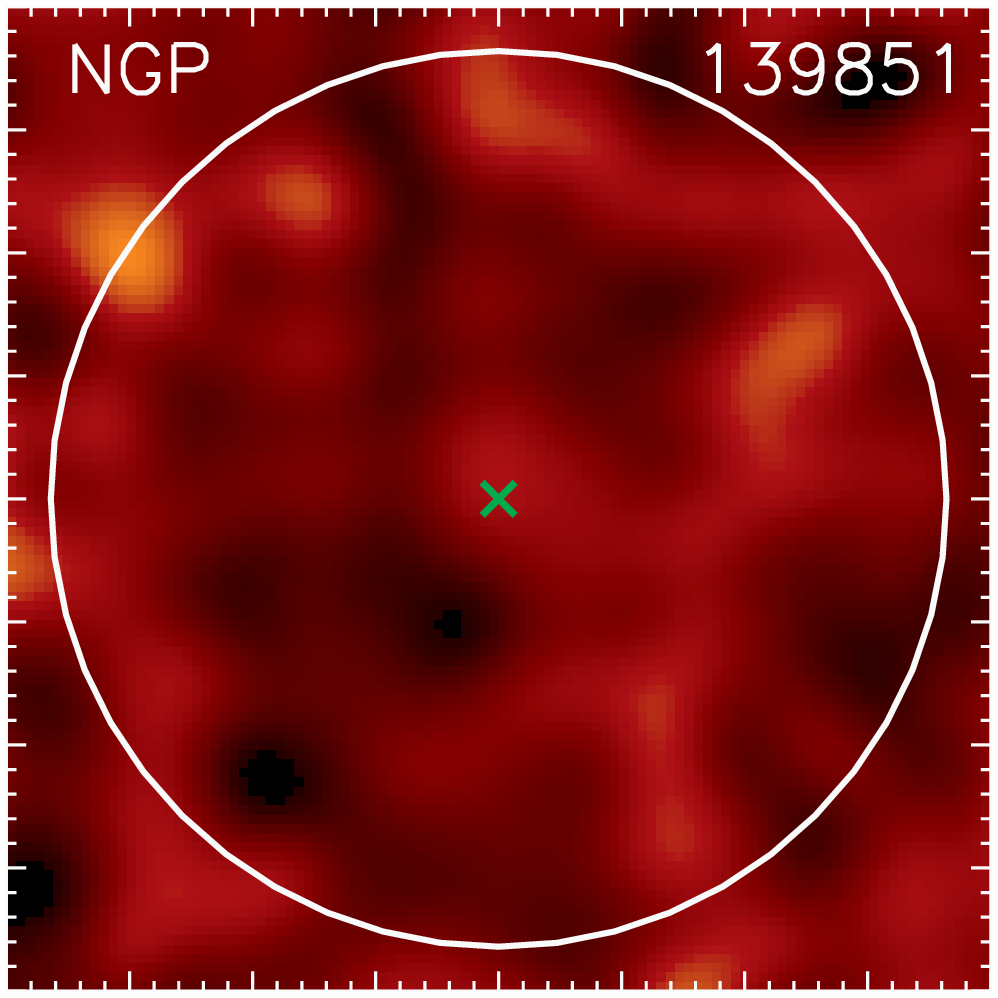}
\includegraphics[totalheight=4.25cm]{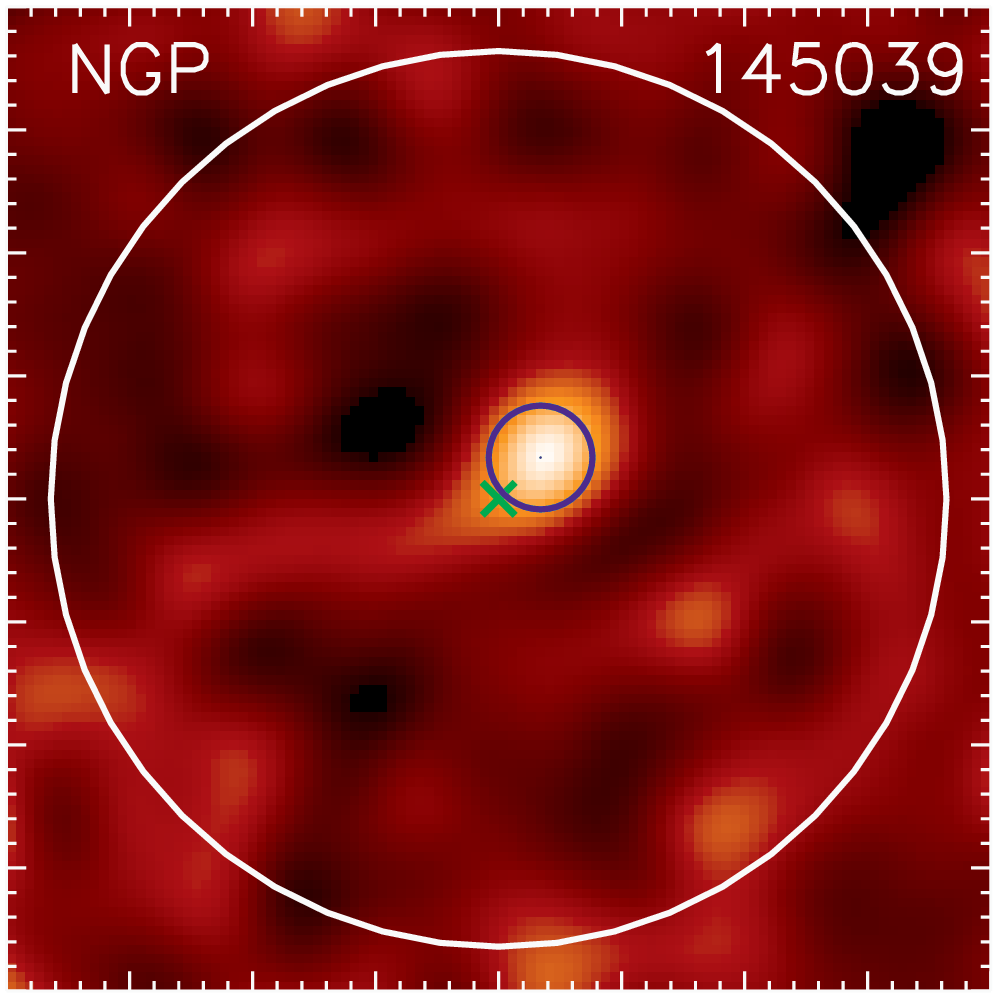}
\includegraphics[totalheight=4.25cm]{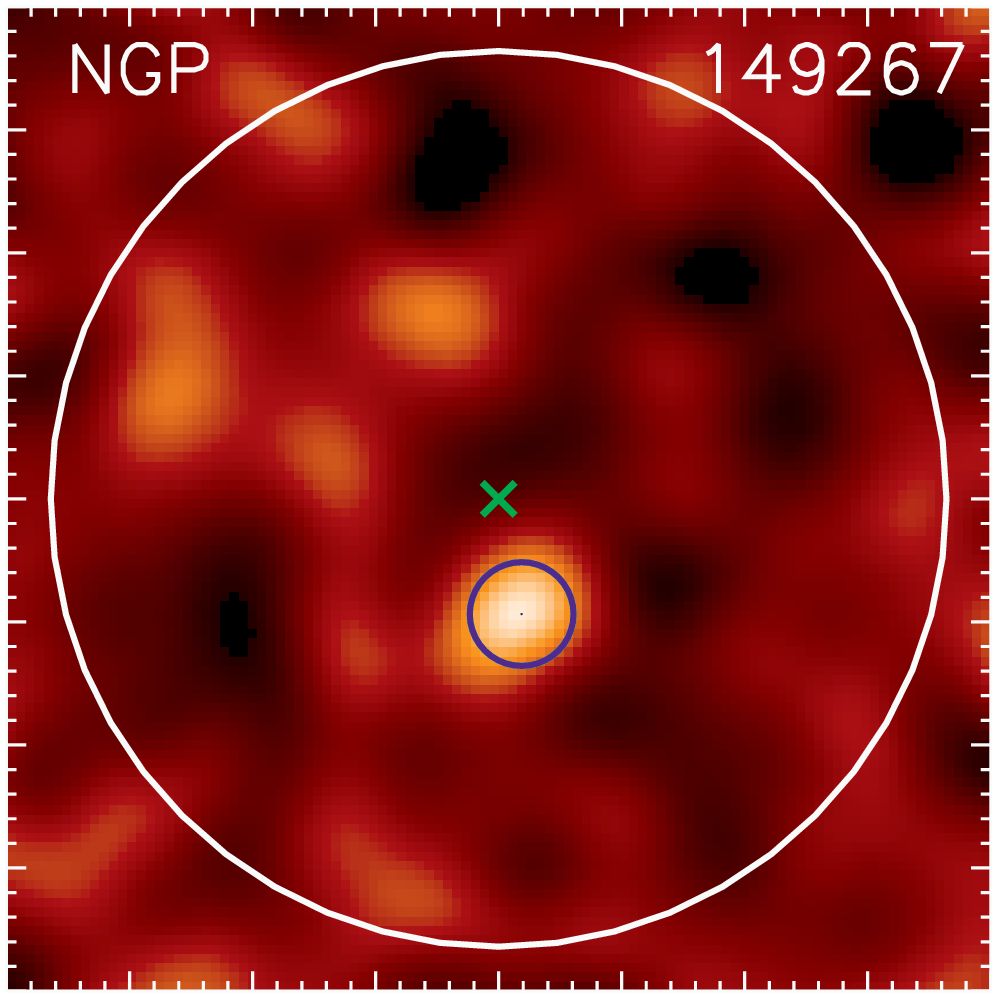}
\includegraphics[totalheight=4.25cm]{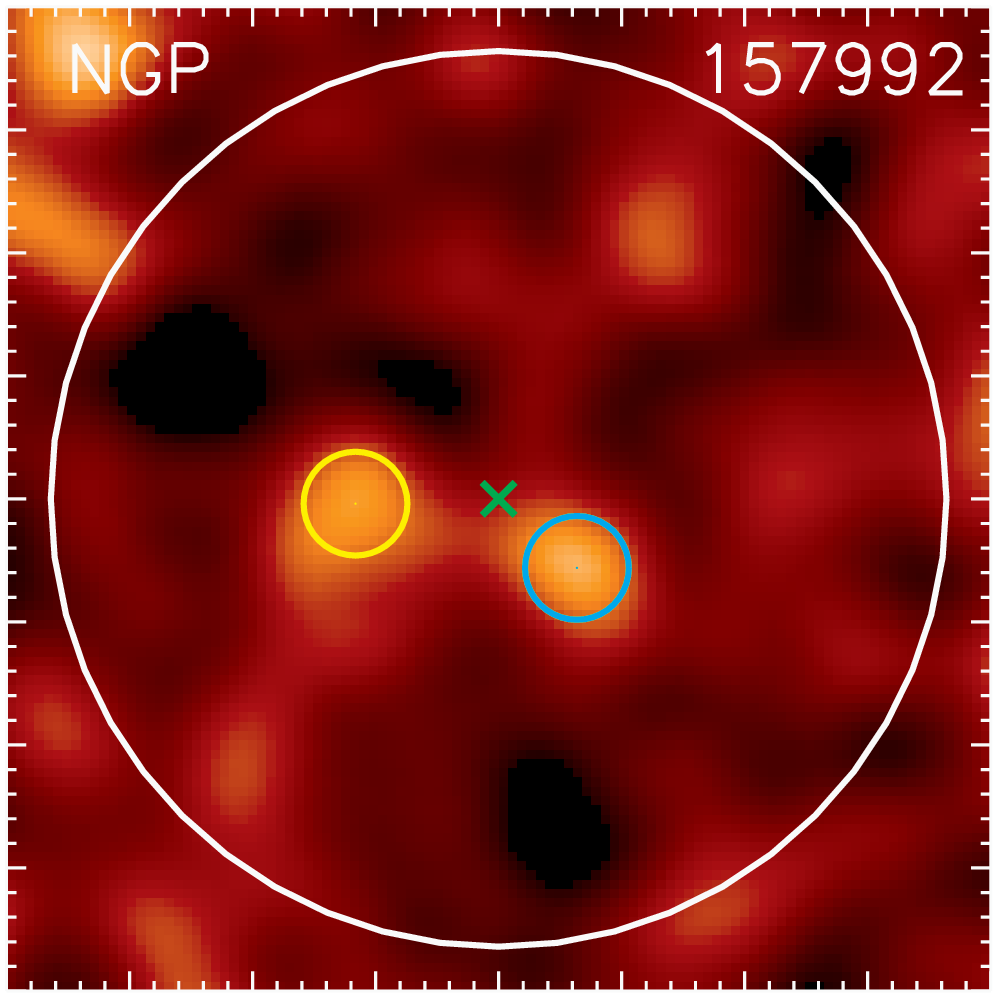} \\
\includegraphics[totalheight=4.25cm]{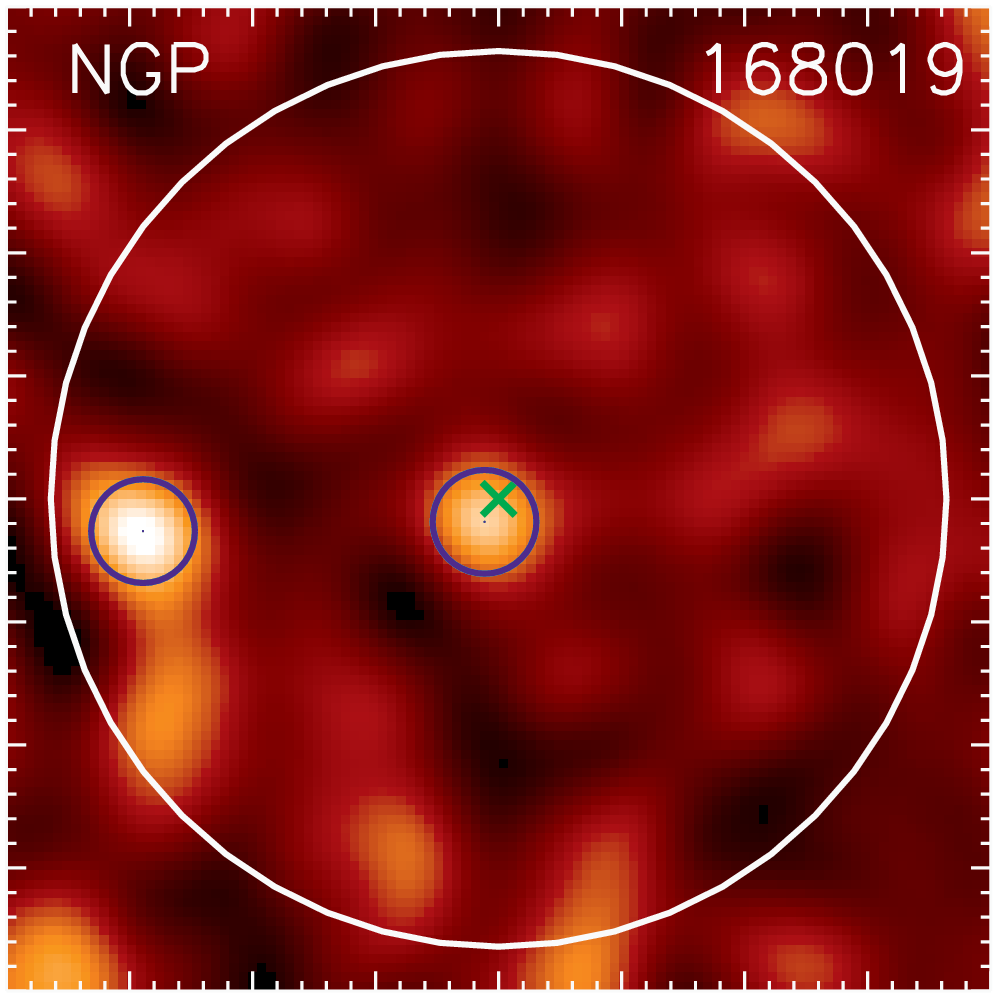}
\includegraphics[totalheight=4.25cm]{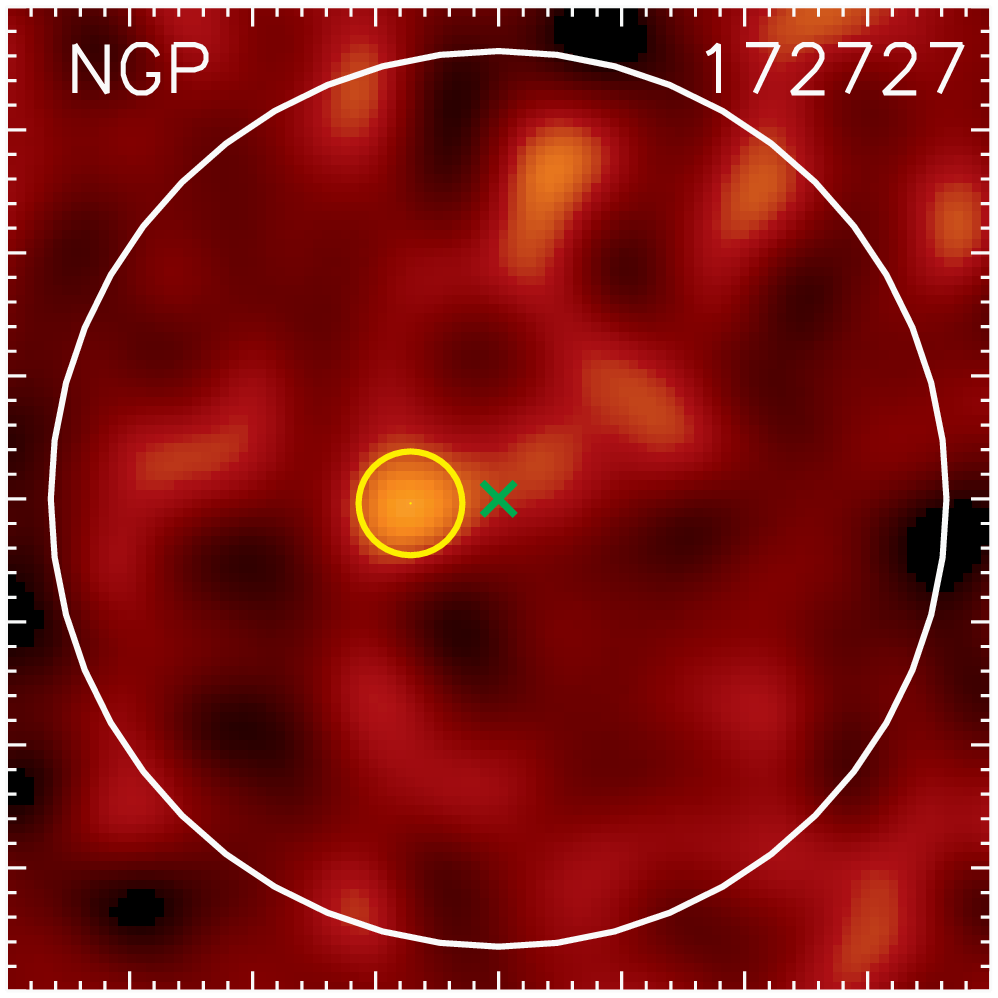}
\includegraphics[totalheight=4.25cm]{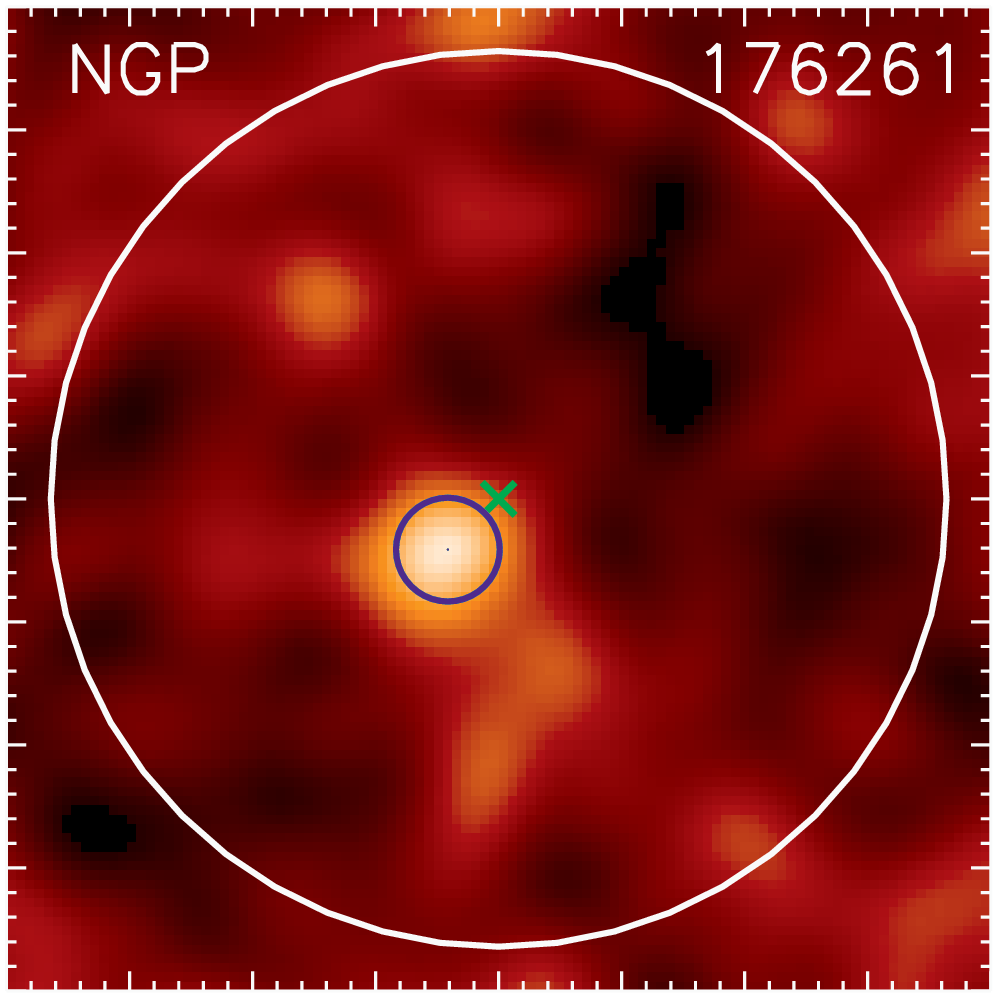}
\includegraphics[totalheight=4.25cm]{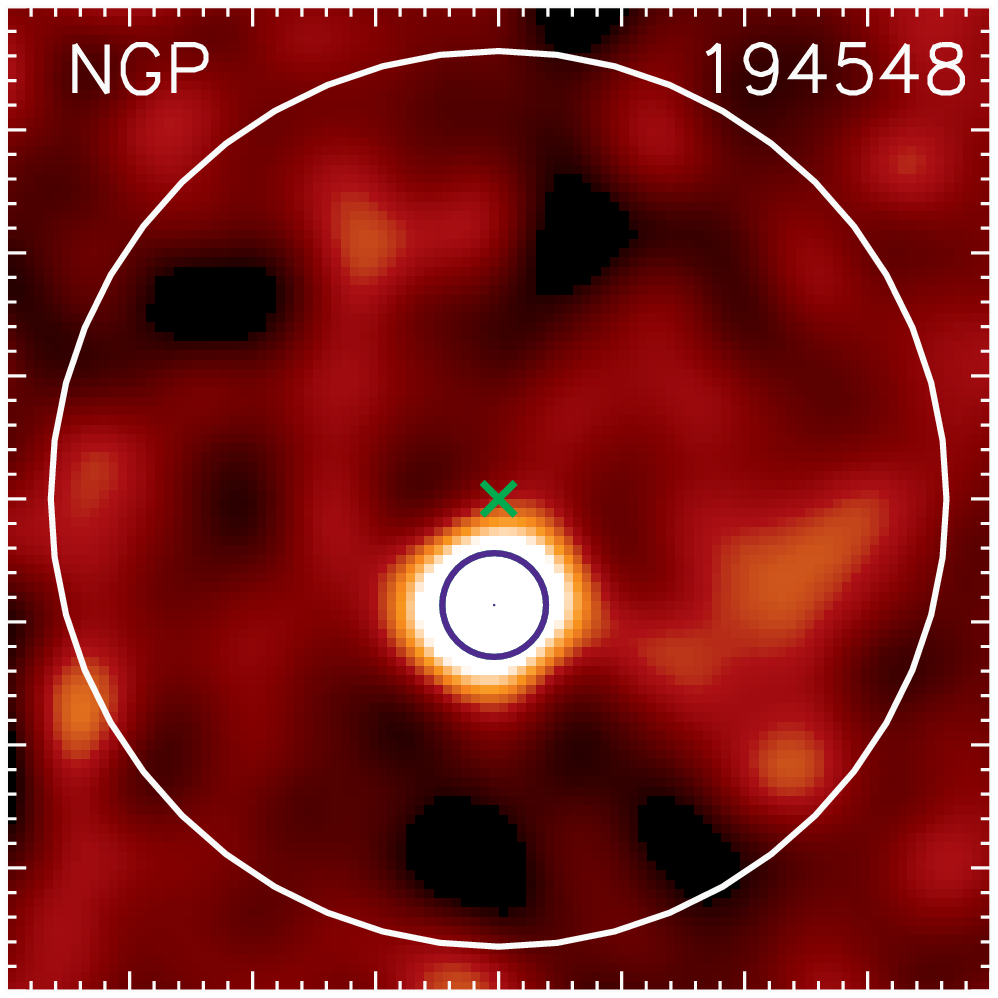} \\
\includegraphics[totalheight=4.25cm]{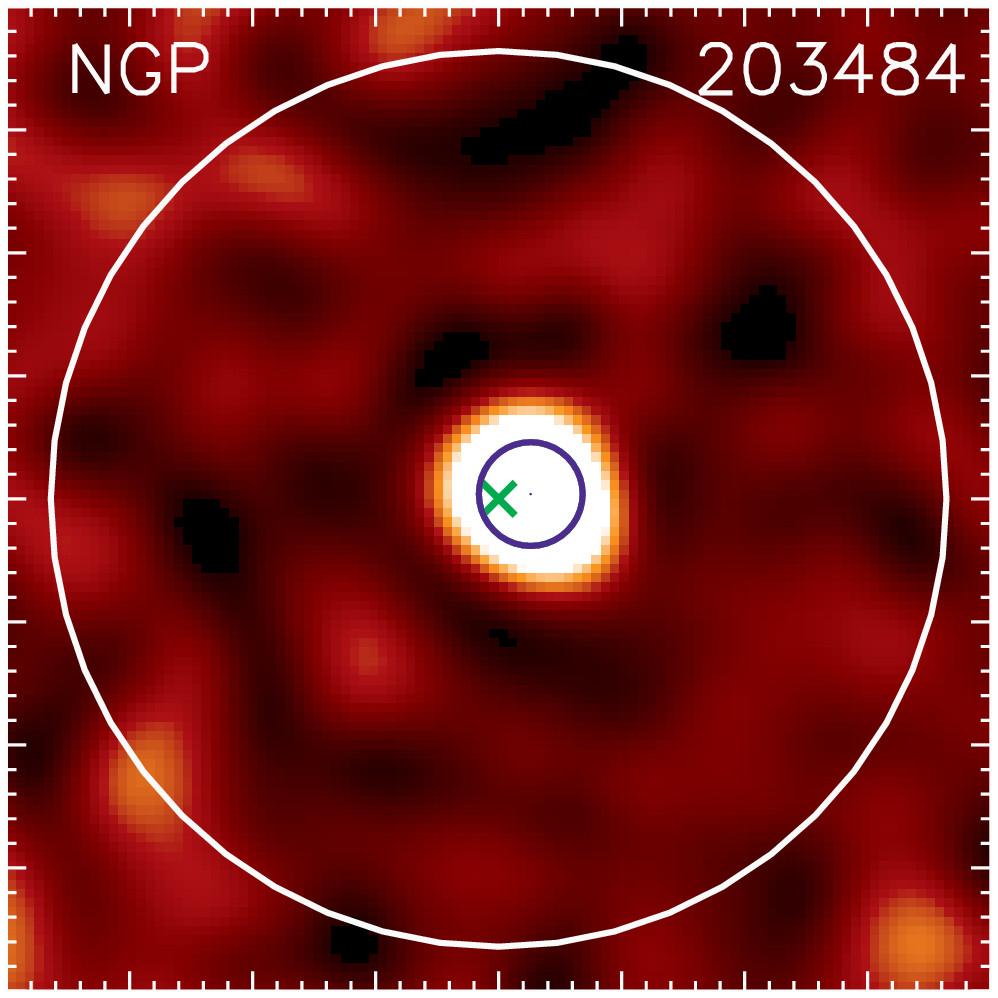}
\includegraphics[totalheight=4.25cm]{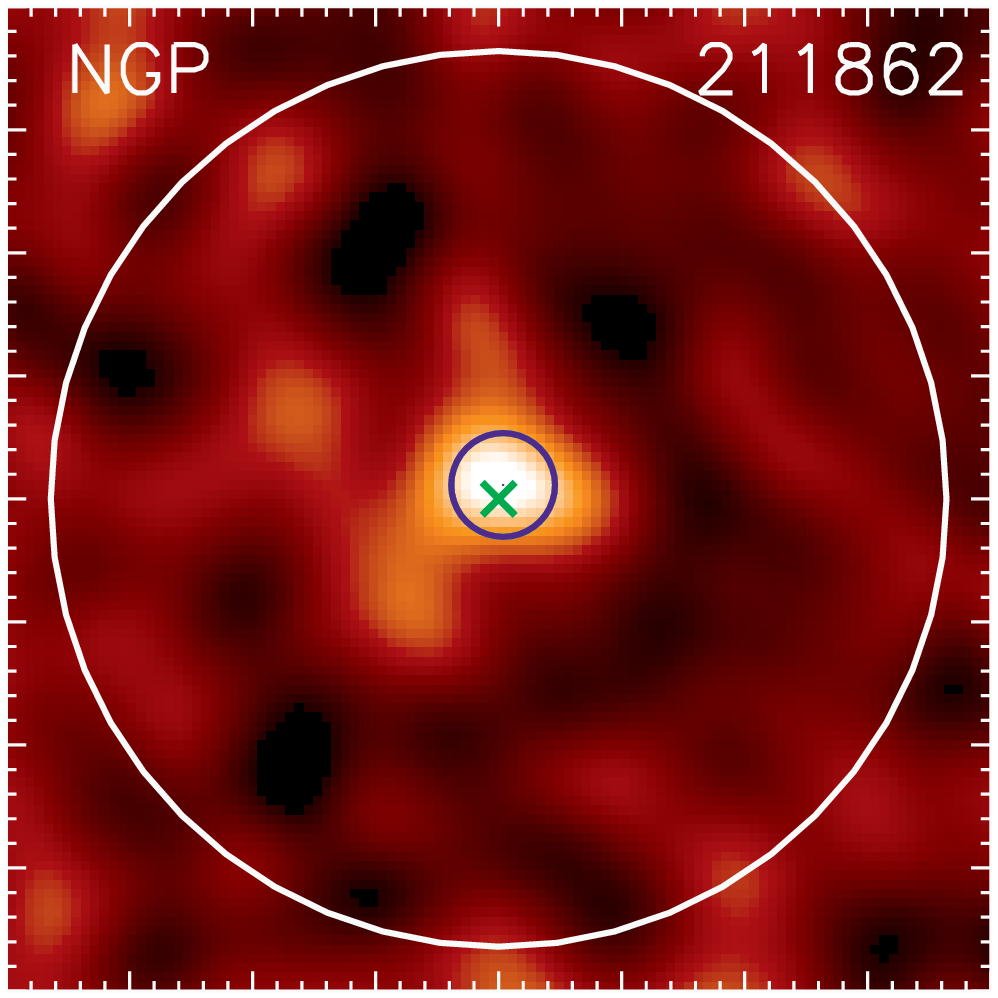}
\includegraphics[totalheight=4.25cm]{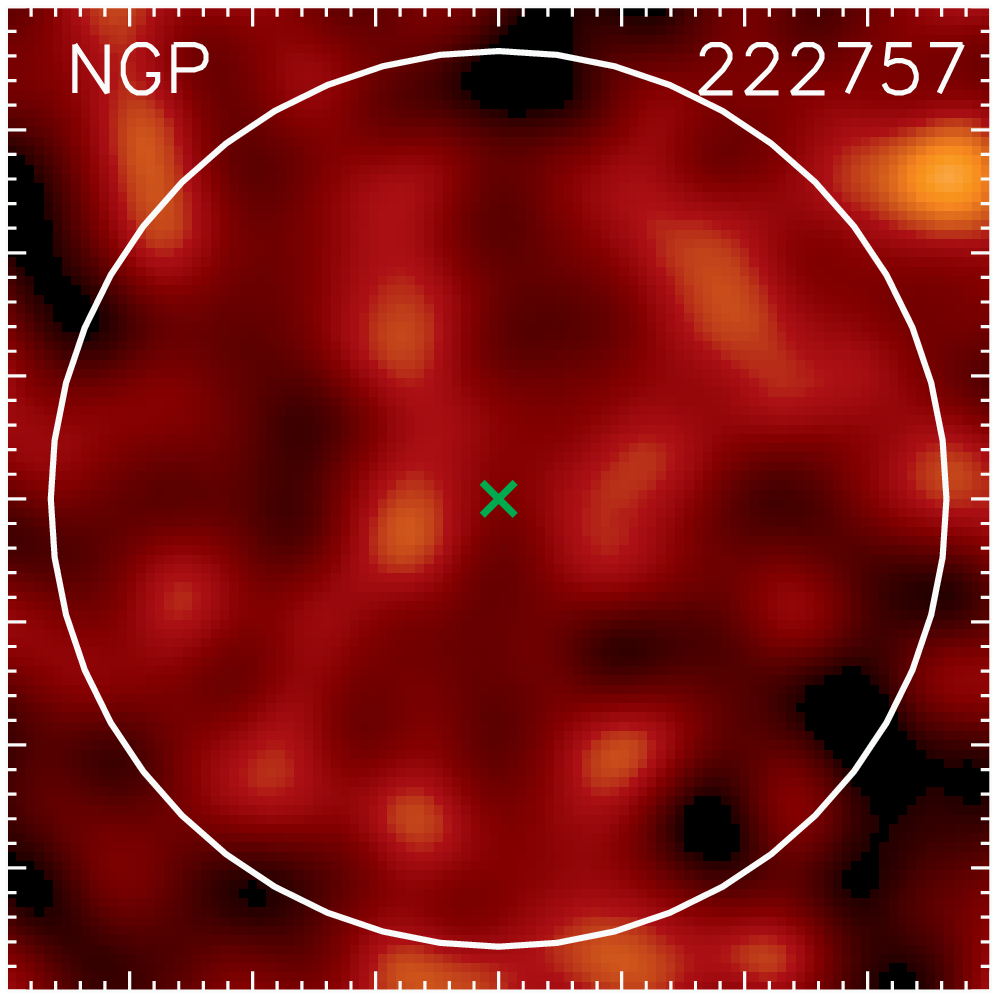}
\includegraphics[totalheight=4.25cm]{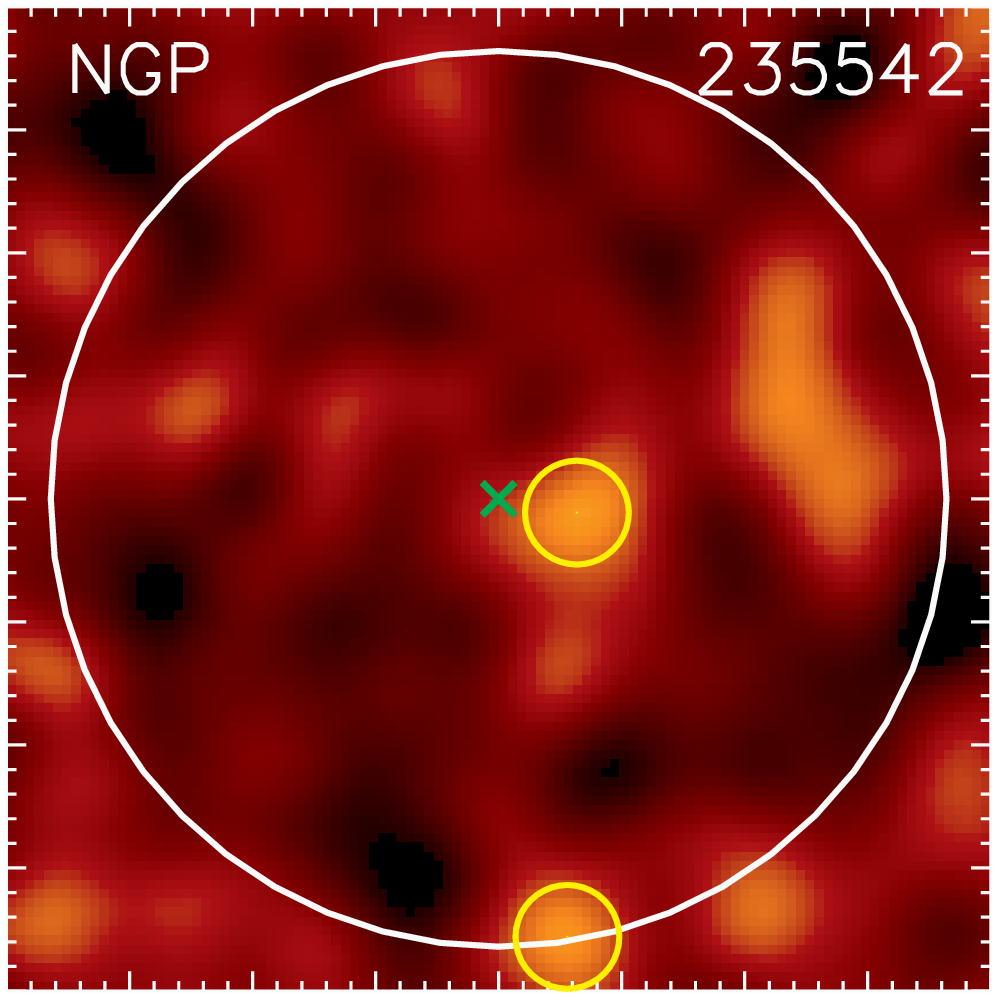} \\
\includegraphics[totalheight=4.25cm]{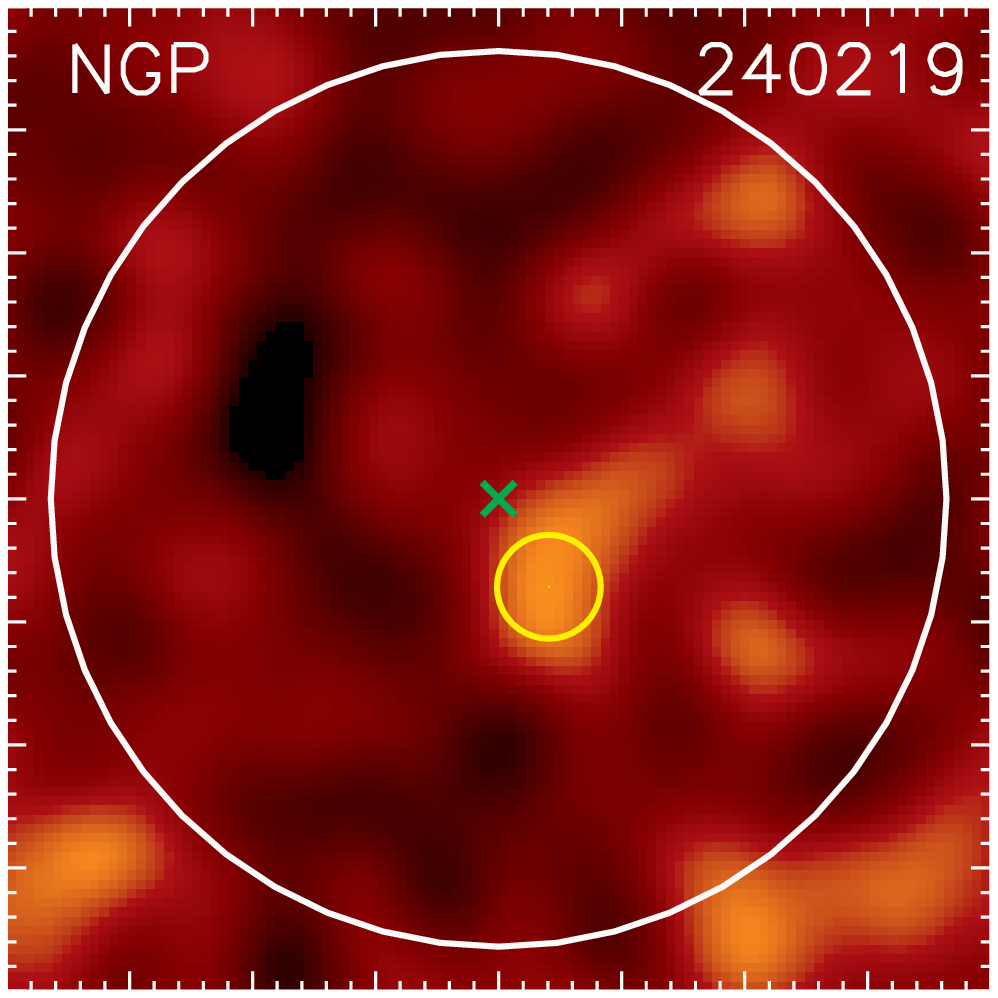}
\includegraphics[totalheight=4.25cm]{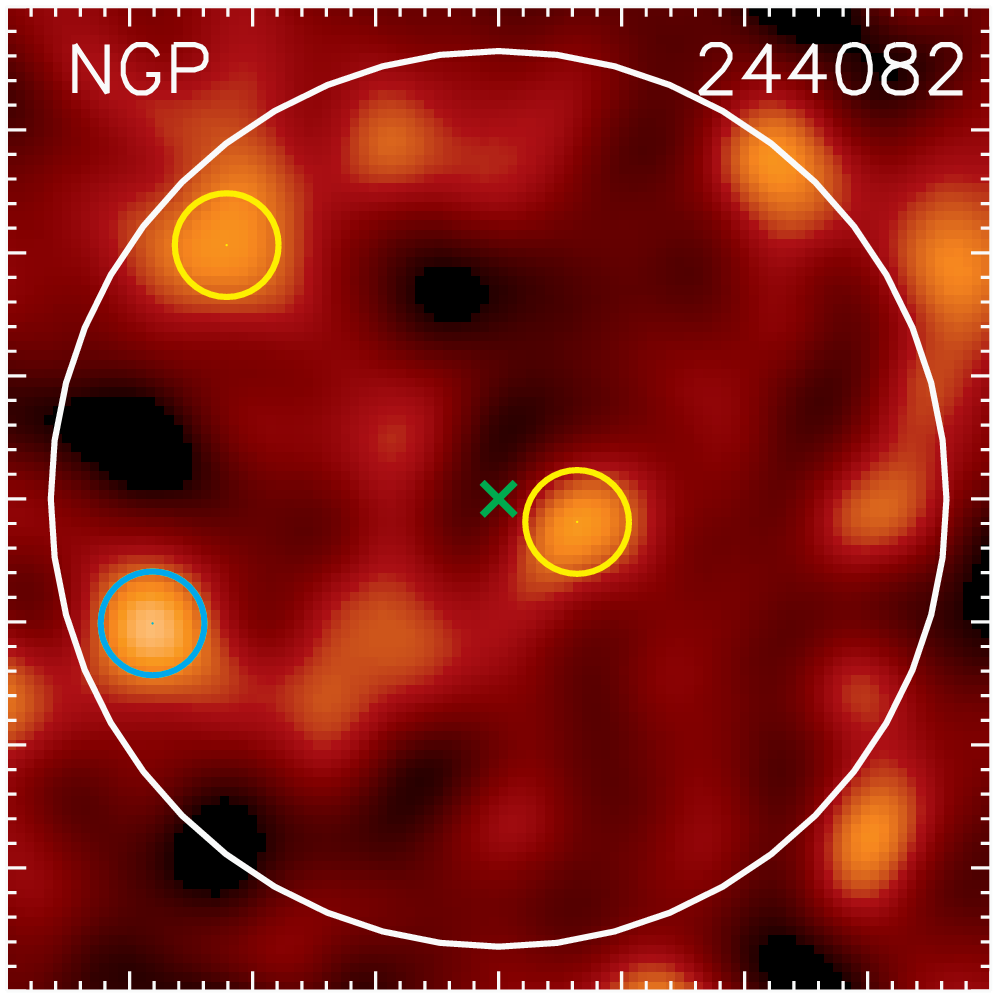}
\includegraphics[totalheight=4.25cm]{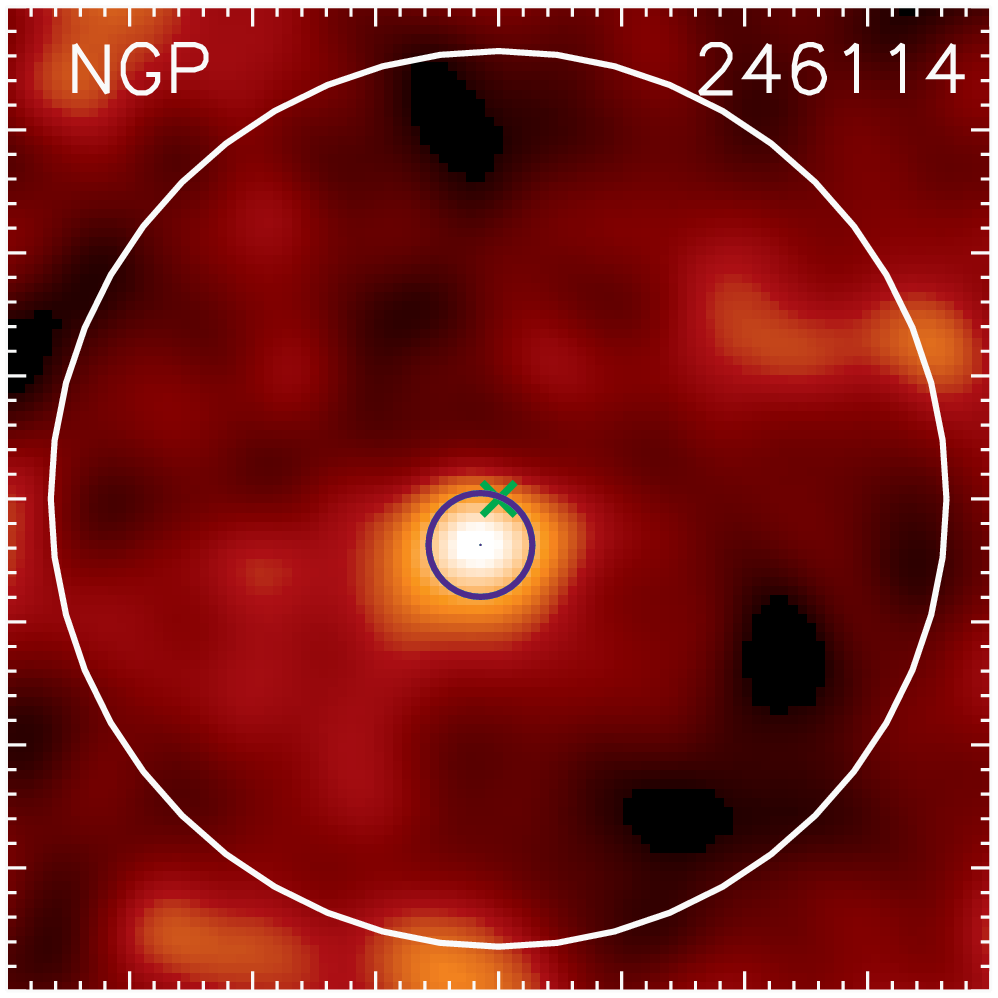}
\includegraphics[totalheight=4.25cm]{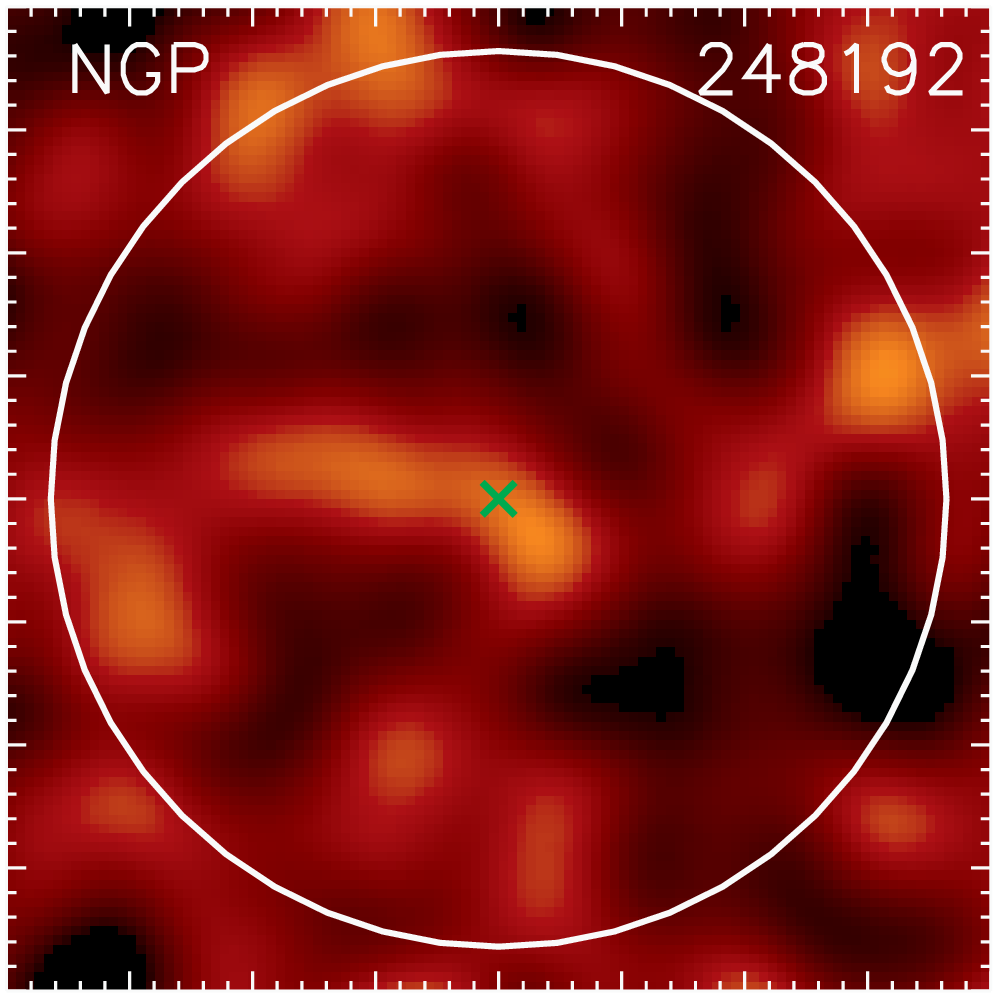} \\
\centering
\contcaption{ }
\end{center}
\end{figure*}
\begin{figure*}\hspace{-0.6cm}
\begin{center}
\includegraphics[totalheight=4.25cm]{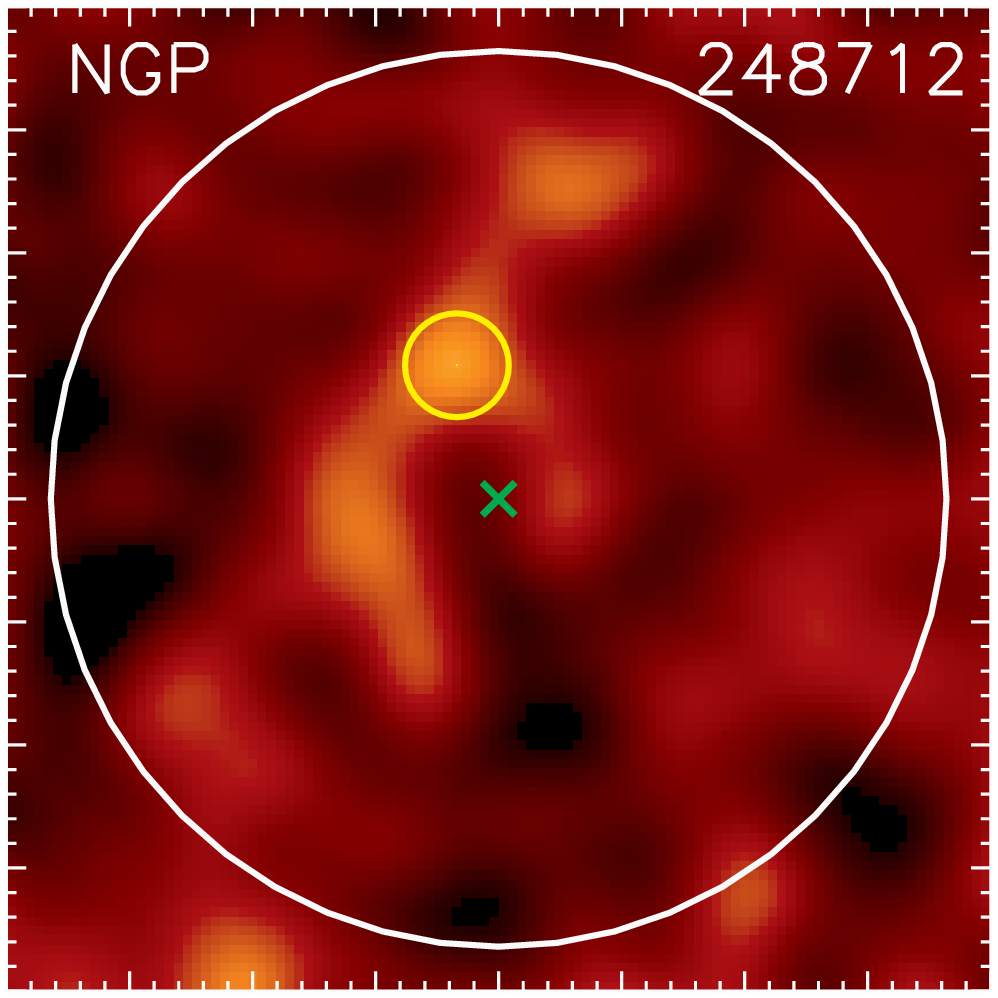}
\includegraphics[totalheight=4.25cm]{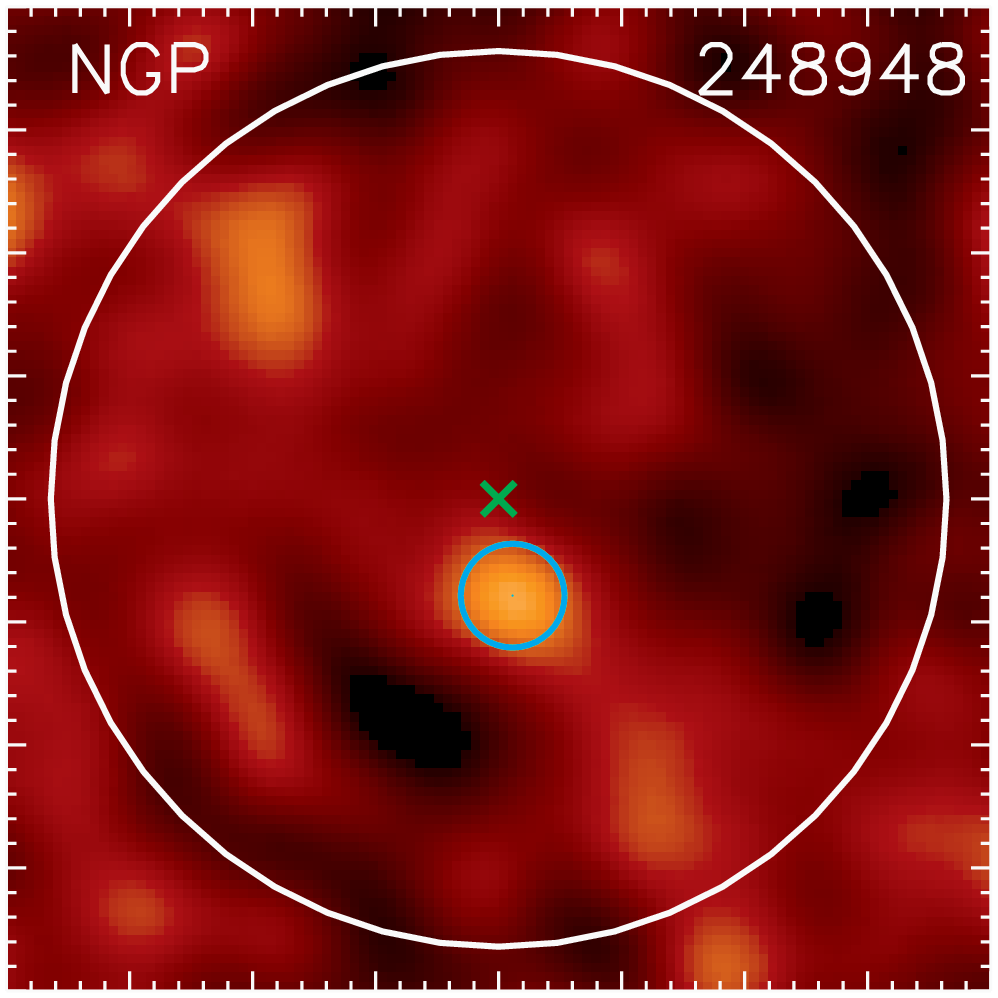}
\includegraphics[totalheight=4.25cm]{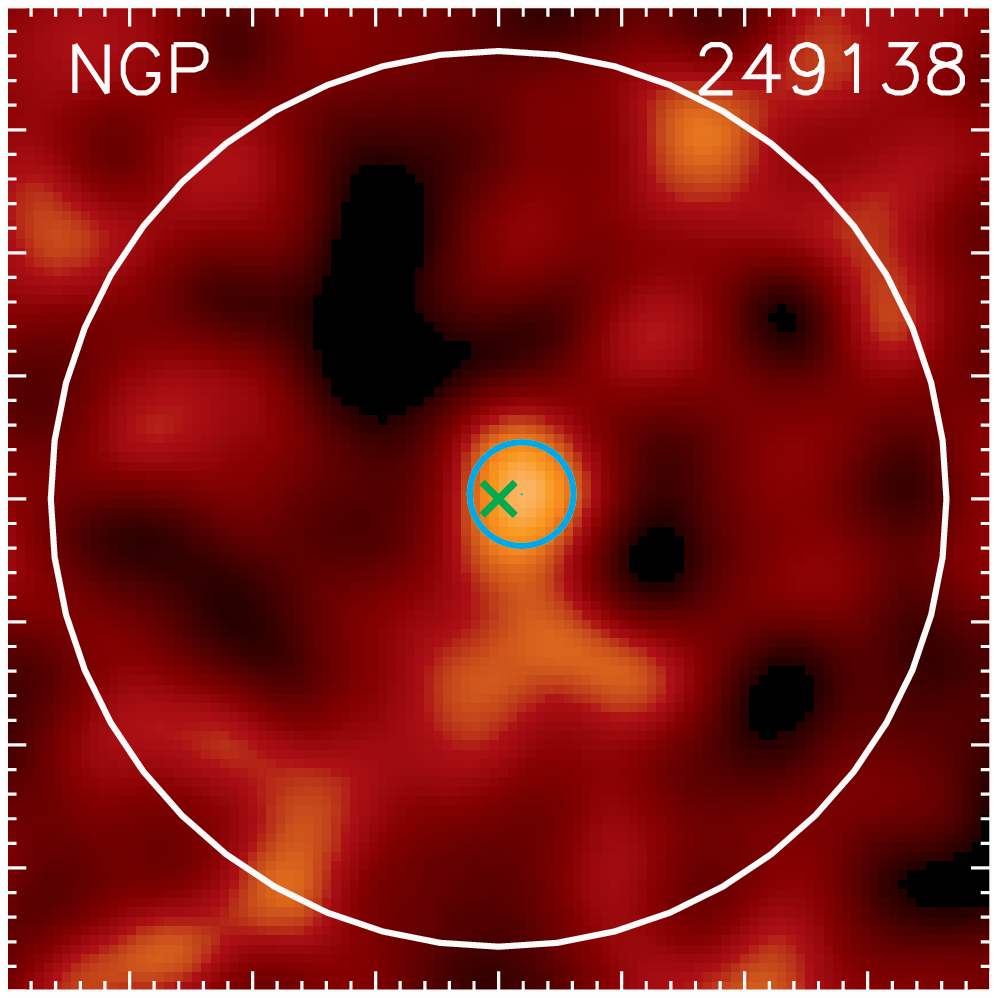}
\includegraphics[totalheight=4.25cm]{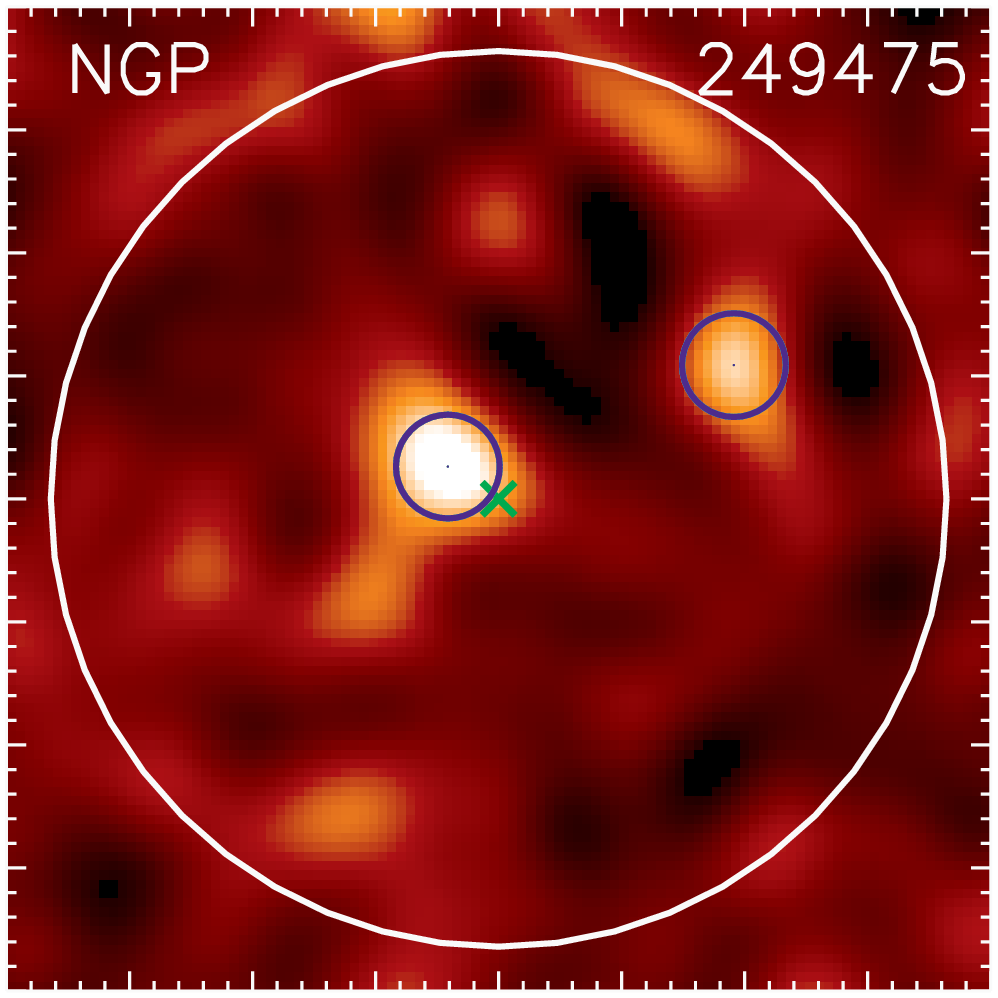} \\
\includegraphics[totalheight=4.25cm]{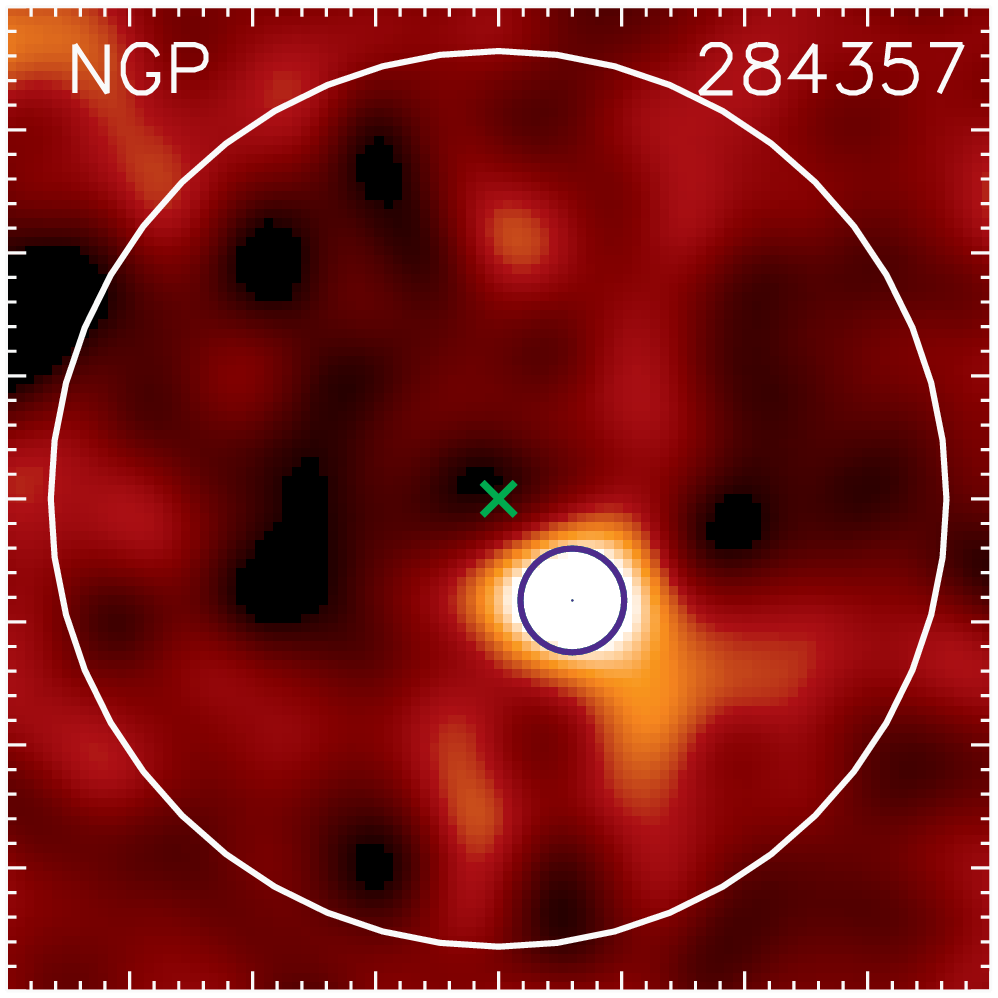}
\includegraphics[totalheight=4.25cm]{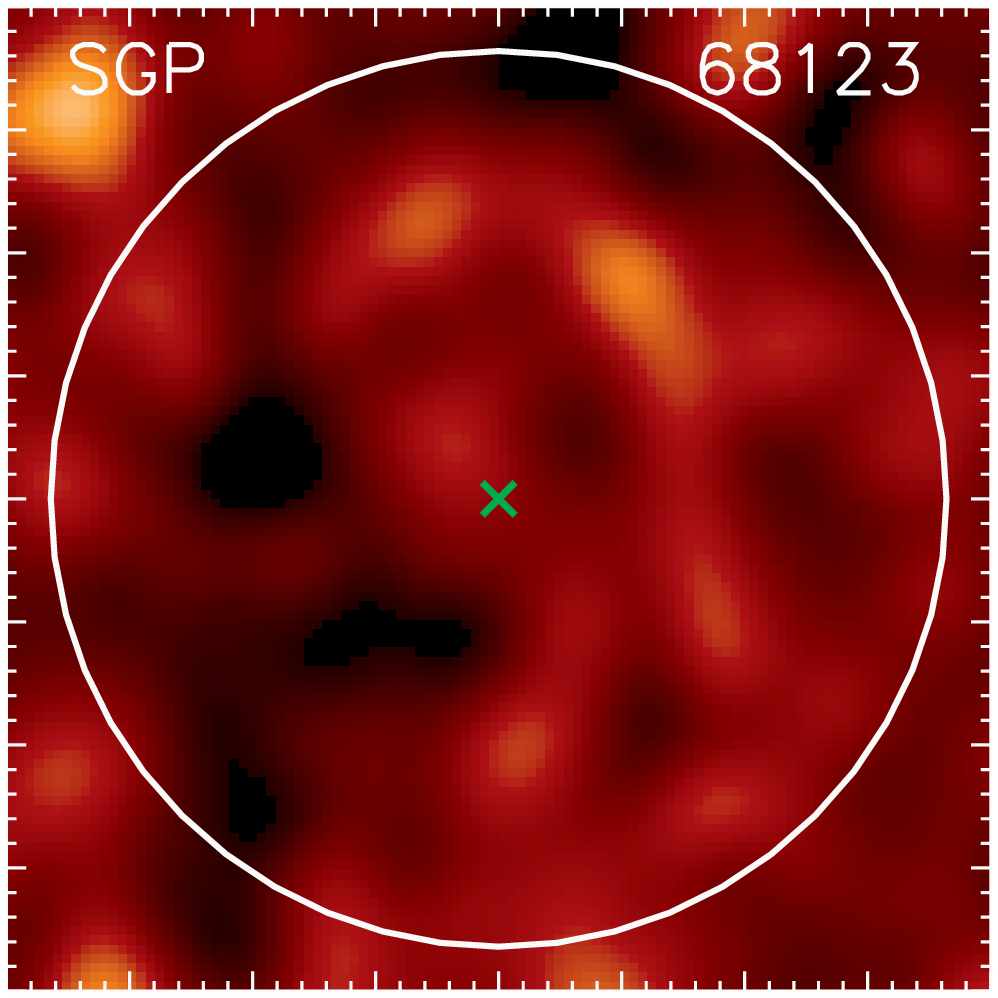}
\includegraphics[totalheight=4.25cm]{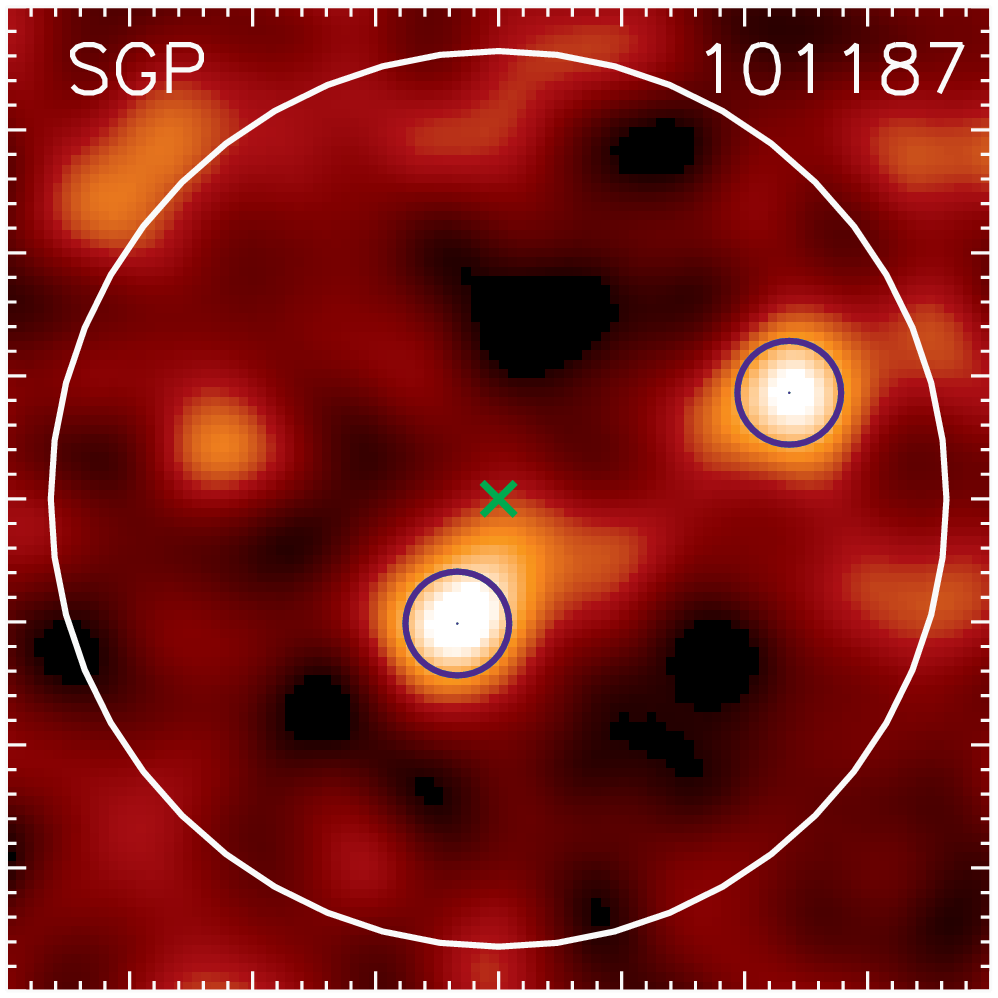}
\includegraphics[totalheight=4.25cm]{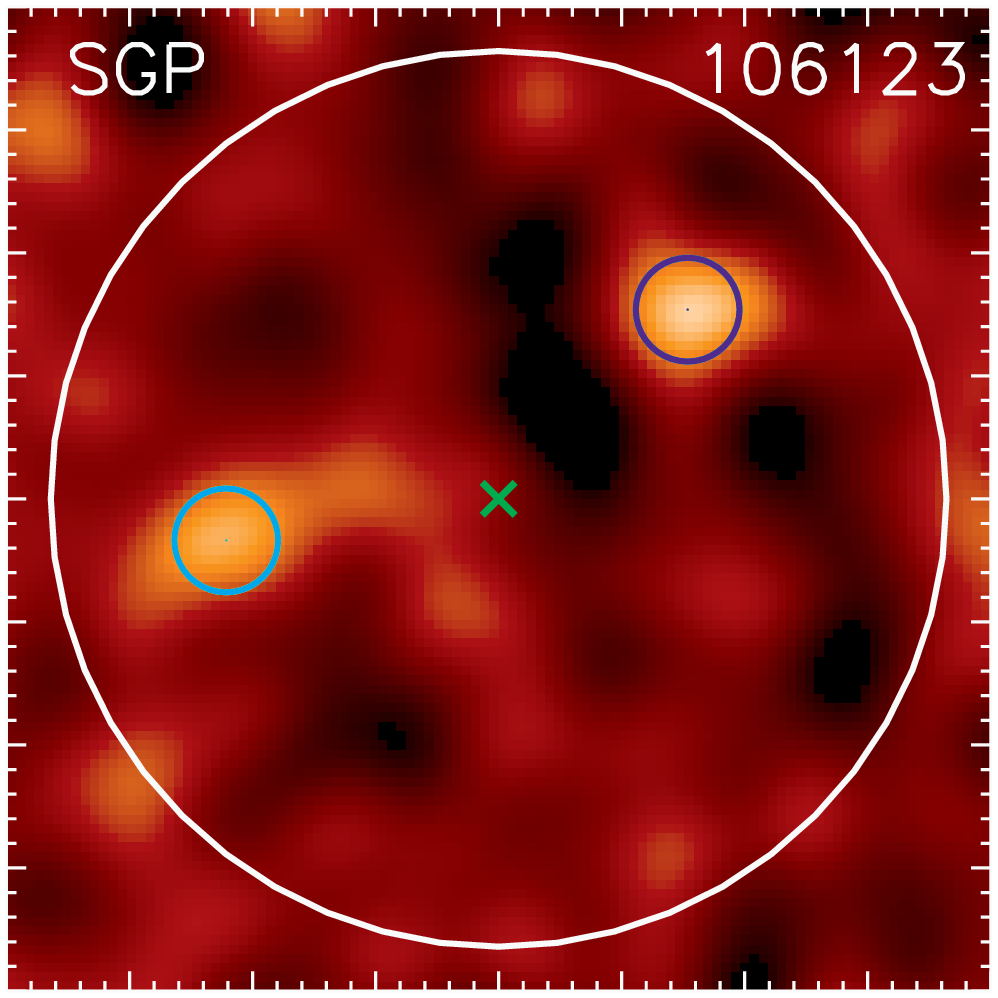} \\
\includegraphics[totalheight=4.25cm]{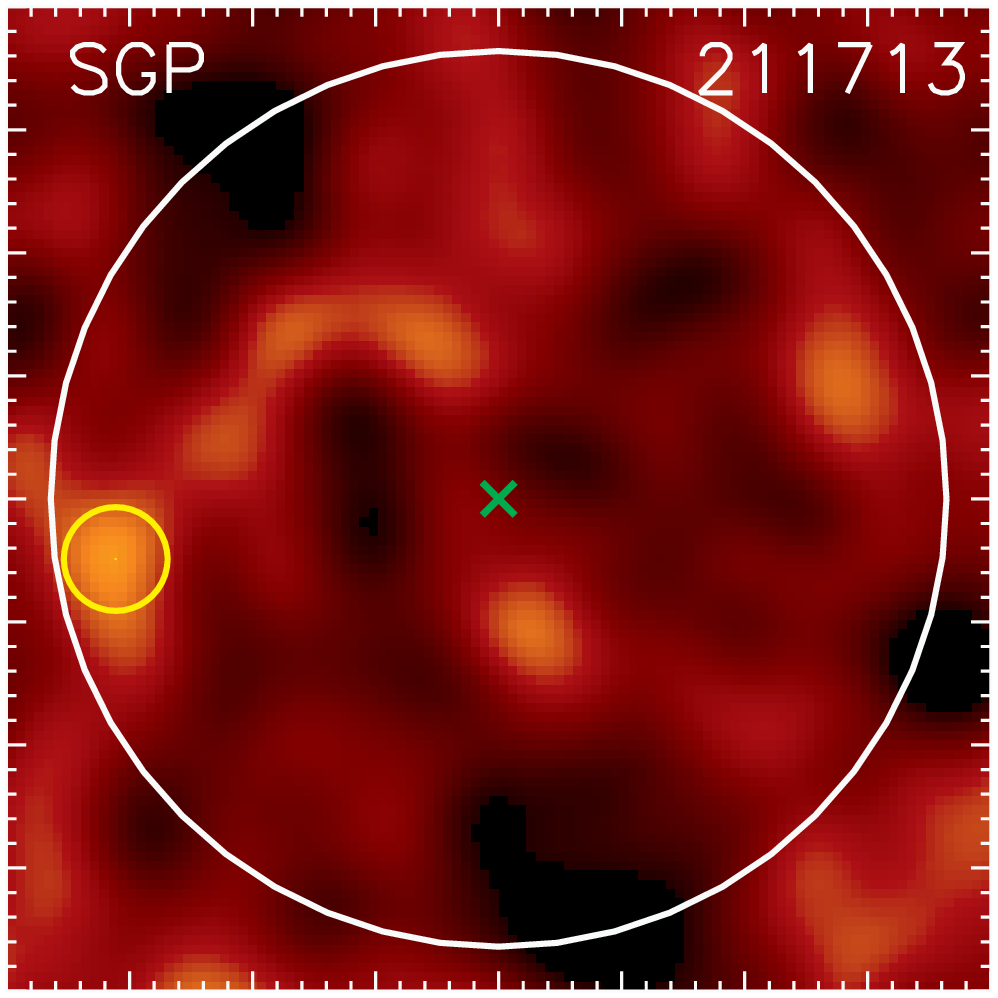}
\includegraphics[totalheight=4.25cm]{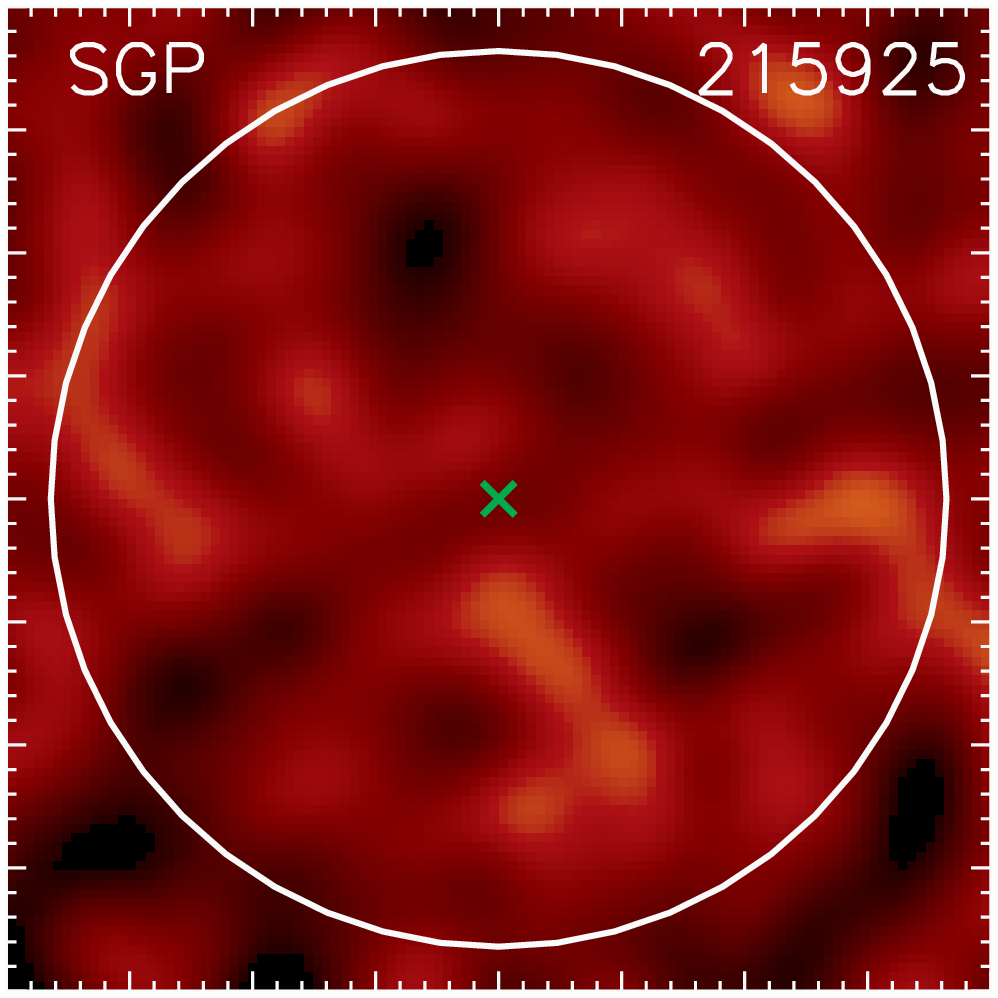}
\includegraphics[totalheight=4.25cm]{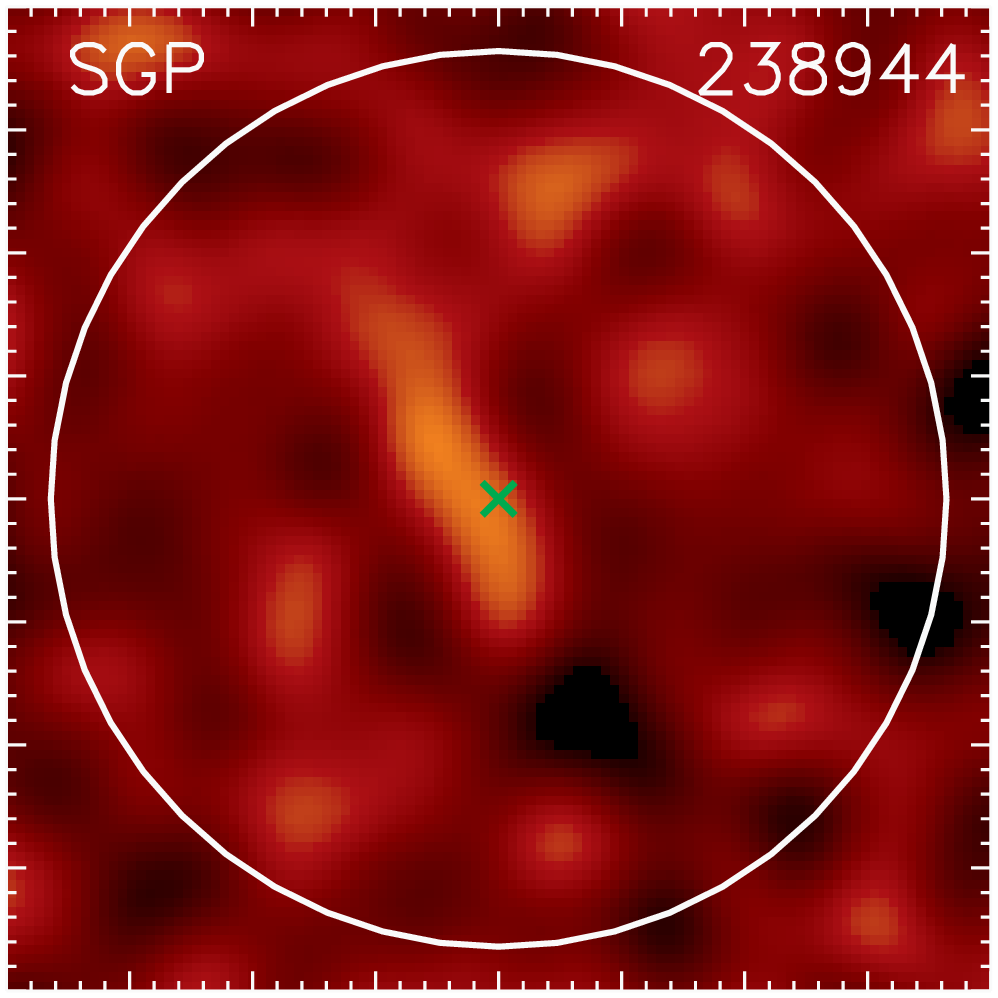}
\includegraphics[totalheight=4.25cm]{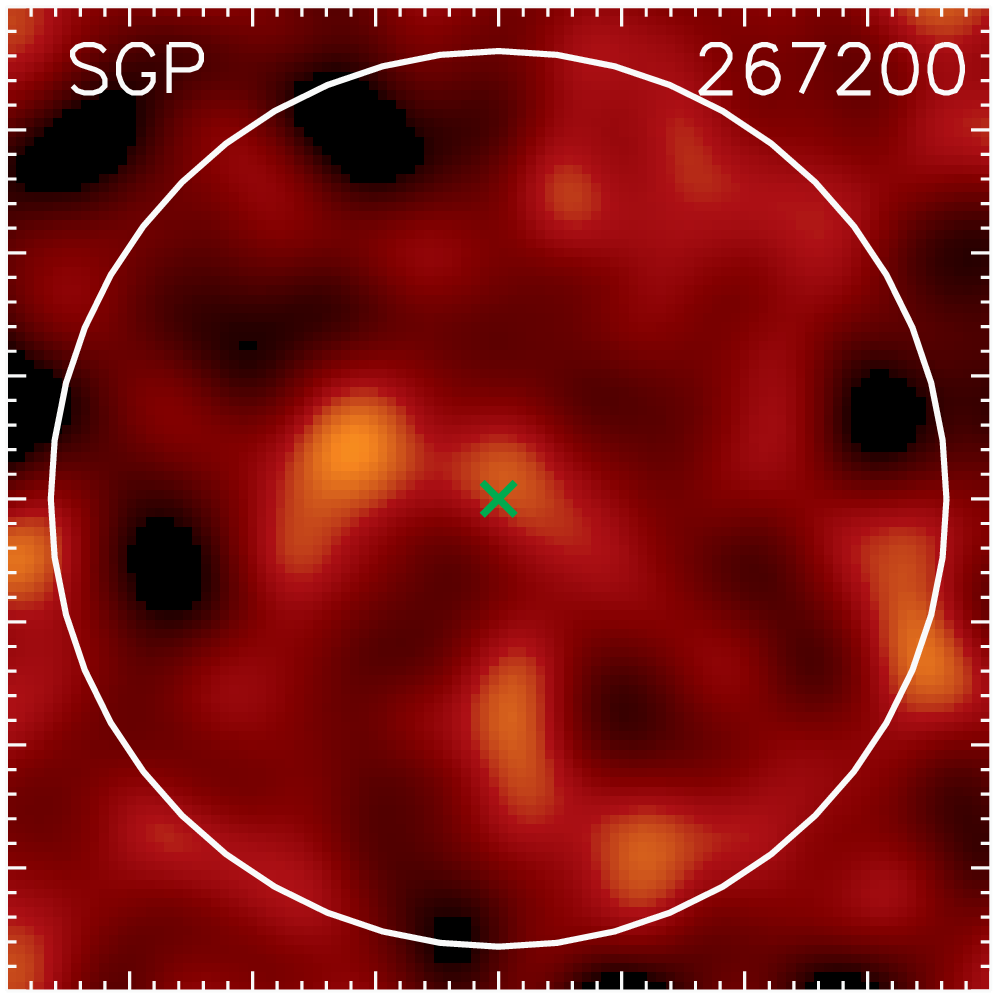} \\
\includegraphics[totalheight=4.25cm]{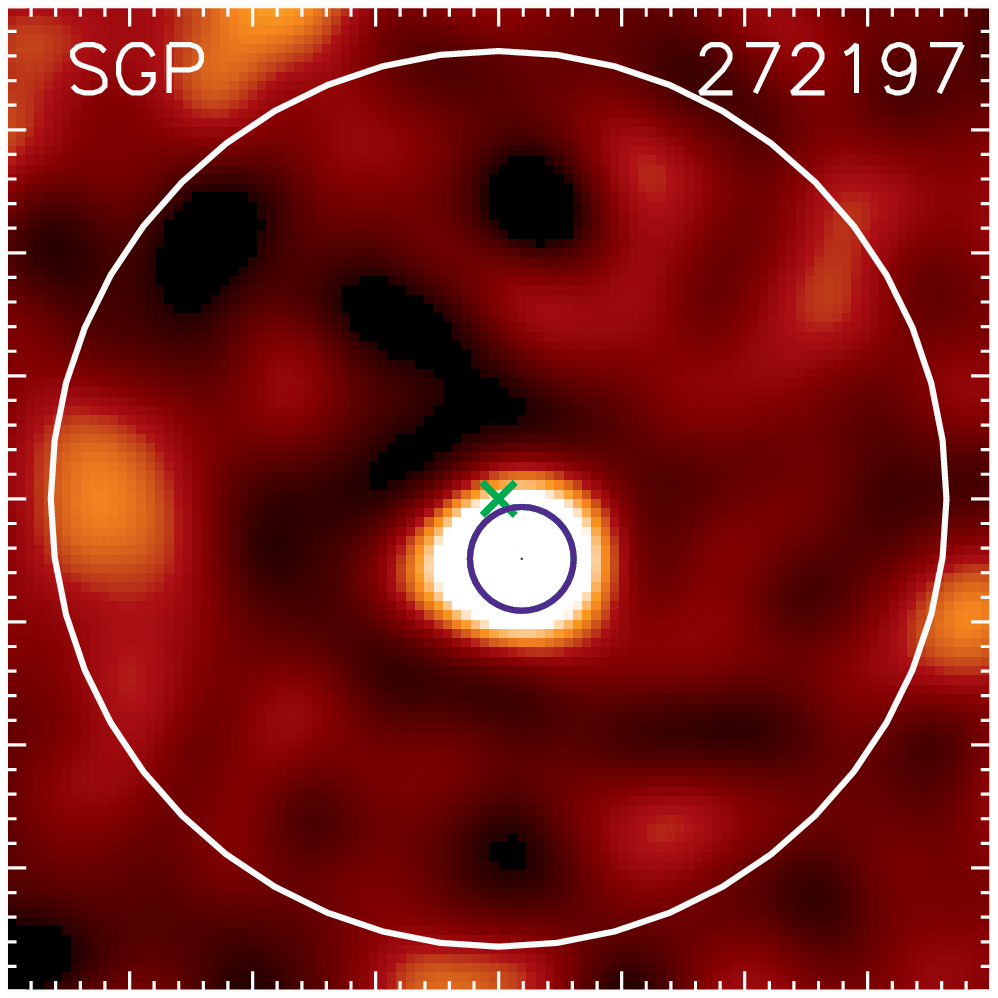}
\includegraphics[totalheight=4.25cm]{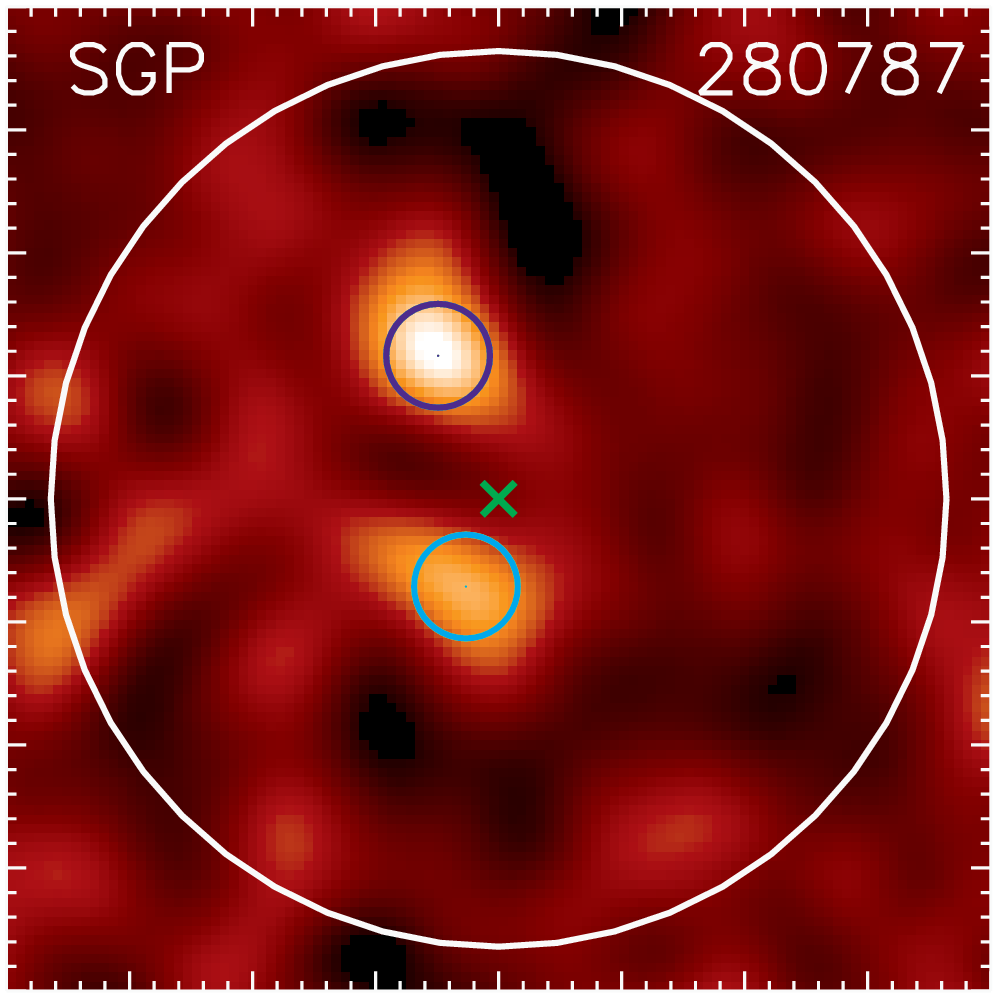}
\includegraphics[totalheight=4.25cm]{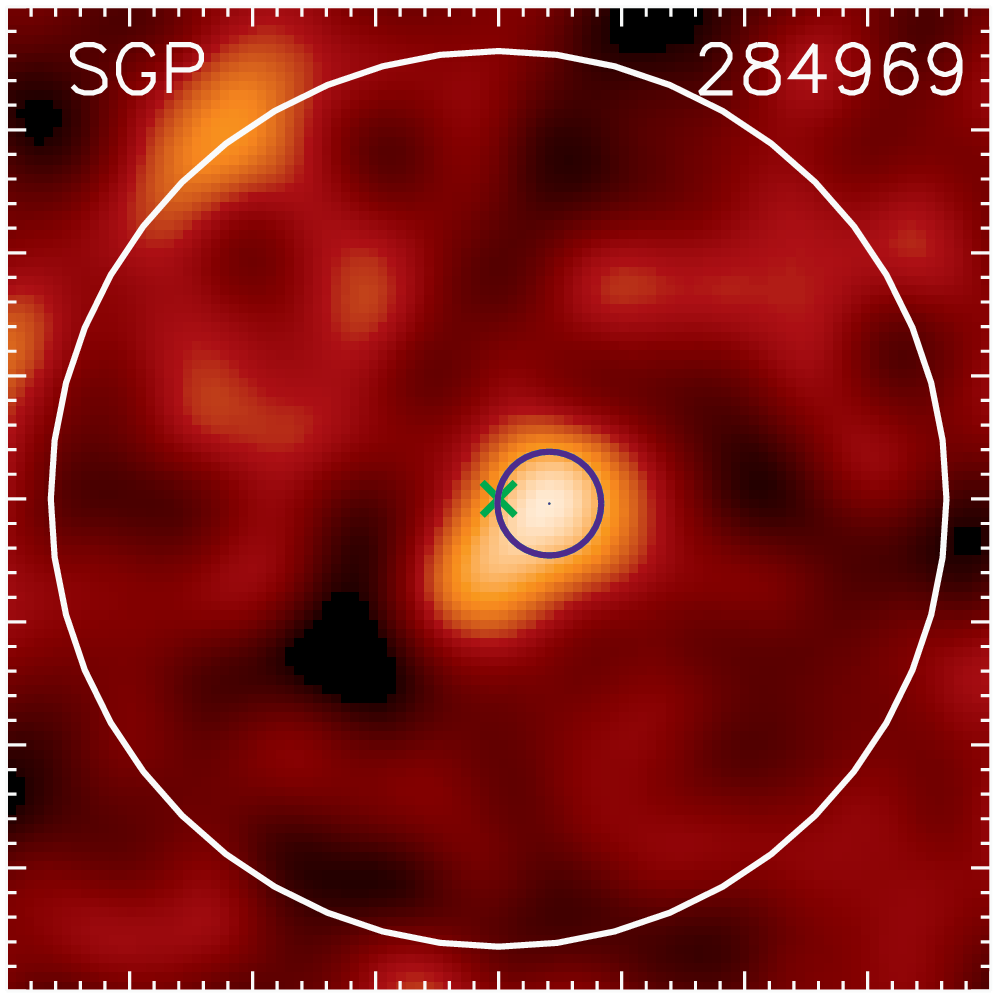}
\includegraphics[totalheight=4.25cm]{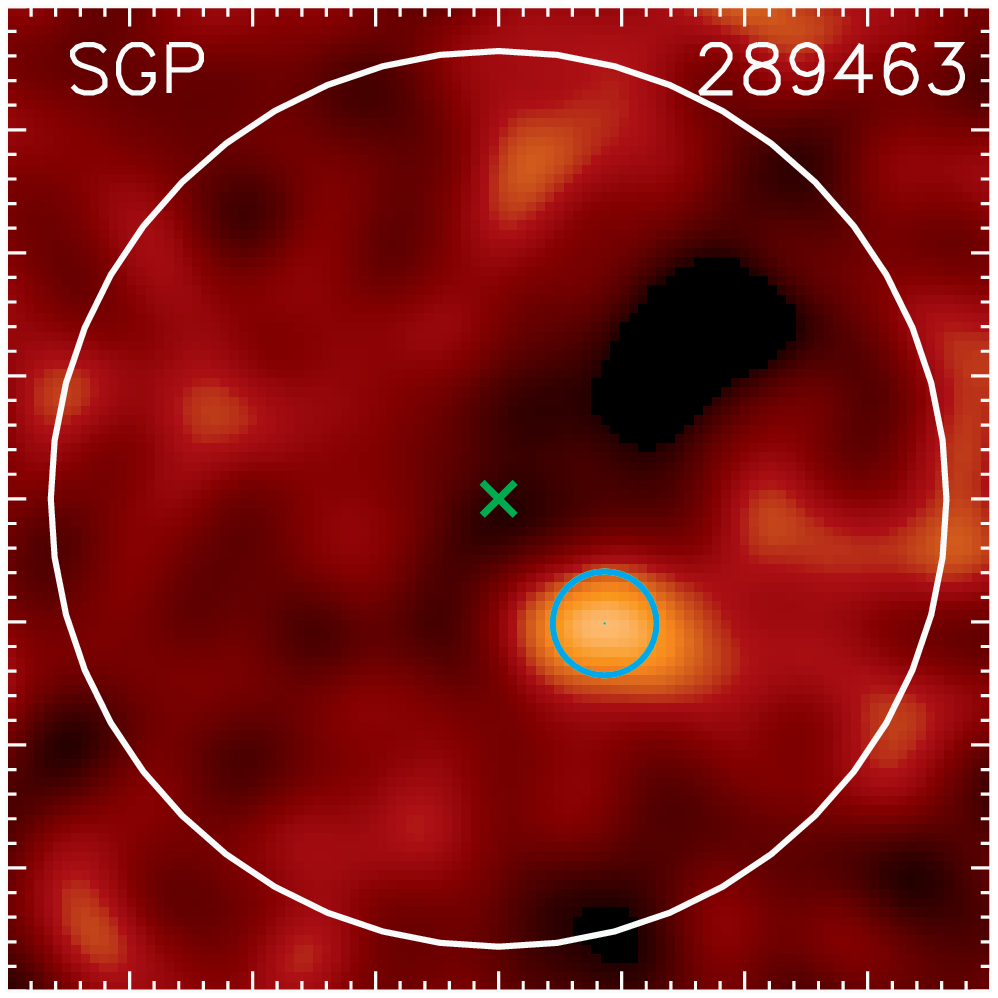} \\
\includegraphics[totalheight=4.25cm]{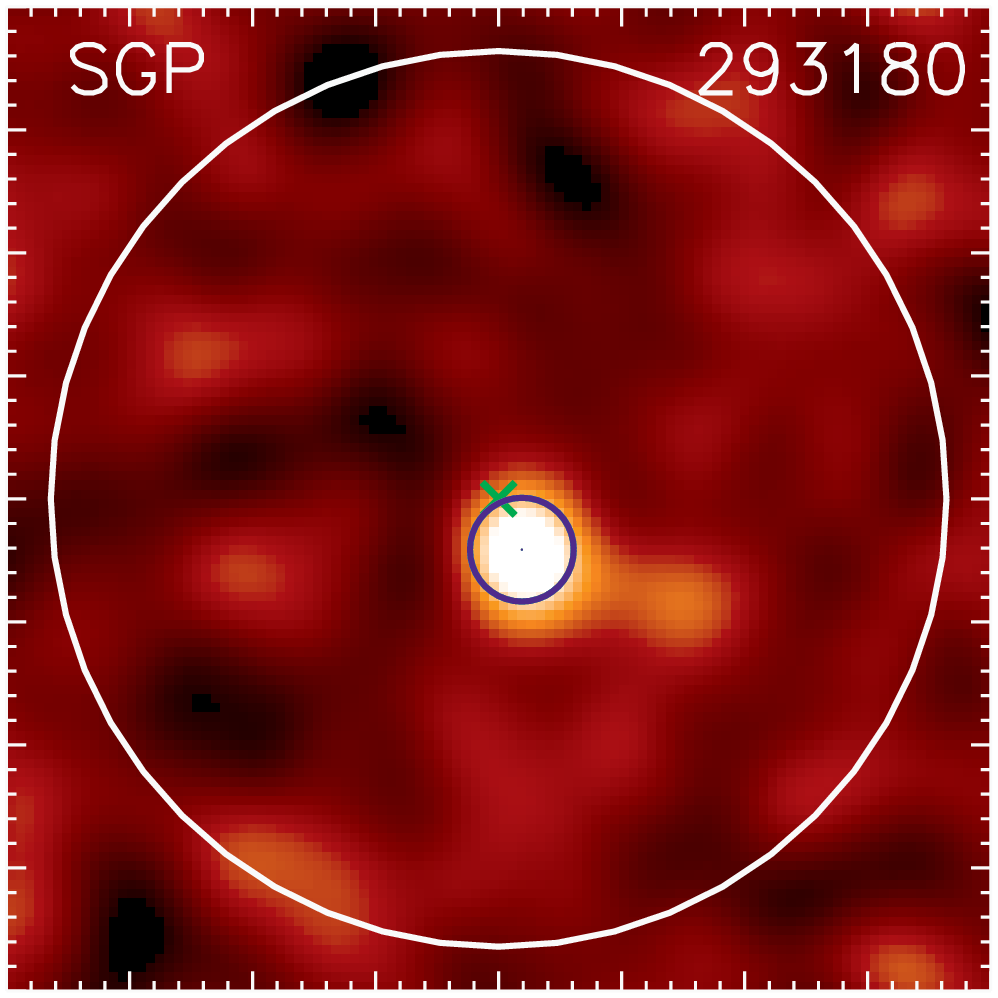}
\includegraphics[totalheight=4.25cm]{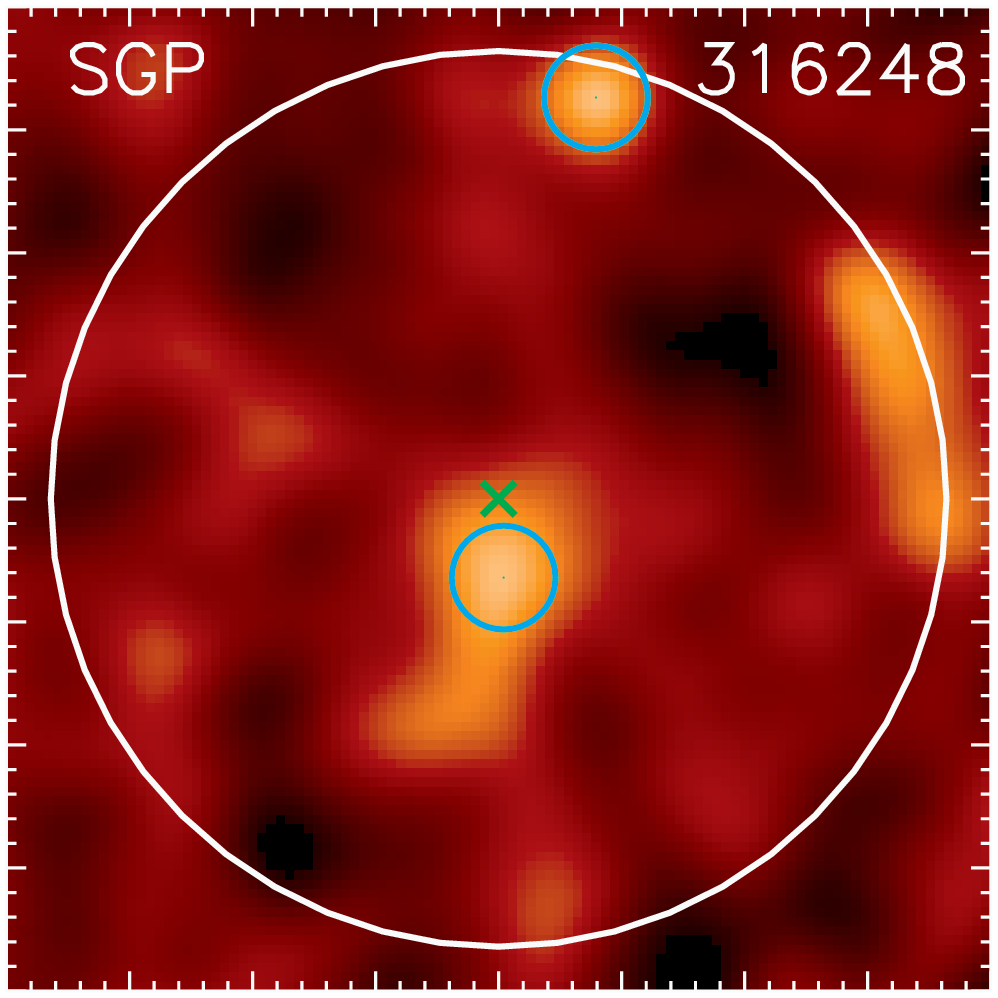}
\includegraphics[totalheight=4.25cm]{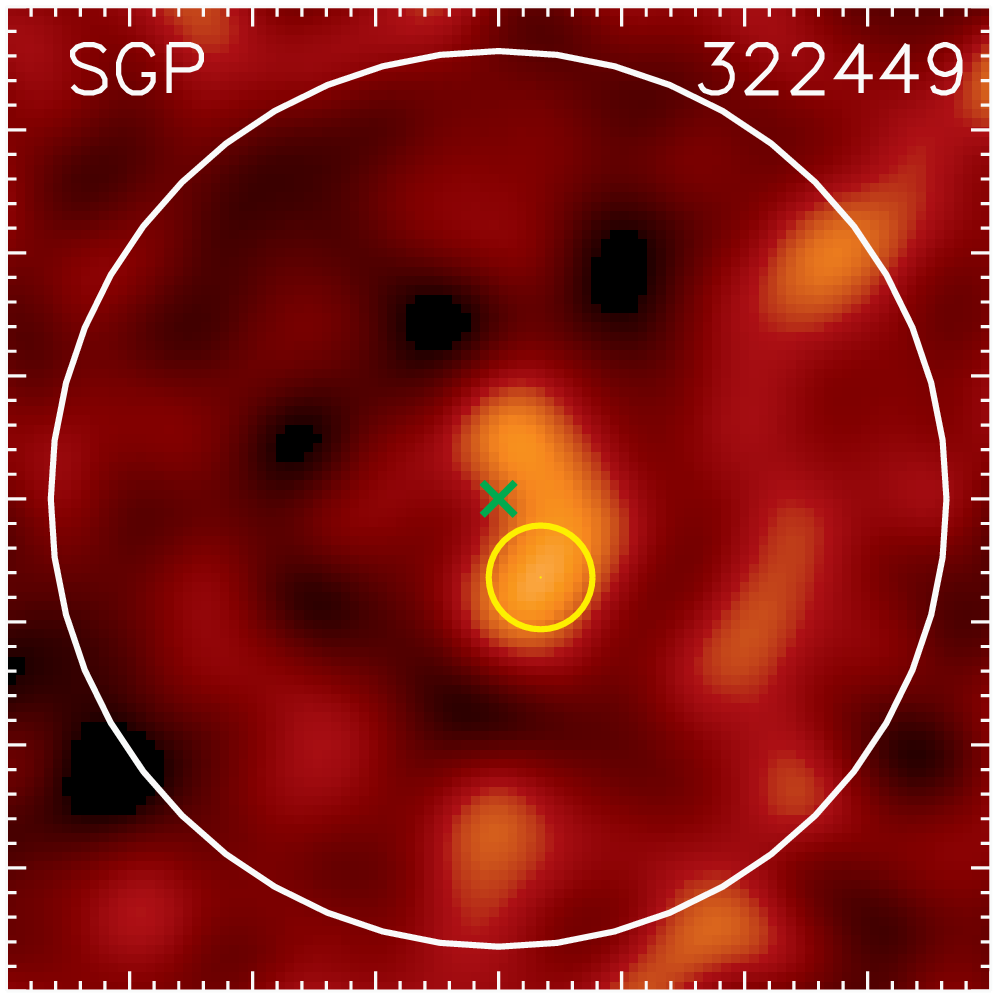}
\includegraphics[totalheight=4.25cm]{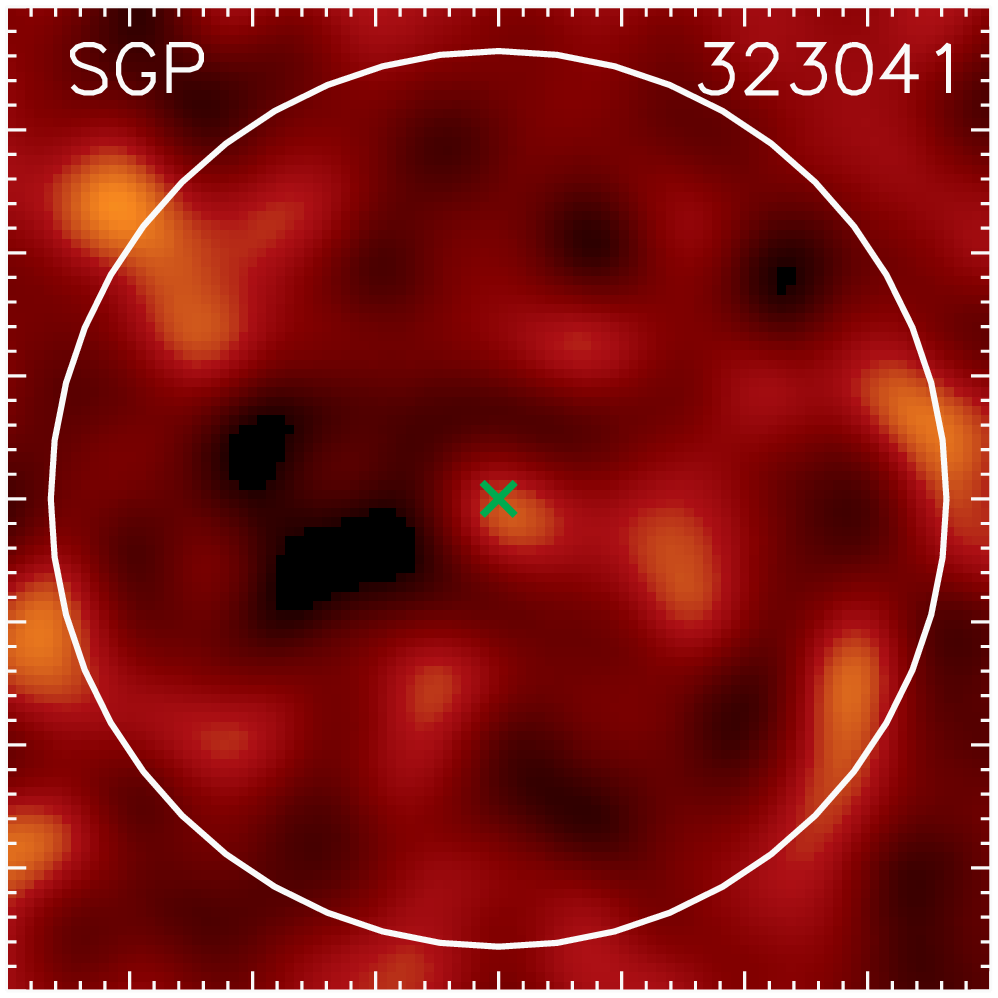} \\
\centering
\contcaption{ }
\end{center}
\end{figure*}
\begin{figure*}\hspace{-0.6cm}
\begin{center}
\includegraphics[totalheight=4.25cm]{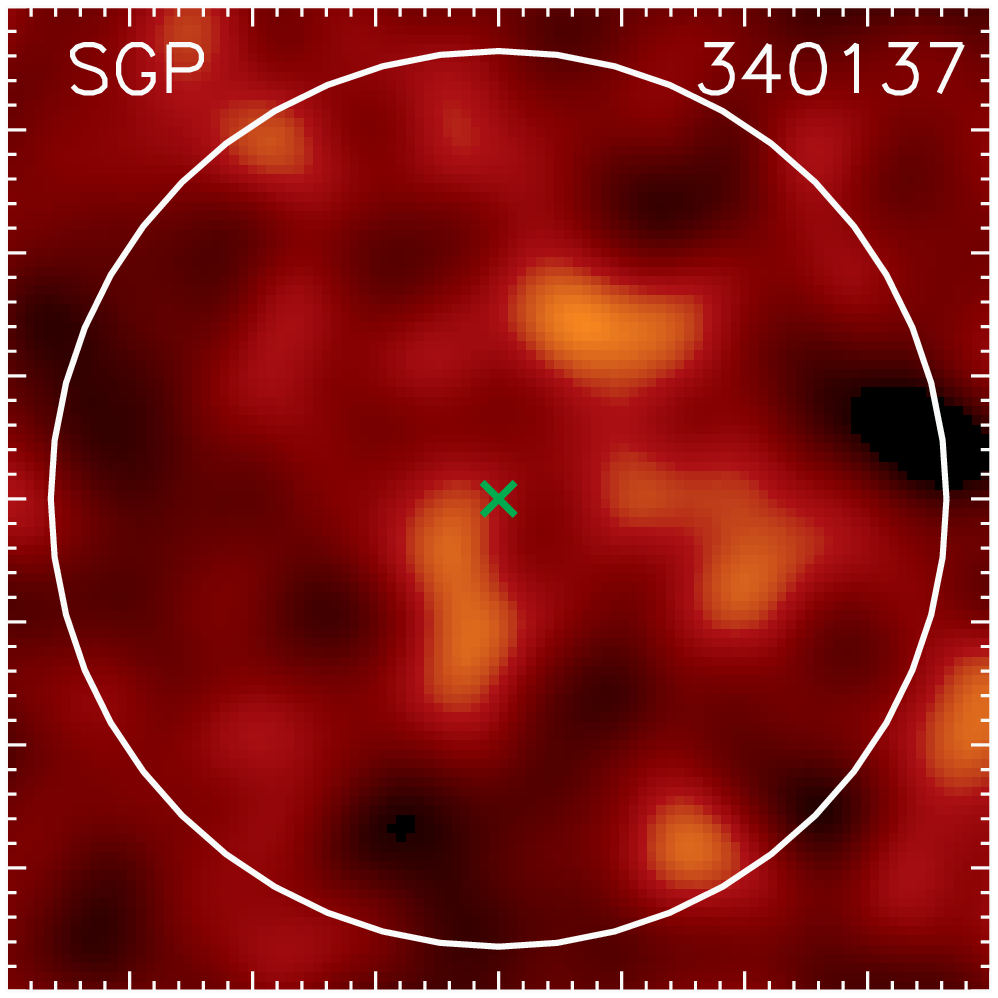}
\includegraphics[totalheight=4.25cm]{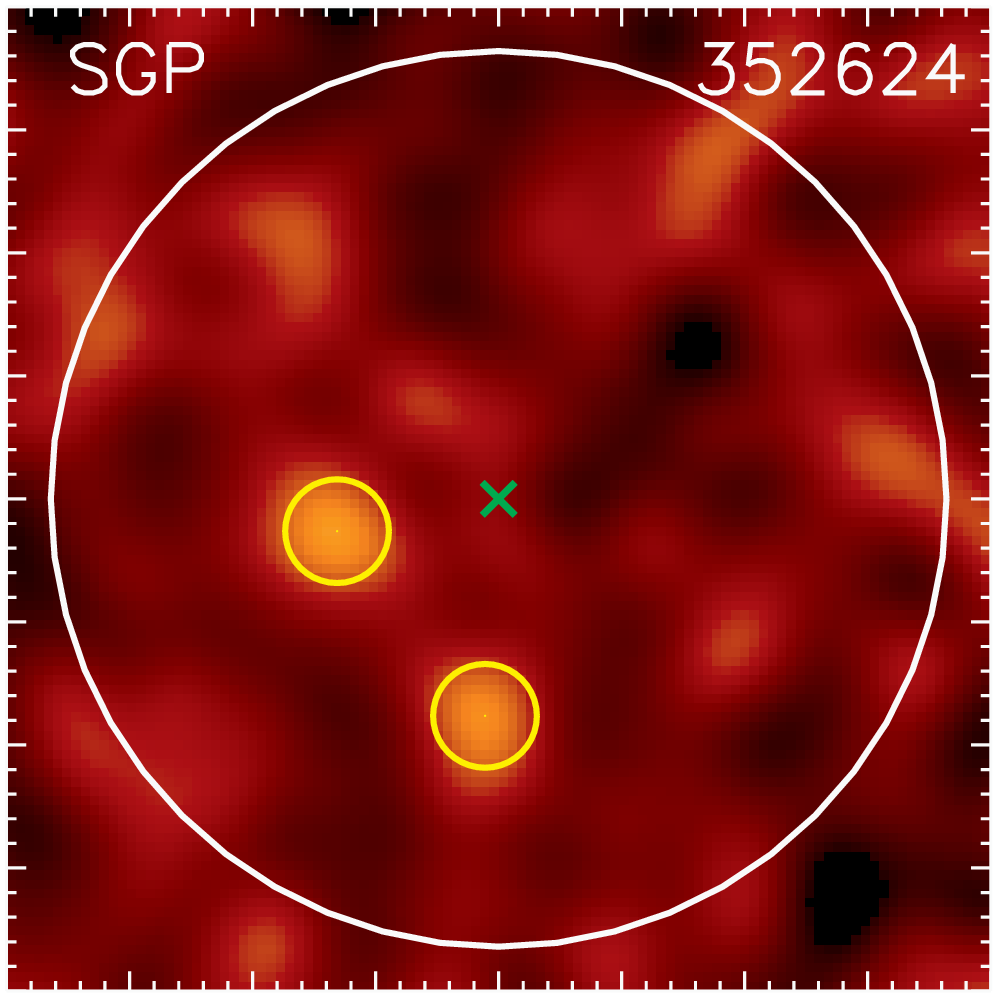}
\includegraphics[totalheight=4.25cm]{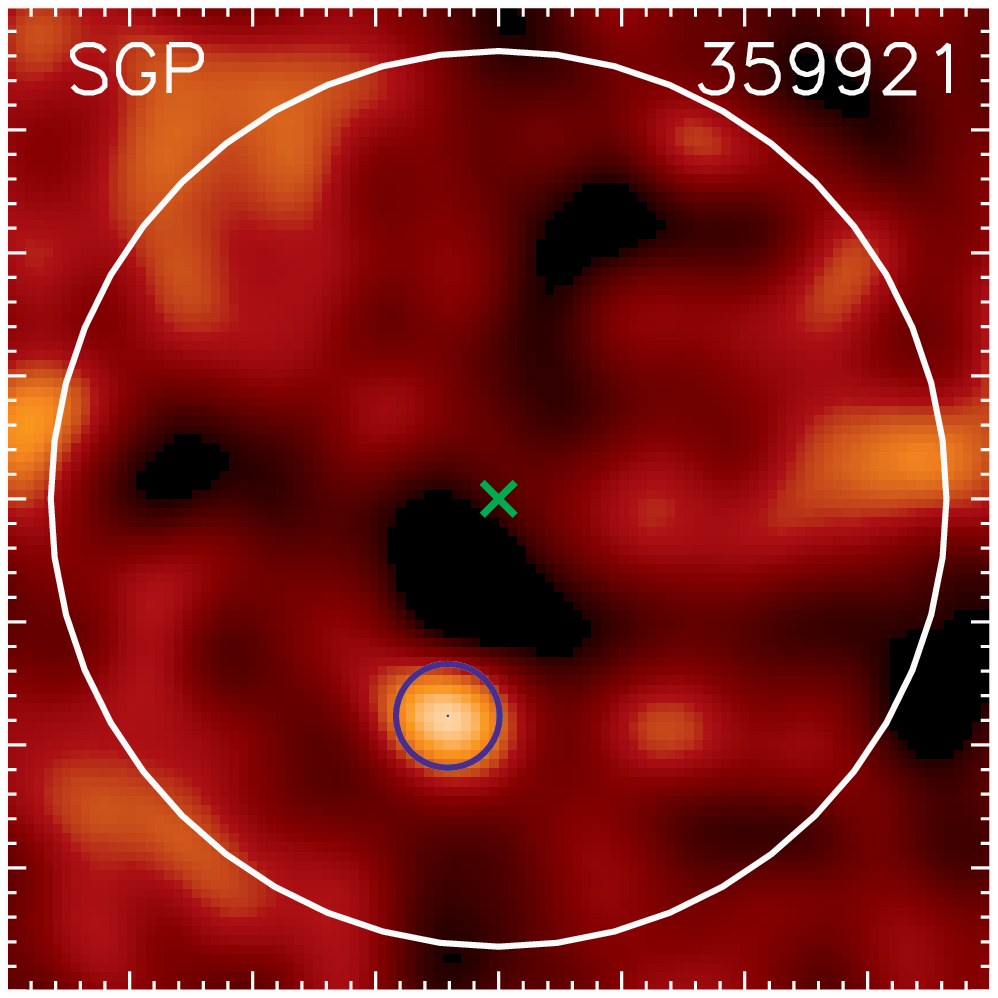}
\includegraphics[totalheight=4.25cm]{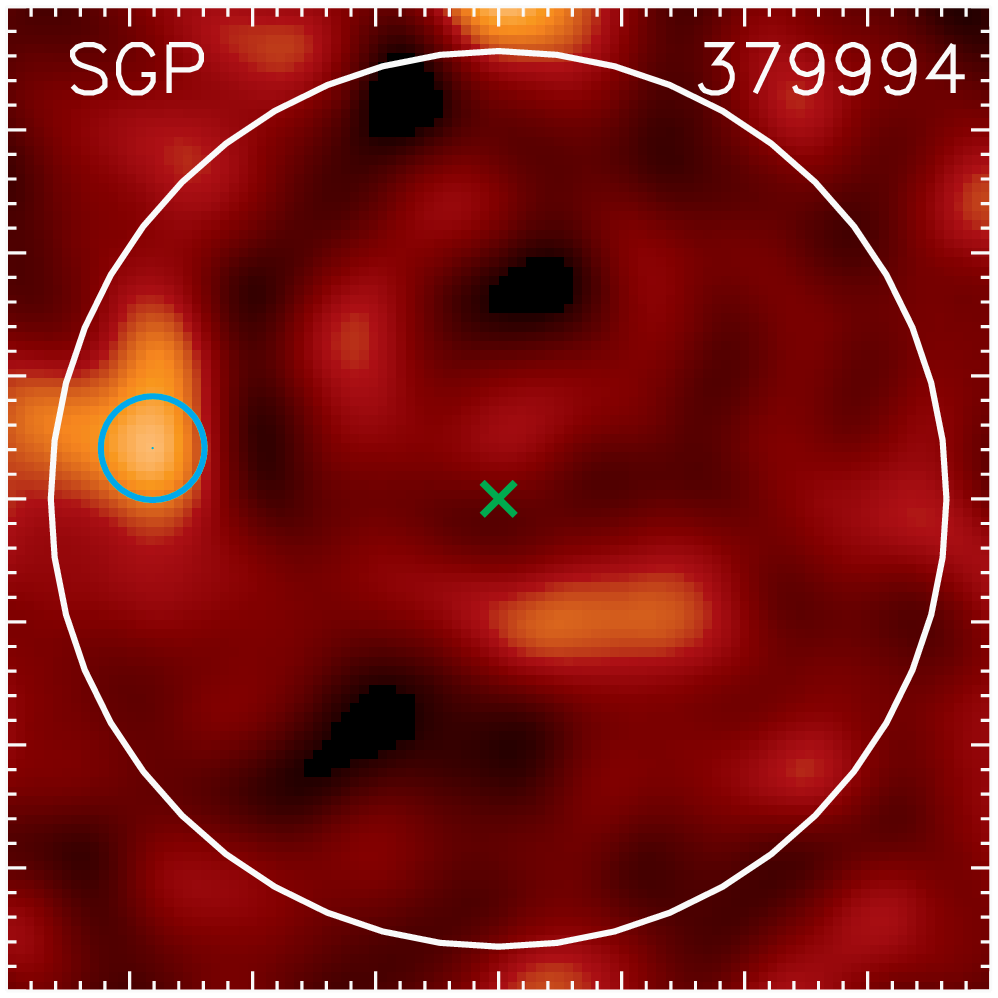} \\
\includegraphics[totalheight=4.25cm]{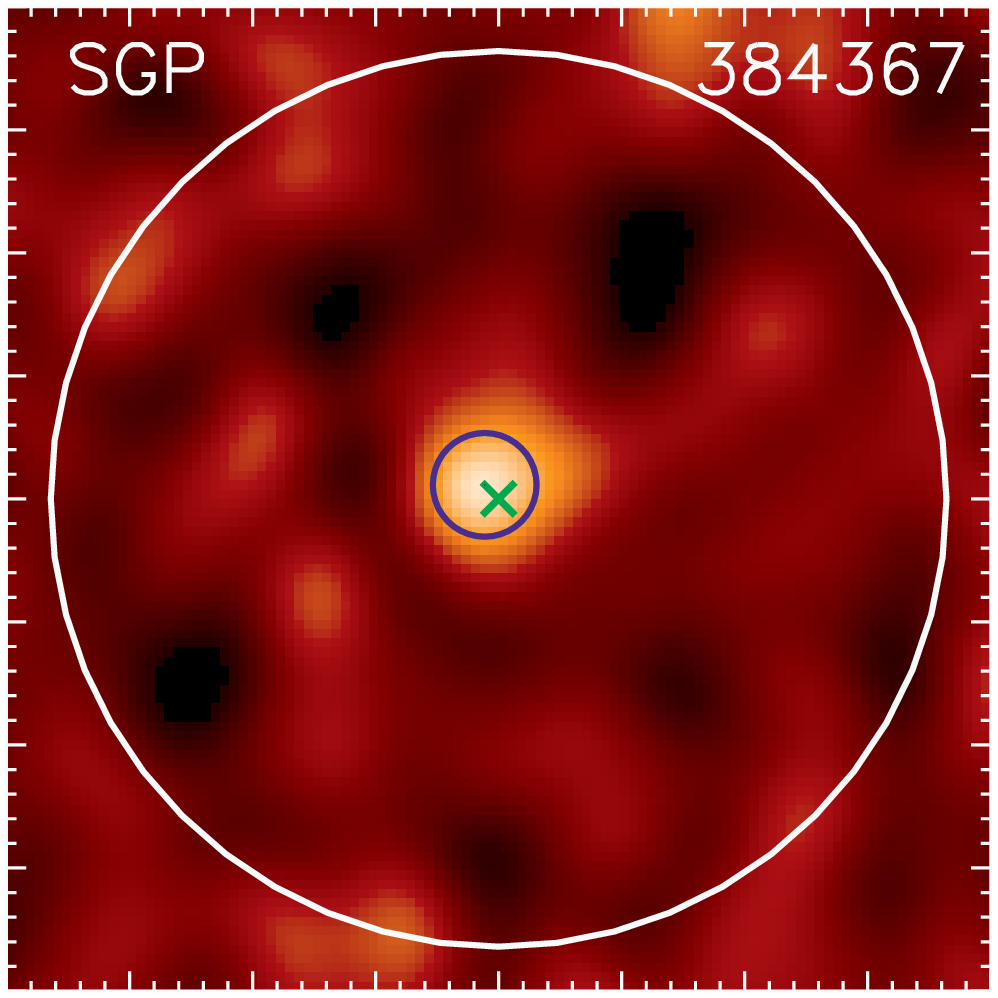}
\includegraphics[totalheight=4.25cm]{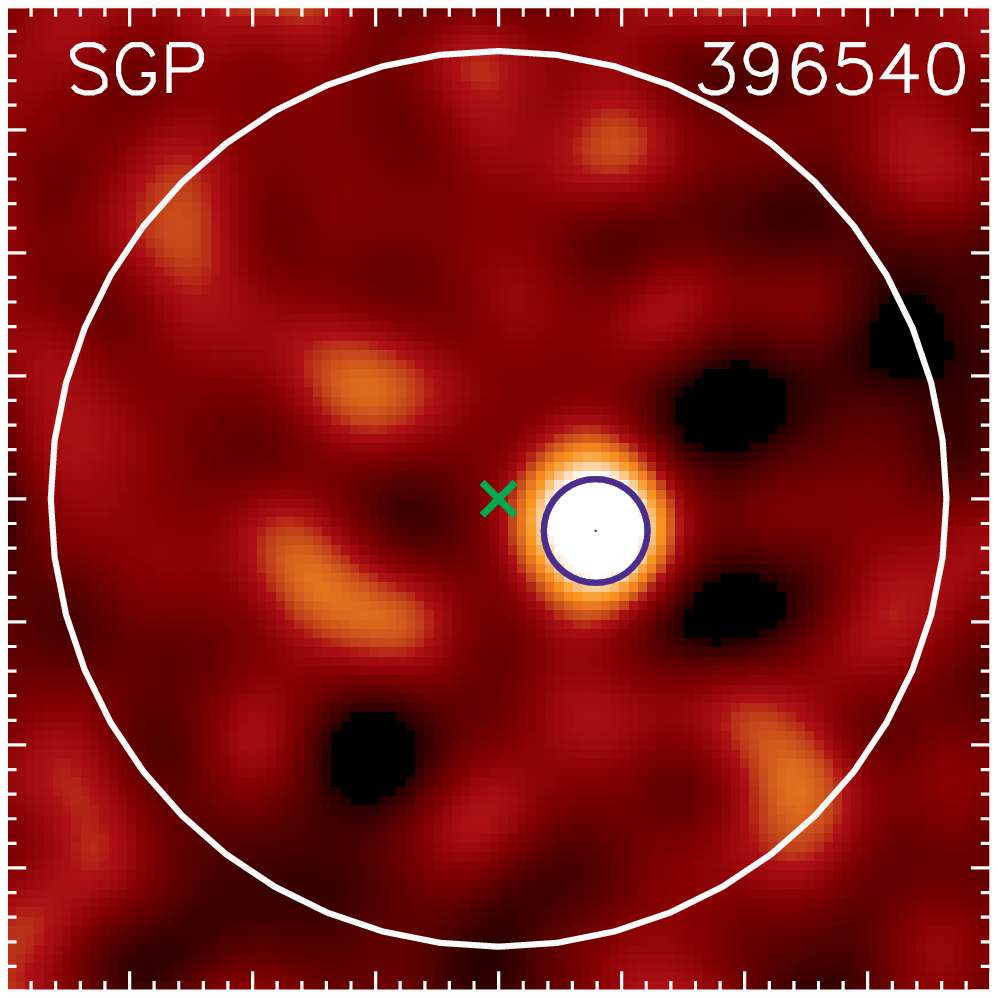}
\includegraphics[totalheight=4.25cm]{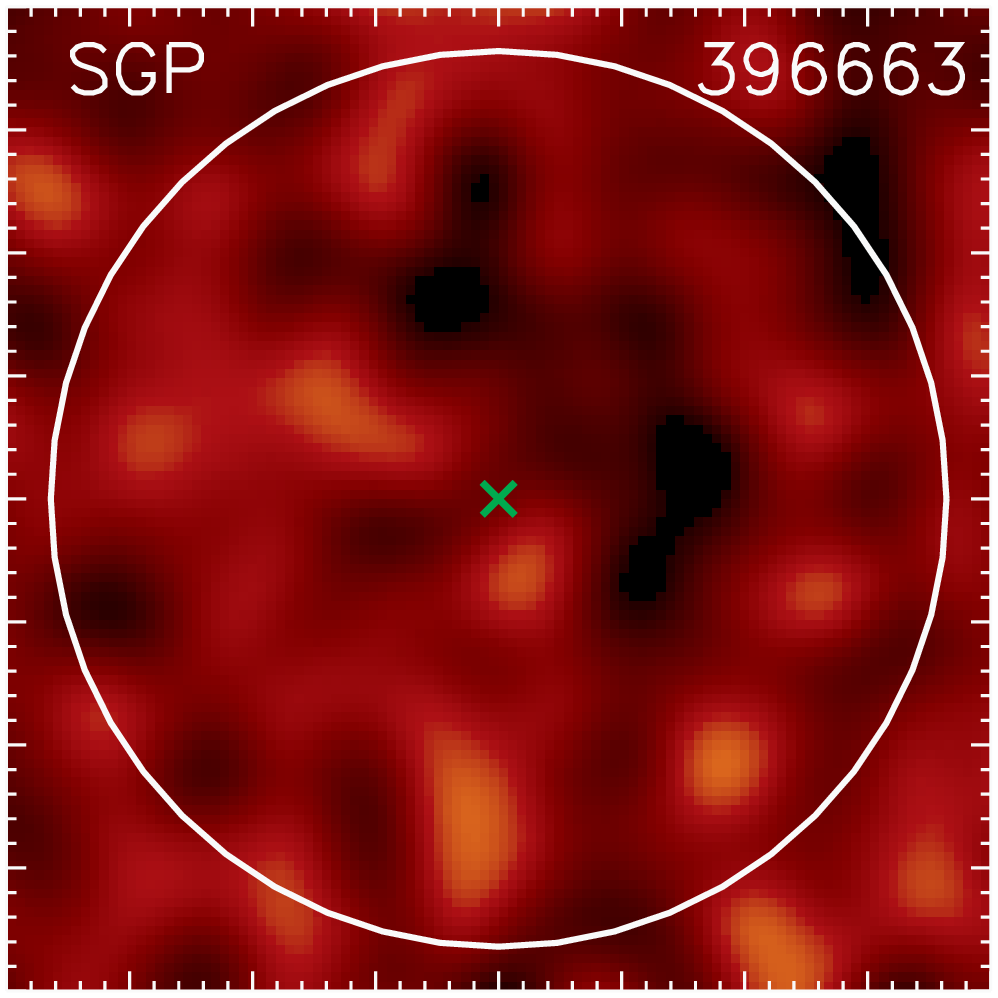}
\includegraphics[totalheight=4.25cm]{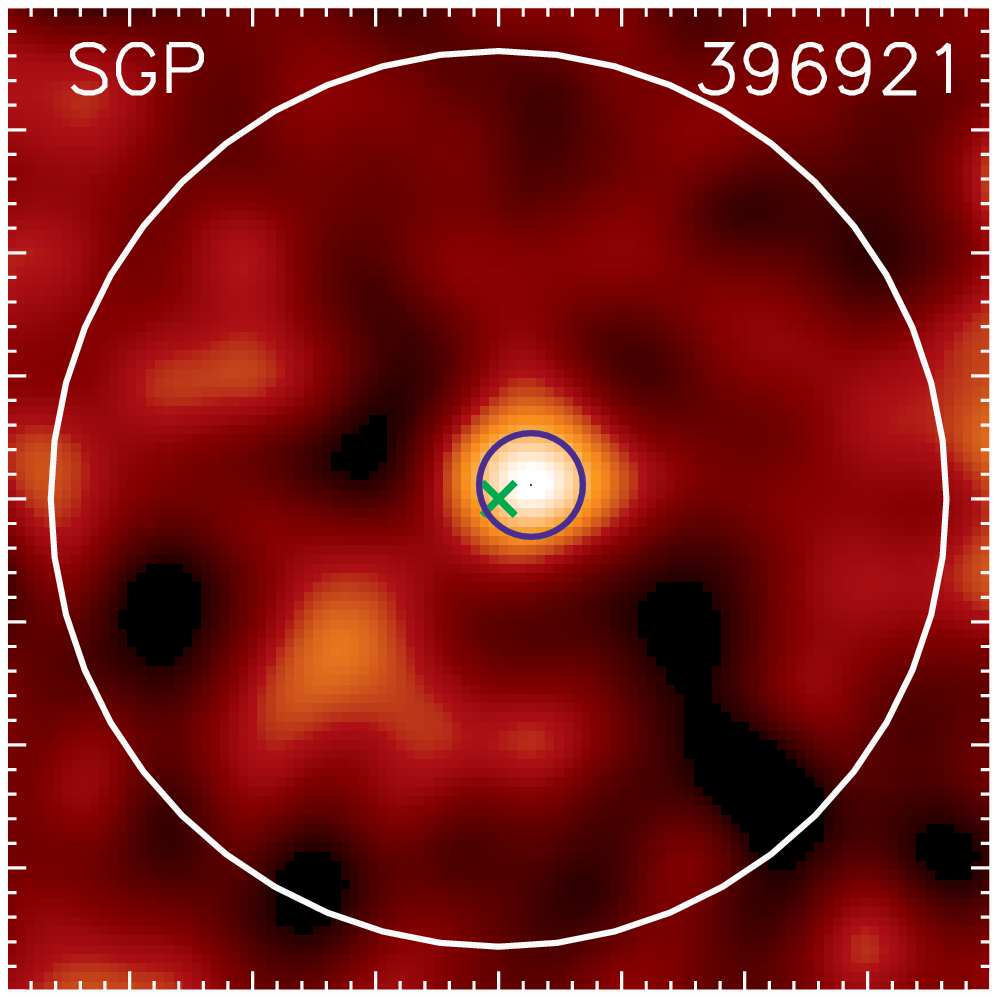} \\
\includegraphics[totalheight=4.25cm]{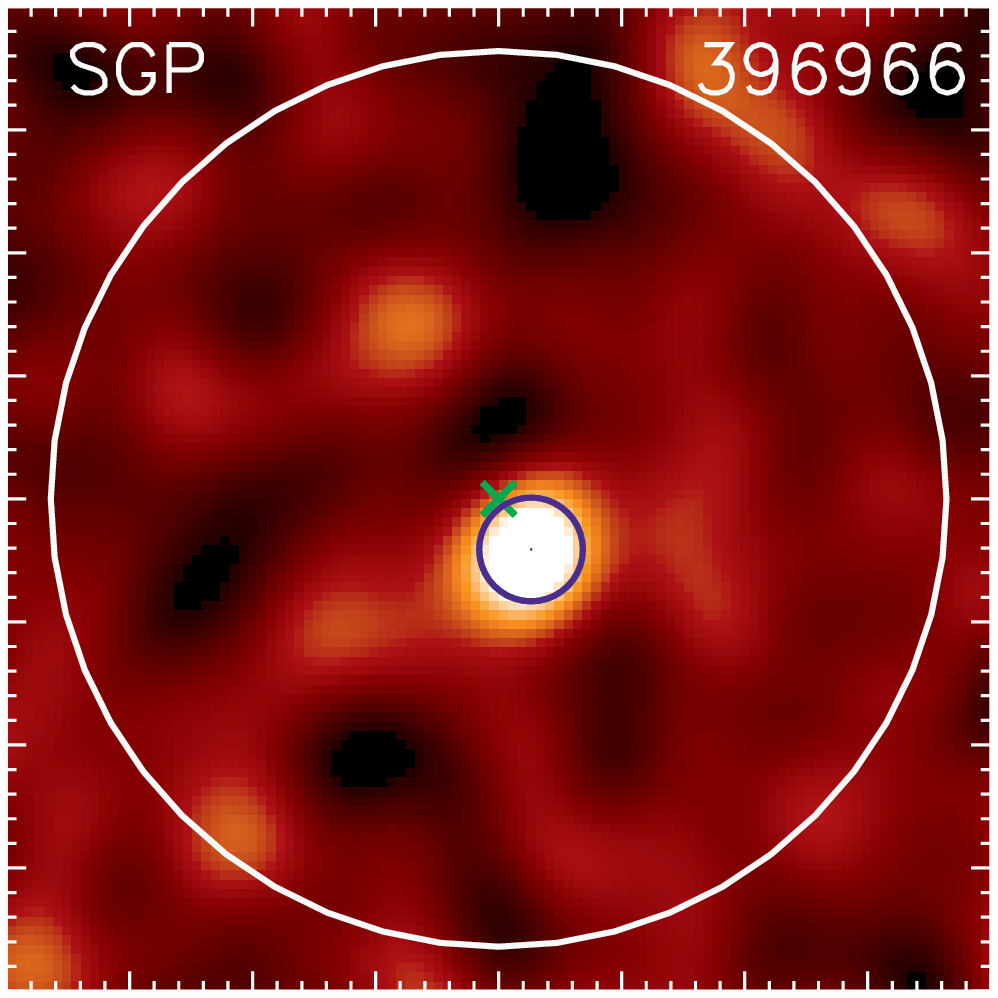}
\includegraphics[totalheight=4.25cm]{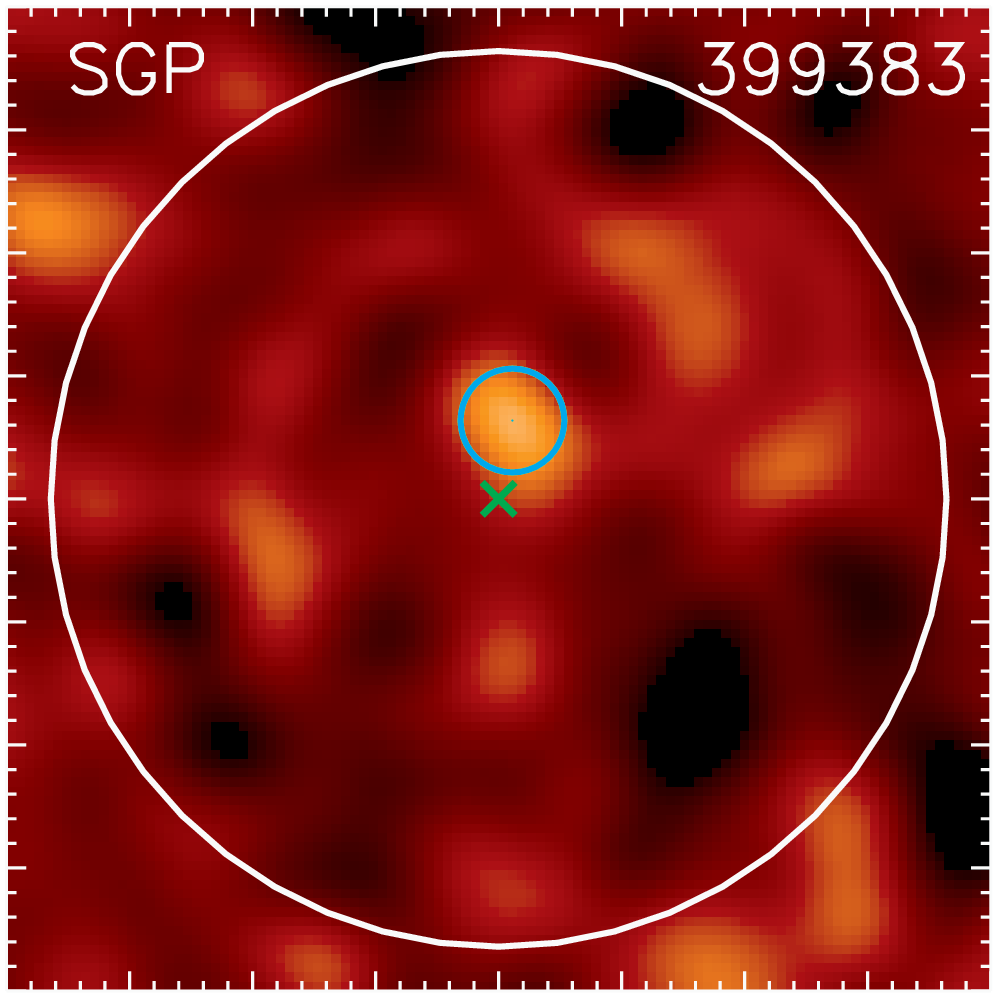}
\includegraphics[totalheight=4.25cm]{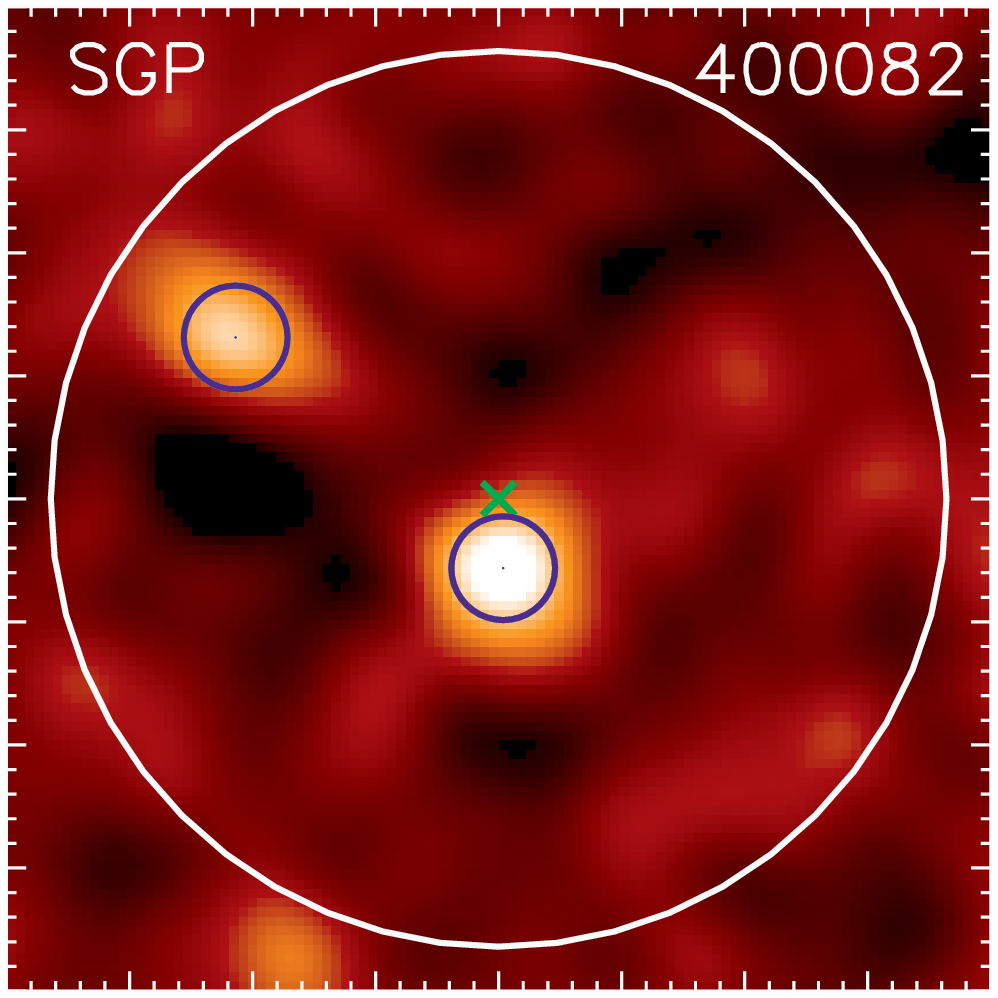}
\includegraphics[totalheight=4.25cm]{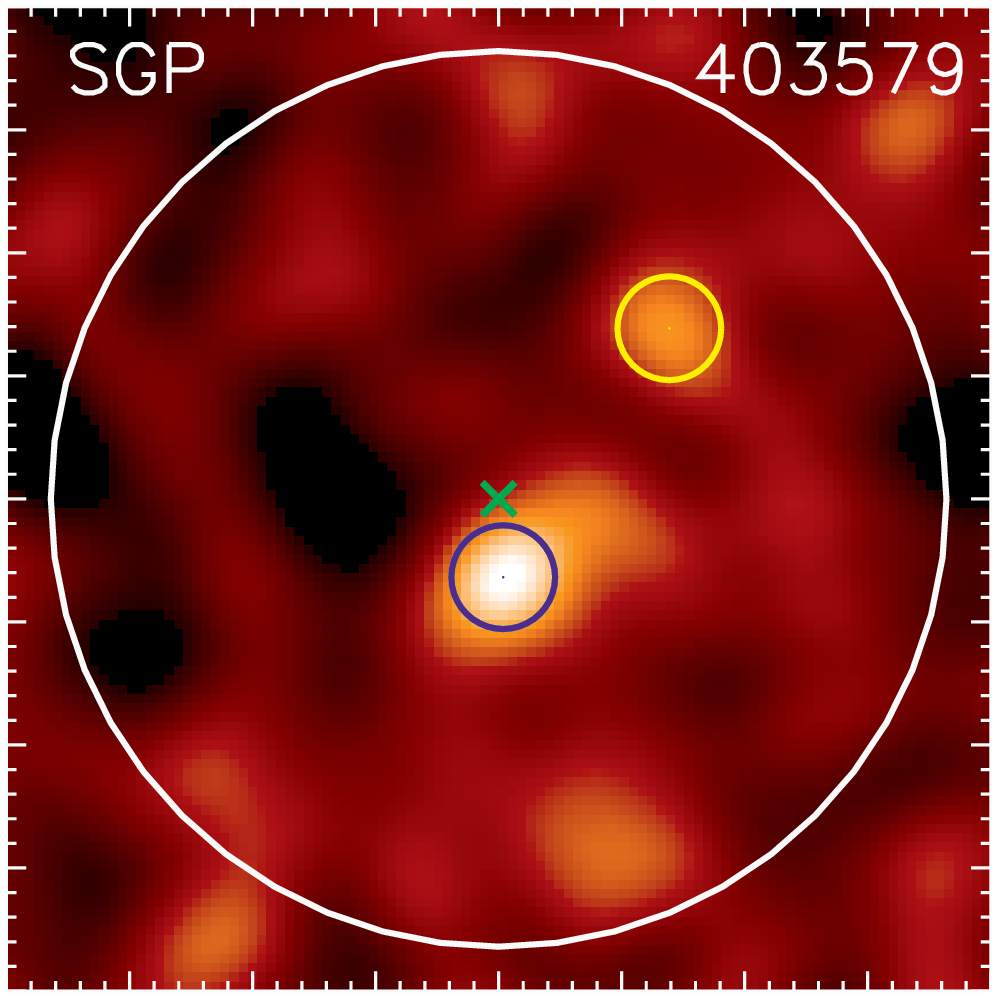}
\includegraphics[totalheight=4.25cm]{./Figures/pstamps/SGP-352624.eps} \\
\centering
\contcaption{ }
\end{center}
\end{figure*}

\begin{landscape}
\begin{table}
\centering
\caption{AzTEC 1.1 mm detections with SNR $\geq 3.5$, including deblended \textit{Herschel}/SPIRE flux densities (columns 4-6). The AzTEC flux density and SNR are listed in columns 7 and 8. SED fitted parameters ($z_\textrm{phot}$, $L_\textrm{IR}$) and SFR are given in columns 9-11 (with dashes indicating that the fitting did not converge). The separation between the AzTEC detection and the \textit{Herschel} position is given in column 12. The last two columns to the right indicate, for SNR detection thresholds $\geq 4.0$ and $\geq 3.5$, if the source is classified as single (S) or as a member of a multiple system (M). An additional P indicates that the sources are physically interacting system candidates. Our list of robust high-$z$ ($> 4$) candidates are indicated with their ID (first column) in boldface font. \label{table:catalogue}}
\begin{tabular}{lrrcccrrrrrrcc}

\hline
ID &
R.A. &
Dec. &
$F_{\rm 250\mu m}$ &
$F_{\rm 350\mu m}$ &
$F_{\rm 500\mu m}$ &
$S_{\rm 1.1 mm}$ &
SNR$_\textrm{Az}$ &
$z_{\rm phot}$ &
$L_{\rm IR(8-1000\mu m)}$ &
SFR &
Separation &
\multicolumn{2}{c}{Class} \\ 
 & 
[deg] &
[deg] & 
[mJy] & 
[mJy] & 
[mJy] & 
[mJy] & 
 & 
 &
[$10^{12}$L$_\odot$] &
[M$_\odot$ yr$^{-1}$] & 
[arcsec] &
$4.0\sigma$ &
$3.5\sigma$ \\
\hline
\textbf{G09-44907}         & $ 129.345104 $ & $   1.427174 $ & $ 17.1 \pm 7.4 $ & $  37.9 \pm  7.8 $ & $  49.8 \pm  9.2 $ &  $ 11.5 \pm  0.9 $ & 12.5 & $ 4.14^{+0.28}_{-0.24} $ & $ 17.7^{+ 5.0}_{-4.0} $ & $ 2614^{+ 734}_{- 586} $ &  7.9   & S  & S  \\
G09-62610                  & $ 137.353750 $ & $   1.926694 $ & $ 17.5 \pm 6.7 $ & $  36.6 \pm  7.3 $ & $  53.7 \pm  8.4 $ &  $  8.1 \pm  0.7 $ & 11.5 & $ 3.58^{+0.24}_{-0.22} $ & $ 13.5^{+ 3.5}_{-3.1} $ & $ 2002^{+ 516}_{- 455} $ &  6.7   & S  & S  \\
G09-64894                  & $ 138.191354 $ & $   1.195965 $ & $ 19.9 \pm 6.8 $ & $  35.5 \pm  7.5 $ & $  55.4 \pm  8.0 $ &  $  4.8 \pm  1.2 $ &  4.1 & $ 3.28^{+0.24}_{-0.24} $ & $ 10.7^{+ 3.2}_{-2.8} $ & $ 1589^{+ 472}_{- 407} $ &  5.9   & S  & S  \\
G09-71054.A                & $ 137.193646 $ & $   1.891493 $ & $ 21.0 \pm 7.0 $ & $  39.2 \pm  7.5 $ & $  44.2 \pm  8.9 $ &  $  5.4 \pm  0.8 $ &  6.6 & $ 3.16^{+0.26}_{-0.22} $ & $  9.9^{+ 3.2}_{-2.5} $ & $ 1464^{+ 468}_{- 373} $ &  2.2   & M  & M  \\
G09-71054.B                & $ 137.185099 $ & $   1.888368 $ & $ 58.9 \pm 7.0 $ & $  50.6 \pm  7.5 $ & $  30.7 \pm  8.9 $ &  $  5.6 \pm  1.0 $ &  5.6 & $ 2.26^{+0.16}_{-0.28} $ & $ 10.8^{+ 2.8}_{-3.2} $ & $ 1602^{+ 411}_{- 475} $ & 34.9   & M  & M  \\
\textbf{G09-81106}$^a$     & $ 132.403229 $ & $   0.247799 $ & $ 14.0 \pm 6.0 $ & $  31.0 \pm  8.2 $ & $  47.6 \pm  8.8 $ &  $ 11.1 \pm  0.8 $ & 13.7 & $ 4.32^{+0.32}_{-0.26} $ & $ 16.8^{+ 4.9}_{-4.1} $ & $ 2482^{+ 721}_{- 604} $ &  6.5   & S  & S  \\
\textbf{G09-83808}$^a$     & $ 135.190729 $ & $   0.689576 $ & $  9.6 \pm 7.2 $ & $  21.8 \pm  7.4 $ & $  41.6 \pm  8.2 $ &  $ 17.5 \pm  0.8 $ & 21.3 & $ 5.34^{+0.62}_{-0.34} $ & $ 24.1^{+ 8.8}_{-5.5} $ & $ 3566^{+1300}_{- 820} $ &  6.2   & S  & S  \\
\textbf{G12-26926}$^a$     & $ 183.489479 $ & $  -1.373146 $ & $ 26.8 \pm 7.0 $ & $  43.1 \pm  7.5 $ & $  51.5 \pm  8.2 $ &  $ 14.9 \pm  1.6 $ &  9.6 & $ 4.30^{+0.22}_{-0.54} $ & $ 25.9^{+ 5.6}_{-7.6} $ & $ 3839^{+ 825}_{-1131} $ &  1.6   & S  & S  \\
G12-31529                  & $ 181.651146 $ & $   1.548312 $ & $ 20.6 \pm 7.4 $ & $  36.8 \pm  7.9 $ & $  56.1 \pm  9.1 $ &  $  8.4 \pm  0.9 $ &  9.1 & $ 3.64^{+0.22}_{-0.22} $ & $ 14.2^{+ 3.7}_{-3.2} $ & $ 2096^{+ 551}_{- 467} $ &  7.5   & S  & S  \\
\textbf{G12-42911}         & $ 175.812813 $ & $   0.479035 $ & $ 18.2 \pm 6.9 $ & $  35.8 \pm  7.2 $ & $  48.8 \pm  8.0 $ &  $ 13.2 \pm  1.5 $ &  8.9 & $ 4.28^{+0.36}_{-0.28} $ & $ 19.8^{+ 5.5}_{-4.3} $ & $ 2931^{+ 820}_{- 637} $ &  5.9   & S  & S  \\
G12-49632.A                & $ 179.712187 $ & $  -0.167757 $ & $ <7.0         $ & $  12.9 \pm  7.5 $ & $  <9.5          $ &  $  5.2 \pm  1.1 $ &  4.7 & $ -                    $ & $ -                   $ & $ -                    $ & 16.4   & M  & M  \\
G12-49632.B                & $ 179.716146 $ & $  -0.164424 $ & $ 21.7 \pm 7.0 $ & $  31.7 \pm  7.5 $ & $  40.1 \pm  9.5 $ &  $  4.6 \pm  1.1 $ &  4.3 & $ 3.14^{+0.34}_{-0.32} $ & $  8.7^{+ 3.4}_{-2.8} $ & $ 1293^{+ 500}_{- 419} $ &  2.7   & M  & M  \\
G12-58719                  & $ 185.308854 $ & $   0.930396 $ & $ 28.2 \pm 7.8 $ & $  45.1 \pm  8.1 $ & $  58.1 \pm  8.9 $ &  $  6.7 \pm  1.3 $ &  5.0 & $ 3.26^{+0.26}_{-0.32} $ & $ 13.3^{+ 3.9}_{-3.7} $ & $ 1971^{+ 578}_{- 544} $ &  1.9   & S  & S  \\
G12-73303                  & $ 183.514691 $ & $  -1.933354 $ & $ <6.9         $ & $  19.9 \pm  7.6 $ & $  34.8 \pm  8.6 $ &  $  4.1 \pm  1.2 $ &  3.5 & $ 3.62^{+0.46}_{-0.62} $ & $  7.7^{+ 4.2}_{-3.7} $ & $ 1137^{+ 614}_{- 544} $ & 10.0   & -  & S  \\
\textbf{G12-77419}         & $ 182.269896 $ & $  -1.090979 $ & $ 13.3 \pm 7.9 $ & $  31.2 \pm  8.0 $ & $  48.1 \pm  9.1 $ &  $  8.7 \pm  1.6 $ &  5.3 & $ 4.00^{+0.36}_{-0.32} $ & $ 14.6^{+ 5.1}_{-4.2} $ & $ 2156^{+ 755}_{- 628} $ &  7.5   & S  & S  \\
\textbf{G12-78868}         & $ 179.059688 $ & $   1.652340 $ & $ 11.9 \pm 7.6 $ & $  34.0 \pm  7.9 $ & $  43.6 \pm  8.5 $ &  $ 10.1 \pm  2.4 $ &  4.2 & $ 4.18^{+0.42}_{-0.40} $ & $ 16.4^{+ 6.1}_{-5.0} $ & $ 2432^{+ 899}_{- 732} $ &  3.2   & S  & S  \\
G15-23358                  & $ 214.860938 $ & $   0.193757 $ & $ 29.3 \pm 6.9 $ & $  51.3 \pm  7.4 $ & $  59.7 \pm  8.3 $ &  $  7.1 \pm  1.0 $ &  6.9 & $ 2.94^{+0.36}_{-0.14} $ & $ 11.9^{+ 2.5}_{-0.9} $ & $ 1758^{+ 366}_{- 137} $ &  3.4   & S  & S  \\
G15-26675                  & $ 221.138229 $ & $   0.278007 $ & $ 23.1 \pm 6.9 $ & $  54.6 \pm  7.3 $ & $  61.9 \pm  7.8 $ &  $ 11.7 \pm  1.8 $ &  6.7 & $ 3.52^{+0.38}_{-0.14} $ & $ 15.9^{+ 2.8}_{-1.0} $ & $ 2356^{+ 416}_{- 149} $ &  2.2   & S  & S  \\
G15-48916.A                & $ 214.619063 $ & $   0.790826 $ & $ 15.8 \pm 7.0 $ & $  27.7 \pm  7.3 $ & $  22.7 \pm  8.3 $ &  $  6.2 \pm  1.1 $ &  5.8 & $ 3.84^{+0.48}_{-0.58} $ & $  9.6^{+ 4.7}_{-4.0} $ & $ 1420^{+ 700}_{- 592} $ & 13.3   & MP & MP \\
G15-48916.B                & $ 214.614271 $ & $   0.788326 $ & $ 18.1 \pm 7.0 $ & $  23.4 \pm  7.3 $ & $  23.4 \pm  8.3 $ &  $  6.0 \pm  1.0 $ &  5.8 & $ 3.50^{+0.64}_{-0.34} $ & $ 10.2^{+ 6.4}_{-3.5} $ & $ 1508^{+ 941}_{- 521} $ &  6.2   & MP & MP \\
G15-57401                  & $ 214.028437 $ & $   1.168215 $ & $ 31.6 \pm 6.8 $ & $  43.3 \pm  7.6 $ & $  39.7 \pm  8.3 $ &  $  4.7 \pm  1.1 $ &  4.3 & $ 2.84^{+0.26}_{-0.26} $ & $  9.4^{+ 3.1}_{-2.5} $ & $ 1387^{+ 460}_{- 375} $ & 32.1   & S  & S  \\
G15-63483                  & $ 221.294479 $ & $   0.016271 $ & $ 19.9 \pm 6.8 $ & $  32.9 \pm  7.5 $ & $  41.1 \pm  8.3 $ &  $  5.1 \pm  1.4 $ &  3.6 & $ 2.92^{+0.44}_{-0.32} $ & $ 10.4^{+ 5.0}_{-3.4} $ & $ 1537^{+ 743}_{- 498} $ &  7.4   & -  & S  \\
G15-82610                  & $ 220.731980 $ & $   1.163035 $ & $ 13.0 \pm 6.8 $ & $  30.1 \pm  7.6 $ & $  43.2 \pm  9.0 $ &  $  7.2 \pm  1.3 $ &  5.3 & $ 3.86^{+0.36}_{-0.32} $ & $ 12.2^{+ 4.5}_{-3.6} $ & $ 1801^{+ 668}_{- 534} $ & 10.5   & S  & S  \\
G15-82660                  & $ 215.636146 $ & $   0.503090 $ & $ 22.6 \pm 7.5 $ & $  52.1 \pm  8.1 $ & $  75.9 \pm  9.0 $ &  $  4.3 \pm  1.0 $ &  4.2 & $ 2.68^{+0.42}_{-0.12} $ & $ 10.2^{+ 2.8}_{-0.8} $ & $ 1507^{+ 411}_{- 112} $ & 10.2   & S  & S  \\
G15-83272                  & $ 213.803646 $ & $  -0.276910 $ & $ 16.3 \pm 7.5 $ & $  31.6 \pm  7.8 $ & $  54.9 \pm  8.8 $ &  $  6.7 \pm  1.7 $ &  4.0 & $ 3.78^{+0.32}_{-0.30} $ & $ 13.2^{+ 4.5}_{-3.8} $ & $ 1952^{+ 671}_{- 561} $ &  4.1   & -  & S  \\
NGP-112775                 & $ 204.170765 $ & $  26.264535 $ & $ 25.2 \pm 7.2 $ & $  33.3 \pm  7.5 $ & $  39.6 \pm  8.4 $ &  $  6.2 \pm  1.3 $ &  4.8 & $ 3.32^{+0.32}_{-0.36} $ & $ 10.8^{+ 3.7}_{-3.4} $ & $ 1603^{+ 551}_{- 500} $ &  6.5   & S  & S  \\
\textbf{NGP-115876.A}      & $ 204.653354 $ & $  27.548868 $ & $ 16.6 \pm 7.7 $ & $  22.2 \pm  8.6 $ & $  15.2 \pm 10.9 $ &  $ 13.8 \pm  1.4 $ & 10.0 & $ 5.02^{+1.00}_{-0.62} $ & $ 21.0^{+14.3}_{-8.9} $ & $ 3104^{+2113}_{-1323} $ & 13.7   & S  & M  \\
NGP-115876.B               & $ 204.648889 $ & $  27.545326 $ & $ 28.1 \pm 7.7 $ & $  43.8 \pm  8.5 $ & $  50.1 \pm 10.4 $ &  $  5.0 \pm  1.4 $ &  3.6 & $ 2.48^{+0.28}_{-0.30} $ & $ 10.4^{+ 4.1}_{-3.4} $ & $ 1543^{+ 610}_{- 499} $ &  6.9   & -  & M  \\
\textbf{NGP-131281}        & $ 193.200871 $ & $  34.403757 $ & $ 26.6 \pm 7.6 $ & $  55.3 \pm  8.4 $ & $  72.7 \pm  9.5 $ &  $ 19.1 \pm  1.3 $ & 15.0 & $ 4.22^{+0.20}_{-0.18} $ & $ 28.8^{+ 5.4}_{-4.6} $ & $ 4264^{+ 798}_{- 676} $ &  2.3   & S  & S  \\
NGP-145039                 & $ 194.371603 $ & $  29.281424 $ & $ 32.5 \pm 7.6 $ & $  42.9 \pm  7.9 $ & $  51.4 \pm  8.7 $ &  $  6.0 \pm  1.2 $ &  4.9 & $ 3.08^{+0.22}_{-0.24} $ & $ 11.7^{+ 3.2}_{-3.0} $ & $ 1736^{+ 477}_{- 442} $ &  6.3   & S  & S  \\
NGP-149267                 & $ 203.000852 $ & $  26.419826 $ & $ 23.2 \pm 7.6 $ & $  46.7 \pm  7.9 $ & $  52.1 \pm  8.4 $ &  $  6.2 \pm  1.3 $ &  4.7 & $ 3.26^{+0.22}_{-0.22} $ & $ 12.4^{+ 3.3}_{-2.9} $ & $ 1838^{+ 486}_{- 435} $ &  9.1   & S  & S  \\
NGP-157992                 & $ 193.515275 $ & $  27.176396 $ & $ 22.9 \pm 7.6 $ & $  46.2 \pm  8.1 $ & $  57.8 \pm  8.6 $ &  $  9.3 \pm  2.5 $ &  3.7 & $ 3.74^{+0.22}_{-0.44} $ & $ 17.9^{+ 4.4}_{-5.3} $ & $ 2642^{+ 655}_{- 784} $ &  9.4   & -  & S  \\
NGP-168019.B               & $ 205.414677 $ & $  32.476257 $ & $ 60.3 \pm 7.5 $ & $  85.7 \pm  8.3 $ & $  82.5 \pm  9.3 $ &  $  8.2 \pm  1.6 $ &  5.2 & $ 2.58^{+0.22}_{-0.14} $ & $ 14.3^{+ 2.1}_{-1.3} $ & $ 2120^{+ 312}_{- 192} $ & 33.4   & M  & M  \\
NGP-168019.A               & $ 205.405540 $ & $  32.476465 $ & $ 28.0 \pm 7.5 $ & $  50.1 \pm  8.2 $ & $  51.2 \pm  9.5 $ &  $  6.5 \pm  1.5 $ &  4.3 & $ 3.20^{+0.26}_{-0.26} $ & $ 12.9^{+ 3.8}_{-3.3} $ & $ 1912^{+ 567}_{- 486} $ &  1.2   & M  & M  \\
NGP-176261                 & $ 199.243213 $ & $  33.915007 $ & $ 24.3 \pm 7.5 $ & $  40.0 \pm  7.9 $ & $  52.7 \pm  8.7 $ &  $  7.3 \pm  1.6 $ &  4.7 & $ 3.48^{+0.24}_{-0.26} $ & $ 13.7^{+ 3.8}_{-3.4} $ & $ 2032^{+ 567}_{- 506} $ &  5.3   & S  & S  \\
NGP-194548$^a$             & $ 203.406969 $ & $  24.259451 $ & $ 17.9 \pm 7.6 $ & $  47.1 \pm  8.0 $ & $  67.9 \pm  8.7 $ &  $ 13.8 \pm  1.3 $ & 11.1 & $ 3.80^{+0.42}_{-0.14} $ & $ 18.0^{+ 3.2}_{-1.0} $ & $ 2661^{+ 474}_{- 154} $ &  7.9   & S  & S  \\
\hline
\multicolumn{13}{l}{$^a$ Redshifts $>4$ have been spectroscopically confirmed (see Table \ref{table:RSR} and \S\ref{secc:highz_candidates}).}\\
\end{tabular}
\end{table}
\end{landscape}

\begin{landscape}
\begin{table}
\centering
\contcaption{}
\begin{tabular}{lrrcccrrrrrrcc}

\hline
ID &
R.A. &
Dec. &
$F_{\rm 250\mu m}$ &
$F_{\rm 350\mu m}$ &
$F_{\rm 500\mu m}$ &
$S_{\rm 1.1 mm}$ &
SNR$_\textrm{Az}$ &
$z_{\rm phot}$ &
$L_{\rm IR(8-1000\mu m)}$ &
SFR &
Separation &
\multicolumn{2}{c}{Class} \\ 
 & 
[deg] &
[deg] & 
[mJy] & 
[mJy] & 
[mJy] & 
[mJy] & 
 & 
 &
[$10^{12}$L$_\odot$] &
[M$_\odot$ yr$^{-1}$] & 
[arcsec] &
$4.0\sigma$ &
$3.5\sigma$ \\
\hline

\textbf{NGP-203484}$^a$    & $ 204.915985 $ & $  31.369396 $ & $ 20.9 \pm 7.7 $ & $  38.5 \pm  7.9 $ & $  51.1 \pm  9.0 $ &  $ 21.3 \pm  1.4 $ & 15.4 & $ 5.18^{+0.26}_{-0.54} $ & $ 35.0^{+ 6.9}_{-8.6} $ & $ 5173^{+1023}_{-1268} $ &   4.1   & S  & S  \\
NGP-211862                 & $ 193.669650 $ & $  26.824187 $ & $ 28.2 \pm 7.5 $ & $  45.7 \pm  8.1 $ & $  45.1 \pm  9.1 $ &  $  8.1 \pm  1.5 $ &  5.4 & $ 3.38^{+0.26}_{-0.26} $ & $ 13.9^{+ 4.0}_{-3.5} $ & $ 2054^{+ 595}_{- 512} $ &   2.3   & S  & S  \\
NGP-244082                 & $ 199.496726 $ & $  34.488118 $ & $ <7.6         $ & $  16.2 \pm  8.2 $ & $  35.7 \pm  9.0 $ &  $  6.6 \pm  1.7 $ &  3.9 & $ 4.36^{+0.72}_{-0.84} $ & $ 10.6^{+ 7.3}_{-5.6} $ & $ 1574^{+1074}_{- 824} $ &  34.5   & -  & S  \\
NGP-246114$^a$             & $ 205.308501 $ & $  33.993556 $ & $ 17.3 \pm 6.5 $ & $  30.4 \pm  8.1 $ & $  33.9 \pm  8.5 $ &  $  7.8 \pm  1.5 $ &  5.2 & $ 3.80^{+0.48}_{-0.34} $ & $ 12.3^{+ 5.1}_{-3.7} $ & $ 1820^{+ 754}_{- 544} $ &   3.1   & S  & S  \\
NGP-248948                 & $ 192.159401 $ & $  29.626687 $ & $  9.1 \pm 7.8 $ & $  31.3 \pm  8.2 $ & $  44.5 \pm  8.8 $ &  $  9.6 \pm  2.7 $ &  3.5 & $ 3.64^{+0.64}_{-0.50} $ & $ 16.6^{+ 8.5}_{-5.9} $ & $ 2451^{+1254}_{- 875} $ &   7.5   & -  & S  \\
NGP-249138                 & $ 193.610448 $ & $  24.625757 $ & $ 21.3 \pm 7.6 $ & $  40.5 \pm  8.0 $ & $  57.7 \pm  8.9 $ &  $  7.2 \pm  1.9 $ &  3.8 & $ 3.60^{+0.26}_{-0.26} $ & $ 14.8^{+ 4.2}_{-3.7} $ & $ 2195^{+ 625}_{- 542} $ &   3.1   & -  & S  \\
NGP-249475.A               & $ 206.236099 $ & $  31.442021 $ & $ 14.8 \pm 7.5 $ & $  33.6 \pm  8.2 $ & $  40.6 \pm  9.8 $ &  $  8.4 \pm  1.4 $ &  5.8 & $ 3.94^{+0.40}_{-0.34} $ & $ 13.5^{+ 5.4}_{-4.2} $ & $ 1994^{+ 792}_{- 615} $ &   5.2   & M  & M  \\
\textbf{NGP-249475.B}               & $ 206.228529 $ & $  31.444312 $ & $ 13.9 \pm 7.5 $ & $  12.6 \pm  8.2 $ & $  <9.8          $ &  $  6.5 \pm  1.5 $ &  4.5 & $ 4.62^{+2.12}_{-0.66} $ & $  9.9^{+19.0}_{-6.9} $ & $ 1471^{+2817}_{-1020} $ &  26.0   & M  & M  \\
\textbf{NGP-284357}$^a$    & $ 203.215088 $ & $  33.394556 $ & $ 12.6 \pm 5.3 $ & $  20.4 \pm  7.8 $ & $  42.4 \pm  8.3 $ &  $ 11.1 \pm  1.4 $ &  7.7 & $ 4.70^{+0.44}_{-0.36} $ & $ 16.5^{+ 5.8}_{-4.5} $ & $ 2442^{+ 859}_{- 671} $ &  11.0   & S  & S  \\
NGP-55628.A                & $ 206.604457 $ & $  34.271562 $ & $ 42.0 \pm 8.0 $ & $  47.9 \pm  8.4 $ & $  60.5 \pm 10.8 $ &  $ 10.8 \pm  1.4 $ &  7.6 & $ 3.52^{+0.16}_{-0.40} $ & $ 20.6^{+ 4.3}_{-5.8} $ & $ 3055^{+ 630}_{- 853} $ &   2.7   & M  & M  \\
NGP-55628.B                & $ 206.599920 $ & $  34.267187 $ & $ 46.0 \pm 8.0 $ & $  59.0 \pm  8.4 $ & $  35.5 \pm 10.7 $ &  $  6.1 \pm  1.5 $ &  4.2 & $ 2.66^{+0.20}_{-0.36} $ & $ 11.1^{+ 3.2}_{-3.6} $ & $ 1645^{+ 466}_{- 527} $ &  21.3   & M  & M  \\
NGP-78659                  & $ 207.218236 $ & $  26.898910 $ & $ 30.8 \pm 7.5 $ & $  53.3 \pm  7.9 $ & $  67.8 \pm  8.4 $ &  $ 14.6 \pm  2.2 $ &  6.8 & $ 3.90^{+0.24}_{-0.22} $ & $ 23.7^{+ 5.1}_{-4.4} $ & $ 3506^{+ 755}_{- 649} $ &   7.2   & S  & S  \\
NGP-94843.A                & $ 204.696040 $ & $  25.669062 $ & $ 33.6 \pm 7.7 $ & $  42.3 \pm  8.2 $ & $  54.8 \pm 10.6 $ &  $  6.7 \pm  1.4 $ &  5.0 & $ 3.16^{+0.24}_{-0.26} $ & $ 12.5^{+ 3.7}_{-3.2} $ & $ 1842^{+ 551}_{- 475} $ &   3.7   & S  & MP \\
NGP-94843.B                & $ 204.698351 $ & $  25.663854 $ & $ 27.5 \pm 7.7 $ & $  31.8 \pm  8.2 $ & $  28.5 \pm 10.6 $ &  $  5.5 \pm  1.4 $ &  3.9 & $ 3.22^{+0.30}_{-0.56} $ & $ 10.5^{+ 4.4}_{-4.4} $ & $ 1554^{+ 646}_{- 651} $ &  22.6   & -  & MP \\
SGP-101187.A               & $  16.970845 $ & $ -30.299799 $ & $ <7.9         $ & $  35.1 \pm  8.1 $ & $  48.6 \pm  9.3 $ &  $  6.0 \pm  1.1 $ &  5.7 & $ 3.48^{+0.30}_{-0.28} $ & $ 10.9^{+ 4.0}_{-3.2} $ & $ 1611^{+ 593}_{- 470} $ &   9.7   & M  & M  \\
SGP-101187.B               & $  16.962158 $ & $ -30.294590 $ & $ 42.7 \pm 7.9 $ & $  68.3 \pm  8.1 $ & $  71.2 \pm  9.3 $ &  $  6.1 \pm  1.1 $ &  5.3 & $ 2.84^{+0.14}_{-0.20} $ & $ 14.0^{+ 2.8}_{-3.0} $ & $ 2073^{+ 415}_{- 449} $ &  29.8   & M  & M  \\
SGP-106123.A               & $  13.846177 $ & $ -28.005632 $ & $ <7.9         $ & $  63.5 \pm  8.6 $ & $  59.3 \pm  9.6 $ &  $  4.3 \pm  1.0 $ &  4.3 & $ 2.94^{+0.22}_{-0.36} $ & $ 10.5^{+ 3.2}_{-3.4} $ & $ 1552^{+ 479}_{- 496} $ &  24.4   & S  & MP \\
SGP-106123.B               & $  13.857975 $ & $ -28.010840 $ & $ <7.9         $ & $  29.8 \pm  8.6 $ & $  38.3 \pm  9.6 $ &  $  3.5 \pm  1.0 $ &  3.7 & $ 2.90^{+0.52}_{-0.30} $ & $  5.8^{+ 1.8}_{-1.0} $ & $  860^{+ 265}_{- 143} $ &  24.4   & -  & MP \\
\textbf{SGP-272197}$^a$    & $   1.530803 $ & $ -32.445340 $ & $ <7.4         $ & $  18.6 \pm  8.2 $ & $  46.1 \pm  8.6 $ &  $ 21.1 \pm  1.5 $ & 13.8 & $ 5.76^{+0.74}_{-0.40} $ & $ 28.6^{+11.3}_{-7.3} $ & $ 4229^{+1672}_{-1073} $ &   5.2   & S  & S  \\
SGP-280787.A               & $ 350.395117 $ & $ -33.077812 $ & $ <7.5         $ & $  11.6 \pm  8.1 $ & $ <13.8          $ &  $  9.6 \pm  1.8 $ &  5.2 & $ -                    $ & $ -                   $ & $ -                    $ &  13.3   & S  & M  \\
SGP-280787.B               & $ 350.394372 $ & $ -33.083021 $ & $ <7.5         $ & $  24.2 \pm  8.1 $ & $  55.4 \pm  8.6 $ &  $  7.0 \pm  1.9 $ &  3.7 & $ 4.00^{+0.36}_{-0.34} $ & $ 13.8^{+ 5.0}_{-4.1} $ & $ 2042^{+ 738}_{- 606} $ &   6.8   & -  & M  \\
SGP-284969                 & $  15.856769 $ & $ -30.058979 $ & $ <7.8         $ & $  29.5 \pm  8.2 $ & $  44.2 \pm  9.5 $ &  $  4.0 \pm  0.8 $ &  4.7 & $ 3.04^{+0.44}_{-0.26} $ & $  6.3^{+ 1.5}_{-0.9} $ & $  929^{+ 222}_{- 128} $ &   5.7   & S  & S  \\
SGP-289463                 & $  25.533576 $ & $ -32.576882 $ & $ <7.8         $ & $  27.8 \pm  8.5 $ & $  59.0 \pm  9.5 $ &  $  7.7 \pm  2.0 $ &  3.9 & $ 4.00^{+0.34}_{-0.34} $ & $ 14.9^{+ 5.2}_{-4.5} $ & $ 2202^{+ 766}_{- 667} $ &  14.6   & -  & S  \\
\textbf{SGP-293180}        & $  18.159985 $ & $ -30.784299 $ & $ 14.4 \pm 7.8 $ & $  23.9 \pm  7.8 $ & $  38.5 \pm  8.7 $ &  $  7.3 \pm  1.1 $ &  6.5 & $ 4.02^{+0.42}_{-0.42} $ & $ 11.7^{+ 4.9}_{-4.1} $ & $ 1734^{+ 720}_{- 607} $ &   4.6   & S  & S  \\
SGP-316248.B               & $ 354.390415 $ & $ -34.829924 $ & $ <11.2        $ & $ <11.8          $ & $ <14.4          $ &  $  5.8 \pm  1.5 $ &  4.0 & $ -                    $ & $ -                   $ & $ -                    $ &  35.0   & -  & M  \\
SGP-316248.A               & $ 354.392953 $ & $ -34.840757 $ & $ 20.4 \pm 7.2 $ & $  24.2 \pm  7.6 $ & $  48.5 \pm  8.9 $ &  $  5.3 \pm  1.3 $ &  4.0 & $ 3.48^{+0.34}_{-0.34} $ & $ 10.1^{+ 3.9}_{-3.2} $ & $ 1501^{+ 569}_{- 472} $ &   5.8   & -  & M  \\
SGP-359921                 & $  16.920650 $ & $ -28.453187 $ & $ 13.0 \pm 8.0 $ & $  24.2 \pm  8.5 $ & $  41.2 \pm  9.3 $ &  $  4.9 \pm  1.1 $ &  4.3 & $ 3.32^{+0.60}_{-0.28} $ & $  7.3^{+ 2.2}_{-1.0} $ & $ 1072^{+ 325}_{- 143} $ &  17.3   & S  & S  \\
SGP-379994                 & $  11.365272 $ & $ -32.553340 $ & $ <7.5         $ & $  18.0 \pm  8.2 $ & $  29.9 \pm  9.3 $ &  $  4.4 \pm  1.1 $ &  3.9 & $ 4.00^{+1.14}_{-0.78} $ & $  7.3^{+ 8.0}_{-4.3} $ & $ 1078^{+1179}_{- 640} $ &  32.9   & -  & S  \\
SGP-384367                 & $  21.078458 $ & $ -32.979812 $ & $ <7.6         $ & $  27.0 \pm  8.1 $ & $  51.0 \pm  9.1 $ &  $  4.8 \pm  1.0 $ &  4.6 & $ 3.32^{+0.44}_{-0.24} $ & $  7.3^{+ 1.6}_{-0.8} $ & $ 1072^{+ 236}_{- 123} $ &   1.9   & S  & S  \\
\textbf{SGP-396540}        & $  10.268114 $ & $ -28.222993 $ & $  9.8 \pm 7.5 $ & $  12.8 \pm  8.0 $ & $  42.5 \pm  8.8 $ &  $  9.6 \pm  1.3 $ &  7.5 & $ 4.66^{+0.54}_{-0.54} $ & $ 14.3^{+ 6.8}_{-5.4} $ & $ 2120^{+ 999}_{- 799} $ &  10.0   & S  & S  \\
SGP-396921                 & $  16.554353 $ & $ -28.230701 $ & $ <7.6         $ & $  23.2 \pm  8.1 $ & $  49.6 \pm  9.2 $ &  $  4.5 \pm  0.9 $ &  5.2 & $ 3.22^{+0.50}_{-0.24} $ & $  6.9^{+ 1.8}_{-0.8} $ & $ 1021^{+ 266}_{- 122} $ &   4.3   & S  & S  \\
SGP-396966                 & $   8.877234 $ & $ -31.505854 $ & $ 20.4 \pm 7.8 $ & $  30.5 \pm  7.8 $ & $  53.4 \pm  8.9 $ &  $  8.6 \pm  1.2 $ &  6.9 & $ 3.76^{+0.28}_{-0.26} $ & $ 13.9^{+ 4.2}_{-3.4} $ & $ 2060^{+ 619}_{- 501} $ &   5.2   & S  & S  \\
SGP-399383                 & $  20.202309 $ & $ -30.974549 $ & $ 24.9 \pm 7.7 $ & $  21.3 \pm  8.3 $ & $  44.4 \pm  8.9 $ &  $  4.6 \pm  1.3 $ &  3.7 & $ 3.24^{+0.36}_{-0.44} $ & $  8.9^{+ 3.9}_{-3.4} $ & $ 1312^{+ 572}_{- 499} $ &   7.5   & -  & S  \\
\textbf{SGP-400082.A}               & $   8.385889 $ & $ -30.081937 $ & $ <7.8         $ & $  20.6 \pm  8.5 $ & $  34.2 \pm 10.2 $ &  $  7.5 \pm  1.3 $ &  5.8 & $ 4.56^{+0.88}_{-0.88} $ & $ 11.2^{+ 8.4}_{-6.0} $ & $ 1658^{+1239}_{- 893} $ &   5.0   & MP & MP \\
\textbf{SGP-400082.B}               & $   8.392871 $ & $ -30.076729 $ & $ 18.1 \pm 7.8 $ & $  14.3 \pm  8.5 $ & $  29.8 \pm 10.2 $ &  $  6.6 \pm  1.5 $ &  4.4 & $ 4.56^{+1.16}_{-0.92} $ & $ 10.1^{+ 9.7}_{-5.9} $ & $ 1489^{+1433}_{- 876} $ &  27.6   & MP & MP \\
\textbf{SGP-403579}        & $ 355.064638 $ & $ -30.446674 $ & $ <7.6         $ & $  <8.0          $ & $  32.2 \pm  8.5 $ &  $  6.1 \pm  1.2 $ &  5.2 & $ 4.30^{+0.76}_{-0.84} $ & $  9.5^{+ 6.9}_{-5.1} $ & $ 1412^{+1023}_{- 756} $ &   5.8   & S  & S  \\
\hline
\multicolumn{13}{l}{$^a$ Redshifts $>4$ have been spectroscopically confirmed (see Table \ref{table:RSR} and \S\ref{secc:highz_candidates}).}\\
\end{tabular}
\end{table}
\end{landscape}

\newpage

\bsp	
\label{lastpage}
\end{document}